\documentclass[apj,twocolappendix,appendixfloats]{emulateapj}

\usepackage{graphicx,amsmath,enumerate}
\usepackage{natbib,times,float}
\usepackage{verbatim}
\usepackage{tabu}

\newcommand{\be}{\begin{equation}}
\newcommand{\ee}{\end{equation}}

\newcommand{\kms}{\,{\rm {km\, s^{-1}}}}
\newcommand{\msun}{{$M_{\odot}$}}

\newcommand{\ergs}{erg~s$^{-1}$}

\newcommand{\ang}{{$\rm{\AA}$}}
 
\newcommand{\gtsima}{$\; \buildrel > \over \sim \;$}
\newcommand{\ltsima}{$\; \buildrel < \over \sim \;$}
\newcommand{\prosima}{$\; \buildrel \propto \over \sim \;$}
\newcommand{\gsim}{\lower.5ex\hbox{\gtsima}}
\newcommand{\lsim}{\lower.5ex\hbox{\ltsima}}
\newcommand{\simgt}{\lower.5ex\hbox{\gtsima}}
\newcommand{\simlt}{\lower.5ex\hbox{\ltsima}}
\newcommand{\simpr}{\lower.5ex\hbox{\prosima}}

\newcommand{\etal}{{et al.~}}

\begin{document}

\title{Multi-epoch spectroscopy of dwarf galaxies with AGN signatures: identifying sources with persistent broad H$\alpha$ emission}

\author{Vivienne F. Baldassare\altaffilmark{1}, Amy E. Reines\altaffilmark{1,2,3}, Elena Gallo\altaffilmark{1}, Jenny E. Greene\altaffilmark{4}, Or Graur \altaffilmark{5,6}, Marla Geha\altaffilmark{7}, Kevin Hainline\altaffilmark{8}, Christopher M. Carroll\altaffilmark{9}, Ryan C. Hickox\altaffilmark{9}}

\altaffiltext{1}{Department of Astronomy, University of Michigan, Ann Arbor, MI 48109}
\altaffiltext{2}{National Optical Astronomy Observatory, 950 N Cherry Ave, Tucson, AZ 85719}
\altaffiltext{3}{Hubble Fellow}
\altaffiltext{4}{Department of Astrophysical Sciences, Princeton University, Princeton, NJ 08544}
\altaffiltext{5}{Center for Cosmology and Particle Physics, New York University, New York, NY 10003}
\altaffiltext{6}{Department of Astrophysics, American Museum of Natural History, New York, NY 10024, USA}
\altaffiltext{7}{Department of Astronomy, Yale University, New Haven, CT 06520}
\altaffiltext{8}{Steward Observatory, Tucson, AZ 85721}
\altaffiltext{9}{Department of Physics \& Astronomy, Dartmouth College, Hanover, NH 03755}


\begin{abstract}

We use time-domain optical spectroscopy to distinguish between broad emission lines powered by accreting black holes (BHs) or stellar processes (i.e., supernovae) for 16 galaxies identified as AGN candidates by Reines \etal (2013). Our study is primarily focused on those objects with narrow emission-line ratios dominated by star formation. Based on follow-up spectra taken with the Magellan Echellette Spectrograph (MagE), the Dual Imaging Spectrograph, and the Ohio State Multi-Object Spectrograph, we find that the broad H$\alpha$ emission has faded or was ambiguous for all of the star-forming objects (14/16) over baselines ranging from 5 to 14 years. For the two objects in our follow-up sample with narrow-line AGN signatures (RGG 9 and RGG 119), we find persistent broad H$\alpha$ emission consistent with an AGN origin. Additionally, we use our MagE observations to measure stellar velocity dispersions for 15 objects in the Reines et al. (2013) sample, all with narrow-line ratios indicating the presence of an AGN.
Stellar masses range from $\sim5\times10^{8}$ to $3\times10^{9}$~\msun, and we measure $\sigma_{\ast}$ ranging from $28-71\kms$.  These $\sigma_{\ast}$ correspond to some of the lowest-mass galaxies with optical signatures of AGN activity. We show that RGG 119, the one object which has both a measured $\sigma_{\ast}$ and persistent broad H$\alpha$ emission, falls near the extrapolation of the $\rm M_{BH}-\sigma_{\star}$ relation to the low-mass end.
 \end{abstract}


\section{Introduction}

While massive black holes (BHs; here defined as $\rm M_{BH} \gtrsim 10^{4-5}$~\msun) are ubiquitous in the centers of galaxies of Milky Way mass and larger \citep{1998AJ....115.2285M}, little is known about the occurrence rate and properties of BHs in dwarf galaxies (i.e., galaxies with $\rm M_{\star} \lesssim 10^{9.5}$~\msun). Studying the population of BHs in dwarf galaxies is useful for understanding BH formation and growth in the early universe. Having had relatively quiet merger histories, their BHs may have masses similar to their ``birth" mass (e.g., \citealt{:fl, 2011ApJ...742...13B}). 

Clues to the formation of BH seeds may reside in the well studied relation between BH mass and stellar velocity dispersion (see e.g., \citealt{2000ApJ...539L...9F, 2000ApJ...539L..13G, 2009ApJ...698..198G}). Using semi-analytic modeling, \cite{2009MNRAS.400.1911V} find that, while the relationship between BH mass and velocity dispersion (or the $\rm M_{BH}-\sigma$ relation) is in place by $z\sim4$ for massive halos, systems with lower mass BHs ($\rm M_{BH} < 10^{6}~M_{\odot}$) evolve onto it at later times. If BH seeds form ``light" ($\sim10^{2}M_{\odot}$; the remnants of Population III stars, see e.g., \citealt{2001ApJ...551L..27M, 2009ApJ...701L.133A, 2014ApJ...784L..38M}), then BHs begin as undermassive with respect to the local $\rm M_{BH}-\sigma_{\star}$ relation, and will evolve towards it from below. Alternatively, if BH seeds form ``heavy" ($10^{4-5}M_{\odot}$; through direct collapse of gas clouds, e.g., \citealt{2006MNRAS.370..289B, 2006MNRAS.371.1813L}), then BHs are initially overmassive and systems evolve rightwards onto the relation (i.e., towards higher velocity dispersions) as the host galaxies grow through mergers. The smallest $z~=~0$ halos ($\sigma_{\star} \sim 20-50~\kms$) have not yet undergone major mergers, and thus may be present-day outliers on the local $\rm M_{BH}-\sigma_{\star}$ relation. 

Finding and weighing relatively low-mass BHs is difficult due to the small gravitational sphere of influence of the BH, which is typically $\lesssim1$ parsec for $\rm M_{BH}\sim10^{5}$~\msun. Thus, in order to identify BHs in dwarf galaxies outside the Local Group, it is necessary to look for signs of BH activity (i.e., active galactic nuclei, or AGN). Large-scale surveys such as the Sloan Digital Sky Survey (SDSS) have made it possible to search for signatures of AGN in samples comprised of tens of thousands of galaxies. Early work by \cite{2004ApJ...610..722G} (see also \citealt{2007ApJ...670...92G}), identified galaxies in the SDSS with low-mass AGN by searching for broad H$\alpha$ emission, which can be produced by virialized gas near the central BH. These studies identified $\sim 200$ low-mass AGN candidates with a median BH mass of $\sim10^{6}$~\msun. 

More recent studies have pushed the search for low-mass AGN into the dwarf galaxy regime. X-ray and radio observations led to the discovery of AGN in the dwarf starburst galaxy Henize 2-10 \citep{Reines:2011fr,2012ApJ...750L..24R}, and in the dwarf galaxy pair Mrk 709 \citep{2014ApJ...787L..30R}. Optical spectroscopic observations revealed the well-studied AGN in NGC 4395 \citep{1989ApJ...342L..11F, 2003ApJ...588L..13F} and POX 52 \citep{2004ApJ...607...90B}, and more recently led to the discovery of a BH with a mass in the range of $27,000-62,000$~\msun~in the dwarf galaxy RGG 118 \citep{2015ApJ...809L..14B}. Additionally, multiwavelength studies using large samples of \textit{bona fide} dwarf galaxies have produced a collective sample of hundreds of AGN candidates in dwarf galaxies (\citealt{Reines:2013fj, 2014AJ....148..136M, 2015ApJ...805...12L,2015MNRAS.454.3722S, 2016arXiv160301622P}).

\citealt{Reines:2013fj} (hereafter R13) searched for dwarf galaxies with spectroscopic signatures of AGN activity based on their narrow and/or broad emission lines. Out of $\sim25,000$ nearby ($z\le0.055$; D $\lesssim250$ Mpc) dwarf galaxies in the SDSS, they identified 136 galaxies with narrow-line ratios indicating photoionization at least partly due to an AGN based on the BPT (Baldwin, Phillips \& Terlevich 1981\nocite{1981PASP...93....5B}) diagram, i.e., they were classified as either AGN or AGN/star-forming composites (see also \citealt{2001ApJ...556..121K}; \citealt{2003MNRAS.346.1055K};  \citealt{2006MNRAS.372..961K}). A fraction of these also had broad H$\alpha$ emission. Additionally, R13 identified 15 galaxies with broad H$\alpha$ emission in their SDSS spectroscopy (FWHM ranging from $\sim600-3700~\kms$), but with narrow-line ratios that placed them in the star forming region of the BPT diagram. We are particularly concerned with the origin of the broad emission in these systems.

Broad Balmer emission with FWHM up to several thousand kilometers per second can be produced by stellar processes such as Type II supernovae \citep{1997ARA&A..35..309F,2012ApJ...750..128P}, luminous blue variables \citep{2011MNRAS.415..773S}, or Wolf-Rayet stars. Follow-up spectroscopy, taken over a sufficiently long baseline, can help distinguish between an AGN and one of the aforementioned transient stellar phenomena. 

Here, we present follow-up optical spectroscopic observations of 14 of the 15 BPT star forming galaxies with broad H$\alpha$ presented in R13, as well as observations of ten BPT AGN (one with broad H$\alpha$) and three BPT composites (one with broad H$\alpha$). Follow-up observations for RGG 118 -- the current record holder for the lowest-mass BH in a galaxy nucleus -- were presented separately in \cite{2015ApJ...809L..14B}. The goals of this work are two-fold. First, for our targets which were identified by R13 to have broad H$\alpha$ emission (broader than narrow lines such as [SII]$\lambda\lambda$6713,6731) with FWHM $>500\kms$, we analyze their spectra and determine whether the broad H$\alpha$ emission is still present and consistent with previous observations. Second, for galaxies with sufficiently high-resolution follow-up spectroscopy, we measure stellar velocity dispersions. 

In Section 2, we describe our observations and data reduction procedures. In Section 3, we discuss our emission line fitting analysis and report stellar velocity dispersion measurements for objects with sufficiently high spectral resolution data. In Section 4, we present results from the follow-up spectroscopy for the broad line objects.


\section{Observations and Data Reduction}

\begin{table*}[t]
\begin{center}
\caption{Summary of Observations}
\begin{tabular}{c c c c c c c c c c}
\hline
\hline
R13 ID & NSAID & SDSS Name & z & R13 BPT class & SDSS obs. & DIS obs. &  MagE obs. &  OSMOS obs. \\
\hline
6 & 105376 & J084025.54+181858.9 & 0.0150 & AGN & 2005-12-07 & -- & 2013-04-20 & -- \\
\textbf 9 & \textbf{10779} &  J090613.75+561015.5 &  0.0466 & AGN & 2000-12-30 &  2013-03-15 &  -- &  -- \\
16 & 30370 & J111319.23+044425.1 & 0.0265 & AGN & 2002-02-21 & -- & 2013-04-19 & -- \\
22 & 77431 & J130434.92+075505.0 & 0.0480 & AGN & 2008-02-08 & -- & 2013-04-20 & -- \\
27 & 78568 & J140228.72+091856.4 & 0.0191 & AGN  & 2007-03-16 & -- & 2013-04-19 & -- \\
28 & 70907 & J140510.39+114616.9 & 0.0174 & AGN  & 2006-03-05 & -- & 2013-04-19 & -- \\
29 & 71023 & J141208.47+102953.8 & 0.0326 & AGN & 2006-04-23 & -- & 2013-04-20 & -- \\
31 & 71565 & J143523.42+100704.2 & 0.0312 & AGN & 2005-06-14 & -- & 2013-04-20 & --  \\
33 & 120870 & J144712.80+133939.2 & 0.0323 & AGN & 2007-05-22  & -- & 2013-04-20 & -- \\
\smallskip
34 & 124249 & J153941.68+171421.8 & 0.0458 & AGN & 2008-04-07 & -- & 2013-04-20 & -- \\ 
118 & 166155 & J152303.80+114546.0 & 0.0234 & Composite & 2007-05-14 & -- & 2013-04-20 & -- \\
\textbf{119} & \textbf{79874} &  J152637.36+065941.6 &  0.0384 & Composite & 2008-03-15 &  2013-03-15 &  2013-04-19 &  -- \\
120 & 72952 & J152913.46+083010.6 & 0.0430 & Composite & 2006-06-19 & -- & 2013-04-20 & --  \\
\smallskip
128 & 73346 & J160544.57+085043.9 & 0.0154 & Composite & 2004-08-11 & -- & 2013-04-19/20 & -- \\ 
\textbf B & \textbf{15952} &  J084029.91+470710.4 &  0.0421 & SF & 2001-03-13 &  2013-02-09  &  -- &  2015-01-20 \\
\textbf C & \textbf{109990} &  J090019.66+171736.9 &  0.0288 &  SF & 2006-11-13 &  2013-02-16 &  -- &  2014-02-02 \\
\textbf D & \textbf{76788} &  J091122.24+615245.5 &  0.0266 & SF &  2007-12-16 &  -- &  -- &  2015-01-21 \\
\textbf E & \textbf{109016} &  J101440.21+192448.9 &  0.0289 &  SF  & 2006-02-02 &  -- &  2013-04-19 &  2015-01-20 \\
\textbf F & \textbf{12793} &  J105100.64+655940.7 &  0.0325 &  SF  & 2001-01-20 &  -- &  -- &  2015-01-22 \\
\textbf G & \textbf{13496} &  J105447.88+025652.4 &  0.0222 &  SF & 2002-03-20 &  -- &  2013-04-20 &  -- \\
\textbf H & \textbf{74914} &  J111548.27+150017.7 &  0.0501 &  SF & 2005-01-09 &  -- &  2013-04-19 &  -- \\
\textbf I & \textbf{112250} &  J112315.75+240205.1 &  0.0250 &  SF & 2007-02-23 &  2013-03-15 &  -- &  2015-01-23 \\ 
\textbf J & \textbf{41331} &  J114343.77+550019.4 &  0.0272 &  SF & 2003-04-05 &  2013-02-16 &  -- &  --  \\
\textbf K & \textbf{91579} &  J120325.66+330846.1 &  0.0349 & SF  & 2005-05-08 &  2013-03-15 &  -- &  2015-01-21 \\
\textbf L & \textbf{33207} &  J130724.64+523715.5 &  0.0262 &  SF & 2002-04-12 &  2013-02-16 &  -- &  -- \\
\textbf M & \textbf{119311} &  J131503.77+223522.7 &  0.0230 &  SF & 2008-02-11 &  2013-02-16 &  -- &  2014-02-02  \\
\textbf N & \textbf{88972} &  J131603.91+292254.0 &  0.0378 &  SF & 2006-06-18 &  -- &  -- &  2015-01-20  \\
\textbf O & \textbf{104565} &  J134332.09+253157.7 &  0.0287 &  SF & 2006-02-01 &  -- &  -- &  2015-01-22  \\
\hline
\end{tabular}
\label{obs}
\end{center}
\textbf{Table \ref{obs}.} Summary of our spectroscopic observations. The first column gives the ID assigned in Reines \etal (2013), the second gives the NASA-Sloan Atlas ID, and the third column gives the SDSS designation. Redshift is given in the fourth column, and the fifth column gives the region of the BPT diagram in which the object fell in R13 based on analysis of the SDSS spectrum. The remaining four columns give the dates of observations for SDSS, DIS, MagE, and MDM when applicable. The R13 ID and NSAID for objects in the SDSS broad H$\alpha$ sample are shown in bold. Note that observations for RGG 118 were presented in a prior work (Baldassare \etal 2015). 
\end{table*}

\vspace{0.5cm}

The sample analyzed here is comprised of 27 total targets. Of these, 16 comprise our ``SDSS broad H$\alpha$ sample" or galaxies identified to have broad H$\alpha$ emission in their SDSS spectroscopy. In the SDSS broad H$\alpha$ sample, 14 objects are BPT star forming, 1 is a BPT composite and 1 is a BPT AGN. Additionally, we have follow-up observations for 11 objects (9 BPT AGN and 2 BPT composites) that did not have broad H$\alpha$ emission in their SDSS spectroscopy.
In addition to the SDSS observations, each galaxy has one or two follow-up spectroscopic observations taken between 5 and 14 years after the SDSS spectrum. Below, we outline our observations and data reduction procedures for each instrument used. See Table~\ref{obs} for a summary of our sample and observations. All spectra were corrected for Galactic extinction.\\

\textit{Clay/MagE.}

We observed 16 targets (including RGG 118) with the Magellan Echellette Spectrograph (MagE; \citealt{2008SPIE.7014E..54M}). MagE is a moderate-resolution (R=4100) spectrograph on the 6.5m Clay Telescope at Las Campanas Observatory. Data were collected on the nights of 2013 April 18-19 using a 1\arcsec~slit. Spectral coverage spans roughly from 3200 \ang~  to 10000 \ang~ across 15 orders. Seeing was measured to be 0.5$''$-1.2\arcsec~over the two nights. Total exposure times per object ranged from 1800 to 4800s. A thorium argon arc lamp was observed for wavelength calibration, and the flux standard $\rm \theta$ Crt was observed for flux calibration. Flux calibrated reference spectra were obtained from the European Southern Observatory library of spectrophotometric standards.

Flat fielding, sky subtraction, extraction, wavelength calibration, and flux calibration were performed with the mage\_reduce pipeline written by George Becker.\footnote{The mage\_reduce reduction pipeline is available for download at: ftp://ftp.ociw.edu/pub/gdb/mage\_reduce/mage\_reduce.tar.gz} One-dimensional spectra were extracted with a 3\arcsec~aperture (the larger signal within 3\arcsec was useful for determining stellar velocity dispersions; see Section 3.2). \\

\textit{APO/DIS.}

Eight targets were observed with the Dual Imaging Spectrograph (DIS\footnote{http://www.astro.princeton.edu/$\sim$rhl/dis.html}) on the 3.5m Astrophysical Research Consortium telescope at Apache Point Observatory. Observations were taken using a 1.5\arcsec~ slit. DIS consists of a red channel and a blue channel, with a transition wavelength of $\rm\sim5350\AA$. We used the B1200 and R1200 gratings (R=1200), which give linear dispersions of 0.62 $\rm\AA$/pix for the blue channel and 0.58 $\rm\AA$/pix for the red channel. DIS targets were observed on the nights of 2013 February 9, 2013 February 16, and 2013 March 15. Each object had three exposures with individual exposure times of 1200s. A Helium-Neon-Argon lamp was observed for wavelength calibration, and standard stars Feige 34 and Feige 66 were observed for flux calibration. The flux calibrated reference spectra were obtained from the \textit{Hubble Space Telescope} CALSPEC Calibration Database. 

Reductions for the two-dimensional images, as well as extraction of the one-dimensional spectra and wavelength calibration were done using standard longslit reduction procedures in \textit{IRAF}\footnote{\textit{IRAF} is distributed by the National Optical Astronomy Observatory, which is operated by the Associate of Universities for Research in Astronomy (AURA) under cooperative agreement with the National Science Foundation.} following Massey \etal 1992 and Massey 1997. \\

\textit{MDM/OSMOS.}

We observed ten targets with the Ohio State Multi-Object Spectrograph (OSMOS; \citealt{2011PASP..123..187M}) on the MDM Observatory 2.4 Hiltner Telescope on 2 February 2014 and 20-23 January 2015. For each target, we used a $1.2''$ by 20$'$ slit, with a VPH grism ($R=1600$, or 0.7 $\rm \AA$ / pixel), which covered an observed wavelength range of $3900 - 6900$ \AA\ . The seeing was measured to be 0.8$''$-1.1$''$ for the February 2 observation, and between 0.8$''$-1.8$''$ for the January 20-23 observations. Each object was observed with multiple exposures of 1800s. For wavelength calibration, we observed an Argon arc lamp, and for flux calibration, we observed the spectrophotometric standards Feige 34 and Feige 66.

The data were reduced following standard \textit{IRAF} routines from the \textit{longslit} package, including bias and dark subtraction, flat fielding, wavelength calibration, and telluric and background subtraction. One-dimensional spectra were then extracted using a $\sim3''$ aperture along the slit, and corrected for heliocentric velocity.\\

\textit{SDSS.}

For purposes of comparison, we also make use of the SDSS observations of the galaxies in our sample. The objects were originally selected by R13 by analyzing SDSS spectroscopy of $\sim25,000$ dwarf emission line galaxies in the NASA-Sloan Atlas \footnote{http://www.nsatlas.org}. 


\section{Analysis}

\subsection{Emission line modeling}

For each galaxy, we modeled the H$\alpha$-[NII] complex in order to ascertain whether broad H$\alpha$ emission was present. We summarize the general procedure here, though more thorough discussions can be found in R13 and \cite{2015ApJ...809L..14B}. In this work, we are mainly interested in whether a broad H$\alpha$ emission line is present. Thus, we simply fit the continuum as a line across the region surrounding the H$\alpha$-[NII] complex. This also ensures that we do not artificially introduce a broad line by over-subtracting an absorption feature. We then create a model for the narrow-line emission using an intrinsically narrow-line (e.g., [SII]$\lambda\lambda$ 6713,6731; forbidden lines are guaranteed not to be produced in the dense broad line region gas). We use the narrow-line model to fit the narrow H$\alpha$ emission and the [NII] lines simultaneously. The width of the narrow H$\alpha$ line is allowed to increase by up to 25\%. The relative amplitudes of the [NII] lines are fixed to their laboratory values. We next add an additional Gaussian component to the model to represent broad H$\alpha$ emission. If the preferred FWHM for the component is $> 500\kms$ and the $\chi^{2}_{\nu}$ is improved by at least 20\%, then we classify the galaxy as having persistent broad H$\alpha$ emission (\citealt{2005AJ....129.1795H}; R13). As noted in R13, this threshold is somewhat arbitrary, but found to be empirically suitable for this work. If the broad model is preferred, we attempt to add an additional broad component, effectively allowing the broad emission to be modeled with up to two Gaussians. Once we have our best-fit model (i.e., adding additional components no longer improves the $\chi^{2}_{\nu}$ by our threshold value), we measure the FWHM and luminosity of broad H$\alpha$, if present. We model and measure emission line fluxes for H$\beta$ and [OIII] $\lambda 5007$. Below, we discuss the narrow-line models used for each instrument.

\subsubsection{Narrow-line models}
\textit{MagE:}
For the MagE spectra, we follow R13 and model the narrow-line emission by fitting Gaussian profiles to the [SII] $\lambda\lambda$6713,6731 doublet. In our modeling, the two lines are required to have the same width.  \\

\textit{OSMOS:}
The wavelength coverage of the OSMOS spectra does not extend to the [SII] doublet, so here we model the narrow emission with the [OIII] $\lambda$5007 line. The profile of [OIII] often has a core component and a wing component \citep{1981ApJ...247..403H, 1984ApJ...286..171D, 1985MNRAS.213....1W, 2005ApJ...627..721G,2013MNRAS.433..622M, 2014MNRAS.442..784Z}. We therefore fit the [OIII] line with two components if the addition of a second component improves the $\chi^{2}_{\nu}$ by at least 20\%. If [OIII] is indeed best fit with both a core and wing component, then we use the width of the core component for our narrow-line model. If it is instead best fit with a single component, we retain the width of the single component for our narrow-line model. In order to determine whether the use of [OIII] as our narrow line would affect our fits to the H$\alpha$-[NII] complex, we compared the profiles of [SII] and [OIII] in the SDSS spectra for our OSMOS targets. We find that, for a given object, the FWHMs of the [OIII] and [SII] lines differ by very little, with percent differences in FWHM ranging from 0.3 to 11\%. Since we allow the width of the H$\alpha$ narrow line to increase by up to 25\%, we do not expect this choice of narrow line to affect our results. \\

\begin{figure}
\centering
\includegraphics[scale=0.44]{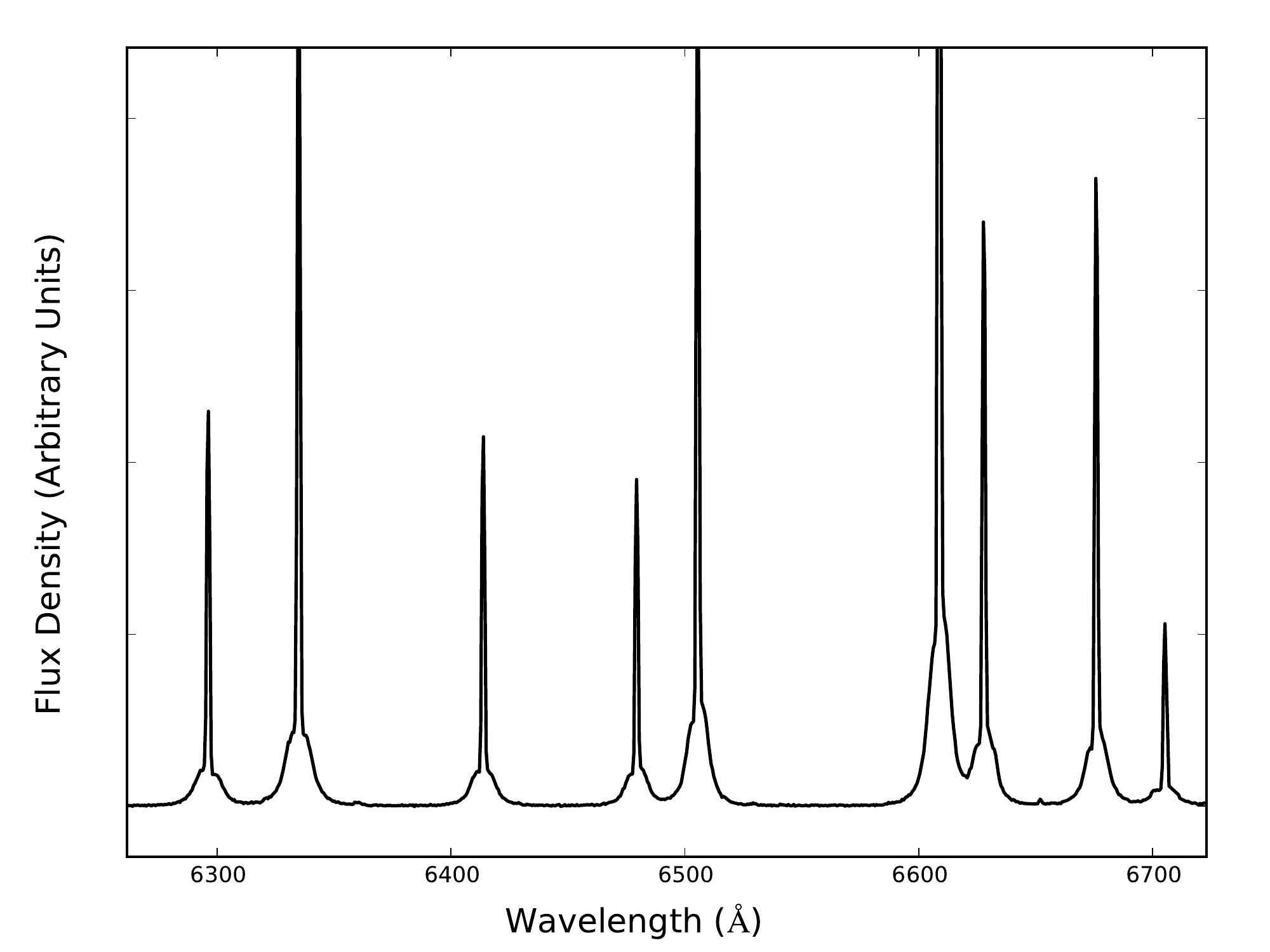}
\caption{Red channel He-Ne-Ar arc lamp spectrum taken as part of our observations with the Dual-Imaging Spectrograph on the ARC 3.5m telescope. Note that the shape of the instrumental broadening has two components -- a narrow core and a broad base.}
\label{comp_lines}
\end{figure}

\begin{figure}
\includegraphics[scale=0.45]{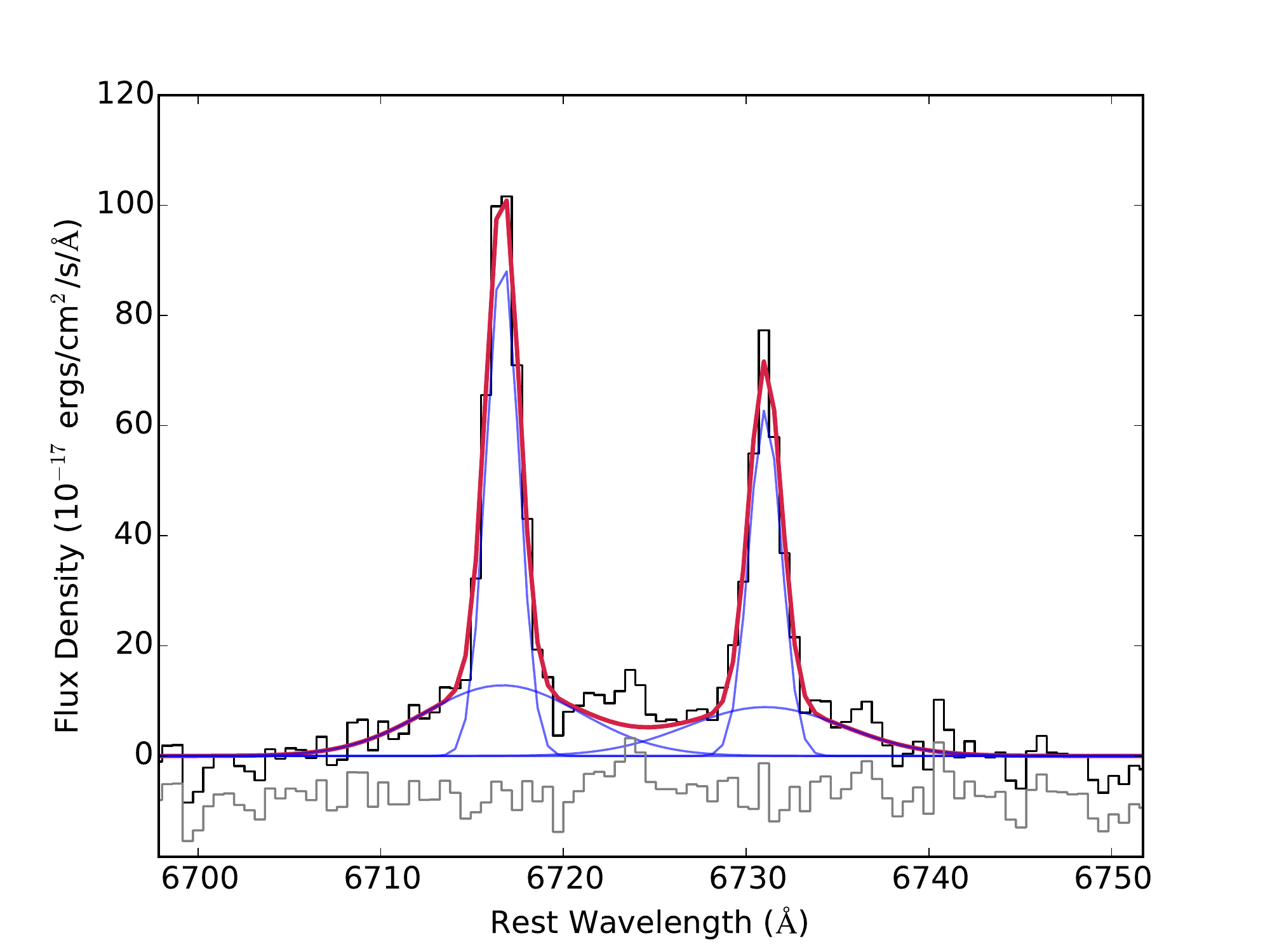}
\caption{[SII]$\lambda\lambda6716,6731$ lines from the DIS spectrum of RGG C (NSA 109990). The black line shows the observed spectrum, the red line is our best fit to the two lines, and the solid blue lines show the various components corresponding to our best fit of the two [SII] lines. We constrain the relative heights and widths of the ``narrow" and ``broad" components and use that model to fit the narrow-lines in the H$\alpha$-[NII] complex for DIS observations. }
\label{SII}
\end{figure}

\textit{DIS:}
For the DIS spectra, we also use the [SII] lines to model the narrow-line emission. However, the emission lines in the DIS spectra display some additional instrumental broadening, which we observe in the red channel arc lamp spectra (see Figure~\ref{comp_lines}). Since this instrumental broadening is not visible in low S/N [SII] lines but can still affect the fit to H$\alpha$, we created a basic model for the narrow-line emission to use in fitting the [SII] lines. Using the galaxy spectrum for which the [SII] lines had the most flux, we generated a narrow-line model consisting of two Gaussians with fixed relative amplitudes and widths. This basic shape was then used to model [SII] lines in all other DIS galaxy spectra, and correspondingly, their narrow H$\alpha$ and [NII] lines. Figure~\ref{SII} shows the [SII] lines from a representative DIS spectrum. 

Given the instrumental broad component seen in the emission lines, we tested the degree to which the narrow-line model affects the presence/properties of broad H$\alpha$ emission to ensure that our results were not influenced by this broadening. To do this, we created two additional narrow-line models: one where the instrumental broadening is less prevalent (i.e., we reduced the relative width and relative amplitude of the ``broad" narrow-line component by 10\%), and one where it is more prevalent (i.e., the relative width and amplitude are increased by 10\%). We then refit the spectra and assessed the security of our broad H$\alpha$ detections. We found this to have little effect on the FWHM and luminosity of broad H$\alpha$ for secure detections (i.e., FWHM varied by less than 10\%), but to greatly influence these measured quantities for more ambiguous detections (with the FWHM varying by as much as 90\% for a slightly different narrow-line model). We err on the side of caution in our classifications; objects which have ambiguous broad H$\alpha$ detections in their DIS spectra are classified as ambiguous.

\subsection{Stellar velocity dispersions}

We measured stellar velocity dispersions for 15 galaxies with MagE observations using pPXF (Penalized Pixel Fitting; \citealt{2004PASP..116..138C}). As a reminder, of these, 9 are classified as BPT AGN, 3 are BPT composites, and 3 are BPT star forming galaxies. Four galaxies (1 composite and 3 star forming) overlap with our SDSS broad H$\alpha$ sample. 

PPXF uses a library of stellar templates to fit the stellar continuum and kinematics of a galaxy spectrum. We used a library of 51 stars from the ELODIE spectral database \citep{2001A&A...369.1048P}. Our library consists of stars covering spectral classes from O through M and luminosity classes from bright giants to dwarfs (see Appendix Table~\ref{sptypes} for a list of stars and spectral types). The ELODIE spectra span from 3900\AA\ to 6800\AA\ in wavelength and have a spectral resolution of R=10,000. 

\begin{figure}
\includegraphics[scale=0.45]{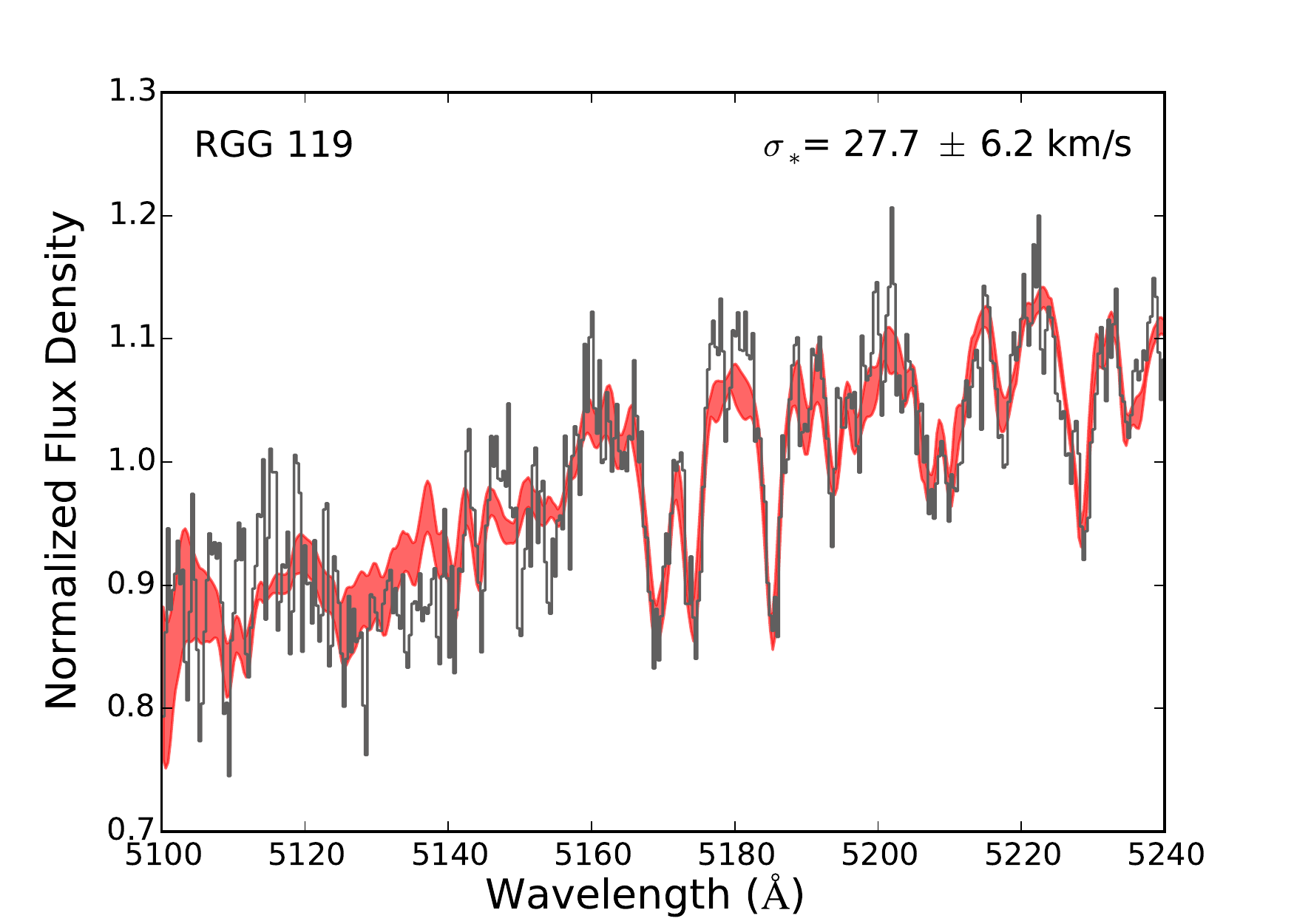}
\caption{\textbf{Spectral region of the MagE spectrum of RGG 119 used to measure the stellar velocity dispersion in pPXF. We use a region of the spectrum encompassing the Mg $b$ triplet.} The observed spectrum is shown in gray, while the shaded red region represents the range of outputs from pPXF for the chosen combinations of additive and multiplicative polynomials.}
\label{rgg119_stel}
\end{figure}

We measured stellar velocity dispersions in region of the spectrum surrounding the Mg \textit{b} triplet, spanning from 5100\AA\ to 5250\AA (see Figure~\ref{rgg119_stel}). We fit each region using combinations of low-order multiplicative and additive polynomials (orders range from 1 to 4; see e.g., \citealt{2010ApJ...716..269W, 2015ApJ...801...38W}) , and report the mean value for all measurements. Uncertainties are reported by PPXF for each individual fit (i.e. each combination of multiplicative and additive polynomial). We take the mean for all velocity dispersion measurements the Mg b band and add the errors in quadrature to obtain a velocity dispersion measurement and corresponding error estimate. We measure $\sigma_{\star}$ values ranging from $28-71 \kms$ with a median $\sigma_{\star}$ of $41 \kms$. These values correspond to galaxies with NASA-Sloan Atlas stellar masses ranging from $\sim5\times10^{8}$ to $3\times10^{9}$~\msun. All stellar velocity dispersion measurements are presented in Table~\ref{veldisp}.                                                                                                                                                                                                 

\begin{table}
{\centering
\caption{Stellar velocity dispersion measurements}
\begin{tabular}{c | c }
\hline
\hline 
R13 ID & $\sigma_{\star}$  \\
 & ($\kms$) \\ 
\hline
6 &  62 $\pm$ 12  \\
16 & 33 $\pm$ 5  \\
22 & 39 $\pm$ 6  \\
27  & 28 $\pm$ 6  \\
28  & 32 $\pm$ 4  \\
29 & 49 $\pm$ 7 \\
31 & 31 $\pm$ 4  \\
33  & 46 $\pm$ 5  \\
34 & 58 $\pm$ 3  \\
119 & 28 $\pm$ 6  \\
120 & 71 $\pm$ 8 \\
128  & 43 $\pm$ 5  \\
E & 37 $\pm$ 12 \\
G & 41 $\pm$ 13 \\
H& 60 $\pm$ 11 \\
\hline
\end{tabular}
\label{veldisp}\\
}
\textbf{Table~\ref{veldisp}.} Stellar velocity dispersion measurements for objects with MagE data. Stellar velocity dispersions were measured in the Mg $b$ region of the spectrum.

\end{table}


\section{Broad line AGN candidates: Results}

\begin{figure*}
\centering
\includegraphics[scale=0.7]{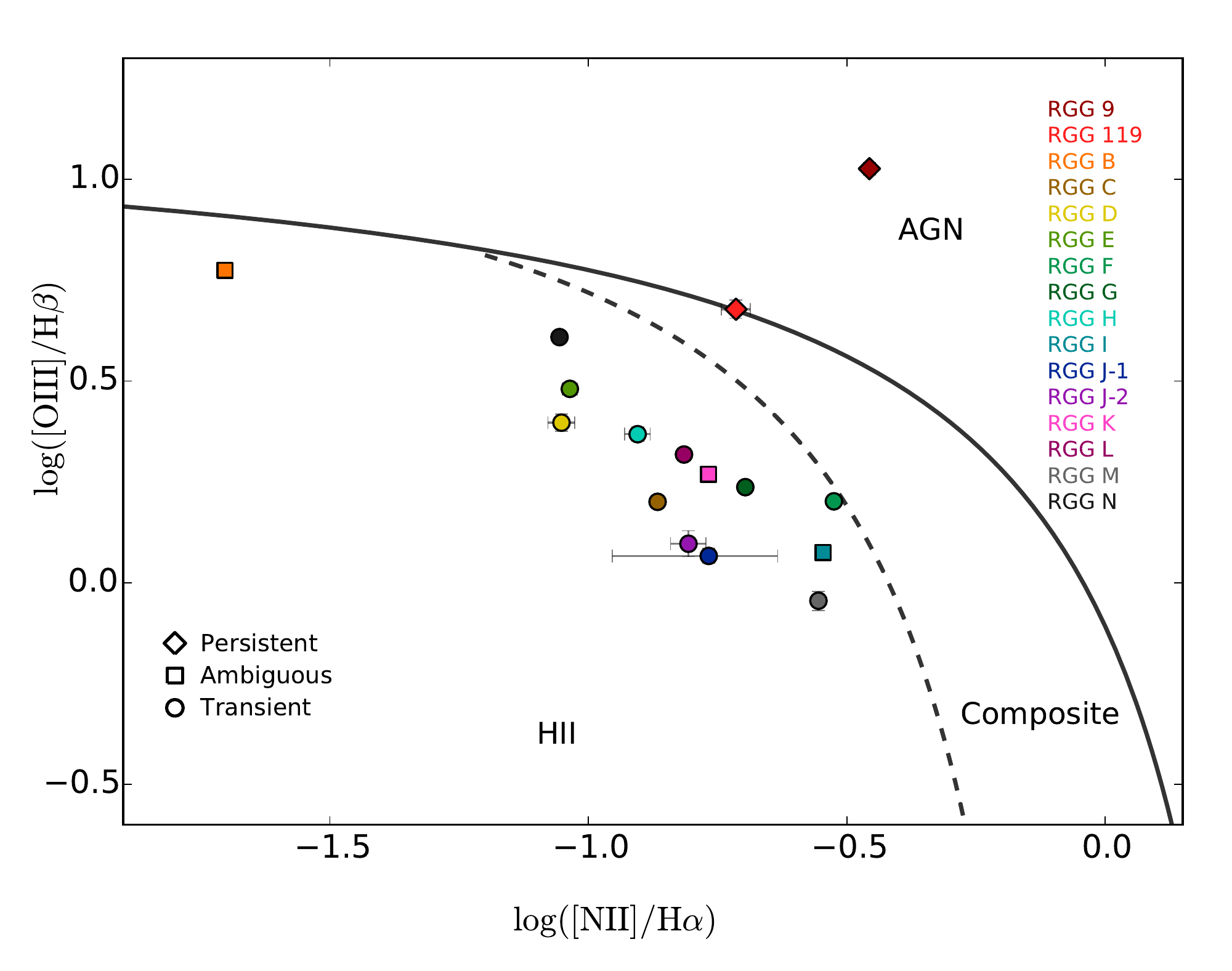}
\caption{Positions of our SDSS broad H$\alpha$ targets on the BPT diagram \citep{1981PASP...93....5B} using the classification of \cite{2006MNRAS.372..961K}. Each target is plotted in a different color. Note that there are two points for RGG J corresponding to the two SDSS observations. A diamond indicates that broad H$\alpha$ was present in follow-up spectroscopy (`persistent'), a square indicates an ambiguous follow-up broad H$\alpha$ detection, and a circle indicates that broad H$\alpha$ was not detected in follow-up observations (`transient').}
\label{bpt}
\end{figure*}

\begin{table*}[t]
{\centering
\caption{Broad H$\alpha$ candidate results}
\begin{tabular}{c c c c c c c c c c}
\hline
\hline
\smallskip
R13 ID & NSAID & R13 $\log L_{\rm H\alpha}$ (\ergs) & R13 FWHM$_{\rm H\alpha}$ ($\kms$) & R13 BPT Class & Follow-up & $\Delta$t (years)& Persistent broad H$\alpha$? \\
\hline
9 &  10779 & 40.15 & 703 & AGN & DIS & 13 & yes \\
\smallskip
119 &  79874 & 40.16 & 1043 & Comp & DIS, MagE & 5 & yes \\ 

C &  109990 & 40.10 & 3690 & SF & DIS, OSMOS & 8 & no \\
D &  76788 & 39.56 & 4124 & SF & OSMOS & 8 & no  \\
E &  109016 & 39.26 & 935 & SF & MagE, OSMOS & 9 & no \\
F &  12793 &  40.09 & 598 & SF & OSMOS & 14 & no  \\
G &  13496 & 39.56 & 2027 & SF & MagE & 11 & no  \\
H &  74914 & 39.97 & 3014 & SF & MagE & 8 & no  \\
J &  41331 & 39.99 & 1521 & SF & DIS & 10 & no  \\
L &  33207 & 39.51 & 3126 & SF & DIS & 11 &no   \\
M &  119311 & 39.75 & 3563 & SF & DIS, OSMOS & 6 & no   \\ 
N &  88972 & 40.42 & 645 & SF & OSMOS & 11 & no   \\
\smallskip
O &  104565 & 38.88 & 1653 & SF & OSMOS & 9 & no \\

B &  15952 & 40.67 & 1245 & SF & DIS, OSMOS & 14 & ambiguous   \\
I &  112250 & 39.39 & 994 & SF & DIS, OSMOS & 8 & ambiguous  \\
K &  91579 & 39.58 & 774 & SF & DIS, OSMOS & 10 & ambiguous  \\
\hline
\end{tabular}
\label{classification}\\}
\textbf{Table~\ref{classification}}. Summary of our analysis for the SDSS broad H$\alpha$ sample. All objects here were observed to have broad H$\alpha$ in their SDSS spectroscopy in R13. Columns 1 \& 2 list the R13 and NASA-Sloan Atlas ID for each object. Columns 3 \& 4 list the luminosity and FWHM of the broad emission measured in R13. Column 5 lists the instrument(s) used for the follow-up spectroscopy, and Column 6 lists the time between the SDSS observation and most recent follow-up observation. Column 7 presents the classifications determined in this work. 
 \end{table*}

We show the locations of all 16 SDSS broad H$\alpha$ targets on the BPT diagram in Figure~\ref{bpt}. Based on our follow-up spectroscopy, we classify our targets as having SDSS broad H$\alpha$ emission that is: transient (broad H$\alpha$ is not present in follow-up spectra), persistent (broad H$\alpha$ is present in follow-up spectroscopy and consistent with prior observations), or ambiguous (state of broad emission is unclear). In the following subsections, we describe our classifications in more detail. Table~\ref{classification} summarizes the results of this analysis. Figures showing the fits to H$\beta$, [OIII], and H$\alpha$ are given for each observation in the Appendix. See Section 6 for a discussion on aperture effects and intrinsic variability as they relate to this work.

\subsection{Transient broad H$\alpha$}
\begin{figure*}[h]
\includegraphics[scale=0.8]{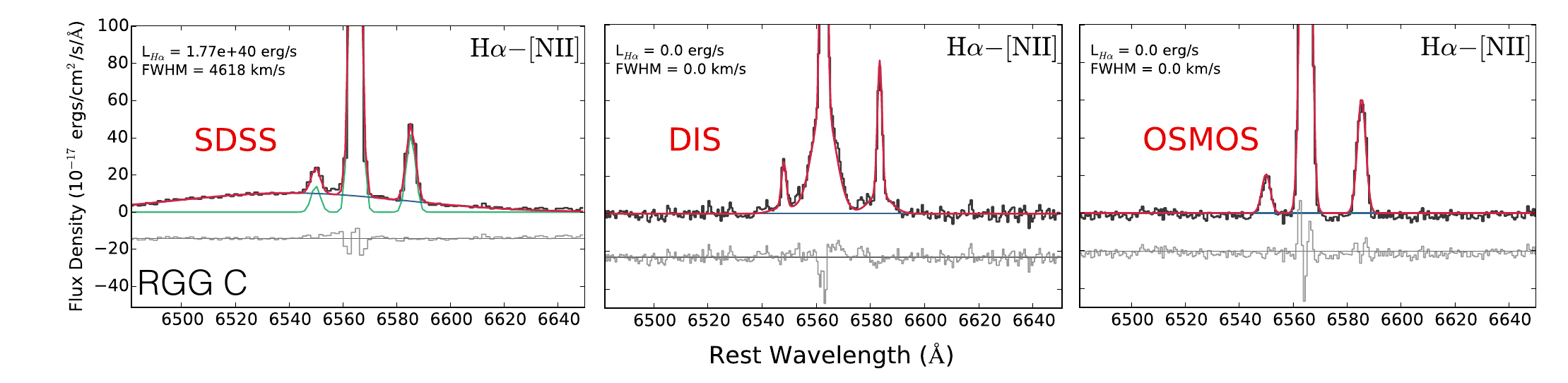}
\caption{The observed H$\alpha$-[NII] complex and corresponding best-fit profile for the SDSS, DIS, and OSMOS spectra of RGG C (NSA 109990). In each, the dark gray line represents the observed spectrum. The narrow emission line fit is plotted in green, and the blue shows the fit to broad H$\alpha$ emission. The overall best-fit model is given in red, and the light gray line below the observed spectrum shows the residual between the observed spectrum and the best fit, offset by an arbitrary amount. This galaxy is classified as having transient broad H$\alpha$ emission.}
\label{rggc}
\end{figure*}

Out of the 16 objects with SDSS broad H$\alpha$, we find 11 to have transient broad H$\alpha$ emission, i.e., they lacked broad emission in their followup spectroscopic observations (see Figure~\ref{rggc} for an example). All 11 fall in the ``HII" region of the BPT diagram, which indicates the presence of recent star formation in the galaxy. We note that several objects (e.g., RGG G, RGG F; see Appendix) had additional components in their H$\alpha$ line models with FWHMs of a few hundred $\kms$; since these are all narrower than the broad emission in the SDSS spectra and fall below our FWHM cut of $500\kms$, we do not classify those objects as having persistent broad emission. 
Given the narrow-emission line ratios and host galaxy properties, we consider Type II supernovae (SNe) to be the most likely origin for the transient broad H$\alpha$ (see Section 5 for more details). We note that, as discussed in detail in R13, luminous blue variables \citep{2011MNRAS.415..773S} and Wolf-Rayet stars can also produce transient broad H$\alpha$ emission. However, Wolf-Rayet stars typically produce other notable spectral features, such as the Wolf-Rayet bump from $\lambda$4650-5690$\rm\AA$. Moreover, their spectra don't typically have strong hydrogen lines \citep{2007ARA&A..45..177C,2011MNRAS.416.1311C}. Finally, recent work has revealed the existence of ``changing look" quasars, i.e., quasars with broad Balmer emission lines that fade significantly over the course of 5-10 years (see e.g., \citealt{2016MNRAS.455.1691R, 2015arXiv150903634R}). In the case of changing look quasar SDSS J101152.98+544206.4, the broad H$\alpha$ luminosity dropped by a factor of 55 on this timescale. This behavior is suspected to be caused by a sudden drop in accretion rate. 

\subsection{Persistent broad H$\alpha$}
\begin{figure*}
\centering
\includegraphics[scale=0.8]{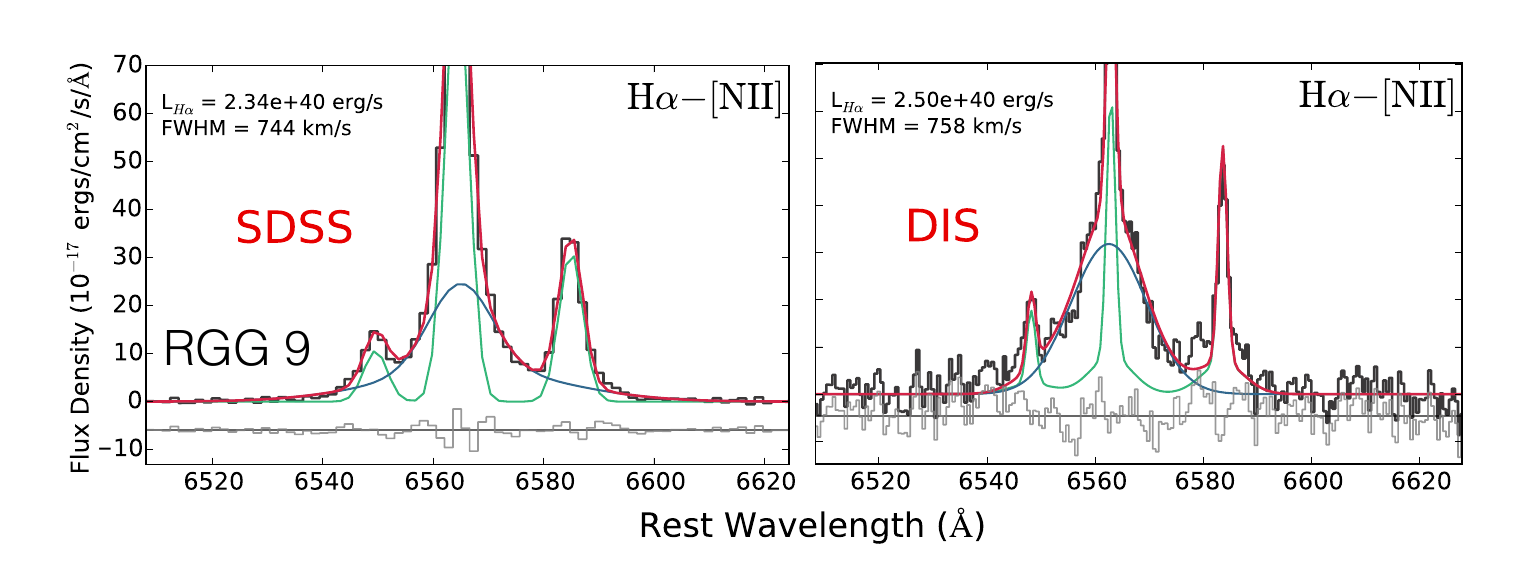}
\caption{The observed H$\alpha$-[NII] complex and corresponding best-fit profile for the SDSS and DIS spectra of RGG 9. The dark gray line represents the observed spectrum. The narrow emission line fit is plotted in green, and the blue shows the fit to broad H$\alpha$ emission. The overall best-fit model is given in red, and the light gray line below the observed spectrum shows the residual between the observed spectrum and the best fit, offset by an arbitrary amount. We classify this galaxy as having persistent broad H$\alpha$.}
\label{nsa10779_main}
\end{figure*}

\begin{figure*}
\includegraphics[scale=0.8]{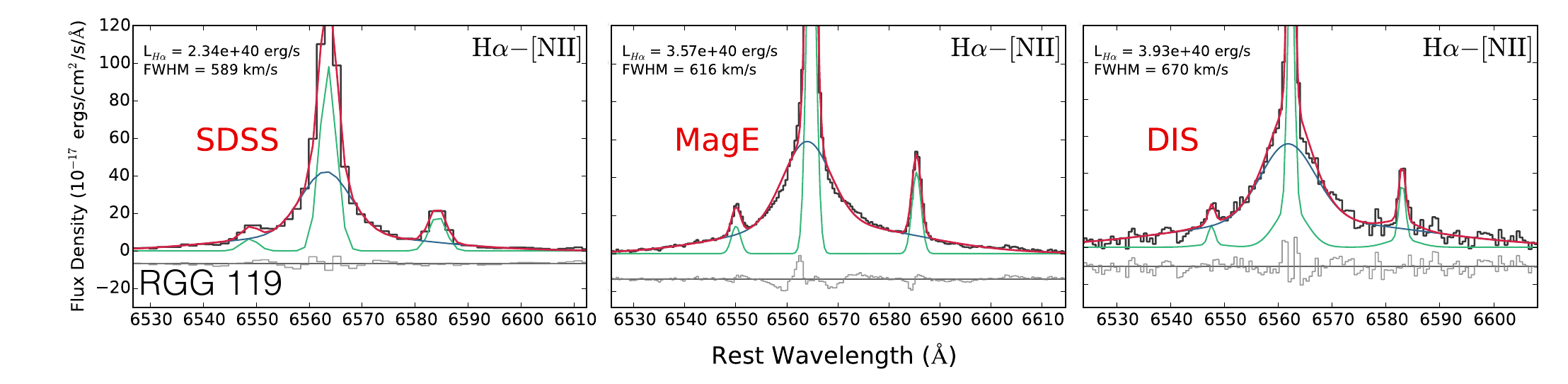}
\caption{The observed H$\alpha$-[NII] complex and corresponding best-fit profile for the SDSS, MagE, and DIS spectra of RGG 119 (NSA 79874). In each, the dark gray line represents the observed spectrum. The narrow emission line fit is plotted in green, and the blue shows the fit to broad H$\alpha$ emission. The overall best-fit model is given in red, and the light gray line below the observed spectrum shows the residual between the observed spectrum and the best fit, offset by an arbitrary amount. We classify this galaxy as having persistent broad H$\alpha$. }
\label{nsa79874_main}
\end{figure*}

We identified 2/16 objects with secure, persistent broad H$\alpha$ emission. RGG 9\footnote{RGG denotes we are using the ID assigned by Reines, Greene, \& Geha (2013).} (NSA 10779), is classified as an AGN on the BPT diagram based on its SDSS spectrum. We observe broad H$\alpha$ emission in both the SDSS spectrum and the DIS spectrum, taken 13 years apart (see Figure~\ref{nsa10779_main} for the best fit to the DIS spectrum). The second secure broad line object, RGG 119 (NSA 79874), is classified as a composite object on the BPT diagram. This object has broad H$\alpha$ emission in the SDSS, MagE, and DIS spectra, spanning five years (see Figure~\ref{nsa79874_main}).

We compute the mass of the central BH using standard virial techniques which employ only the FWHM and luminosity of the broad H$\alpha$ emission \citep{2005ApJ...630..122G,2009ApJ...705..199B, 2013ApJ...767..149B}. The calculation also includes a scale factor $\epsilon$ intended to account for the unknown geometry of the broad line region; we adopt $\epsilon=1$ (see Equation 5 in R13). This method makes use of multiple scaling relations between parameters (e.g., the relation between FWHM$_{\rm H\alpha}$ and FWHM$_{\rm H\beta}$), each of which has its own intrinsic scatter, giving a systematic uncertainty on the measured BH mass of 0.42 dex \citep{2015ApJ...809L..14B}. To ensure that our choice of continuum subtraction method would not affect our BH mass estimates, we measured BH masses for the SDSS spectra using both R13 and our continuum subtraction, and found the masses to be consistent with one another.

For RGG 9, we measure a BH mass from the SDSS observation of $\rm 3.5(\pm0.7)\times10^{5}~M_{\odot}$. From the DIS observation, we measure a BH mass of $\rm 3.7(\pm0.3)\times10^{5}~M_{\odot}$. Using these two measurements, we compute a mean mass of $\rm M_{BH} = 3.6(\pm0.8)\times10^{5}M_{\odot}$. Taking into account the systematic uncertainty of 0.42 dex, this gives us a final estimate of $\rm M_{BH}=3.6^{+5.9}_{-2.3}\times10^{5}M_{\odot}$. Using the stellar mass of RGG 9 from R13 ($\rm M_{\star}=2.3\times10^{9}~M_{\odot}$), we compute a BH mass-to-stellar mass ratio of $\rm M_{BH}/M_{\star} = 1.6\times10^{-4}$. 

For RGG 119, we measure a BH mass of $\rm 2.1(\pm0.5)\times10^{5}~M_{\odot}$ from the SDSS spectroscopy. The MagE observation yields a BH mass of $\rm 2.9(\pm0.2)\times10^{5}~M_{\odot}$, and the DIS observation gives a BH mass of $\rm 3.6(\pm0.3)\times10^{5}~M_{\odot}$. 
We compute a mean BH mass of $\rm M_{BH} = 2.9(\pm0.6)\times10^{5}M_{\odot}$ using the three spectroscopic observations. With the additional systematic uncertainty, our final BH mass estimate is $\rm M_{BH} = 2.9^{+4.9}_{-1.8}\times10^{5}M_{\odot}$. R13 reports a stellar mass of $\rm M_{\star} = 2.1\times10^{9}M_{\odot}$ for RGG 119, giving a BH mass to stellar mass ratio of $\rm M_{BH}/M_{\star} = 1.3\times10^{-4}$. 
See Table~\ref{broadparams} for the measured broad H$\alpha$ FWHM and luminosities, and corresponding BH masses for each observation. We note that our measured BH masses are consistent with those measured in R13 ($2.5\times10^{5}M_{\odot}$ and $5.0\times10^{5}M_{\odot}$ for RGG 9 and RGG 119, respectively). We also refer the reader to Section 6 for a discussion of the potential effect of instrumentation/aperture on the measured properties of the broad H$\alpha$ emission. 

\begin{figure}
\includegraphics[scale=0.65]{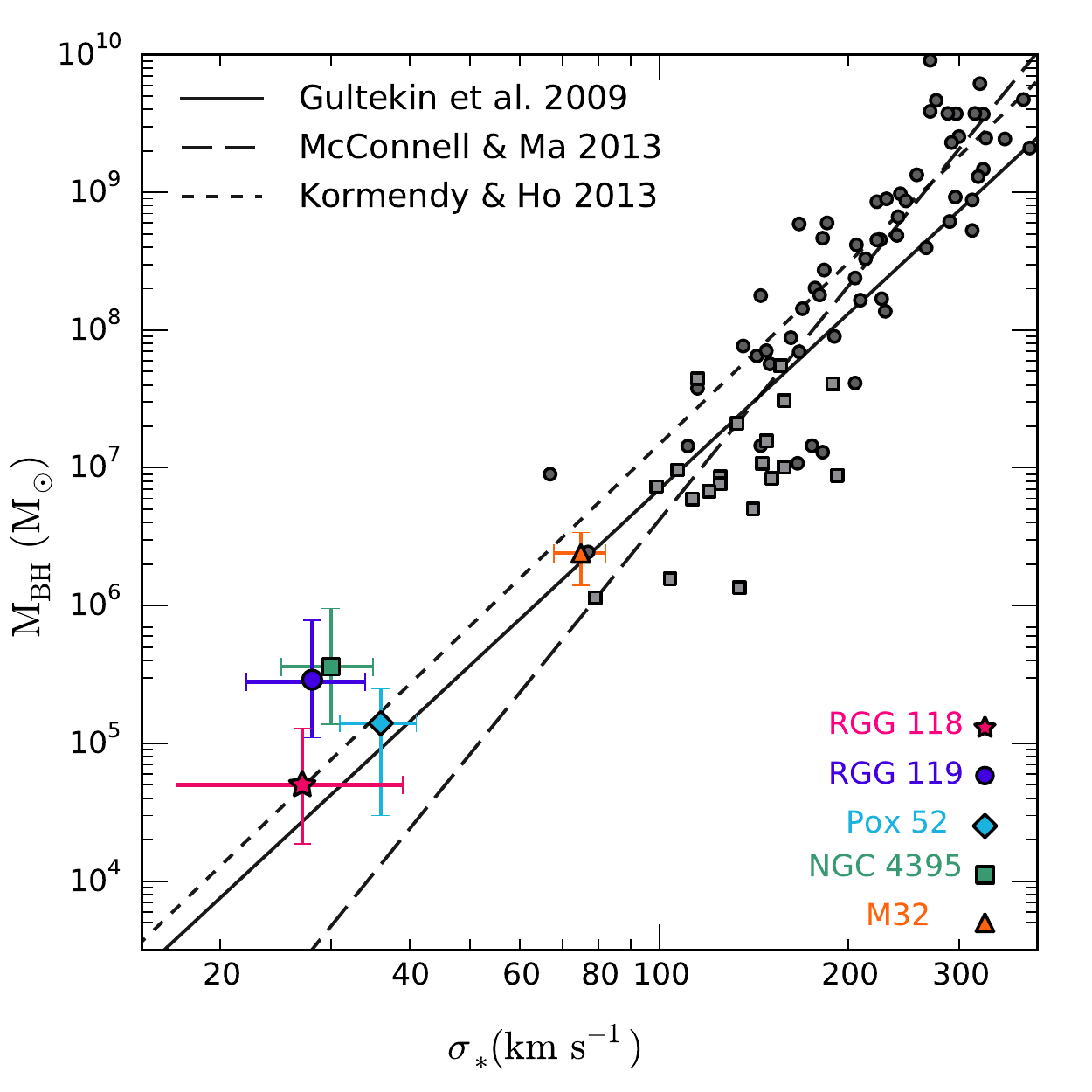}
\caption{The relation between black hole mass and stellar velocity dispersion. The $\rm M_{BH}-\sigma_{\star}$ relations determined in \cite{2009ApJ...698..198G}, \cite{2013ApJ...764..184M}, and \cite{Kormendy:2013ve} are plotted. We also show points for systems with dynamical BH mass measurements as compiled by \cite{Kormendy:2013ve} (pseudobulges are shown as gray squares, while classical bulges are plotted as gray circles).  The black hole mass and stellar velocity dispersion for RGG 119 (NSA 79874; shown as a purple circle) were measured in this work. Also plotted are the low-mass AGN RGG 118 \citep{2015ApJ...809L..14B}, NGC 4395 \citep{1989ApJ...342L..11F, 2003ApJ...588L..13F}, and Pox 52 \citep{2004ApJ...607...90B}, and the dwarf elliptical M32 \citep{1980ApJ...242...53W, 2010MNRAS.401.1770V}. }
\label{msig}
\end{figure}

Using the MagE spectrum of RGG 119 we measure a stellar velocity dispersion of $28 \pm 6~\kms$ from the Mg \textit{b} triplet, and place this galaxy on the M$_{BH}$-$\sigma_{\star}$ relation (Figure~\ref{msig}). RGG 119 sits close to the extrapolation of M$_{BH}$-$\sigma_{\star}$ to low BH masses, similar to well-studied low-mass AGN NGC 4395 \citep{2003ApJ...588L..13F} and Pox 52 \citep{2004ApJ...607...90B}. Our BH-to-galaxy stellar mass ratios are also consistent with other low-mass AGNs \citep{2015ApJ...813...82R}.

\begin{table*}[t]
{\centering
\caption{Broad H$\alpha$ parameters for galaxies with persistent broad H$\alpha$.}
\begin{tabular}{c c c c c}
\hline
\hline
R13 ID & Obs. & $\log L_{\rm H\alpha}$ (\ergs) & FWHM$_{\rm H\alpha}$ ($\kms$) & $\rm M_{BH} (10^{5}~M_{\odot})$  \\
\hline
9 & SDSS & $40.36\pm0.01$ & $744\pm69$ & $3.5 \pm 0.7$ \\
\smallskip
9 & DIS & $40.40^{+0.01}_{-0.02}$ & $758\pm27$ & $3.7 \pm 0.3$ \\

119 & SDSS & $40.36\pm0.01$ & $589\pm69$ & $2.1 \pm 0.5$ \\
119 & MagE & $40.55\pm0.01$ &  $616\pm23$ & $2.9 \pm 0.2$ \\
119 & DIS & $40.59\pm0.01$ & $670\pm27$ & $3.6 \pm 0.3$ \\
\hline
\end{tabular}
\label{broadparams}\\}
\textbf{Table~\ref{broadparams}.} Best fit broad H$\alpha$ FWHM and luminosity from each observation for objects found to have persistent broad H$\alpha$ emission. We also present the corresponding BH mass calculated using the FWHM and luminosity. These values are all measured using the fitting code described in this paper, but the SDSS spectra are continuum subtracted using continuum fits obtained by R13, while the continua in the DIS and MagE spectra were modeled with a straight line. \vspace{0.5cm}
\end{table*}

\subsection{Ambiguous galaxies}
In three cases, we were unable to definitively state whether broad H$\alpha$ emission was still present and thus classify these objects as ambiguous. All three ambiguous galaxies fall in the HII region of the BPT diagram.

RGG B (NSA 15952) was observed to have broad H$\alpha$ emission in its SDSS spectrum and OSMOS spectrum. Moreover, if we fit the broad H$\alpha$ emission from each spectrum with one Gaussian component, we get FWHM$_{\rm H\alpha, SDSS} = 1036 \pm 69 \kms$, and FWHM$_{\rm H\alpha, OSMOS} = 977 \pm 32 \kms$ (i.e., the broad components are consistent with one another). However, the DIS spectrum -- taken in between the SDSS and OSMOS spectra -- does not require a broad component according to our criteria. We cannot discount that we do not detect a broad component in the DIS spectrum due to the atypical narrow-line shapes. Nevertheless, additional follow-up spectroscopy of RGG B would be useful for determining the true nature of the SDSS broad H$\alpha$ emission. We note that RGG B is of particular interest since its position on the BPT diagram is consistent with a low-metallicity AGN \citep{2007ApJ...671.1297I}. Line ratios involving [NII] are affected by the metallicity of the AGN host galaxy \citep{2006MNRAS.371.1559G} and AGN hosts with sub-solar metallicities can migrate left-ward into the star forming region of the BPT diagram.

RGG I (NSA 112250) has broad emission detected in the SDSS, DIS, and OSMOS spectra, but the FWHM measured for the DIS spectrum is highly dependent on the narrow-line model used, i.e., changing the parameters of the narrow-line model changes the measured FWHM by up to $\sim90\%$. The FWHM is also not consistent between the three spectra ($1237 \pm 69 \kms$, $316 \pm 27 \kms$, and $1717 \pm 32 \kms$ for SDSS, DIS, and OSMOS, respectively). 

Finally, RGG K (NSA 91579) has a highly asymmetric broad line observed in the SDSS and OSMOS spectra. However, the DIS spectrum, taken in between the SDSS and OSMOS observations, does not require a broad component. We do observe asymmetric H$\alpha$ emission by eye in the all three spectra. Similarly to with RGG B, we cannot rule out that there is indeed persistent broad H$\alpha$ emission, and that the DIS spectrum does not require it because of issues relating to the fitting of the unusually shaped narrow lines.  Additional observations of all three ambiguous broad H$\alpha$ targets will be necessary to determine whether these objects do indeed host AGN.  See the Appendix for figures showing fits for all observations of the ambiguous objects.


\section{Type II supernovae in dwarf galaxies}

We consider Type II supernovae the most-likely explanation for the fading broad H$\alpha$ emission observed in 11/16 objects in our SDSS broad H$\alpha$ sample. In this section, we further examine our SNe II candidates, and discuss this hypothesis in more detail. All 11 objects with transient broad H$\alpha$ have narrow line ratios consistent with recent star formation. Moreover, their broad emission properties and galaxy properties are offset from the remainder of the SDSS broad line objects in two important ways.

\begin{figure}
\centering
\includegraphics[scale=0.5]{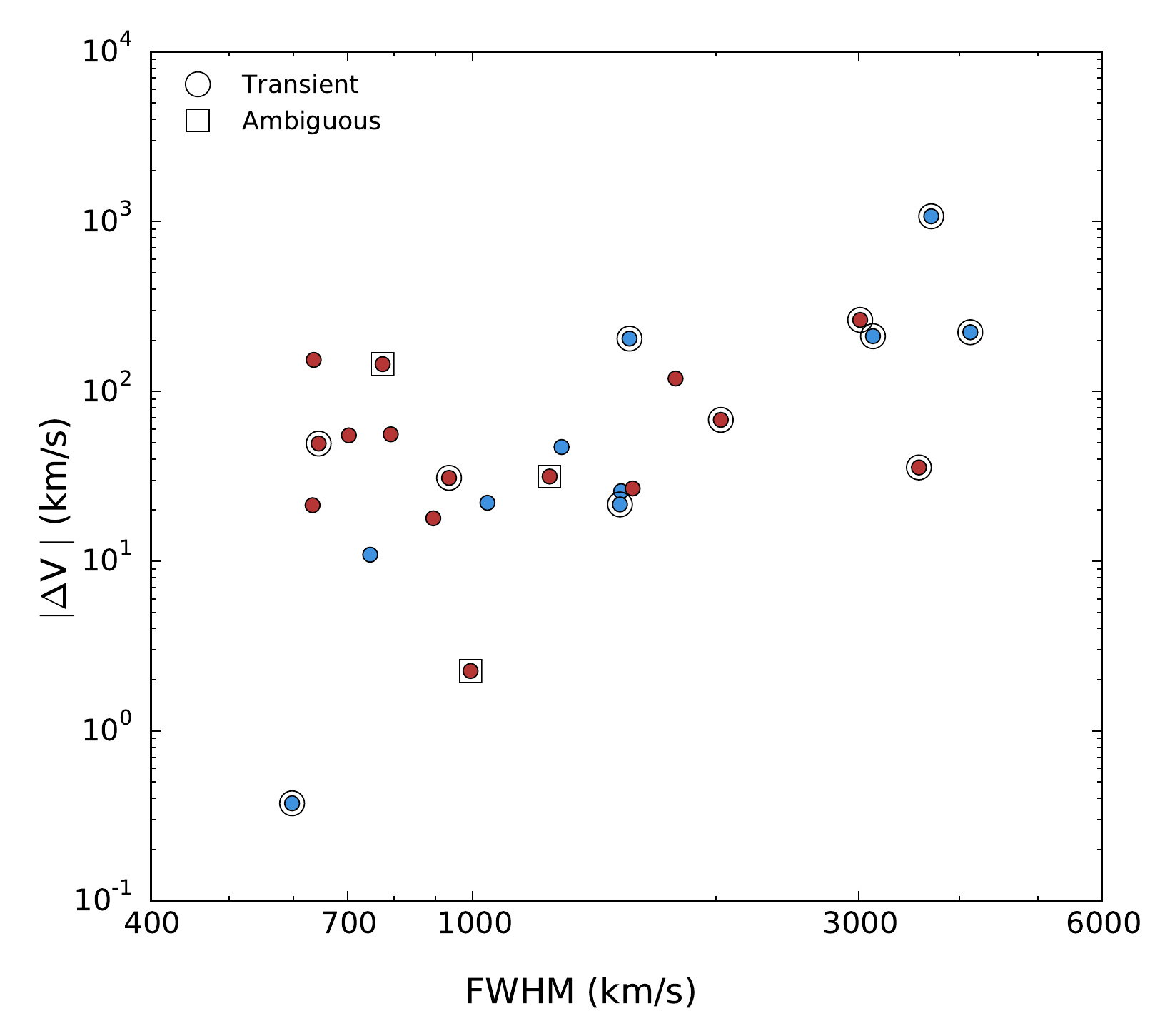}
\caption{Velocity offset from the center of the H$\alpha$ emission line in $\kms$ versus the FWHM of the broad H$\alpha$ emission from R13. A red dot indicates the broad line is red-shifted in velocity space, blue indicates the broad line is blue-shifted. The dots with circles around them are objects with transient broad H$\alpha$ (i.e., likely SNe II hosts). Large FWHM and/or velocity shifts are more characteristic of galaxies with transient broad H$\alpha$ emission.}
\label{offset}
\end{figure}

As seen in Figure~\ref{offset}, SDSS broad emission that we found to be transient tended to have larger FWHM and broad lines that were more offset from the H$\alpha$ line center in velocity space. All SDSS broad H$\alpha$ emission lines with FWHM $\gtrsim2000\kms$ in our sample were found to be transient, as well as all those with absolute velocity offsets of more than $\sim200\kms$. 

\begin{figure}
\centering
\includegraphics[scale=0.5]{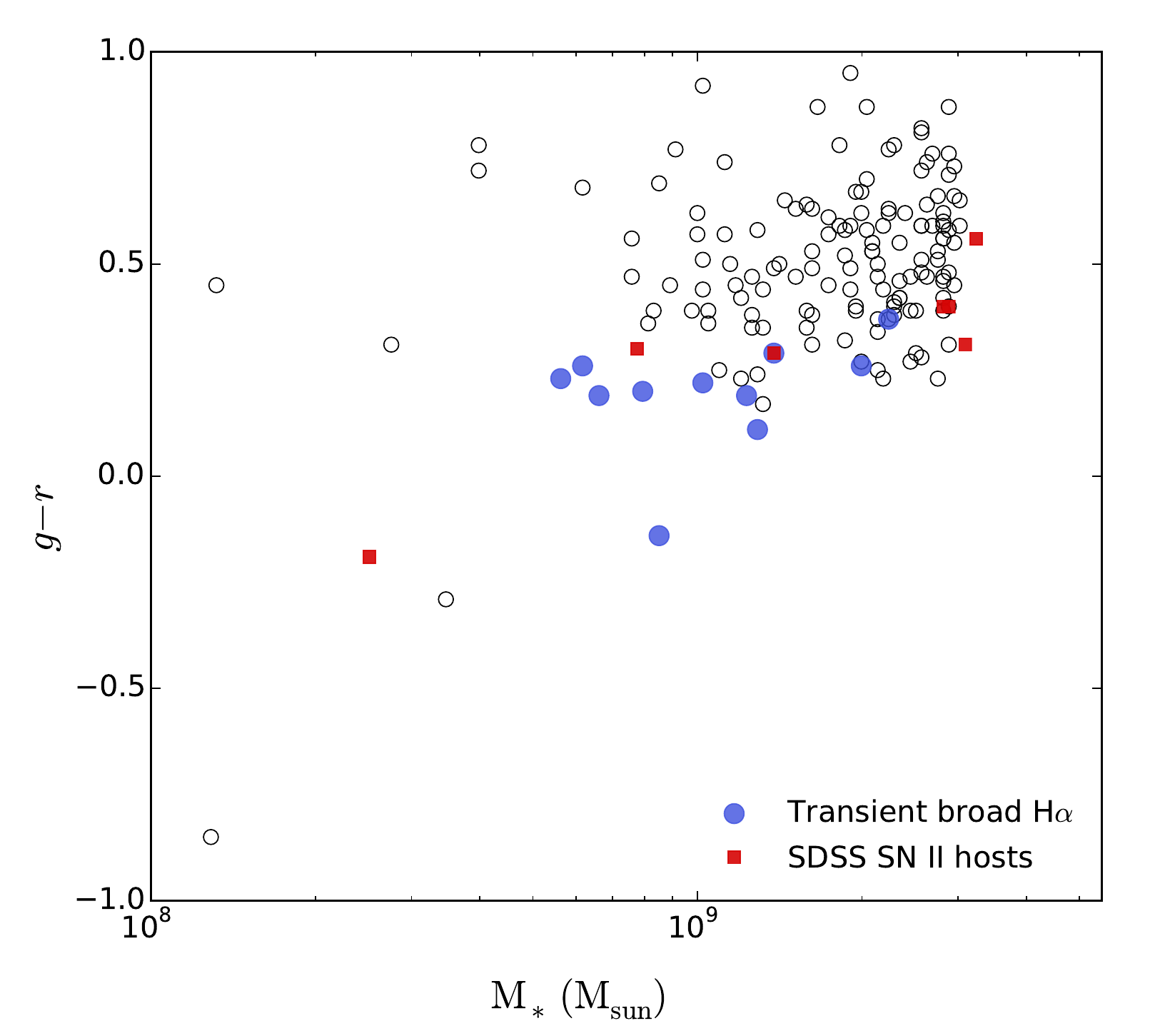}
\caption{Galaxy $g-r$ color versus host galaxy stellar mass for narrow and broad line AGN candidates from R13. Stellar masses and $g-r$ colors are from the NASA-Sloan Atlas database. Objects with transient broad H$\alpha$ are plotted as the larger blue circles. These galaxies tend to have bluer colors with respect to the rest of the sample. We also include SNe II hosts identified in the SDSS \citep{2013MNRAS.430.1746G, 2015MNRAS.450..905G} in galaxies below our mass cut-off. The masses and colors for these also come from the NASA-Sloan Atlas database (except for the lowest mass SDSS SN II host, for which the mass/color are derived from the MPA-JHU Galspec pipeline). }
\label{colors}
\end{figure}

Additionally, galaxies with transient broad H$\alpha$ also tend to be bluer with respect to the sample of broad and narrow-line candidate AGN host galaxies from R13 (Figure~\ref{colors}). The narrow-line AGN candidates were found to have a median galaxy color of $g-r = 0.51$~(R13), while the transient broad H$\alpha$ galaxies have a median host galaxy color of $g-r = 0.22$~(with all transient broad H$\alpha$ galaxies having $g-r<0.4$). This suggests the transient broad H$\alpha$ galaxies have, in general, younger stellar populations more likely to produce SNe II. 

Using time-domain spectroscopy, we have empirically identified targets most likely to be SNe based on the disappearance of broad emission lines over long ($\sim5-10$ year) time scales. A complementary analysis was performed by Graur et al. (2013, 2015)\nocite{2013MNRAS.430.1746G, 2015MNRAS.450..905G}. They searched through most SDSS galaxy spectra for SN-like spectra, and used SN template matching, rather than broad emission line fitting, to identify potential SNe. Here, we compare the results of the two methods. 

Similar to Graur et al. (2013, 2015), R13 did identify 9 SN candidates (in a sample of $\sim25,000$ objects) directly from the single-epoch SDSS spectra (Table 5 in R13; none of those have been followed up in this work). Those objects were identified because they exhibited P Cygni profiles in H$\alpha$. 
Of the 15 BPT star forming galaxies with broad H$\alpha$ identified in R13 as AGN candidates (14 of which are considered in this paper), one was detected by Graur et al. (2015) and classified as a SN II (RGG M, identified in Table 1 of Graur et al. 2015 as SDSS J131503.77+223522.7 or 2651-54507-488). The remainder did not meet all of the detection criteria set by Graur et al. (2013, 2015) and thus were not detected by Graur et al. (2015). The Graur et al. (2013, 2015) method is superior at detecting bona fide SNe II near maximum light, when these objects display prominent P-Cygni profiles. The nature of the objects considered in this paper is generally more ambiguous, as no P-Cygni profiles are seen and the emission line profiles are relatively symmetric (although generally broader and more asymmetric than the bona fide AGN; Figure~\ref{offset}).

During their survey, Graur et al. (2015) also identified five ambiguous broad-line objects in dwarf galaxies, which their pipeline classified as either SNe or AGNs. Of these, three were not in the R13 parent sample due to the stellar mass cut in two cases and a glitch in the NSA catalog in the third. Two are in R13; one was identified as a transient \citep{2014MNRAS.445..515K}, and the other (RGG J) is included in this work.

\begin{figure*}
\centering
\includegraphics[scale=0.53]{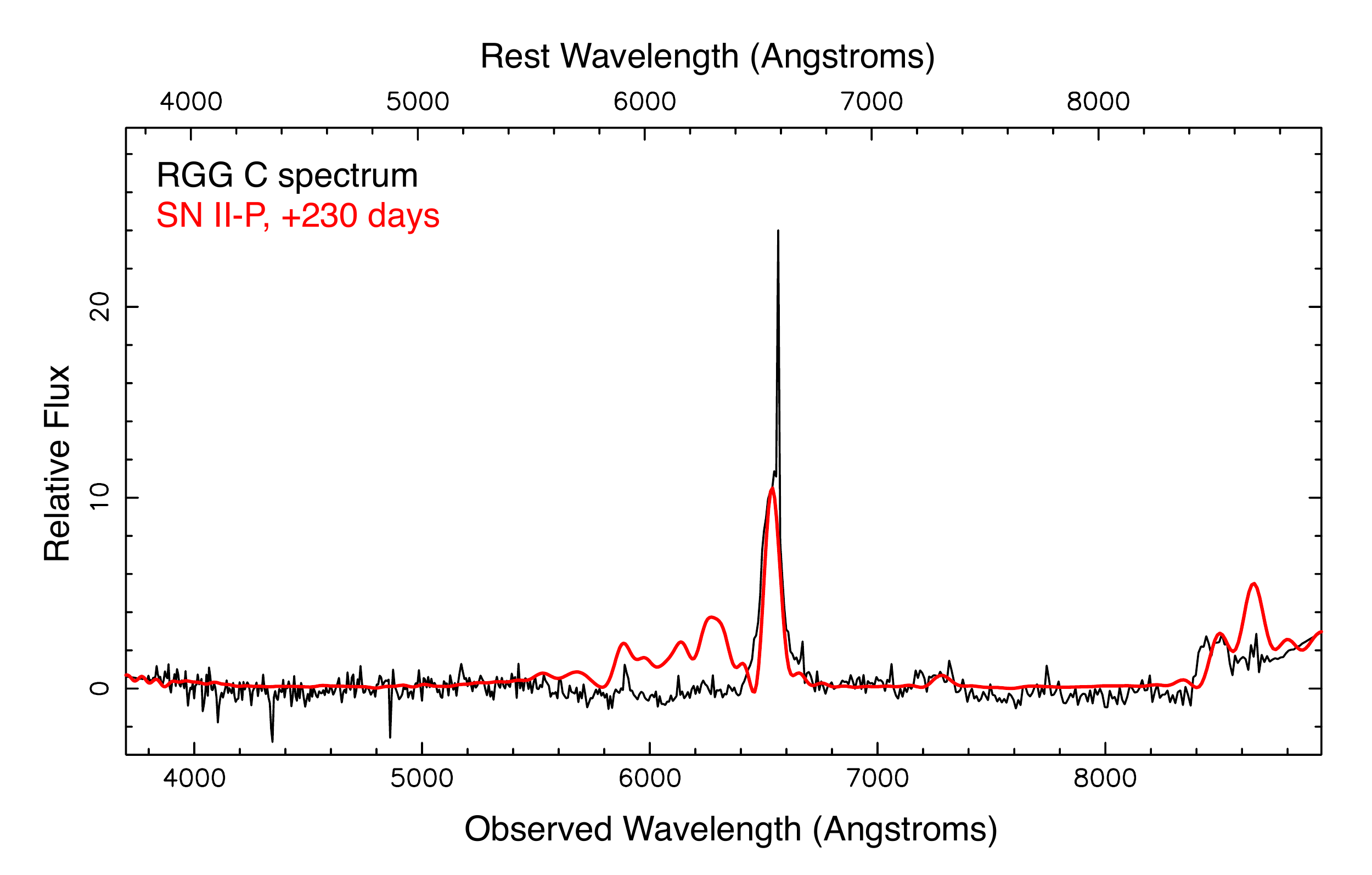}
\caption{The Supernova Identification (SNID) code fit to the SDSS spectrum of RGG C. This fit had an rlap value of 7.1. The observed spectrum is shown in black, while the SNID fit is shown in red. SNID finds that the spectrum is best fit by a post-plateau SN II-P. }
\label{SNID}
\end{figure*}

To understand why the majority of the transient sources presented here were not identified by Graur et al. (2013, 2015), and to try and learn something about their nature, we pass our continuum-subtracted SDSS spectra through the same supernova identification scheme used by Graur et al. First, the stellar continuum and narrow emission line fits are removed. We employ a different procedure from that used by Graur et al.\ to fit the continuum, but the overall result should be similar. Second, we de-redshift the spectra, remove their continua, and manually ran them through the Supernova Identification code (SNID;\footnote{http://people.lam.fr/blondin.stephane/software/snid/} \citealt{2007ApJ...666.1024B}). Four objects were classified as old SNe II, i.e., when their spectra are dominated by a broad H$\alpha$ feature, but with relatively low rlap values. The SNID `rlap' parameter is used to assess the quality of the fit, and is equal to the height-to-noise ratio of the normalized correlation function times the overlap between the input and template spectra in $\rm{ln}~\lambda$ space (see \citealt{2007ApJ...666.1024B} for details). A value of rlap=5 is the default minimum and fits with rlap values close to 5 are regarded as
suspect (see \citealt{2007ApJ...666.1024B} for details). Of the four objects identified as SNe II, two had rlap values $<5$ and two had rlap $\sim 7$. One object was classified as an old SN Ia at z $\sim$0.1. As the spectra were de-redshifted, this classification is not trustworthy. The other three objects failed to be classified by SNID. See Figure~\ref{SNID} for an example of the fit obtained by SNID to RGG C. We will discuss the special case of RGG J in some detail in Section 5.1. 

Since we consider SNe II the best explanation for the transient broad emission, in the following we compute the expected number of SNe II in the R13 parent sample, and check for consistency with our results. 
R13 identified and removed 9 galaxies with SNe II from their sample based on P Cygni profiles in the broad H$\alpha$ emission. Considering the 11 objects with transient broad H$\alpha$ and three ambiguous objects identified in this work, we have $20-23$ SNe II candidates. 
The Lick Observatory Supernova Search \citep{2000AIPC..522..103L} calculated the rate of SNe II in the local universe as a function of color and Hubble Type by observing more than 10,000 galaxies over the course of 12 years \citep{2011MNRAS.412.1473L}. Using representative SDSS \textit{gri} magnitudes and $(g-r)$ and $(r-i)$ colors for the star forming galaxies from R13, we use filter transformations \citep{2005AJ....130..873J, 2008MNRAS.384.1178B} to determine a typical $(B-K)$ color in order to select the correct SN rate from \cite{2011MNRAS.412.1473L}. Choosing a galaxy stellar mass of $1.2\times10^{9}$~\msun and using the equation for $(B-K) < 2.3$, we find the expected rate of SNe II in a given galaxy over the course of 6 months is $0.002^{+0.003}_{-0.001}$. We then convert this to a probability using Poisson statistics. We choose to compute the rate over 6 months since SNe II typically begin to fade at $100-150$ days post explosion (see e.g., \citealt{2009ApJ...703.2205K}). 

Similarly, Graur et al. (2015) measured a mass-normalized SN II rate of $5.5^{+3.7}_{-2.4}~{\rm statistical}~^{+1.2}_{-0.7}~{\rm systematic~\times10^{-12}~yr^{-1}~M_\odot^{-1}}$ in galaxies with stellar masses of $0.08^{+0.06}_{-0.05}~{\times10^{10}~\rm M_\odot}$ (see their table 2). Thus, for a galaxy of $10^9~{\rm M_\odot}$, we would expect $0.0026^{+0.0025}_{-0.0016}$ SNe II in a period of six months, consistent with the rate derived from Li et al. (2011).

R13 analyzed spectroscopy of $\sim25,000$ dwarf galaxies. For a sample of this size, the rate derived from \cite{2011MNRAS.412.1473L} suggests we can expect $50^{+75}_{-25}$ SNe II. Using the rate from Graur et al. (2015), we expect $65^{+63}_{-40}$ SNe II in a sample of 25,000 galaxies. Both rates are roughly consistent with the number we detect in our sample. 
Using either rate, we compute that in a sample of 25,000 galaxies, we expect less than 1 to have an observed SN signature in the original SDSS observation \textit{and} a signature from a new SN in the follow-up observation.

\subsection{A Possible SN IIn in RGG J} 

In the following, we discuss the interesting case of RGG J. This particular source, classified in this work as having transient broad H$\alpha$ emission, has fooled many of us. It was identified as a broad line AGN in \cite{2007ApJ...670...92G}, \cite{2012MNRAS.423..600S}, and \cite{2012ApJ...755..167D}. Thus, it serves as an excellent cautionary tale that there may well be transient sources lurking in our broad-line AGN samples even at higher mass.

\begin{figure*}
\centering
\includegraphics[scale=0.7]{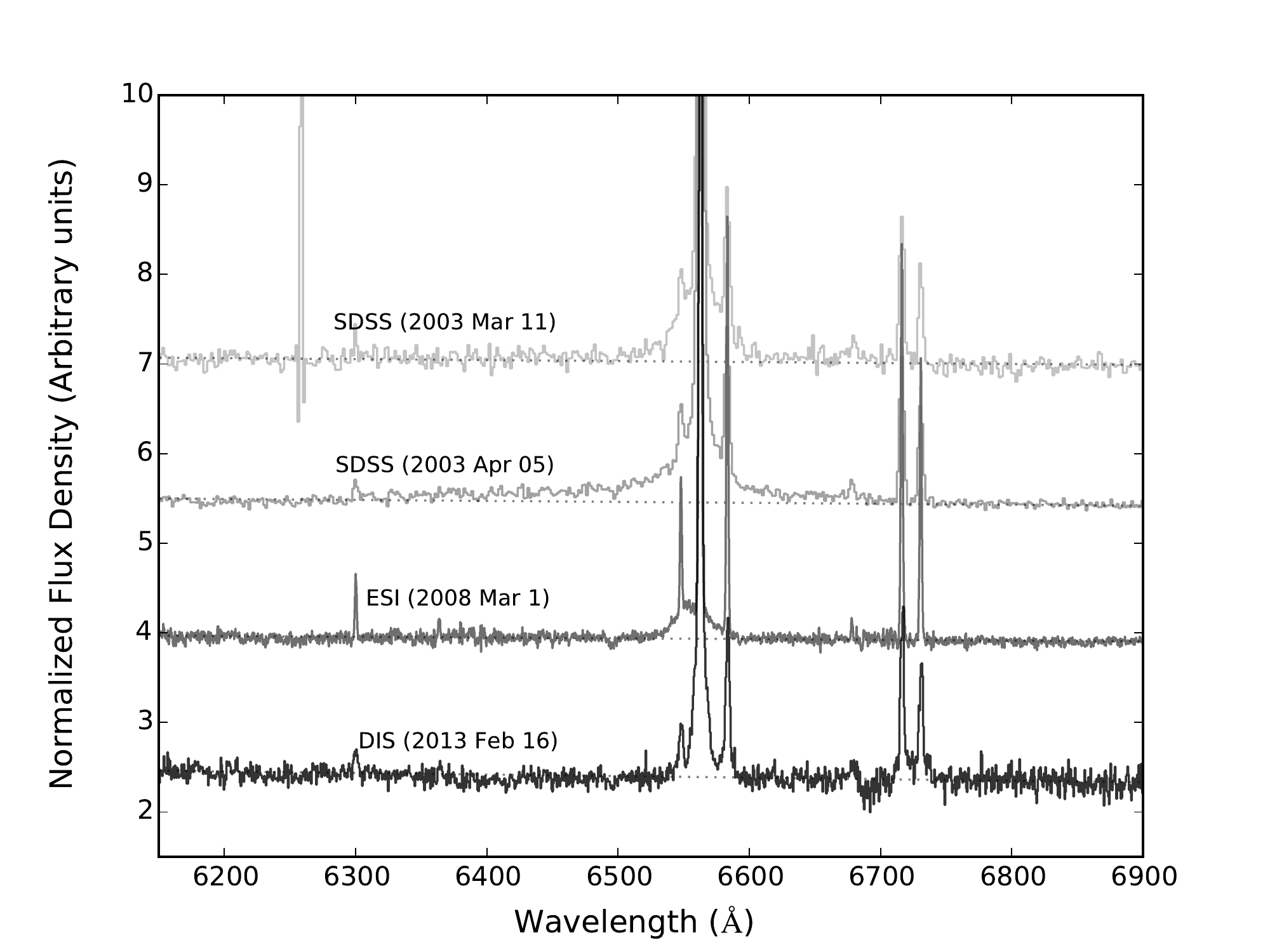}
\caption{The H$\alpha$ regime of the four observed spectra of RGG J. Broad H$\alpha$ is clearly visibile in the two SDSS spectra taken in 2003. An Keck II ESI spectrum was taken in 2008 (\citealt{2011ApJ...739...28X}; spectrum provided by Aaron Barth, private communication). This spectrum shows some broad H$\alpha$ emission as well, though it is distinctly blue-shifted with respect to the narrow H$\alpha$ line center. The DIS spectrum, taken in 2013 and analyzed in this work, does not show evidence for broad H$\alpha$ emission.  }
\label{sdss_j}
\end{figure*}

We also have more epochs for this target than others. It was observed
twice by the SDSS, once by \cite{2011ApJ...739...28X} (spectrum taken in
2008), and once by us in 2013. The SDSS spectra were taken on 11 March 2003
and 05 April 2003 (see Figure~\ref{sdss_j}). Thus, our observations span a full decade. The SDSS flagged
the second epoch as the ``primary'' spectrum, and so the fits in the
literature are most likely made to that later epoch. During
their survey, Graur et al. (2015) classified this source as either a
SN IIn or an AGN based on a fit to the April 2003 spectrum. By 2008, when \cite{2011ApJ...739...28X} observed this source with the Echellette Spectrograph and Imager on Keck II, there was still a (rather marginal) detection of broad Ha. The broad line is undetectable in our DIS spectrum, so we conclude the broad line may be produced by a SN IIn (Figure~\ref{sdss_j}).

Unlike SNe II-P and II-L, which exhibit broad H emission lines with P-Cygni profiles, Type IIn supernovae (SNe IIn) are characterized by narrow lines (FWHM $\lsim 200$ km/s; hence the `n' in `IIn') with intermediate-width bases (FWHM $\sim 1000-2000$ km/s) and bluer continua (see \citealt{1997ARA&A..35..309F} for a review). SNe IIn comprise $\approx 5 \pm 2$\% of all SNe and show a wide spread of peak luminosity \citep{2011MNRAS.412.1441L}.  The narrow lines are thought to be the result of strong interaction between the SN ejecta and a dense circumstellar medium. Though some SNe IIn have been associated with luminous blue variables (e.g., SN 2005gl; \citealt{2009Natur.458..865G}, SN 2010jl; \citealt{2011ApJ...732...63S}, and SN 2009ip; \citealt{2011ApJ...732...32F}), the exact nature of their progenitors is still debated (see \citealt{2014ARA&A..52..487S} for a review). 

\section{Discussion}

We analyzed follow-up spectroscopy of 27 dwarf galaxies with AGN signatures identified in a sample of $\sim25,000$ dwarf galaxies in the SDSS (see R13). Of our follow-up targets, 16 were found by R13 to have broad H$\alpha$ emission in their SDSS spectra, and 14 had narrow emission line ratios indicative of AGN activity (there is some overlap between the broad and narrow-line objects).

Of the 16 SDSS broad H$\alpha$ objects for which we have follow-up spectroscopy, we found that the broad H$\alpha$ emission faded for eleven, all of which fall in the star forming region of the BPT diagram. As stated above, we consider SNe II to be the most likely cause of the original broad emission for these objects, though luminous blue variables and changing look quasars also produce transient broad H$\alpha$ (see R13 for a more complete discussion of potential contaminants). 
We also find two of the SDSS broad H$\alpha$ objects, both of which had narrow-line AGN signatures, to have persistent broad H$\alpha$ emission, suggesting it is emission from virialized gas around a central BH. This also suggests that the broad emission for the remainder of the R13 narrow-line AGN/composite objects is due to an AGN. R13 calculates the type 1 fraction (or the fraction with detectable broad H$\alpha$ emission) for BPT AGN to be $\sim17\%$ (6/35 objects). The type 1 fraction for BPT composites is lower; including RGG 118 \citep{2015ApJ...809L..14B}, $\sim5\%$ of the composites (5/101) are type 1.
Finally, three SDSS broad H$\alpha$ objects (all BPT star forming) are ambiguous and will require further observations to determine the source of broad emission.

In the case of RGG J, (the expected Type II SN; see Section 5.1), the broad emission was persistent over at least 5 years, but the FWHM and velocity shift varied considerably. Thus, in addition to determining whether broad emission is \textit{persistent}, we also check for consistency between epochs. We note that differences in aperture could slightly affect broad emission measurements since the measured luminosity and FWHM of broad H$\alpha$ depend on the fit to the narrow emission lines. Additionally, the instruments used all have different spectral resolutions, which can affect the measured FWHM. Finally, broad line regions can vary intrinsically on these timescales \citep{2015A&A...580A.113T}, making it difficult to determine the degree to which aperture effects play a role. Nevertheless, for RGG 9, we find the two measured FWHM to be consistent with one another within the errors. For RGG 119, the measured FWHM of H$\alpha$ increases by $\sim13\%$ over the course of our three observations (though the first and last measurements are consistent within the errors).

\vspace{0.4cm}
In summary, we were able to securely detect broad H$\alpha$ arising from an AGN in the follow-up observations of two SDSS broad H$\alpha$ galaxies; both of these objects also have narrow-line ratios which support the presence of an AGN. Both objects (RGG 9 and RGG 119) have BH masses of a few times $\rm10^{5}~M_{\odot}$. \textit{Chandra X-ray Observatory} observations (Cycle 16, PI: Reines) will provide further valuable information about the accretion properties of these AGN. Three SDSS broad line objects (all BPT star forming) are classified as ambiguous and will require additional observations to determine whether an AGN is present.  For the remainder of the star forming SDSS broad H$\alpha$ dwarf galaxies, we find that transient stellar phenomena are likely to be the source of detected broad H$\alpha$ emission. Given the relatively low FWHMs, faint flux levels, and the prevalence of young stars, we confirm that, in dwarf galaxies (and perhaps more massive galaxies) broad H$\alpha$ alone should not be taken as evidence for an AGN -- another piece of evidence (such as narrow line ratios) is required (see also discussions in R13 and \citealt{2014AJ....148..136M}).

We also measure stellar velocity dispersions for 12 galaxies with narrow-line AGN signatures. One of these -- RGG 119 -- also has persistent broad H$\alpha$ emission. In this case, we use the broad emission to measure the BH mass and find that RGG 119 lies near the extrapolation of the M$_{\rm BH}$--$\sigma_{\star}$ relation to low BH masses. 
The measured stellar velocity dispersions for these targets range from $28-71\kms$ with a median value of $41\kms$. With galaxy stellar masses of just a few times $10^{9}$, these stellar velocity dispersions correspond to some of the lowest-mass galaxies with accreting BHs. If the remaining narrow-line AGN/composites similarly follow the M$_{\rm BH}$--$\sigma_{\star}$ relation, their BH masses would range from $\sim6\times10^{4}-2\times10^{6}~M_{\odot}$ (using the relation reported in \citealt{2009ApJ...698..198G}). 
%


\begin{center}\textbf{Acknowledgments}\end{center}
VFB is supported by the National Science Foundation Graduate Research Fellowship Program grant DGE 1256260. Support for AER was provided by NASA through Hubble Fellowship grant HST-HF2-51347.001-A awarded by the Space Telescope Science Institute, which is operated by the Association of Universities for Research in Astronomy, Inc., for NASA, under contract NAS 5-26555. RCH acknowledges support from a Alfred P. Sloan Research Fellowship and a Dartmouth Class of 1962 Faculty Fellowship.

This work makes use of: observations obtained at the MDM Observatory, operated by Dartmouth College, Columbia University, Ohio State University, Ohio University, and the University of Michigan; observations obtained with the Apache Point Observatory 3.5-meter telescope, which is owned and operated by the Astrophysical Research Consortium; data gathered with the 6.5 meter Magellan Telescopes located at Las Campanas Observatory, Chile. We thank Aaron Barth for providing us with the Keck ESI spectrum of RGG J. 


\bibliographystyle{apj}

\begin{thebibliography}{79}
\expandafter\ifx\csname natexlab\endcsname\relax\def\natexlab#1{#1}\fi

\bibitem[{{Alvarez} {et~al.}(2009){Alvarez}, {Wise}, \&
  {Abel}}]{2009ApJ...701L.133A}
{Alvarez}, M.~A., {Wise}, J.~H., \& {Abel}, T. 2009, \apjl, 701, L133

\bibitem[{{Baldassare} {et~al.}(2015){Baldassare}, {Reines}, {Gallo}, \&
  {Greene}}]{2015ApJ...809L..14B}
{Baldassare}, V.~F., {Reines}, A.~E., {Gallo}, E., \& {Greene}, J.~E. 2015,
  \apjl, 809, L14

\bibitem[{{Baldwin} {et~al.}(1981){Baldwin}, {Phillips}, \&
  {Terlevich}}]{1981PASP...93....5B}
{Baldwin}, J.~A., {Phillips}, M.~M., \& {Terlevich}, R. 1981, \pasp, 93, 5

\bibitem[{{Barth} {et~al.}(2004){Barth}, {Ho}, {Rutledge}, \&
  {Sargent}}]{2004ApJ...607...90B}
{Barth}, A.~J., {Ho}, L.~C., {Rutledge}, R.~E., \& {Sargent}, W.~L.~W. 2004,
  \apj, 607, 90

\bibitem[{{Begelman} {et~al.}(2006){Begelman}, {Volonteri}, \&
  {Rees}}]{2006MNRAS.370..289B}
{Begelman}, M.~C., {Volonteri}, M., \& {Rees}, M.~J. 2006, \mnras, 370, 289

\bibitem[{{Bellovary} {et~al.}(2011){Bellovary}, {Volonteri}, {Governato},
  {Shen}, {Quinn}, \& {Wadsley}}]{2011ApJ...742...13B}
{Bellovary}, J., {Volonteri}, M., {Governato}, F., {et~al.} 2011, \apj, 742, 13

\bibitem[{{Bentz} {et~al.}(2009){Bentz}, {Walsh}, {Barth}, {Baliber},
  {Bennert}, {Canalizo}, {Filippenko}, {Ganeshalingam}, {Gates}, {Greene},
  {Hidas}, {Hiner}, {Lee}, {Li}, {Malkan}, {Minezaki}, {Sakata}, {Serduke},
  {Silverman}, {Steele}, {Stern}, {Street}, {Thornton}, {Treu}, {Wang}, {Woo},
  \& {Yoshii}}]{2009ApJ...705..199B}
{Bentz}, M.~C., {Walsh}, J.~L., {Barth}, A.~J., {et~al.} 2009, \apj, 705, 199

\bibitem[{{Bentz} {et~al.}(2013){Bentz}, {Denney}, {Grier}, {Barth},
  {Peterson}, {Vestergaard}, {Bennert}, {Canalizo}, {De Rosa}, {Filippenko},
  {Gates}, {Greene}, {Li}, {Malkan}, {Pogge}, {Stern}, {Treu}, \&
  {Woo}}]{2013ApJ...767..149B}
{Bentz}, M.~C., {Denney}, K.~D., {Grier}, C.~J., {et~al.} 2013, \apj, 767, 149

\bibitem[{{Bilir} {et~al.}(2008){Bilir}, {Ak}, {Karaali}, {Cabrera-Lavers},
  {Chonis}, \& {Gaskell}}]{2008MNRAS.384.1178B}
{Bilir}, S., {Ak}, S., {Karaali}, S., {et~al.} 2008, \mnras, 384, 1178

\bibitem[{{Blondin} \& {Tonry}(2007)}]{2007ApJ...666.1024B}
{Blondin}, S., \& {Tonry}, J.~L. 2007, \apj, 666, 1024

\bibitem[{{Cappellari} \& {Emsellem}(2004)}]{2004PASP..116..138C}
{Cappellari}, M., \& {Emsellem}, E. 2004, \pasp, 116, 138

\bibitem[{{Crowther}(2007)}]{2007ARA&A..45..177C}
{Crowther}, P.~A. 2007, \araa, 45, 177

\bibitem[{{Crowther} \& {Walborn}(2011)}]{2011MNRAS.416.1311C}
{Crowther}, P.~A., \& {Walborn}, N.~R. 2011, \mnras, 416, 1311

\bibitem[{{De Robertis} \& {Osterbrock}(1984)}]{1984ApJ...286..171D}
{De Robertis}, M.~M., \& {Osterbrock}, D.~E. 1984, \apj, 286, 171

\bibitem[{{Dong} {et~al.}(2012){Dong}, {Ho}, {Yuan}, {Wang}, {Fan}, {Zhou}, \&
  {Jiang}}]{2012ApJ...755..167D}
{Dong}, X.-B., {Ho}, L.~C., {Yuan}, W., {et~al.} 2012, \apj, 755, 167

\bibitem[{{Ferrarese} \& {Merritt}(2000)}]{2000ApJ...539L...9F}
{Ferrarese}, L., \& {Merritt}, D. 2000, \apjl, 539, L9

\bibitem[{{Filippenko}(1997)}]{1997ARA&A..35..309F}
{Filippenko}, A.~V. 1997, \araa, 35, 309

\bibitem[{{Filippenko} \& {Ho}(2003)}]{2003ApJ...588L..13F}
{Filippenko}, A.~V., \& {Ho}, L.~C. 2003, \apjl, 588, L13

\bibitem[{{Filippenko} \& {Sargent}(1989)}]{1989ApJ...342L..11F}
{Filippenko}, A.~V., \& {Sargent}, W.~L.~W. 1989, \apjl, 342, L11

\bibitem[{{Foley} {et~al.}(2011){Foley}, {Berger}, {Fox}, {Levesque},
  {Challis}, {Ivans}, {Rhoads}, \& {Soderberg}}]{2011ApJ...732...32F}
{Foley}, R.~J., {Berger}, E., {Fox}, O., {et~al.} 2011, \apj, 732, 32

\bibitem[{{Gal-Yam} \& {Leonard}(2009)}]{2009Natur.458..865G}
{Gal-Yam}, A., \& {Leonard}, D.~C. 2009, \nat, 458, 865

\bibitem[{{Gebhardt} {et~al.}(2000){Gebhardt}, {Bender}, {Bower}, {Dressler},
  {Faber}, {Filippenko}, {Green}, {Grillmair}, {Ho}, {Kormendy}, {Lauer},
  {Magorrian}, {Pinkney}, {Richstone}, \& {Tremaine}}]{2000ApJ...539L..13G}
{Gebhardt}, K., {Bender}, R., {Bower}, G., {et~al.} 2000, \apjl, 539, L13

\bibitem[{{Graur} {et~al.}(2015){Graur}, {Bianco}, \&
  {Modjaz}}]{2015MNRAS.450..905G}
{Graur}, O., {Bianco}, F.~B., \& {Modjaz}, M. 2015, \mnras, 450, 905

\bibitem[{{Graur} \& {Maoz}(2013)}]{2013MNRAS.430.1746G}
{Graur}, O., \& {Maoz}, D. 2013, \mnras, 430, 1746

\bibitem[{{Greene} \& {Ho}(2004)}]{2004ApJ...610..722G}
{Greene}, J.~E., \& {Ho}, L.~C. 2004, \apj, 610, 722

\bibitem[{{Greene} \& {Ho}(2005{\natexlab{a}})}]{2005ApJ...627..721G}
---. 2005{\natexlab{a}}, \apj, 627, 721

\bibitem[{{Greene} \& {Ho}(2005{\natexlab{b}})}]{2005ApJ...630..122G}
---. 2005{\natexlab{b}}, \apj, 630, 122

\bibitem[{{Greene} \& {Ho}(2007)}]{2007ApJ...670...92G}
---. 2007, \apj, 670, 92

\bibitem[{{Groves} {et~al.}(2006){Groves}, {Heckman}, \&
  {Kauffmann}}]{2006MNRAS.371.1559G}
{Groves}, B.~A., {Heckman}, T.~M., \& {Kauffmann}, G. 2006, \mnras, 371, 1559

\bibitem[{{G{\"u}ltekin} {et~al.}(2009){G{\"u}ltekin}, {Richstone}, {Gebhardt},
  {Lauer}, {Tremaine}, {Aller}, {Bender}, {Dressler}, {Faber}, {Filippenko},
  {Green}, {Ho}, {Kormendy}, {Magorrian}, {Pinkney}, \&
  {Siopis}}]{2009ApJ...698..198G}
{G{\"u}ltekin}, K., {Richstone}, D.~O., {Gebhardt}, K., {et~al.} 2009, \apj,
  698, 198

\bibitem[{{Hao} {et~al.}(2005){Hao}, {Strauss}, {Fan}, {Tremonti}, {Schlegel},
  {Heckman}, {Kauffmann}, {Blanton}, {Gunn}, {Hall}, {Ivezi{\'c}}, {Knapp},
  {Krolik}, {Lupton}, {Richards}, {Schneider}, {Strateva}, {Zakamska},
  {Brinkmann}, \& {Szokoly}}]{2005AJ....129.1795H}
{Hao}, L., {Strauss}, M.~A., {Fan}, X., {et~al.} 2005, \aj, 129, 1795

\bibitem[{{Heckman} {et~al.}(1981){Heckman}, {Miley}, {van Breugel}, \&
  {Butcher}}]{1981ApJ...247..403H}
{Heckman}, T.~M., {Miley}, G.~K., {van Breugel}, W.~J.~M., \& {Butcher}, H.~R.
  1981, \apj, 247, 403

\bibitem[{{Izotov} {et~al.}(2007){Izotov}, {Thuan}, \&
  {Guseva}}]{2007ApJ...671.1297I}
{Izotov}, Y.~I., {Thuan}, T.~X., \& {Guseva}, N.~G. 2007, \apj, 671, 1297

\bibitem[{{Jester} {et~al.}(2005){Jester}, {Schneider}, {Richards}, {Green},
  {Schmidt}, {Hall}, {Strauss}, {Vanden Berk}, {Stoughton}, {Gunn},
  {Brinkmann}, {Kent}, {Smith}, {Tucker}, \& {Yanny}}]{2005AJ....130..873J}
{Jester}, S., {Schneider}, D.~P., {Richards}, G.~T., {et~al.} 2005, \aj, 130,
  873

\bibitem[{{Kasen} \& {Woosley}(2009)}]{2009ApJ...703.2205K}
{Kasen}, D., \& {Woosley}, S.~E. 2009, \apj, 703, 2205

\bibitem[{{Kauffmann} {et~al.}(2003){Kauffmann}, {Heckman}, {Tremonti},
  {Brinchmann}, {Charlot}, {White}, {Ridgway}, {Brinkmann}, {Fukugita}, {Hall},
  {Ivezi{\'c}}, {Richards}, \& {Schneider}}]{2003MNRAS.346.1055K}
{Kauffmann}, G., {Heckman}, T.~M., {Tremonti}, C., {et~al.} 2003, \mnras, 346,
  1055

\bibitem[{{Kewley} {et~al.}(2001){Kewley}, {Dopita}, {Sutherland}, {Heisler},
  \& {Trevena}}]{2001ApJ...556..121K}
{Kewley}, L.~J., {Dopita}, M.~A., {Sutherland}, R.~S., {Heisler}, C.~A., \&
  {Trevena}, J. 2001, \apj, 556, 121

\bibitem[{{Kewley} {et~al.}(2006){Kewley}, {Groves}, {Kauffmann}, \&
  {Heckman}}]{2006MNRAS.372..961K}
{Kewley}, L.~J., {Groves}, B., {Kauffmann}, G., \& {Heckman}, T. 2006, \mnras,
  372, 961

\bibitem[{{Kormendy} \& {Ho}(2013)}]{Kormendy:2013ve}
{Kormendy}, J., \& {Ho}, L.~C. 2013, \araa, 51, 511

\bibitem[{{Koss} {et~al.}(2014){Koss}, {Blecha}, {Mushotzky}, {Hung},
  {Veilleux}, {Trakhtenbrot}, {Schawinski}, {Stern}, {Smith}, {Li}, {Man},
  {Filippenko}, {Mauerhan}, {Stanek}, \& {Sanders}}]{2014MNRAS.445..515K}
{Koss}, M., {Blecha}, L., {Mushotzky}, R., {et~al.} 2014, \mnras, 445, 515

\bibitem[{{Lemons} {et~al.}(2015){Lemons}, {Reines}, {Plotkin}, {Gallo}, \&
  {Greene}}]{2015ApJ...805...12L}
{Lemons}, S.~M., {Reines}, A.~E., {Plotkin}, R.~M., {Gallo}, E., \& {Greene},
  J.~E. 2015, \apj, 805, 12

\bibitem[{{Li} {et~al.}(2011{\natexlab{a}}){Li}, {Chornock}, {Leaman},
  {Filippenko}, {Poznanski}, {Wang}, {Ganeshalingam}, \&
  {Mannucci}}]{2011MNRAS.412.1473L}
{Li}, W., {Chornock}, R., {Leaman}, J., {et~al.} 2011{\natexlab{a}}, \mnras,
  412, 1473

\bibitem[{{Li} {et~al.}(2011{\natexlab{b}}){Li}, {Leaman}, {Chornock},
  {Filippenko}, {Poznanski}, {Ganeshalingam}, {Wang}, {Modjaz}, {Jha}, {Foley},
  \& {Smith}}]{2011MNRAS.412.1441L}
{Li}, W., {Leaman}, J., {Chornock}, R., {et~al.} 2011{\natexlab{b}}, \mnras,
  412, 1441

\bibitem[{{Li} {et~al.}(2000){Li}, {Filippenko}, {Treffers}, {Friedman},
  {Halderson}, {Johnson}, {King}, {Modjaz}, {Papenkova}, {Sato}, \&
  {Shefler}}]{2000AIPC..522..103L}
{Li}, W.~D., {Filippenko}, A.~V., {Treffers}, R.~R., {et~al.} 2000, in American
  Institute of Physics Conference Series, Vol. 522, American Institute of
  Physics Conference Series, ed. S.~S. {Holt} \& W.~W. {Zhang}, 103--106

\bibitem[{{Lodato} \& {Natarajan}(2006)}]{2006MNRAS.371.1813L}
{Lodato}, G., \& {Natarajan}, P. 2006, \mnras, 371, 1813

\bibitem[{{Madau} {et~al.}(2014){Madau}, {Haardt}, \&
  {Dotti}}]{2014ApJ...784L..38M}
{Madau}, P., {Haardt}, F., \& {Dotti}, M. 2014, \apjl, 784, L38

\bibitem[{{Madau} \& {Rees}(2001)}]{2001ApJ...551L..27M}
{Madau}, P., \& {Rees}, M.~J. 2001, \apjl, 551, L27

\bibitem[{{Magorrian} {et~al.}(1998){Magorrian}, {Tremaine}, {Richstone},
  {Bender}, {Bower}, {Dressler}, {Faber}, {Gebhardt}, {Green}, {Grillmair},
  {Kormendy}, \& {Lauer}}]{1998AJ....115.2285M}
{Magorrian}, J., {Tremaine}, S., {Richstone}, D., {et~al.} 1998, \aj, 115, 2285

\bibitem[{{Marshall} {et~al.}(2008){Marshall}, {Burles}, {Thompson},
  {Shectman}, {Bigelow}, {Burley}, {Birk}, {Estrada}, {Jones}, {Smith},
  {Kowal}, {Castillo}, {Storts}, \& {Ortiz}}]{2008SPIE.7014E..54M}
{Marshall}, J.~L., {Burles}, S., {Thompson}, I.~B., {et~al.} 2008, in Society
  of Photo-Optical Instrumentation Engineers (SPIE) Conference Series, Vol.
  7014, Society of Photo-Optical Instrumentation Engineers (SPIE) Conference
  Series, 54

\bibitem[{{Martini} {et~al.}(2011){Martini}, {Stoll}, {Derwent}, {Zhelem},
  {Atwood}, {Gonzalez}, {Mason}, {O'Brien}, {Pappalardo}, {Pogge}, {Ward}, \&
  {Wong}}]{2011PASP..123..187M}
{Martini}, P., {Stoll}, R., {Derwent}, M.~A., {et~al.} 2011, \pasp, 123, 187

\bibitem[{{McConnell} \& {Ma}(2013)}]{2013ApJ...764..184M}
{McConnell}, N.~J., \& {Ma}, C.-P. 2013, \apj, 764, 184

\bibitem[{{Moran} {et~al.}(2014){Moran}, {Shahinyan}, {Sugarman}, {V{\'e}lez},
  \& {Eracleous}}]{2014AJ....148..136M}
{Moran}, E.~C., {Shahinyan}, K., {Sugarman}, H.~R., {V{\'e}lez}, D.~O., \&
  {Eracleous}, M. 2014, \aj, 148, 136

\bibitem[{{Mullaney} {et~al.}(2013){Mullaney}, {Alexander}, {Fine}, {Goulding},
  {Harrison}, \& {Hickox}}]{2013MNRAS.433..622M}
{Mullaney}, J.~R., {Alexander}, D.~M., {Fine}, S., {et~al.} 2013, \mnras, 433,
  622

\bibitem[{{Pardo} {et~al.}(2016){Pardo}, {Goulding}, {Greene}, {Somerville},
  {Gallo}, {Hickox}, {Miller}, {Reines}, \& {Silverman}}]{2016arXiv160301622P}
{Pardo}, K., {Goulding}, A.~D., {Greene}, J.~E., {et~al.} 2016, ArXiv e-prints

\bibitem[{{Pritchard} {et~al.}(2012){Pritchard}, {Roming}, {Brown}, {Kuin},
  {Bayless}, {Holland}, {Immler}, {Milne}, \& {Oates}}]{2012ApJ...750..128P}
{Pritchard}, T.~A., {Roming}, P.~W.~A., {Brown}, P.~J., {et~al.} 2012, \apj,
  750, 128

\bibitem[{{Prugniel} \& {Soubiran}(2001)}]{2001A&A...369.1048P}
{Prugniel}, P., \& {Soubiran}, C. 2001, \aap, 369, 1048

\bibitem[{{Reines} \& {Deller}(2012)}]{2012ApJ...750L..24R}
{Reines}, A.~E., \& {Deller}, A.~T. 2012, \apjl, 750, L24

\bibitem[{{Reines} {et~al.}(2013){Reines}, {Greene}, \& {Geha}}]{Reines:2013fj}
{Reines}, A.~E., {Greene}, J.~E., \& {Geha}, M. 2013, \apj, 775, 116

\bibitem[{{Reines} {et~al.}(2014){Reines}, {Plotkin}, {Russell}, {Mezcua},
  {Condon}, {Sivakoff}, \& {Johnson}}]{2014ApJ...787L..30R}
{Reines}, A.~E., {Plotkin}, R.~M., {Russell}, T.~D., {et~al.} 2014, \apjl, 787,
  L30

\bibitem[{{Reines} {et~al.}(2011){Reines}, {Sivakoff}, {Johnson}, \&
  {Brogan}}]{Reines:2011fr}
{Reines}, A.~E., {Sivakoff}, G.~R., {Johnson}, K.~E., \& {Brogan}, C.~L. 2011,
  \nat, 470, 66

\bibitem[{{Reines} \& {Volonteri}(2015)}]{2015ApJ...813...82R}
{Reines}, A.~E., \& {Volonteri}, M. 2015, \apj, 813, 82

\bibitem[{{Ruan} {et~al.}(2015){Ruan}, {Anderson}, {Cales}, {Eracleous},
  {Green}, {Morganson}, {Runnoe}, {Shen}, {Wilkinson}, {Blanton}, {Dwelly},
  {Georgakakis}, {Greene}, {LaMassa}, {Merloni}, \&
  {Schneider}}]{2015arXiv150903634R}
{Ruan}, J.~J., {Anderson}, S.~F., {Cales}, S.~L., {et~al.} 2015, ArXiv e-prints

\bibitem[{{Runnoe} {et~al.}(2016){Runnoe}, {Cales}, {Ruan}, {Eracleous},
  {Anderson}, {Shen}, {Green}, {Morganson}, {LaMassa}, {Greene}, {Dwelly},
  {Schneider}, {Merloni}, {Georgakakis}, \&
  {Roman-Lopes}}]{2016MNRAS.455.1691R}
{Runnoe}, J.~C., {Cales}, S., {Ruan}, J.~J., {et~al.} 2016, \mnras, 455, 1691

\bibitem[{{Sartori} {et~al.}(2015){Sartori}, {Schawinski}, {Treister},
  {Trakhtenbrot}, {Koss}, {Shirazi}, \& {Oh}}]{2015MNRAS.454.3722S}
{Sartori}, L.~F., {Schawinski}, K., {Treister}, E., {et~al.} 2015, \mnras, 454,
  3722

\bibitem[{{Smith}(2014)}]{2014ARA&A..52..487S}
{Smith}, N. 2014, \araa, 52, 487

\bibitem[{{Smith} {et~al.}(2011{\natexlab{a}}){Smith}, {Li}, {Silverman},
  {Ganeshalingam}, \& {Filippenko}}]{2011MNRAS.415..773S}
{Smith}, N., {Li}, W., {Silverman}, J.~M., {Ganeshalingam}, M., \&
  {Filippenko}, A.~V. 2011{\natexlab{a}}, \mnras, 415, 773

\bibitem[{{Smith} {et~al.}(2011{\natexlab{b}}){Smith}, {Li}, {Miller},
  {Silverman}, {Filippenko}, {Cuillandre}, {Cooper}, {Matheson}, \& {Van
  Dyk}}]{2011ApJ...732...63S}
{Smith}, N., {Li}, W., {Miller}, A.~A., {et~al.} 2011{\natexlab{b}}, \apj, 732,
  63

\bibitem[{{Stern} \& {Laor}(2012)}]{2012MNRAS.423..600S}
{Stern}, J., \& {Laor}, A. 2012, \mnras, 423, 600

\bibitem[{{Tremou} {et~al.}(2015){Tremou}, {Garcia-Marin}, {Zuther}, {Eckart},
  {Valencia-Schneider}, {Vitale}, \& {Shan}}]{2015A&A...580A.113T}
{Tremou}, E., {Garcia-Marin}, M., {Zuther}, J., {et~al.} 2015, \aap, 580, A113

\bibitem[{{van den Bosch} \& {de Zeeuw}(2010)}]{2010MNRAS.401.1770V}
{van den Bosch}, R.~C.~E., \& {de Zeeuw}, P.~T. 2010, \mnras, 401, 1770

\bibitem[{Volonteri(2010)}]{:fl}
Volonteri, M. 2010, A\&A Rev., 18

\bibitem[{{Volonteri} \& {Natarajan}(2009)}]{2009MNRAS.400.1911V}
{Volonteri}, M., \& {Natarajan}, P. 2009, \mnras, 400, 1911

\bibitem[{{Wenger} {et~al.}(2000){Wenger}, {Ochsenbein}, {Egret}, {Dubois},
  {Bonnarel}, {Borde}, {Genova}, {Jasniewicz}, {Lalo{\"e}}, {Lesteven}, \&
  {Monier}}]{2000A&AS..143....9W}
{Wenger}, M., {Ochsenbein}, F., {Egret}, D., {et~al.} 2000, \aaps, 143, 9

\bibitem[{{Whitmore}(1980)}]{1980ApJ...242...53W}
{Whitmore}, B.~C. 1980, \apj, 242, 53

\bibitem[{{Whittle}(1985)}]{1985MNRAS.213....1W}
{Whittle}, M. 1985, \mnras, 213, 1

\bibitem[{{Woo} {et~al.}(2015){Woo}, {Yoon}, {Park}, {Park}, \&
  {Kim}}]{2015ApJ...801...38W}
{Woo}, J.-H., {Yoon}, Y., {Park}, S., {Park}, D., \& {Kim}, S.~C. 2015, \apj,
  801, 38

\bibitem[{{Woo} {et~al.}(2010){Woo}, {Treu}, {Barth}, {Wright}, {Walsh},
  {Bentz}, {Martini}, {Bennert}, {Canalizo}, {Filippenko}, {Gates}, {Greene},
  {Li}, {Malkan}, {Stern}, \& {Minezaki}}]{2010ApJ...716..269W}
{Woo}, J.-H., {Treu}, T., {Barth}, A.~J., {et~al.} 2010, \apj, 716, 269

\bibitem[{{Xiao} {et~al.}(2011){Xiao}, {Barth}, {Greene}, {Ho}, {Bentz},
  {Ludwig}, \& {Jiang}}]{2011ApJ...739...28X}
{Xiao}, T., {Barth}, A.~J., {Greene}, J.~E., {et~al.} 2011, \apj, 739, 28

\bibitem[{{Zakamska} \& {Greene}(2014)}]{2014MNRAS.442..784Z}
{Zakamska}, N.~L., \& {Greene}, J.~E. 2014, \mnras, 442, 784

\end{thebibliography}

\clearpage


\begin{appendix}
\smallskip

\section{Stellar velocity dispersion library}

\begin{table*}[h]
{\centering
\caption{Spectral library}
\begin{tabular}{c c}
\hline
\hline
Star & Sp. Type \\
\hline
HD061366 & K0III\\ 
HD062161 & F3V\\ 
HD062437 & A5IV\\ 
HD062968 & G8\\ 
HD063108 & F0\\ 
HD064649 & G5\\ 
HD065123 & F7V\\ 
HD065604 & K5\\ 
HD071155 & A0Va\\ 
HD071557 & A0\\ 
HD072722 & K5\\ 
HD073667 & K2V\\ 
HD079210 & M0V\\ 
HD090361 & A1V\\ 
HD042217 & K5\\ 
HD042250 & G9V\\ 
HD042548 & F0\\ 
HD042618 & G4V\\ 
HD043318 & F5V\\ 
HD043318 & F5V\\ 
HD043338 & F0\\ 
HD044285 & K3III\\ 
HD044418 & G8III\\ 
HD044638 & K0/1III\\ 
HD044770 & F5V\\ 
HD044771 & G8Ib/II\\ 
HD046784 & M0III\\ 
HD047072 & Am\\ 
HD047309 & G0\\ 
HD048144 & K5\\ 
HD048279 & O8V+F2V\\ 
HD048433 & K1III\\ 
HD060503 & G8III\\ 
HD027340 & A3IV/V\\ 
HD028343 & M0.5V\\ 
HD028946 & G9V\\ 
HD029139 & K5III\\ 
HD029310 & G0\\ 
HD036395 & M1.5V\\ 
HD037828 & G8III\\ 
HD037958 & B9III/IV\\ 
HD038145 & F0V\\ 
HD038237 & A5IV\\ 
HD039833 & G3V\\ 
HD040573 & B9.5V\\ 
HD040259 & F0V\\ 
HD040460 & K1III\\ 
HD040573 & B9.5V\\ 
HD040616 & G3V\\ 
HD041079 & K2\\
\hline
\end{tabular}
\label{sptypes}\\}
\textbf{Table~\ref{sptypes}}. Stars used in our spectral library for stellar velocity dispersion measurements (see main text Section 3.2). Spectra were taken from the ELODIE spectral database \citep{2001A&A...369.1048P} and spectral types are taken from the SIMBAD database \citep{2000A&AS..143....9W}.  \end{table*}

\clearpage

\section{Emission line fits}

\begin{figure*}[h]
\centering
\includegraphics[scale=0.44]{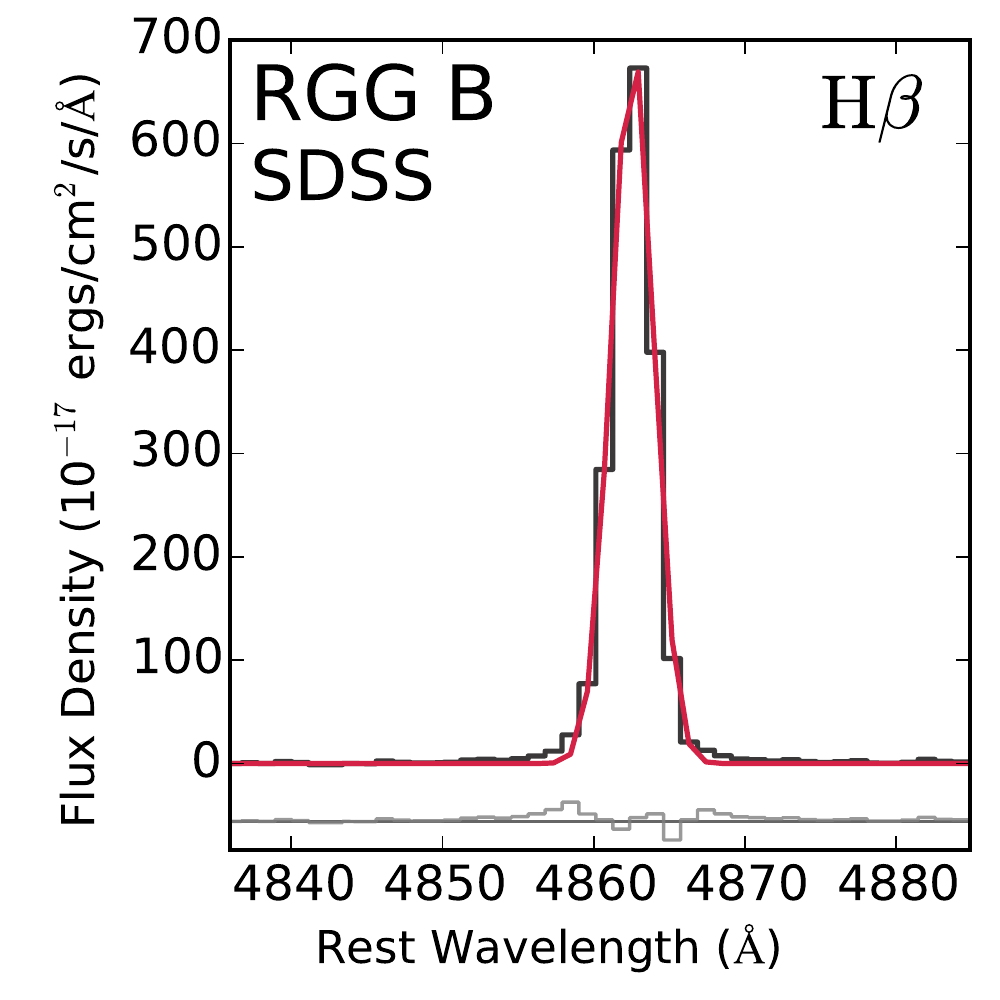}
\includegraphics[scale=0.44]{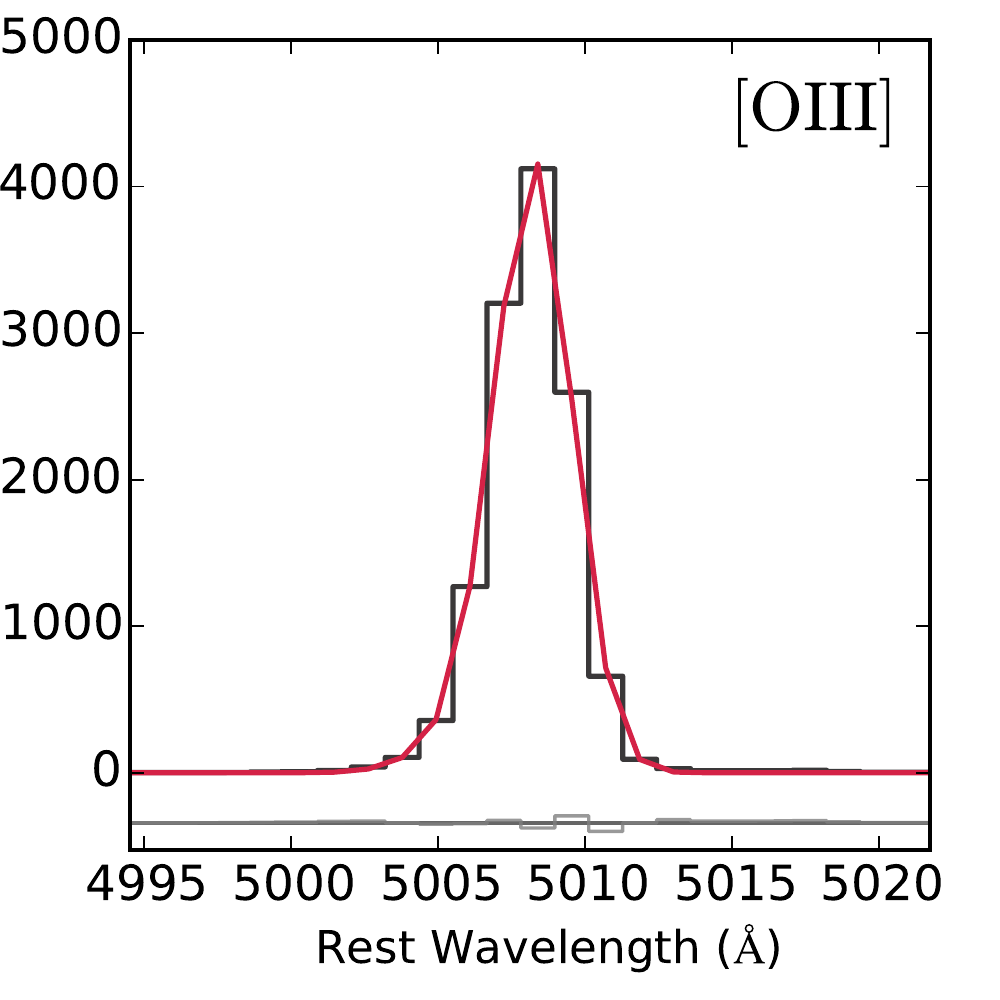}
\includegraphics[scale=0.44]{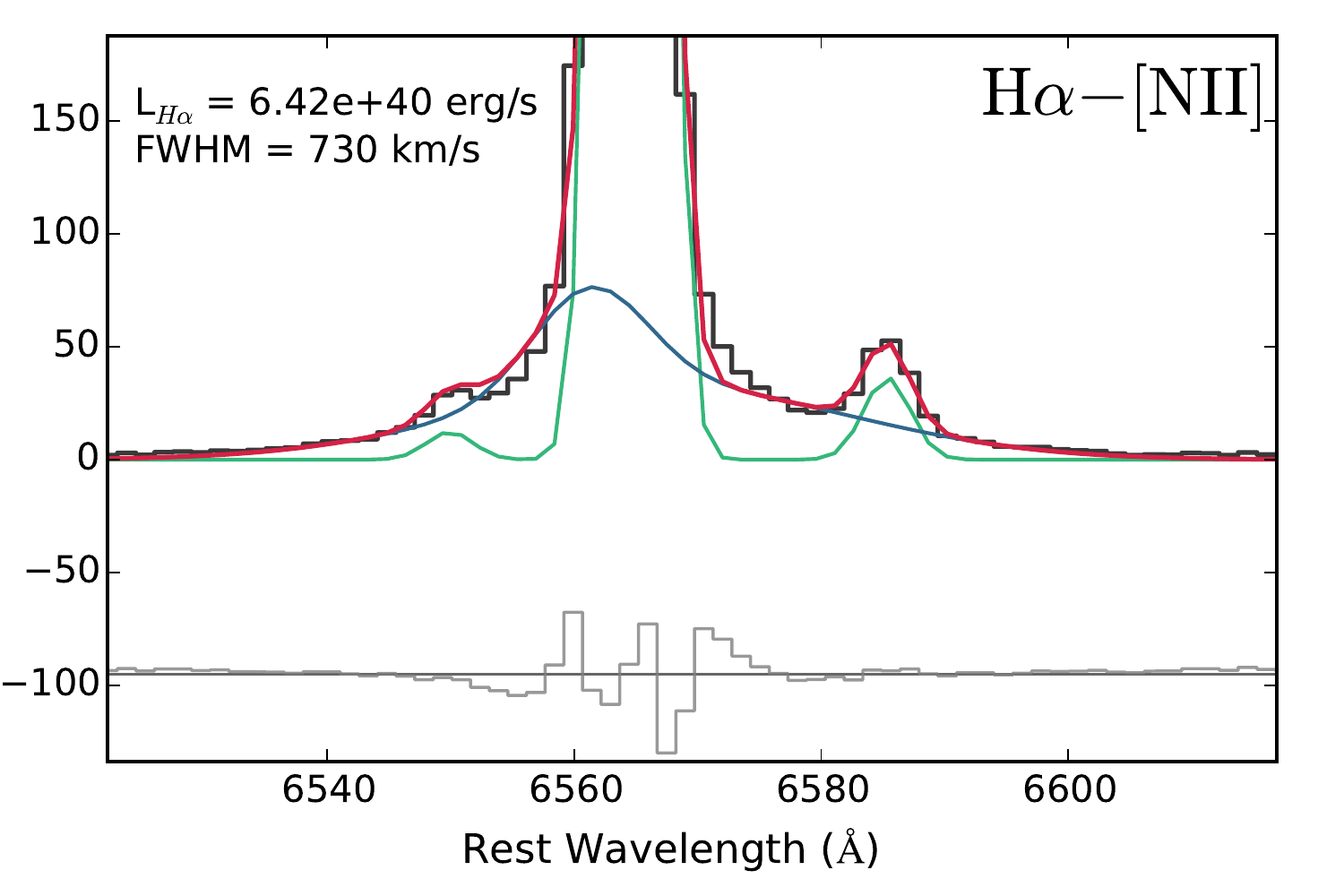}\\

\includegraphics[scale=0.44]{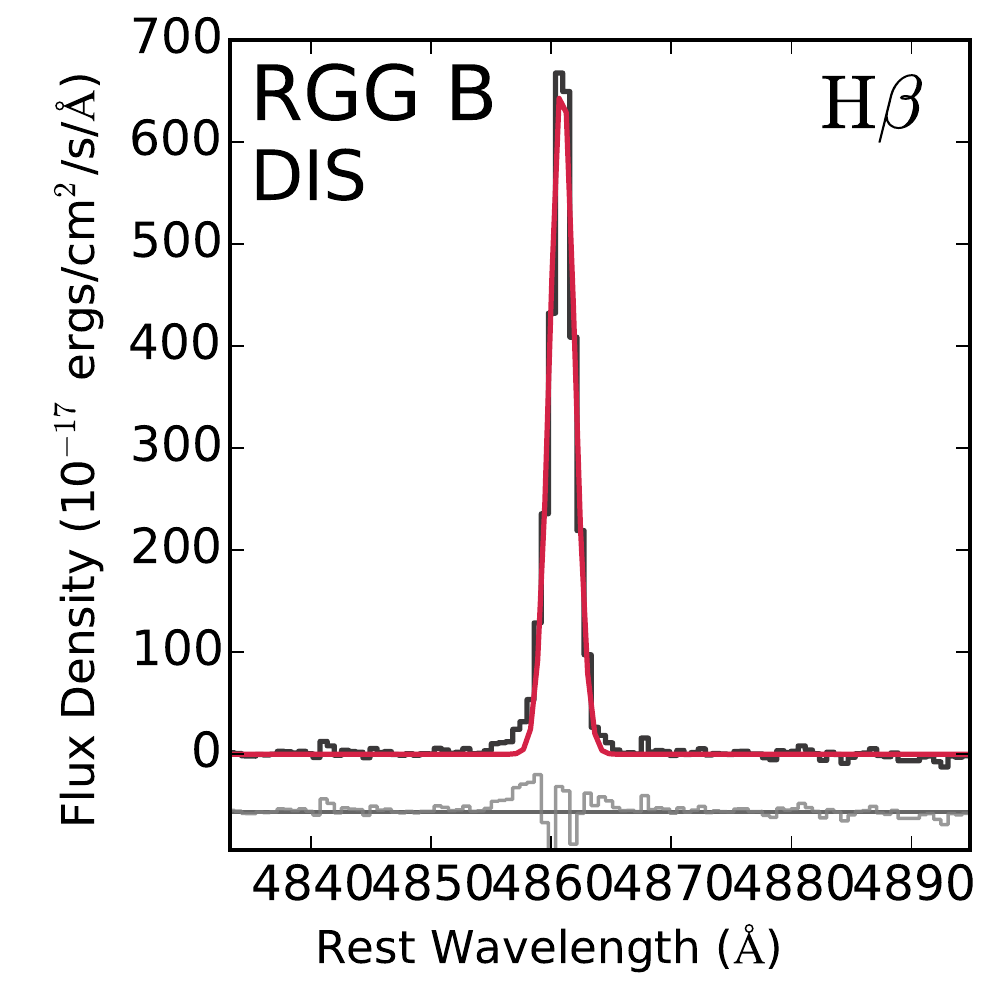}
\includegraphics[scale=0.44]{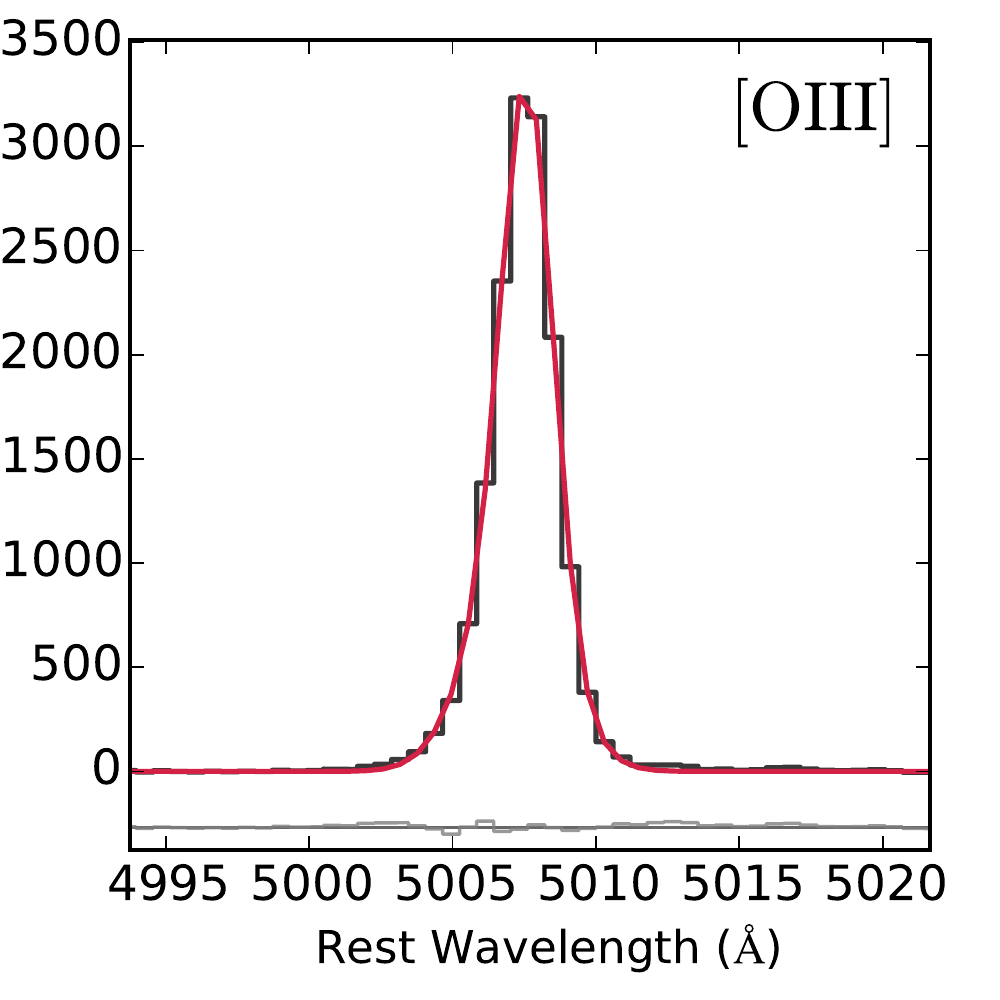}
\includegraphics[scale=0.44]{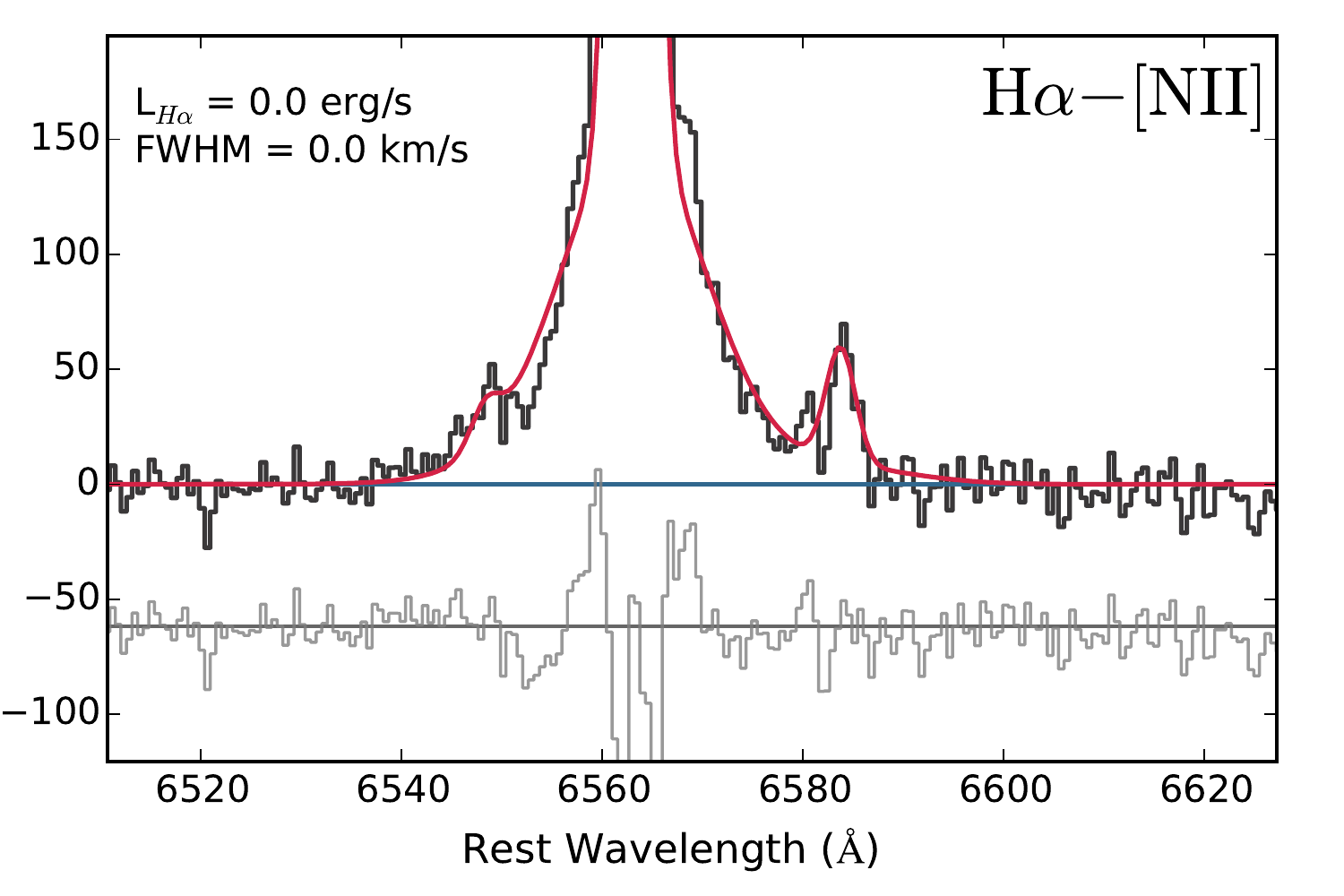}\\

\includegraphics[scale=0.44]{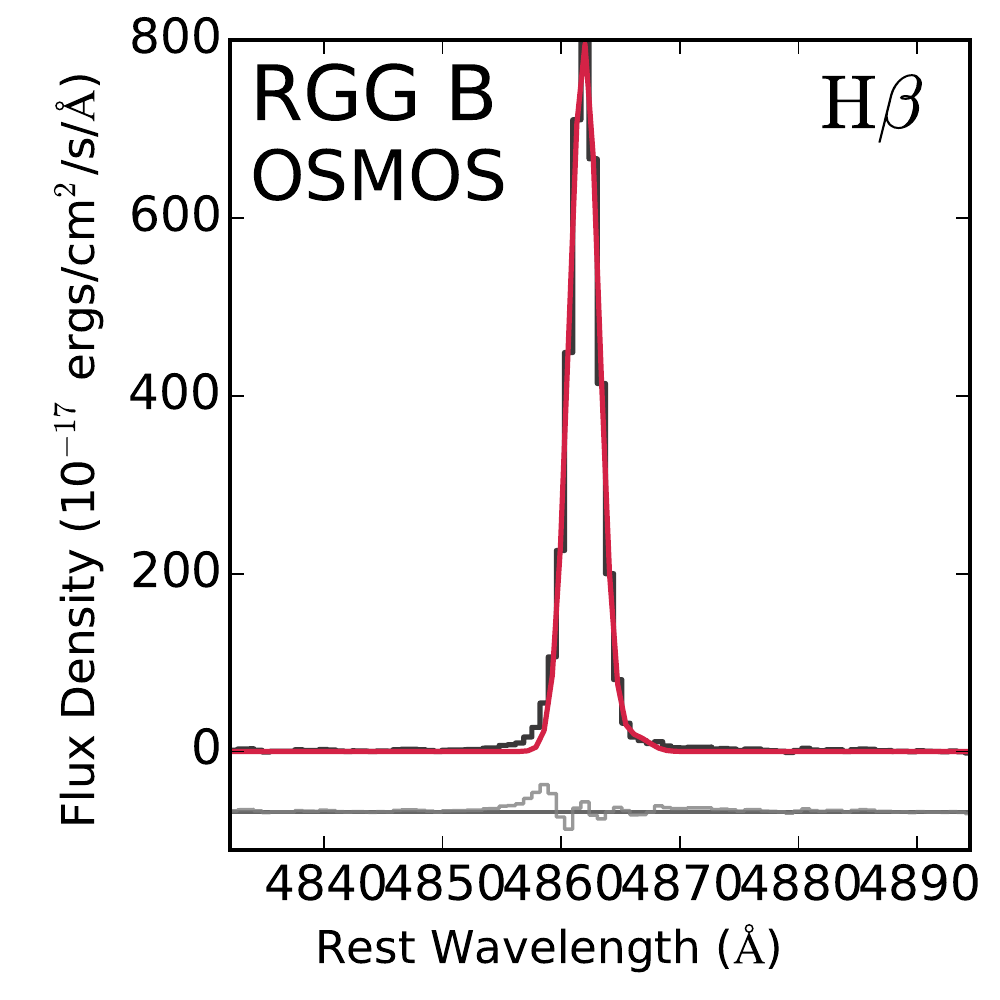}
\includegraphics[scale=0.44]{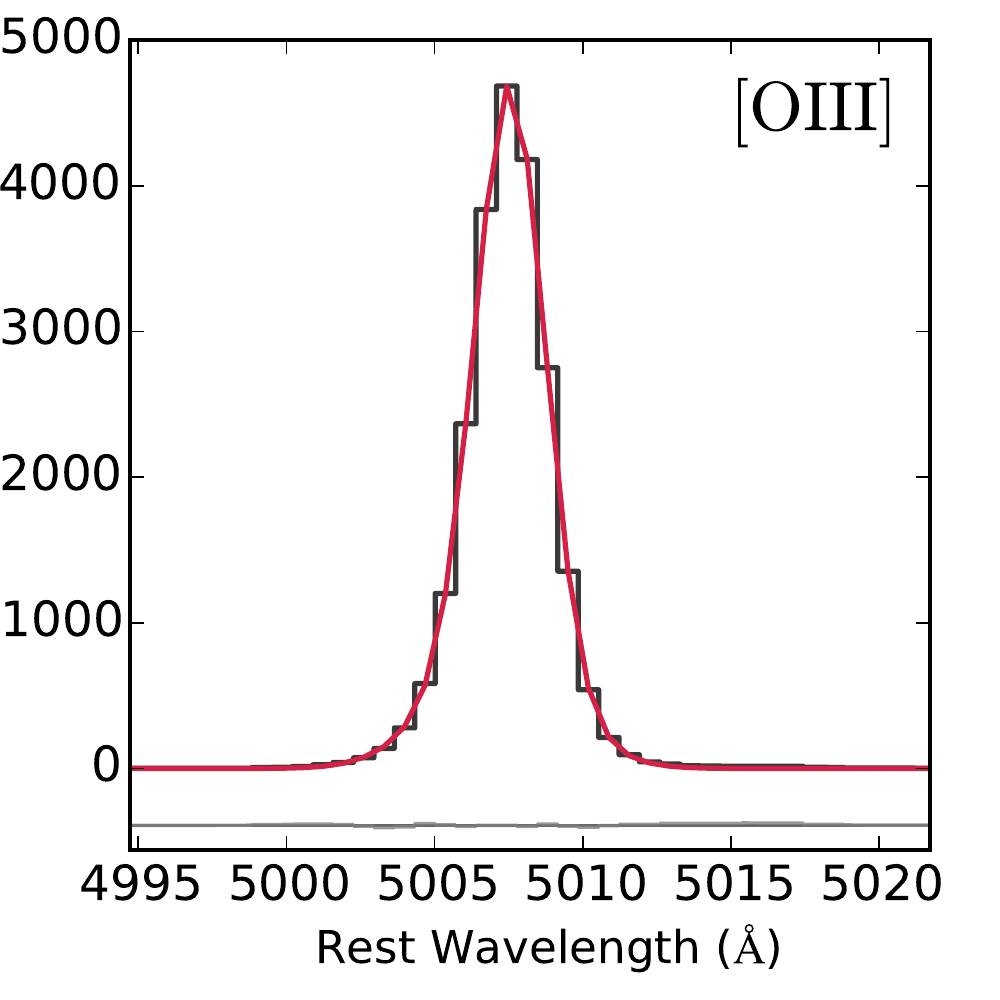}
\includegraphics[scale=0.44]{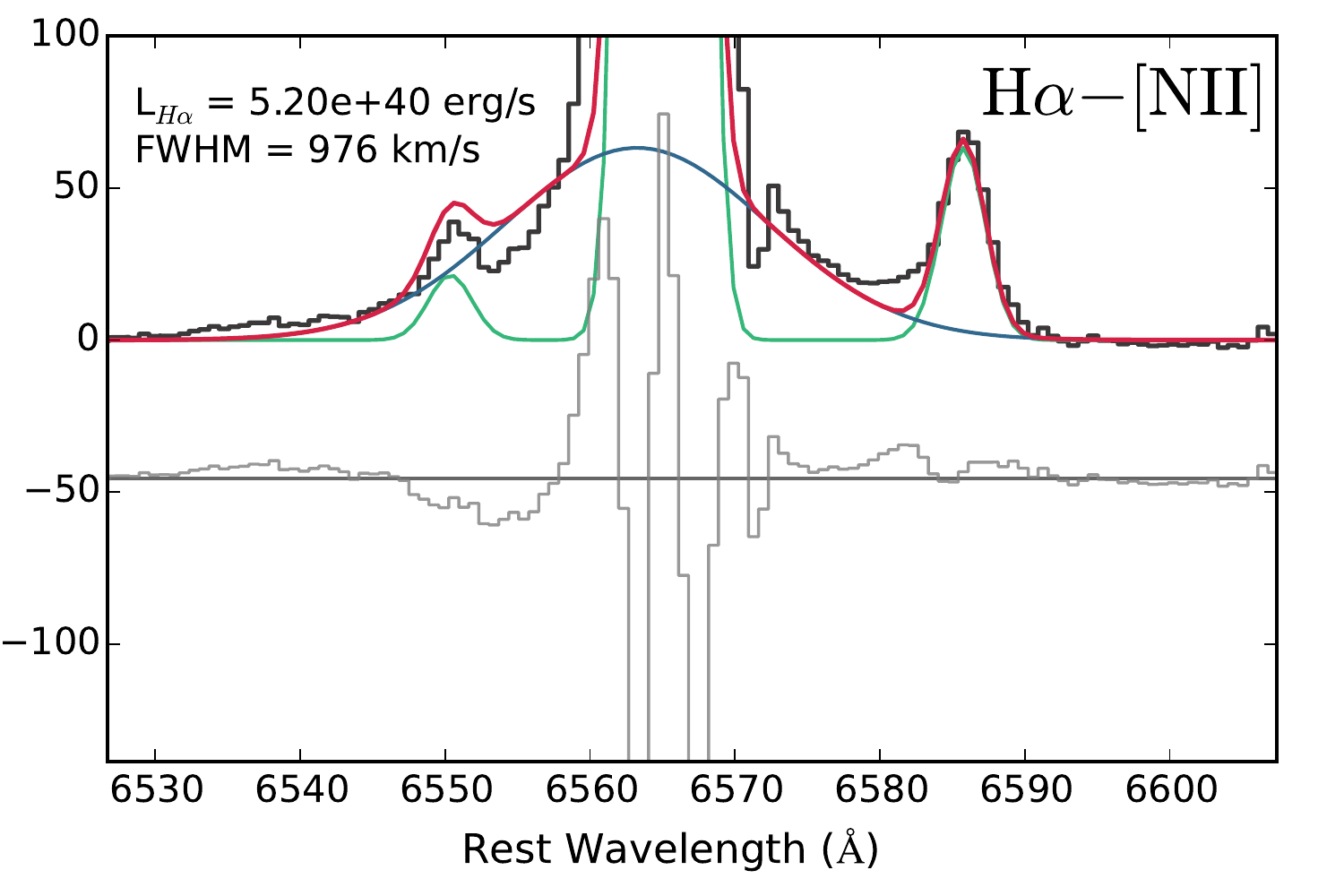}
\caption{These plots show the $\rm H\beta$, [OIII]$\lambda$5007, H$\alpha$, and [NII]$\lambda\lambda6718,6731$ lines for each observation taken of RGG B (NSA 15952). We place this object in the ``ambiguous" category (see main paper). In each plot, the observed spectrum is shown as the solid, dark gray line. The best fit line profile is given by the red line, and the residual is plotted at the bottom of each panel in light gray, offset by an arbitrary amount. In the H$\alpha$-[NII] panel, the blue and green lines give the best fit broad and narrow components, respectively. If there was no broad component required, the blue line representing the broad component will look like a constant line at $F_{\lambda}=0$.}
\label{nsa15952}
\end{figure*}

\begin{figure*}
\centering
\includegraphics[scale=0.44]{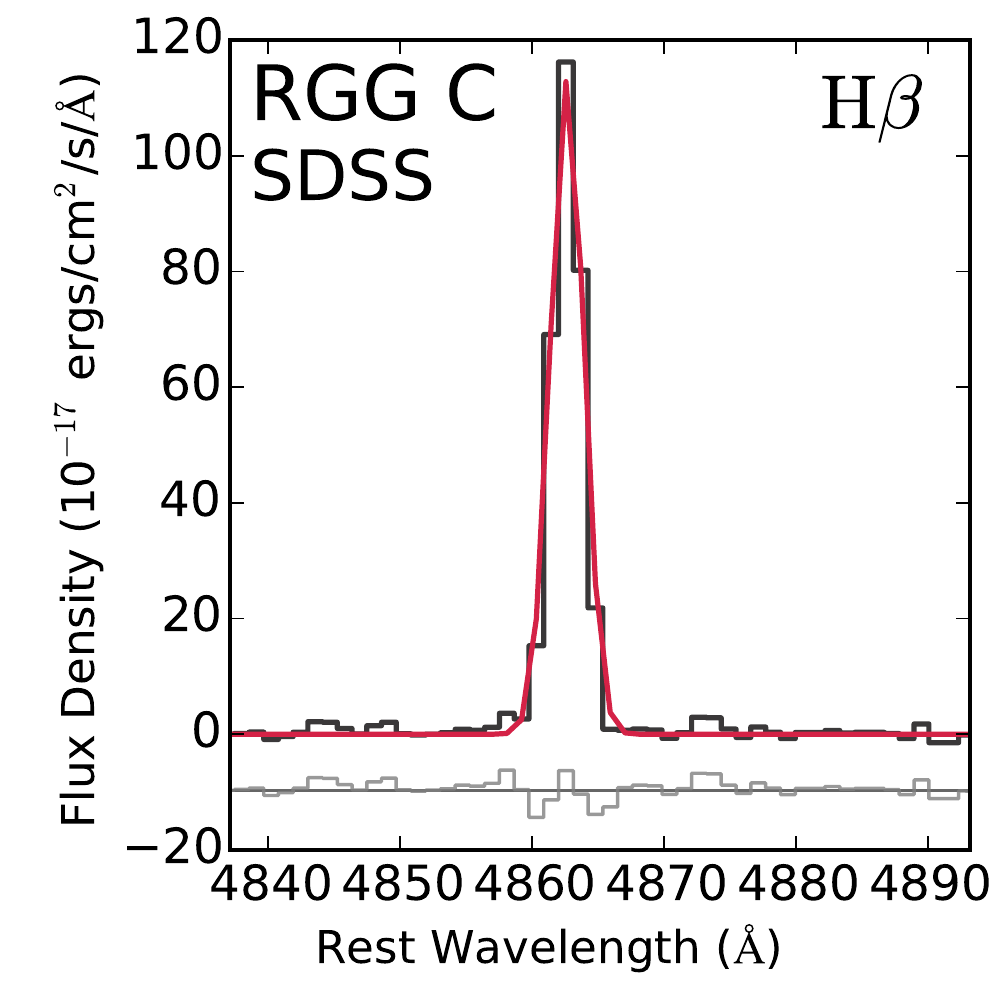}
\includegraphics[scale=0.44]{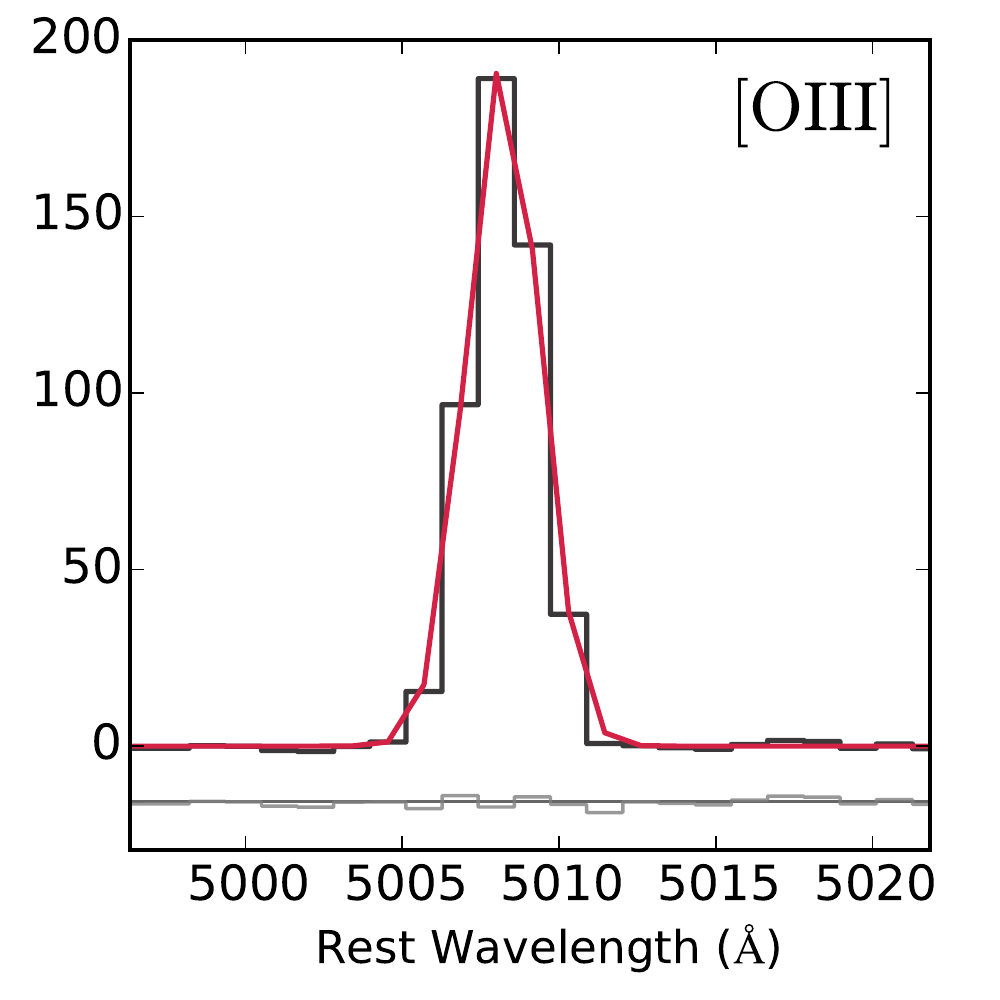}
\includegraphics[scale=0.44]{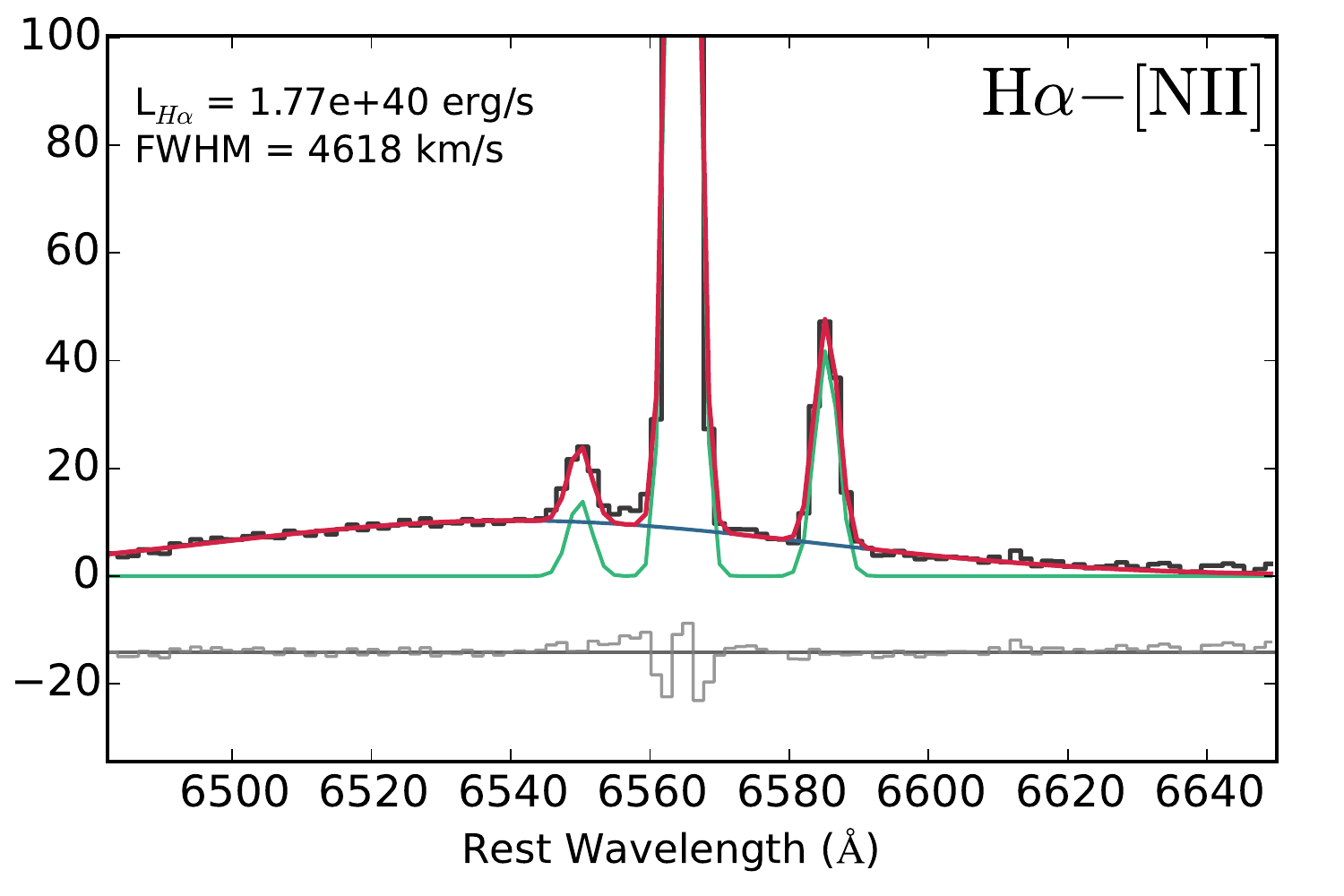}\\

\includegraphics[scale=0.44]{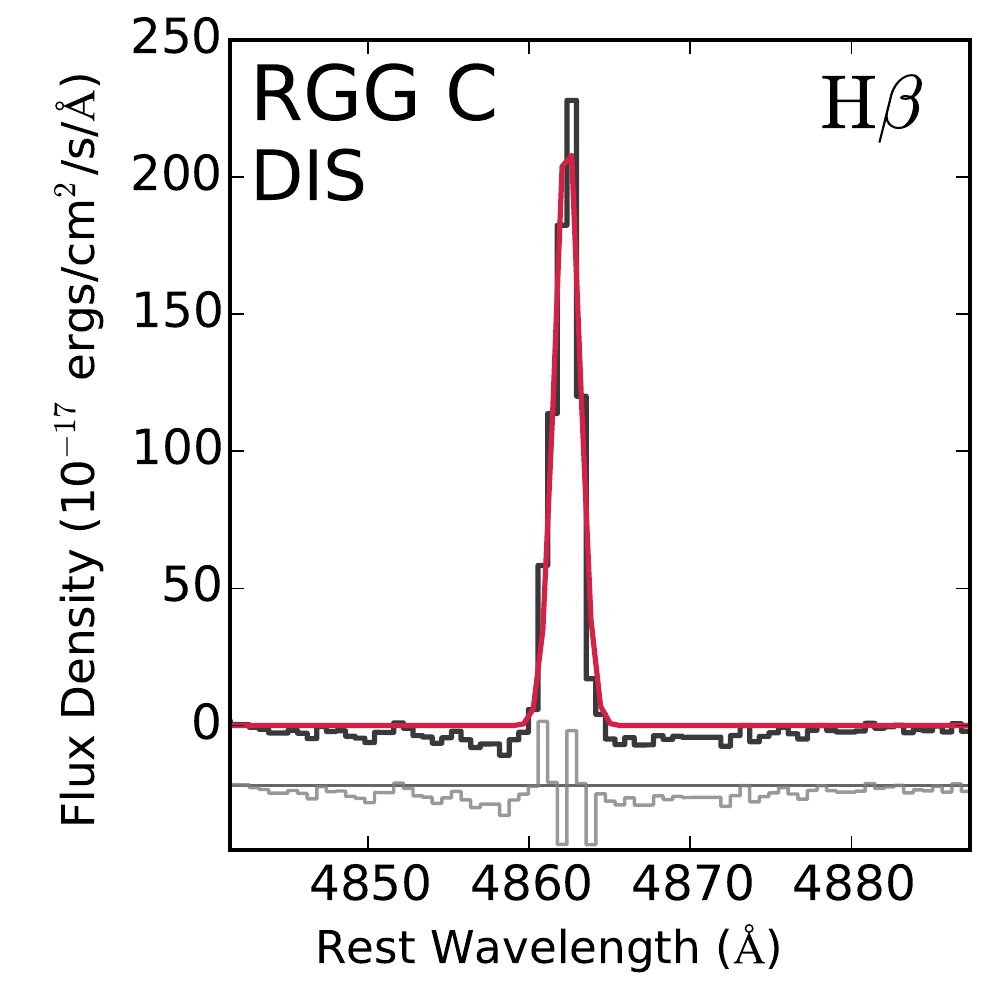}
\includegraphics[scale=0.44]{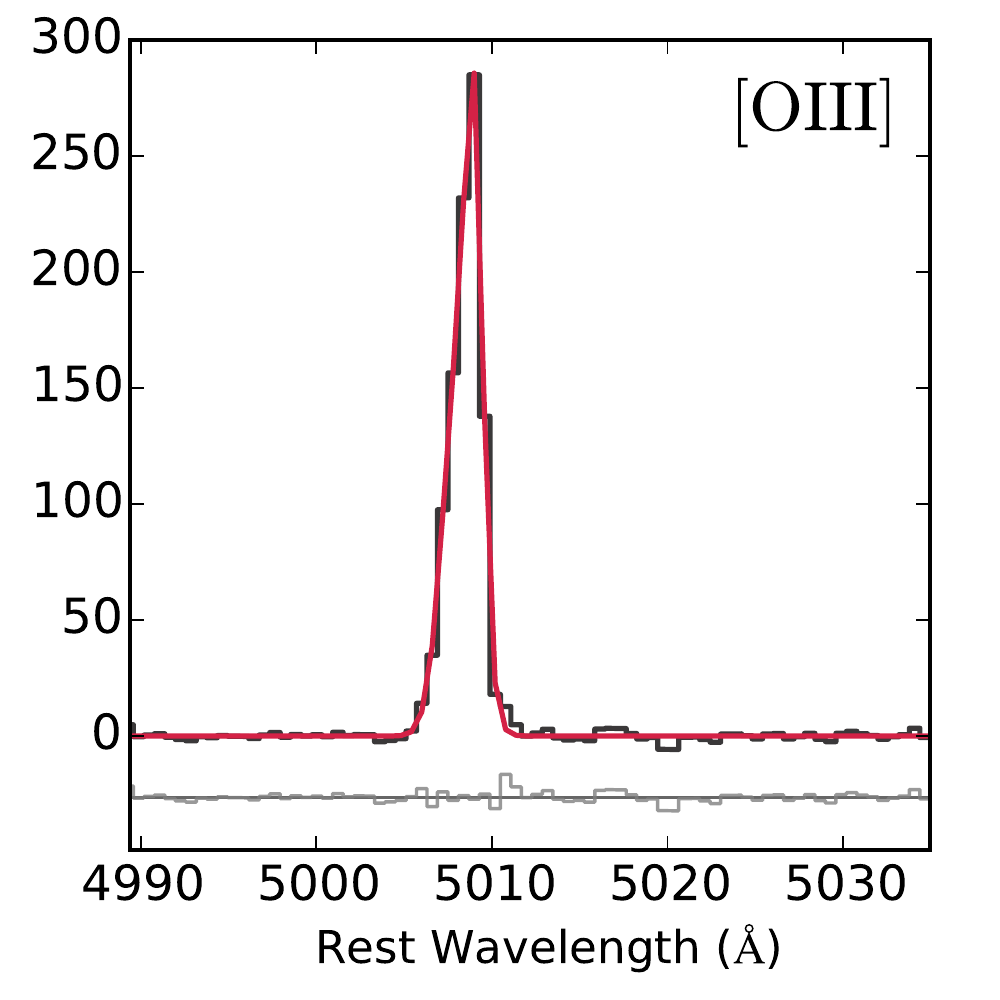}
\includegraphics[scale=0.44]{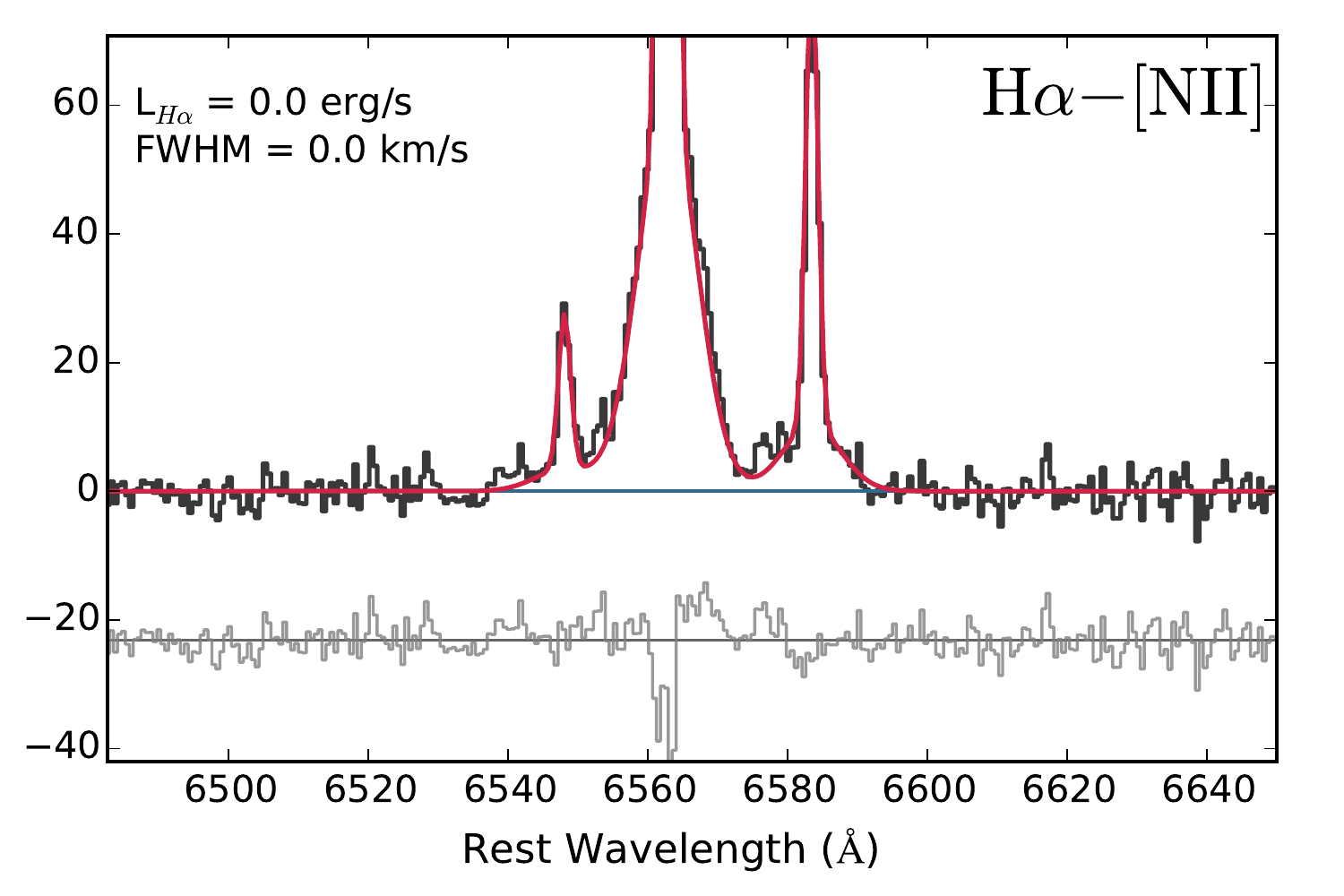}

\includegraphics[scale=0.44]{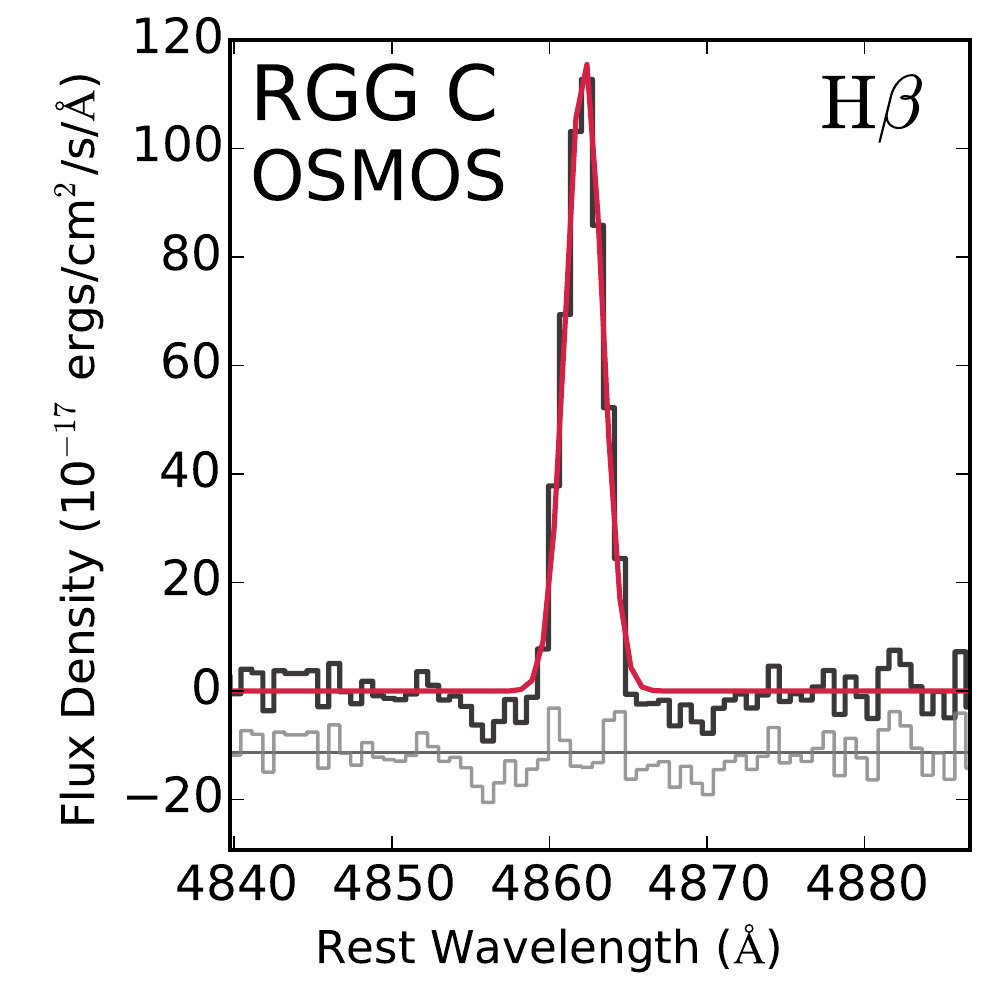}
\includegraphics[scale=0.44]{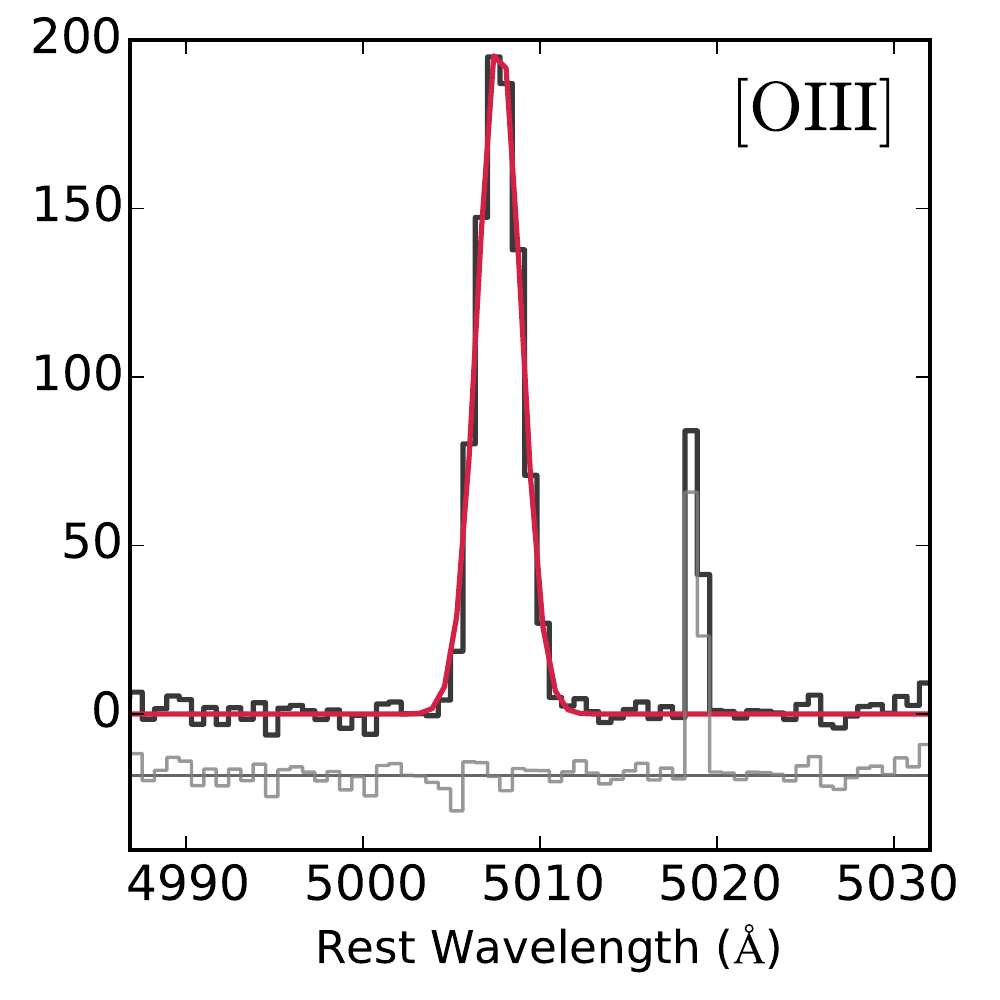}
\includegraphics[scale=0.44]{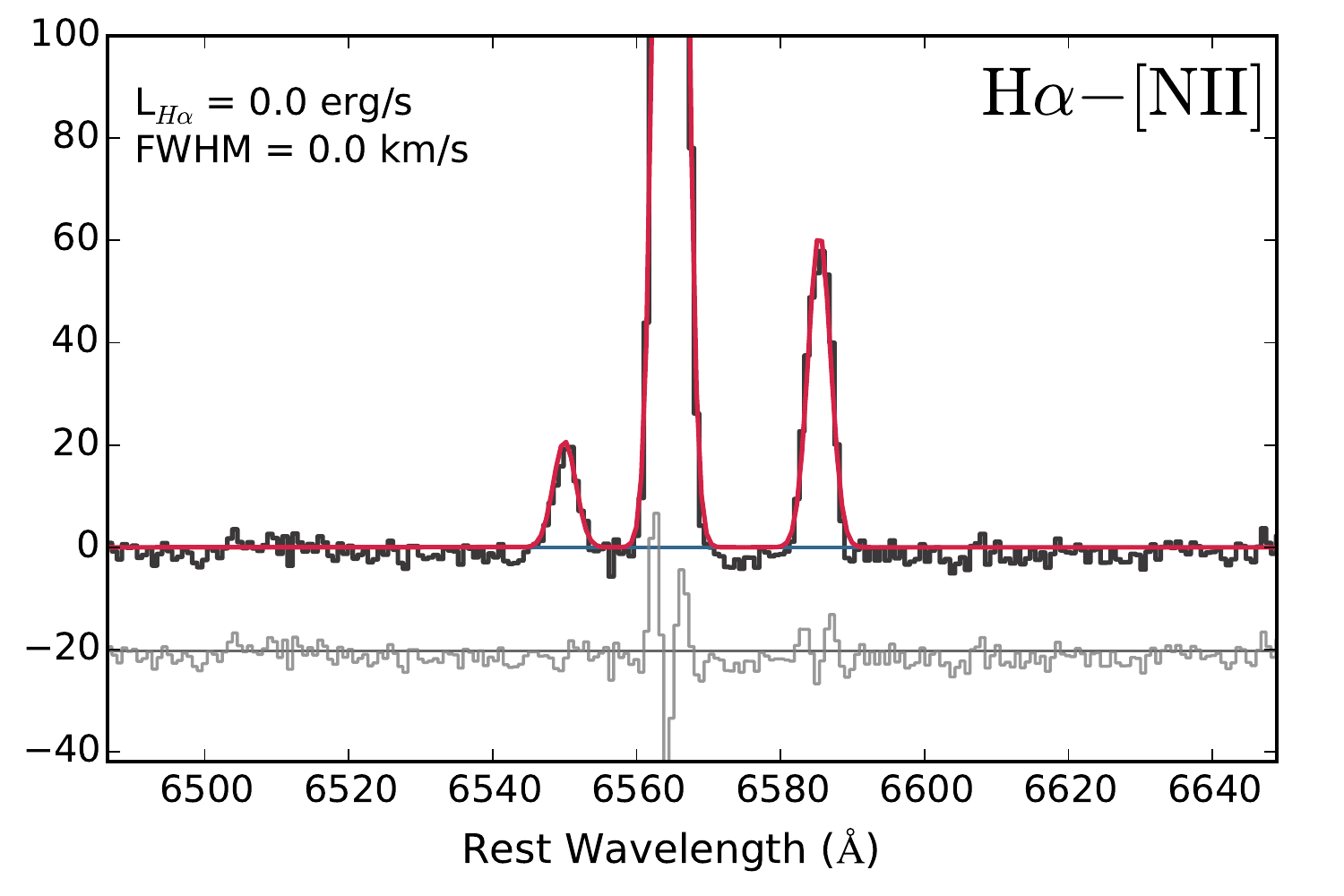}
\caption{These plots show the $\rm H\beta$, [OIII]$\lambda$5007, H$\alpha$, and [NII]$\lambda\lambda6718,6731$ lines for each observation taken of RGG C (NSA 109990). Description is same as for Figure~\ref{nsa15952}. We place this object in the ``transient" category.}
\label{nsa109990}
\end{figure*}

\begin{figure*}
\centering
\includegraphics[scale=0.44]{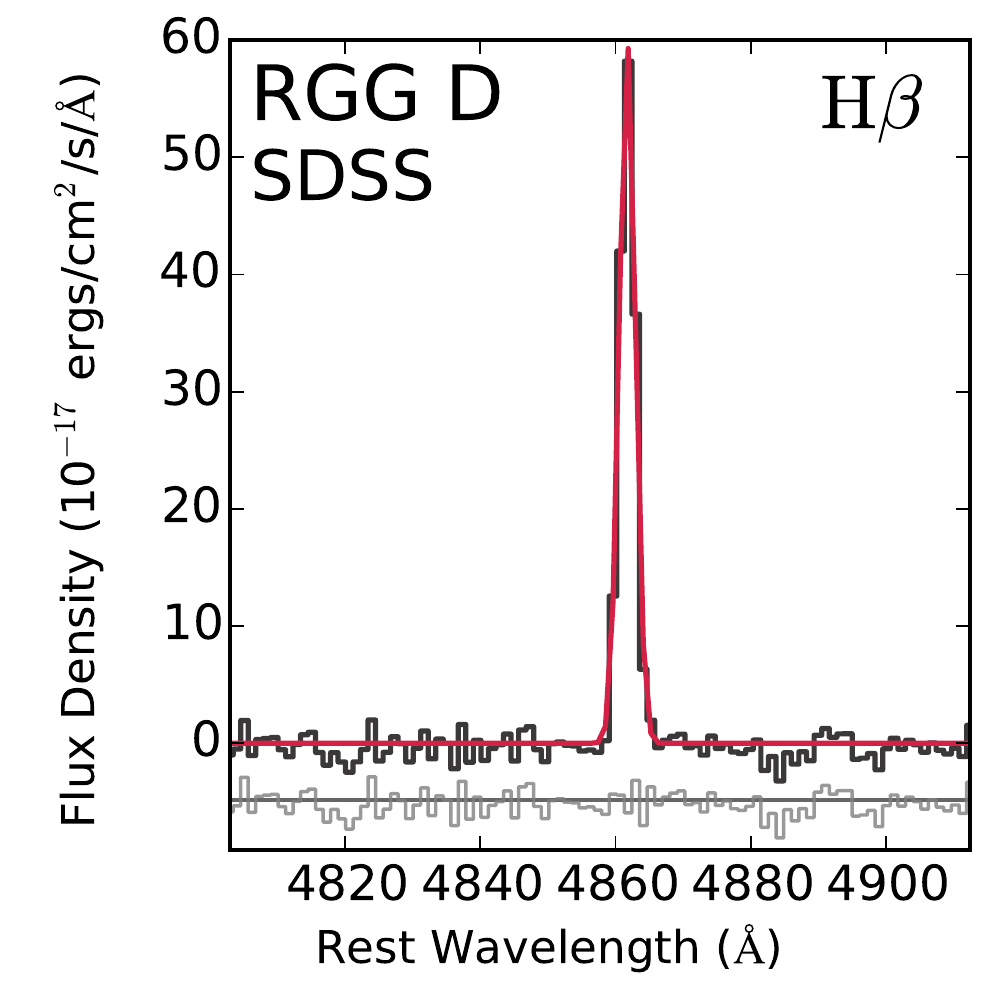}
\includegraphics[scale=0.44]{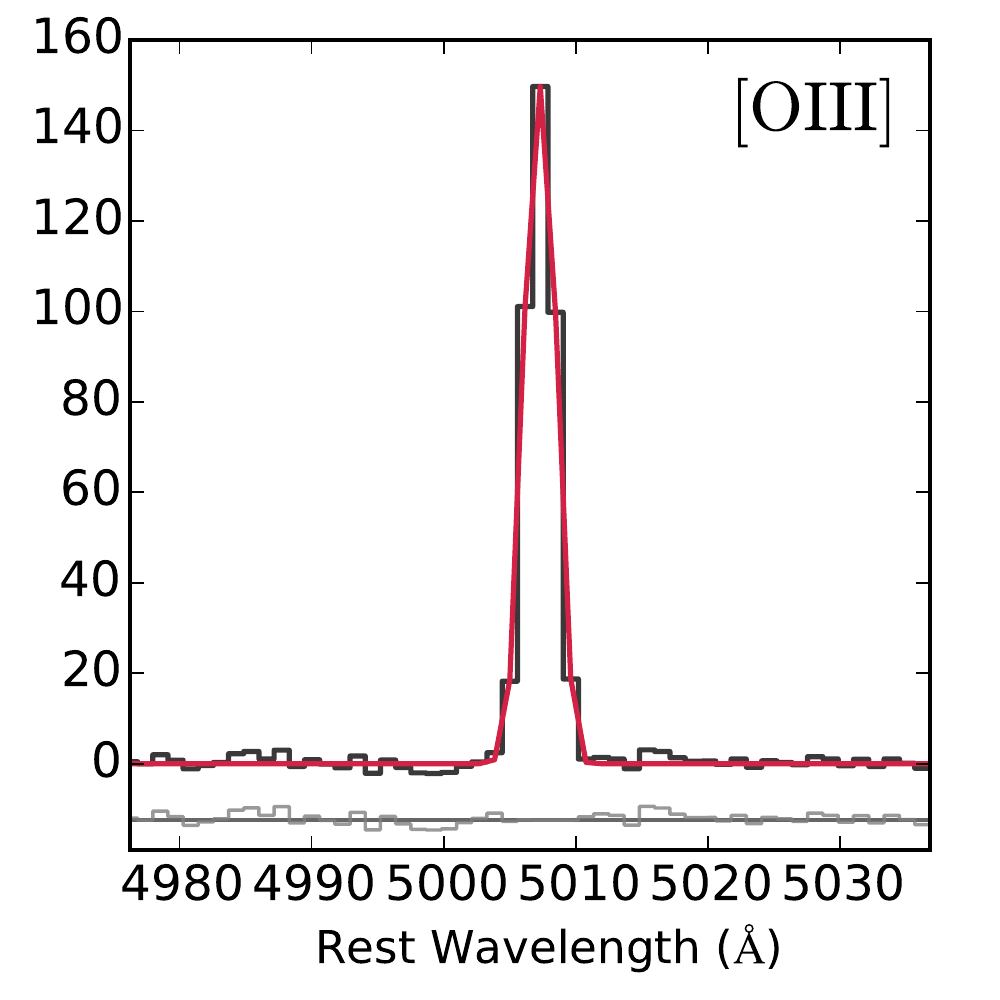}
\includegraphics[scale=0.44]{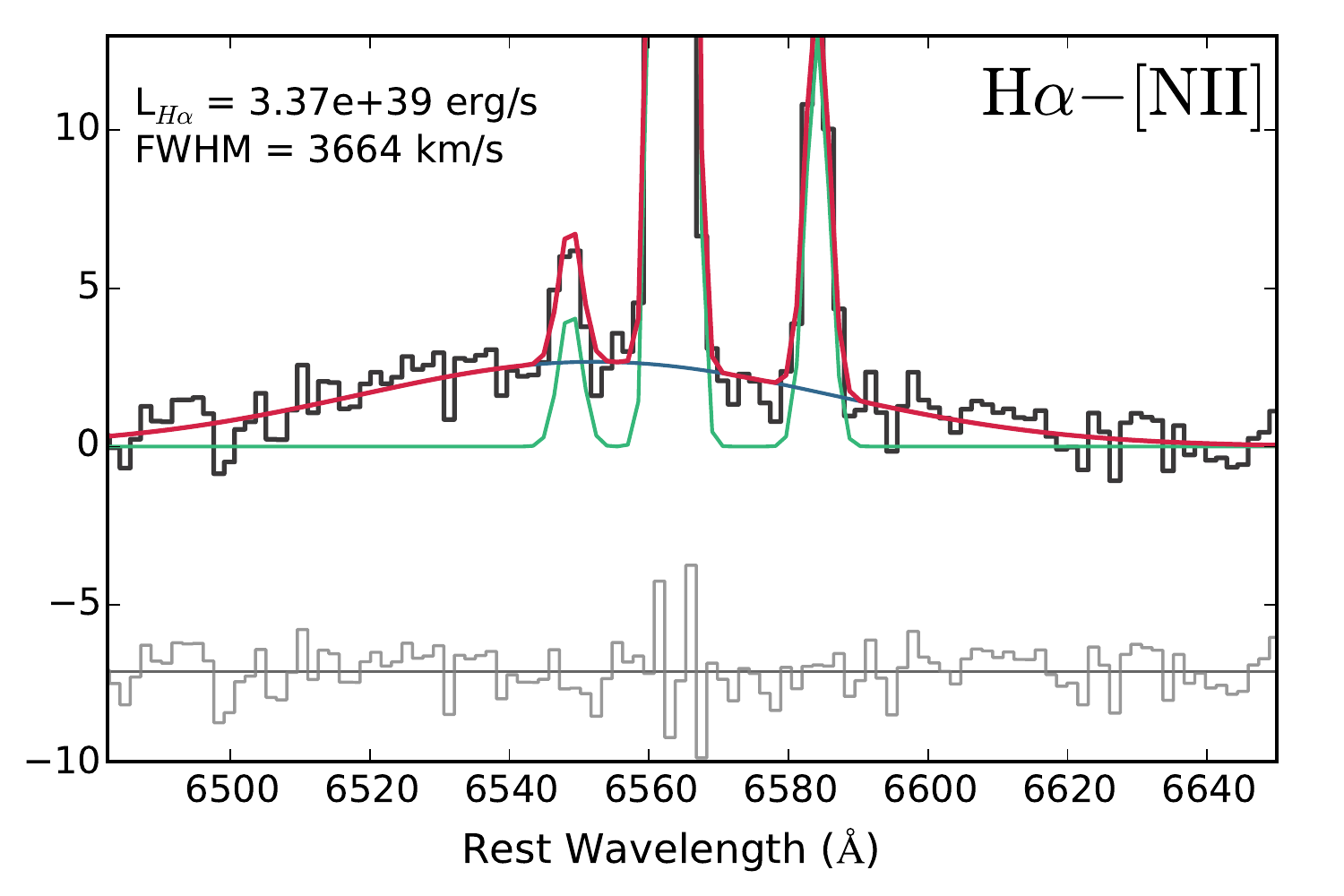}\\

\includegraphics[scale=0.44]{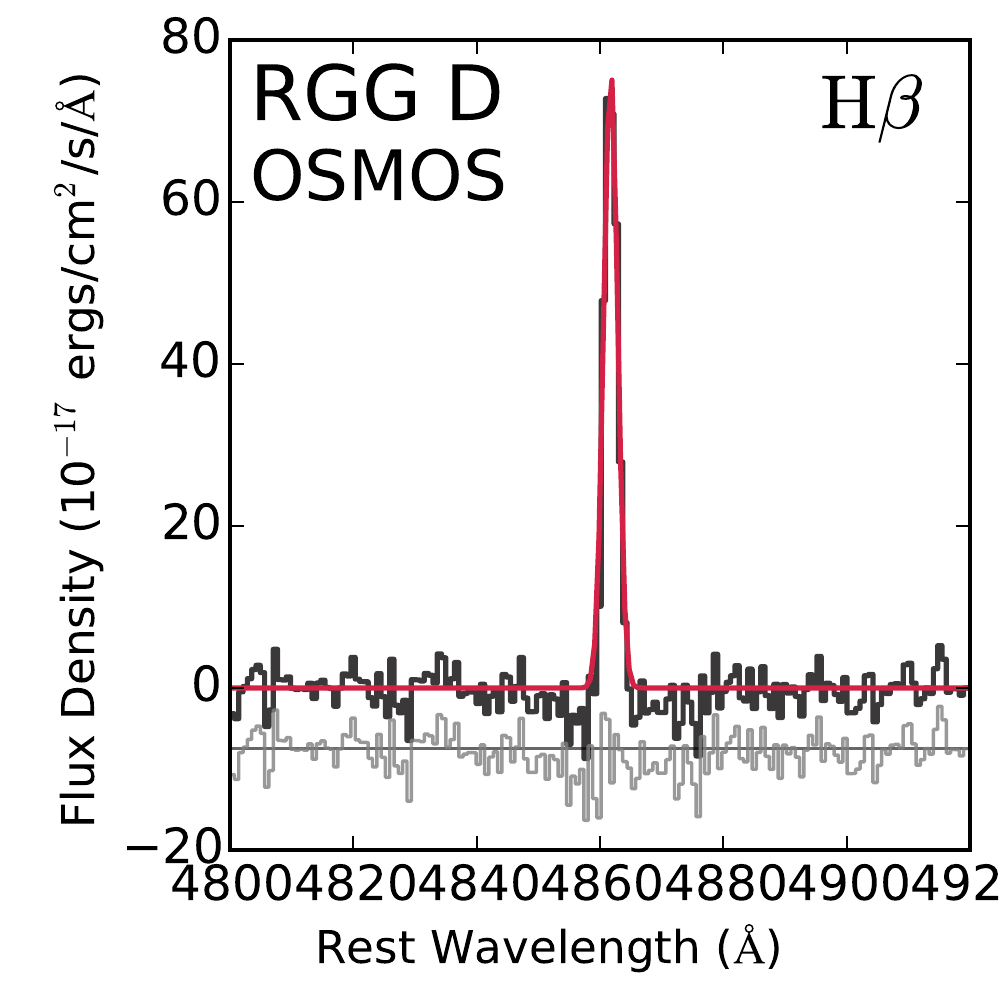}
\includegraphics[scale=0.44]{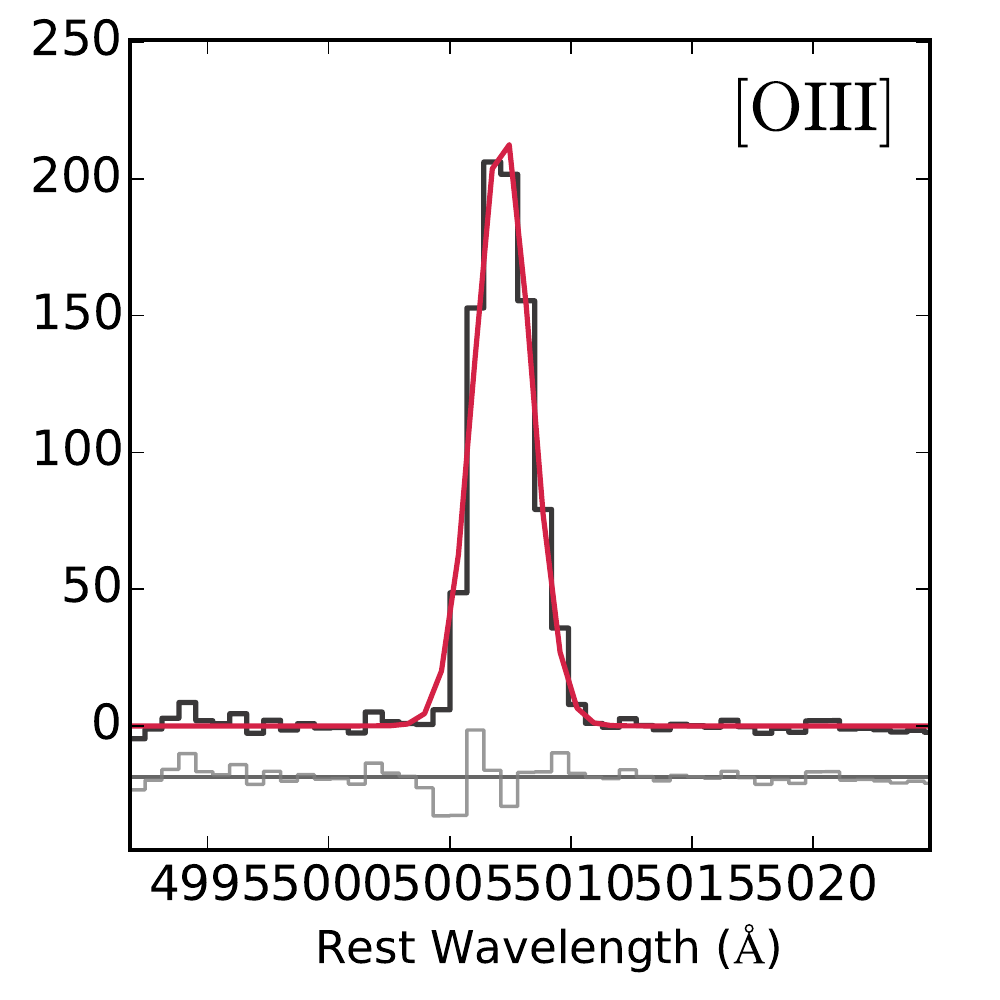}
\includegraphics[scale=0.44]{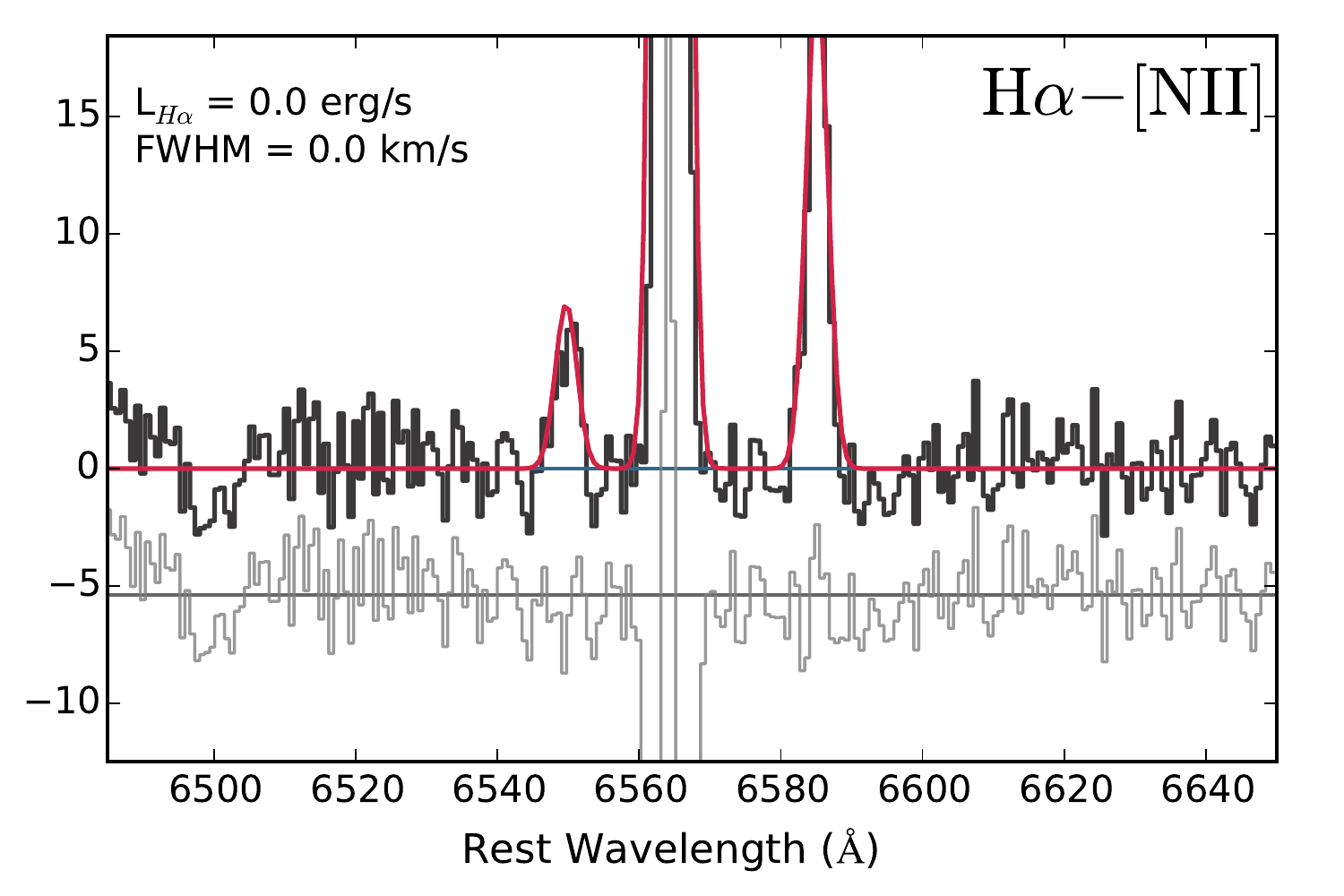}
\caption{These plots show the $\rm H\beta$, [OIII]$\lambda$5007, H$\alpha$, and [NII]$\lambda\lambda6718,6731$ lines for each observation taken of RGG D (NSA 76788). Description is same as for Figure~\ref{nsa15952}. We place this object in the ``transient" category.}
\label{nsa76788}
\end{figure*}

\begin{figure*}
\centering
\includegraphics[scale=0.44]{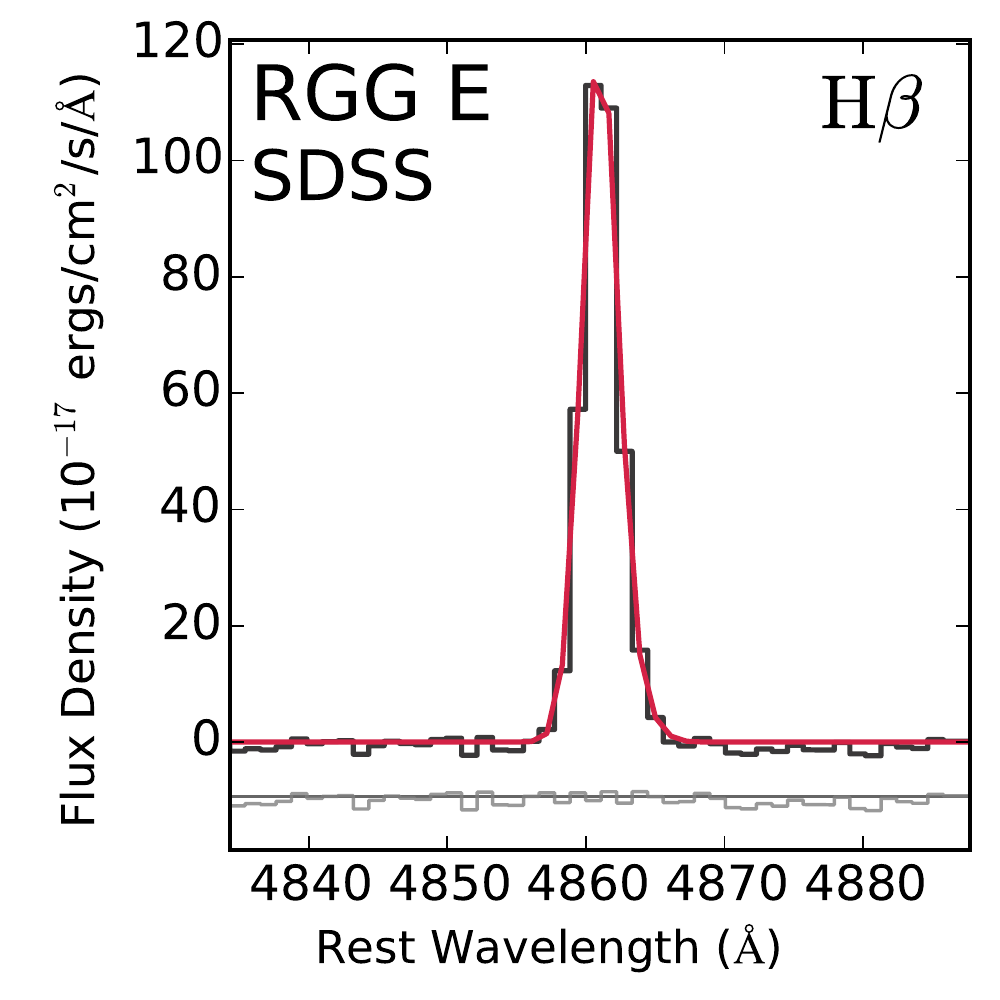}
\includegraphics[scale=0.44]{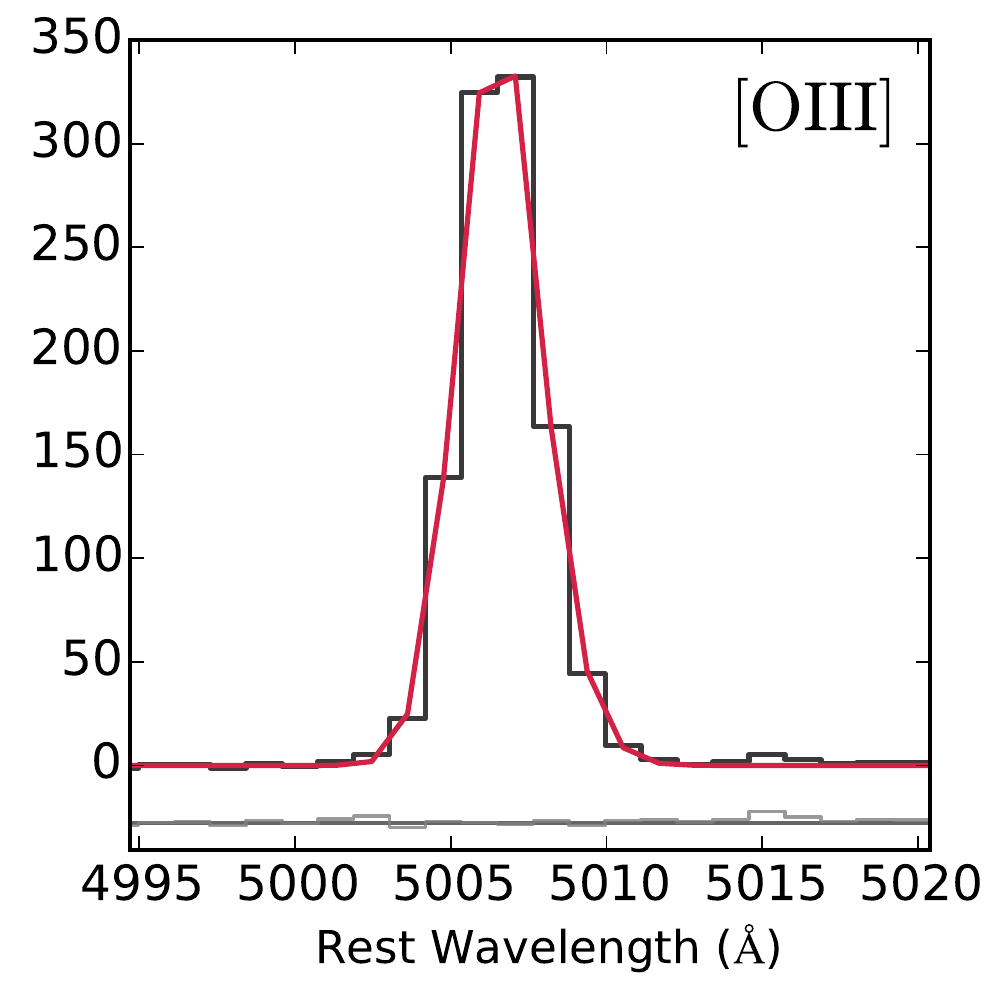}
\includegraphics[scale=0.44]{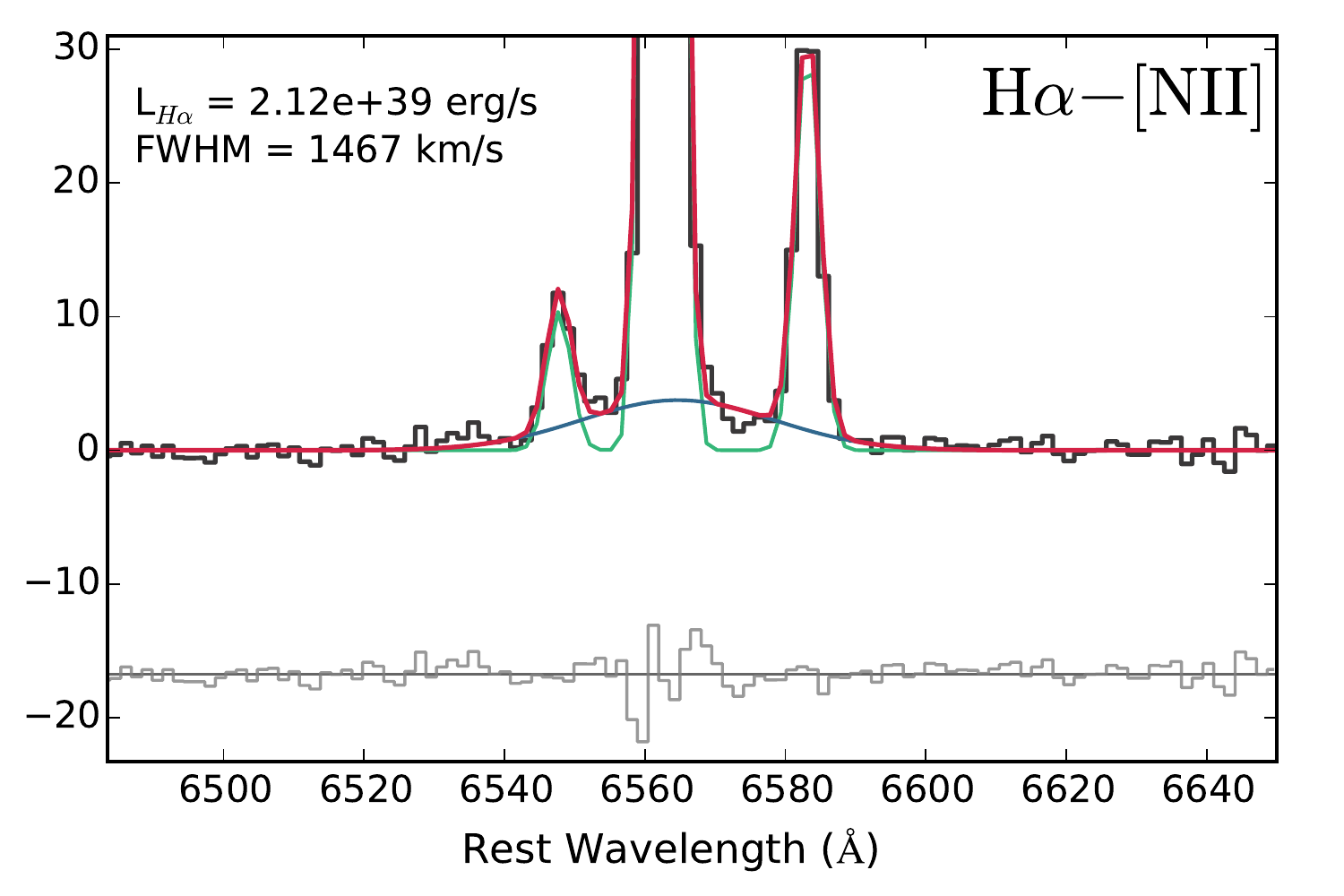}\\

\includegraphics[scale=0.44]{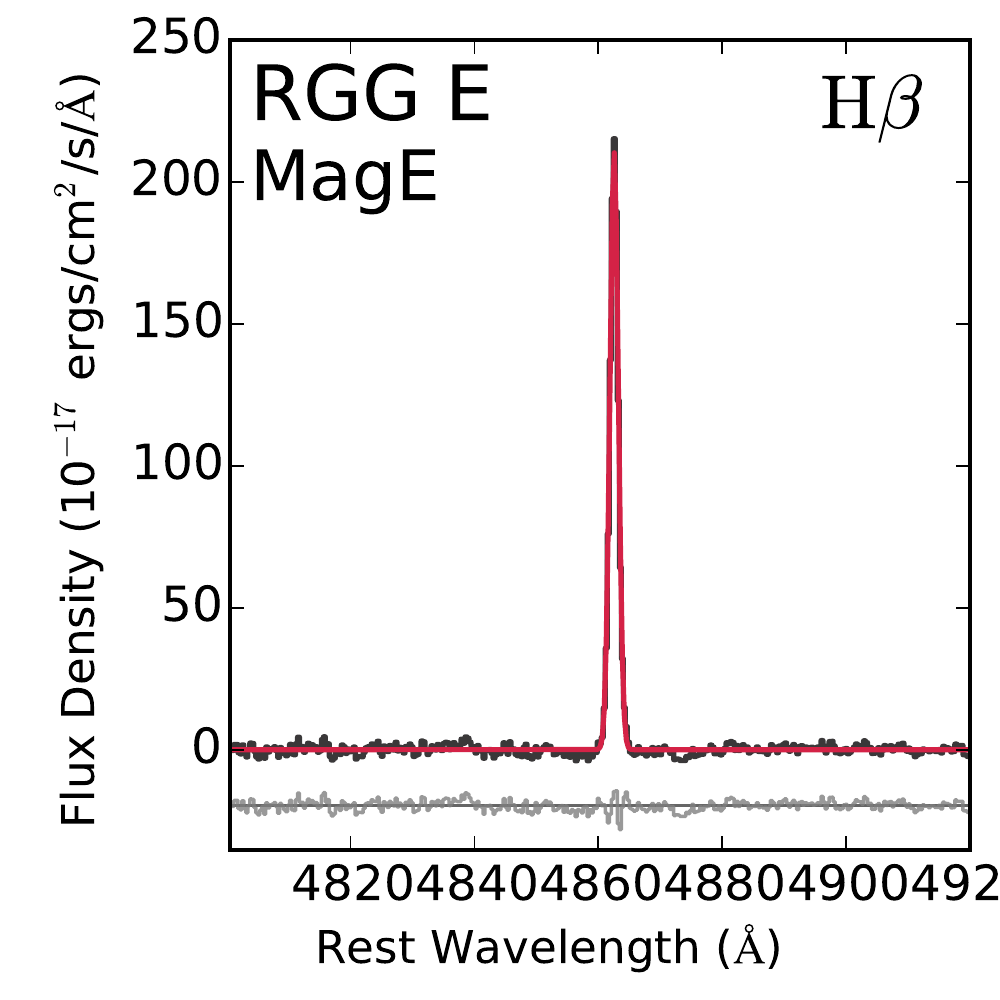}
\includegraphics[scale=0.44]{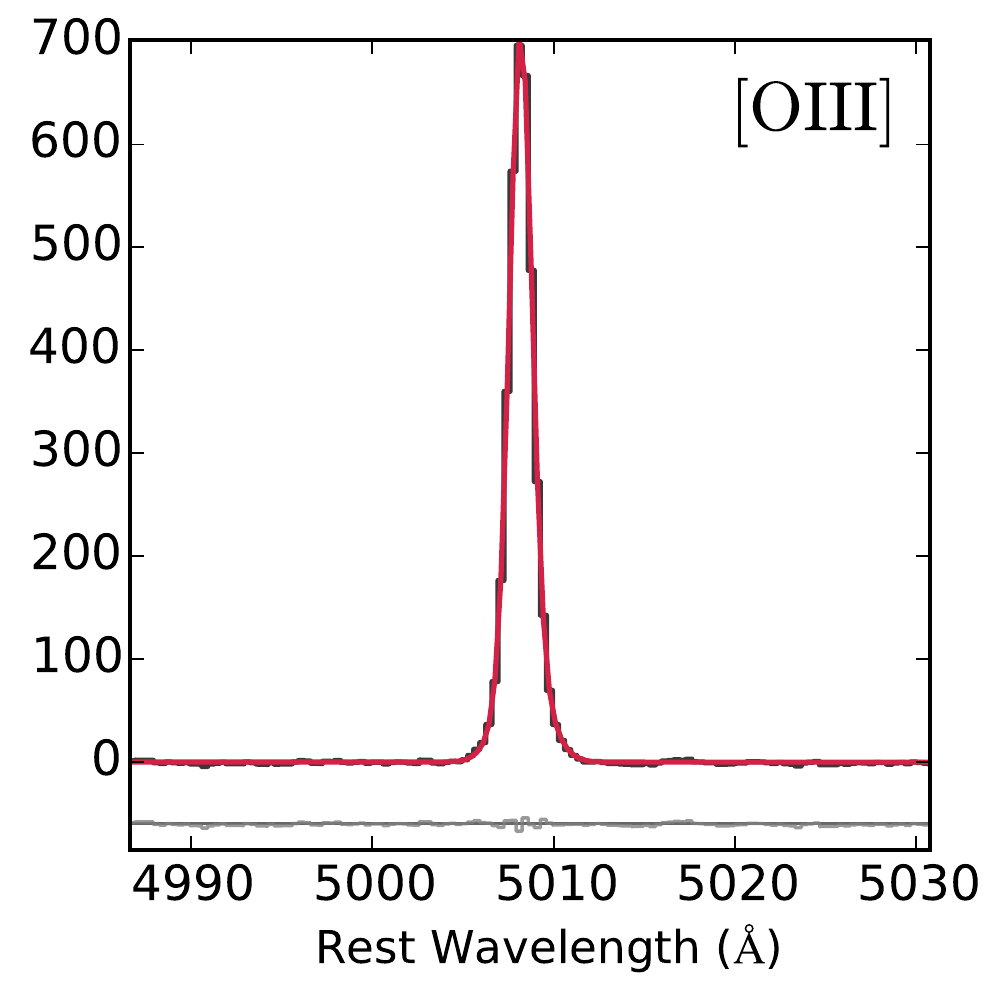}
\includegraphics[scale=0.44]{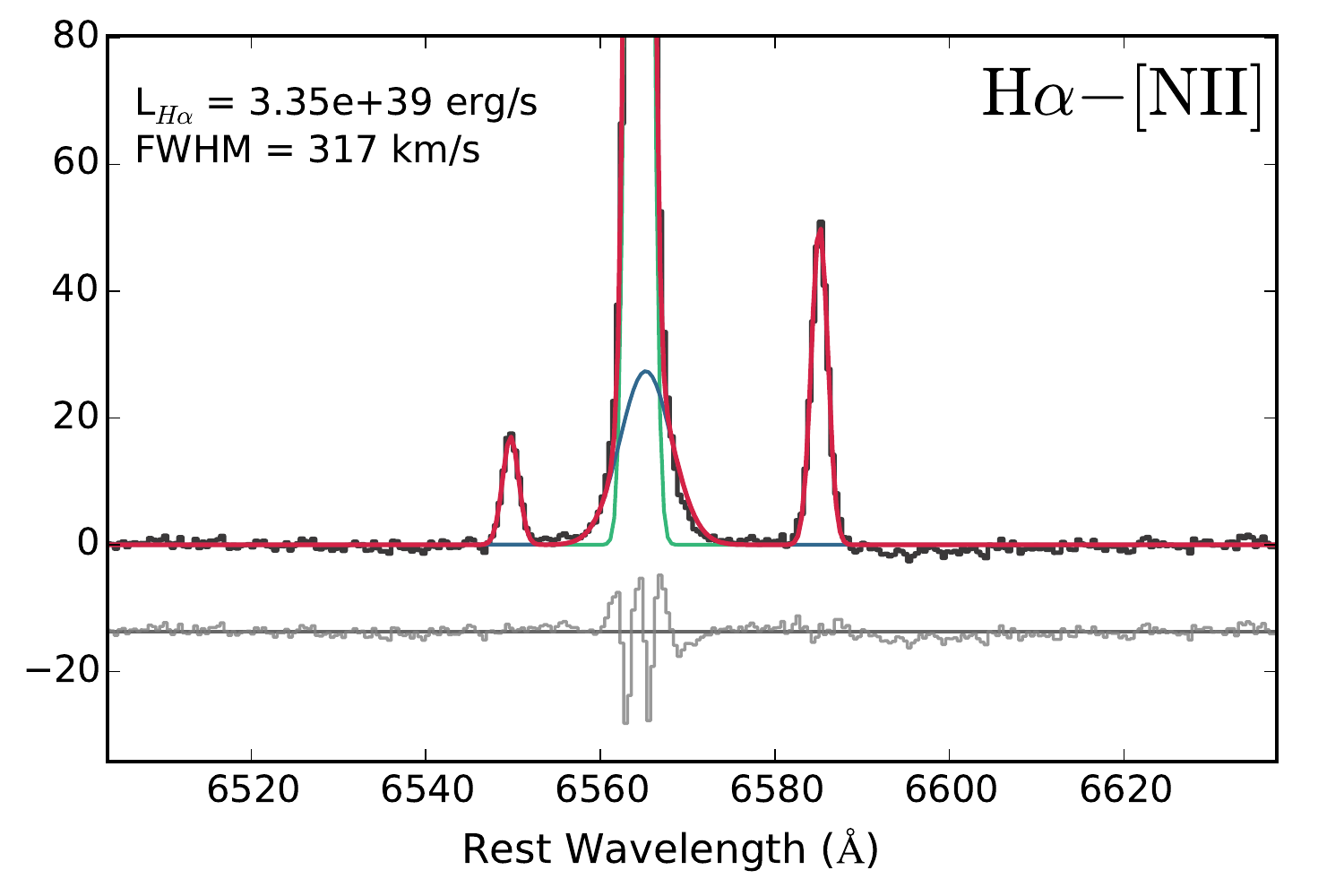}
\caption{These plots show the $\rm H\beta$, [OIII]$\lambda$5007, H$\alpha$, and [NII]$\lambda\lambda6718,6731$ lines for each observation taken of RGG E (NSA 109016). Description is same as for Figure~\ref{nsa15952}. We place this object in the ``transient" category.}
\label{nsa109016}
\end{figure*}

\begin{figure*}
\centering
\includegraphics[scale=0.44]{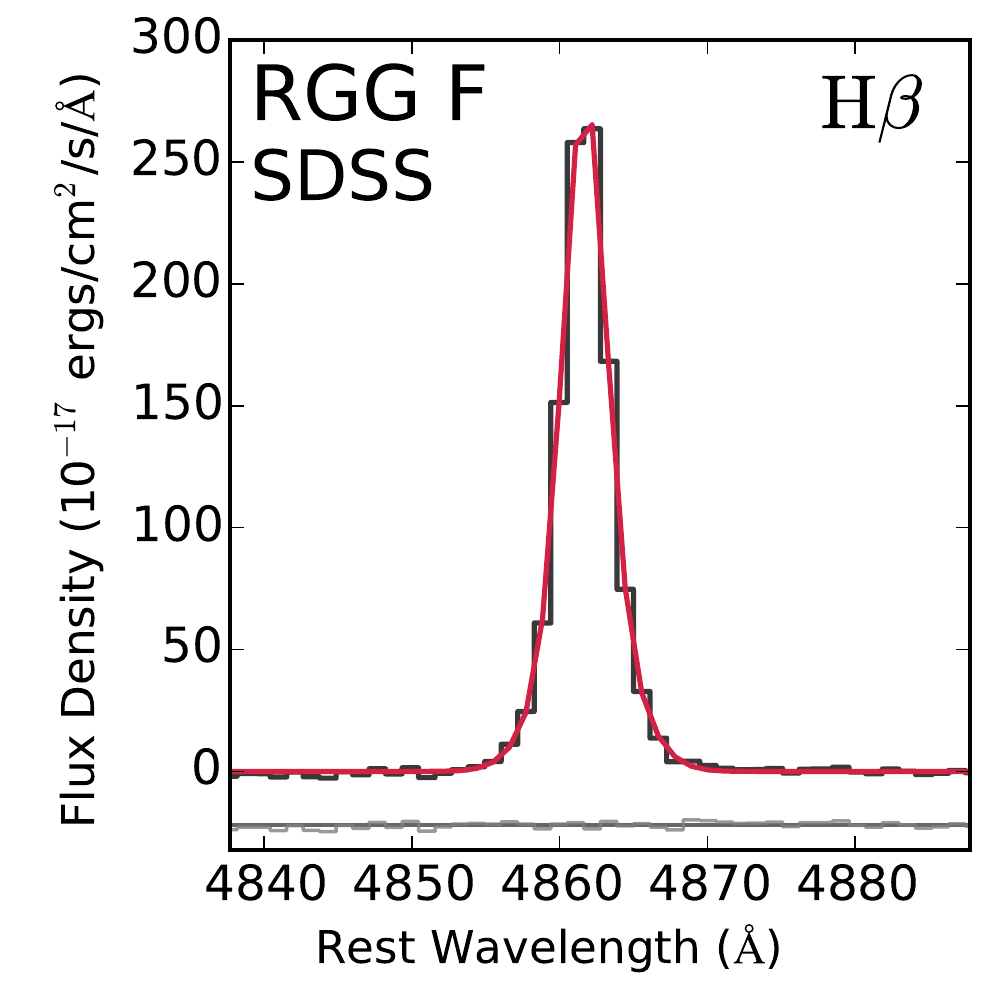}
\includegraphics[scale=0.44]{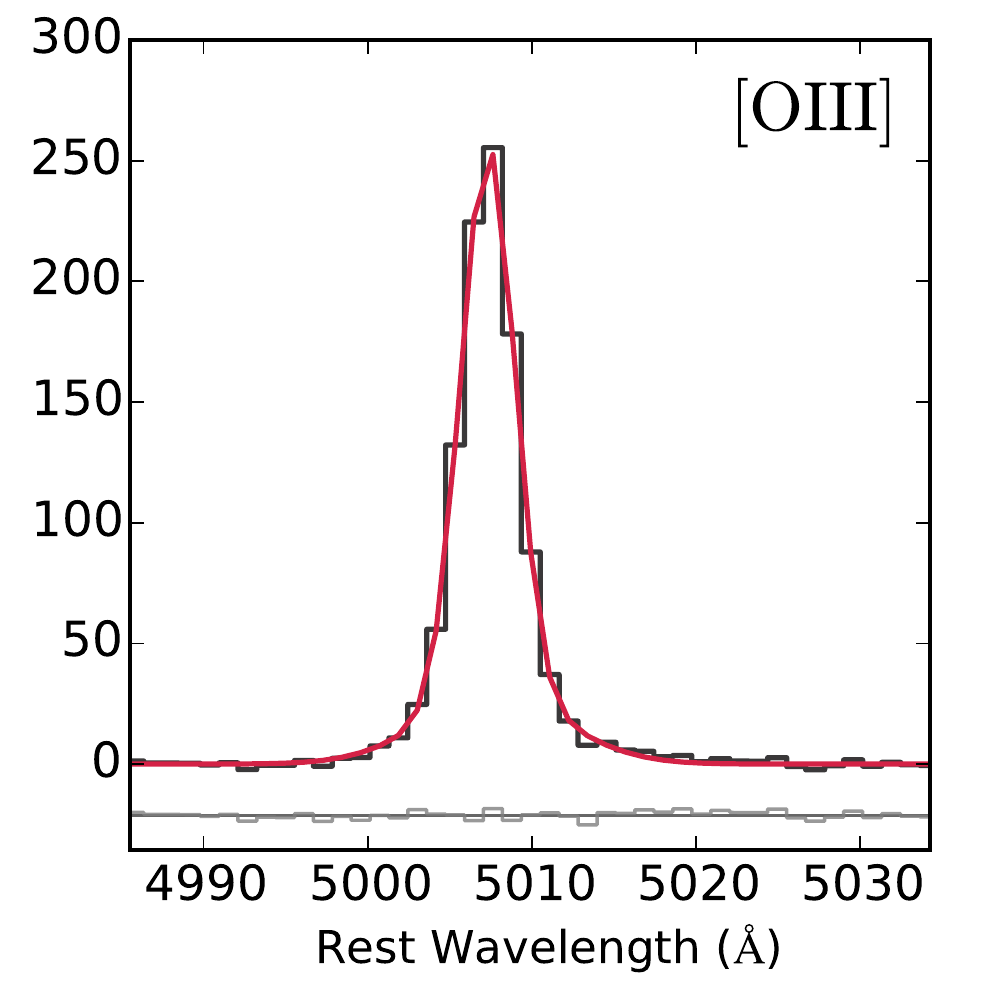}
\includegraphics[scale=0.44]{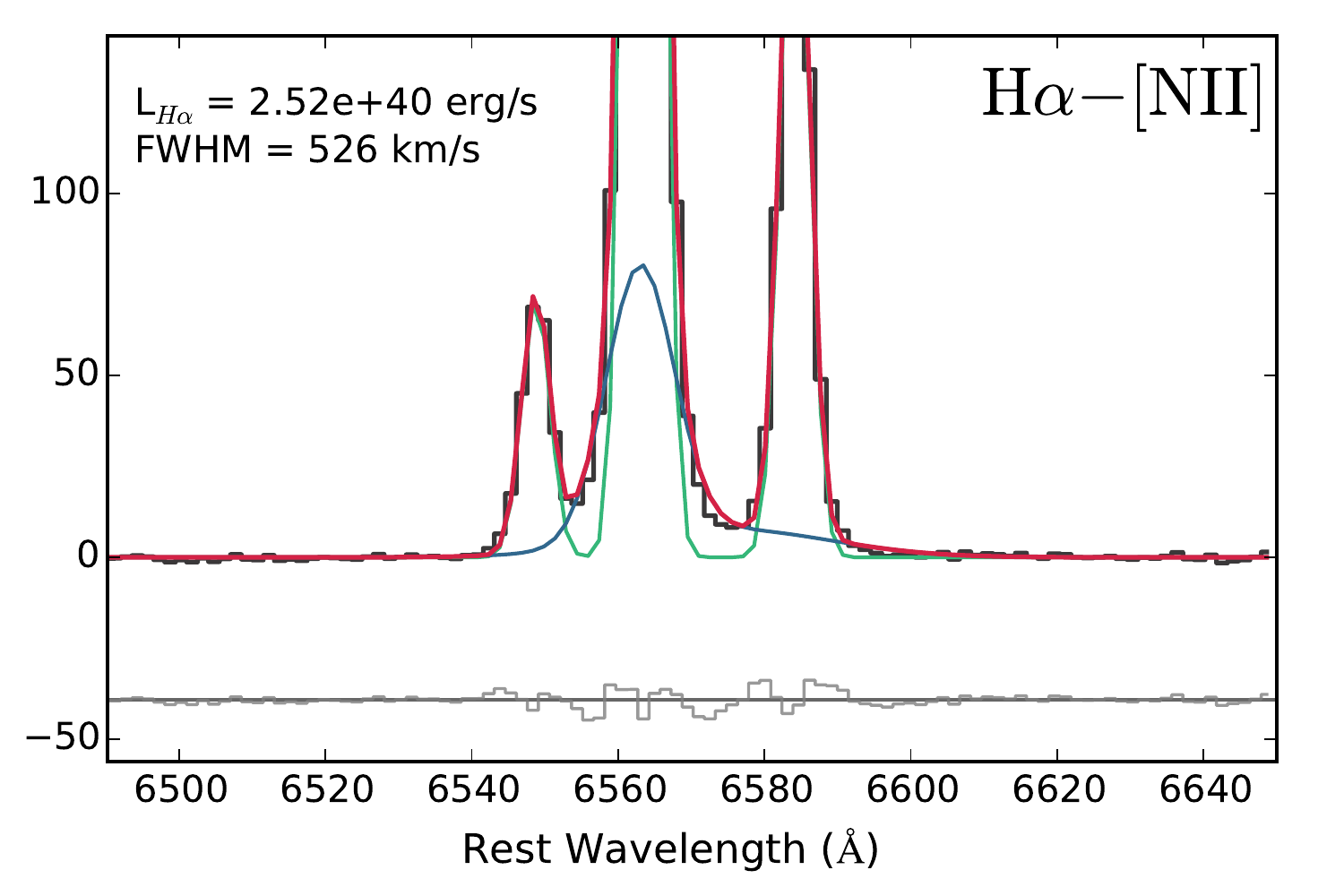}\\

\includegraphics[scale=0.44]{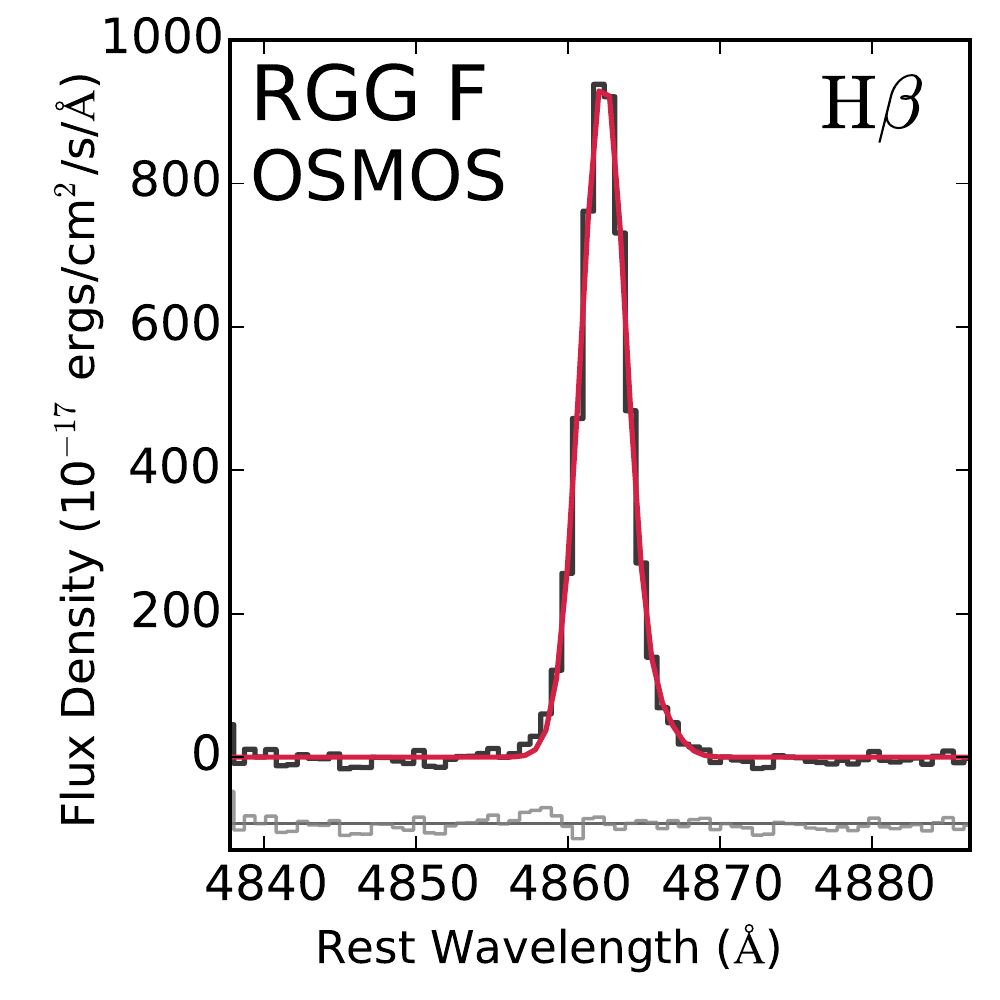}
\includegraphics[scale=0.44]{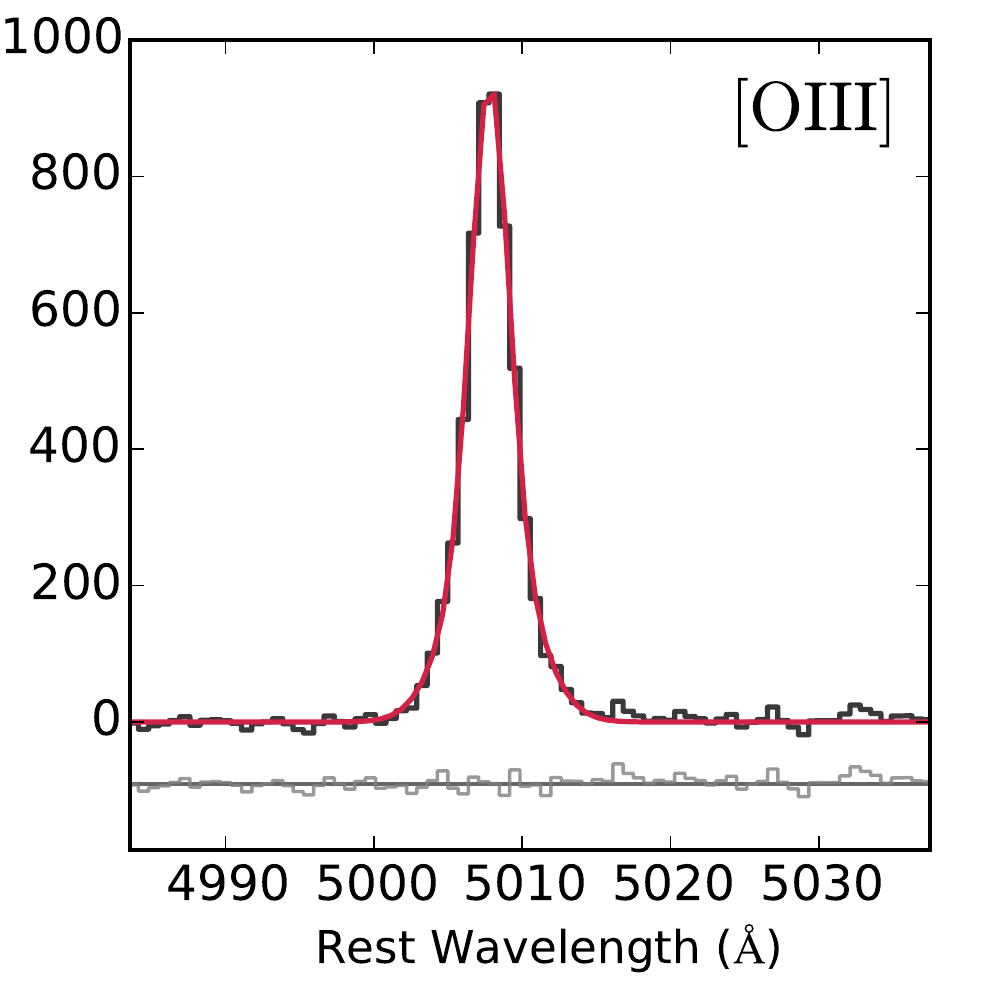}
\includegraphics[scale=0.44]{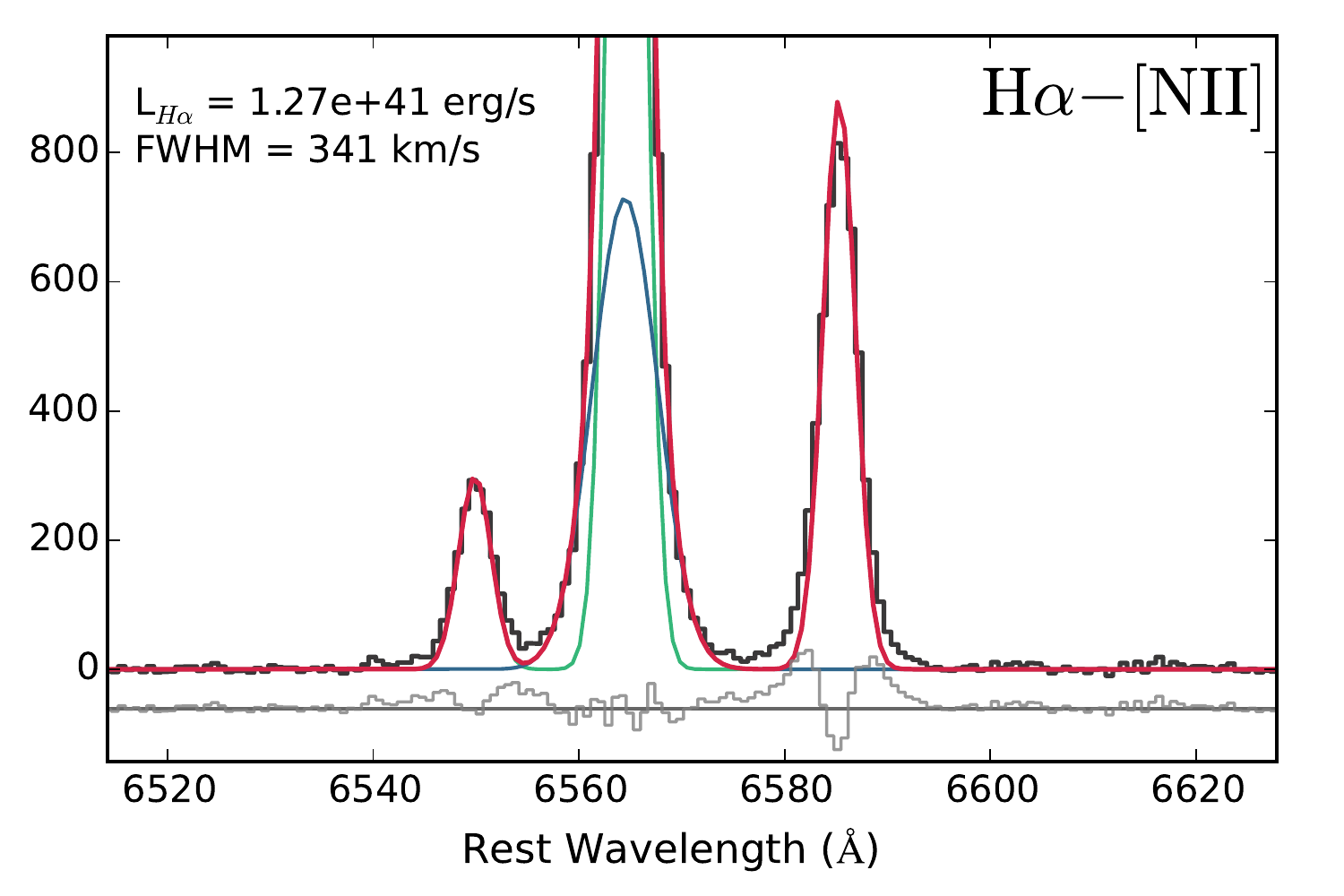}
\caption{These plots show the $\rm H\beta$, [OIII]$\lambda$5007, H$\alpha$, and [NII]$\lambda\lambda6718,6731$ lines for each observation taken of RGG F (NSA 12793). Description is same as for Figure~\ref{nsa15952}. We place this object in the ``transient" category.}
\label{nsa12793}
\end{figure*}

\begin{figure*}
\centering
\includegraphics[scale=0.44]{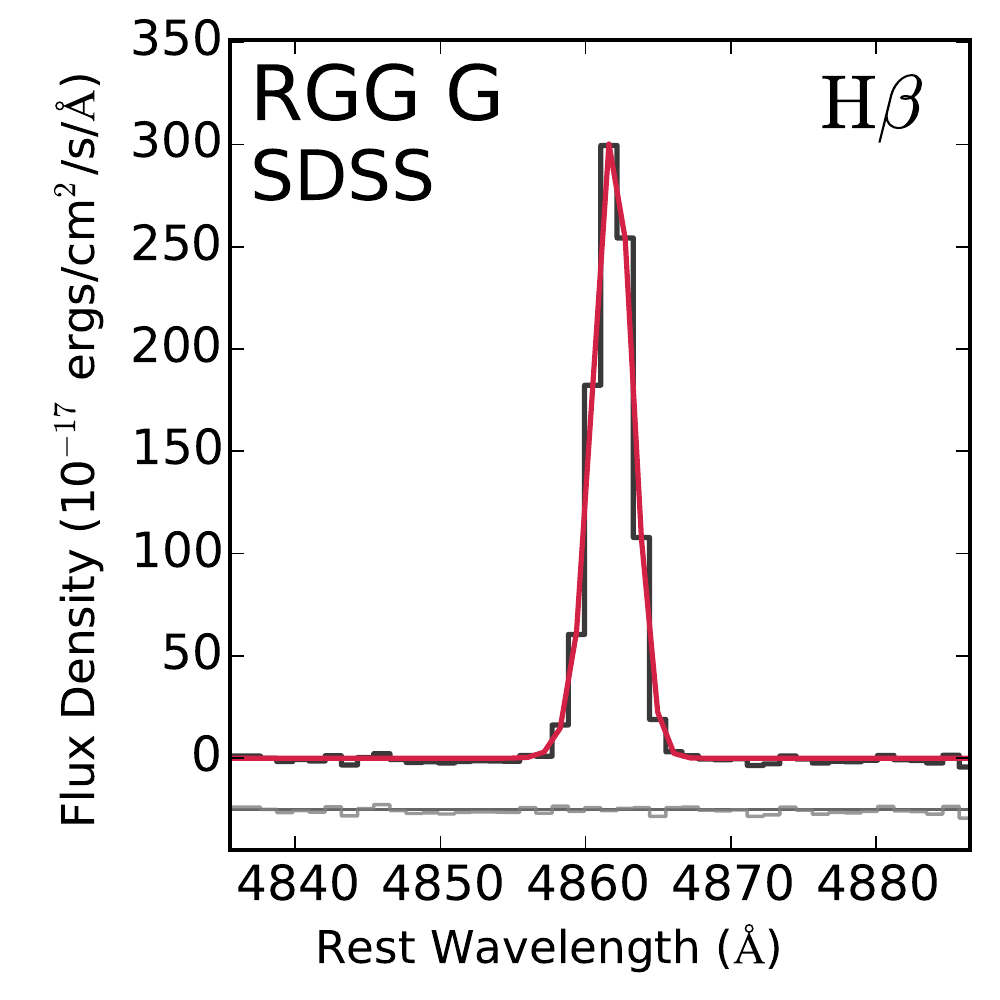}
\includegraphics[scale=0.44]{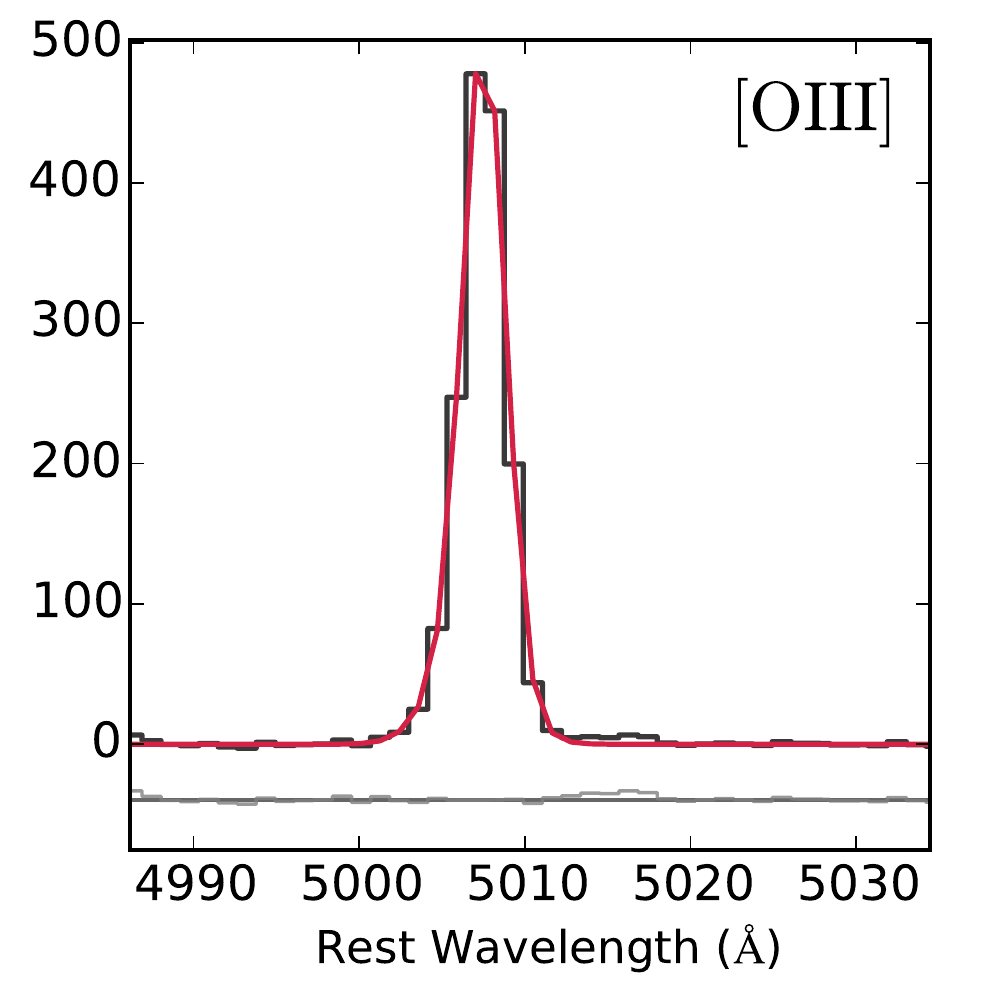}
\includegraphics[scale=0.44]{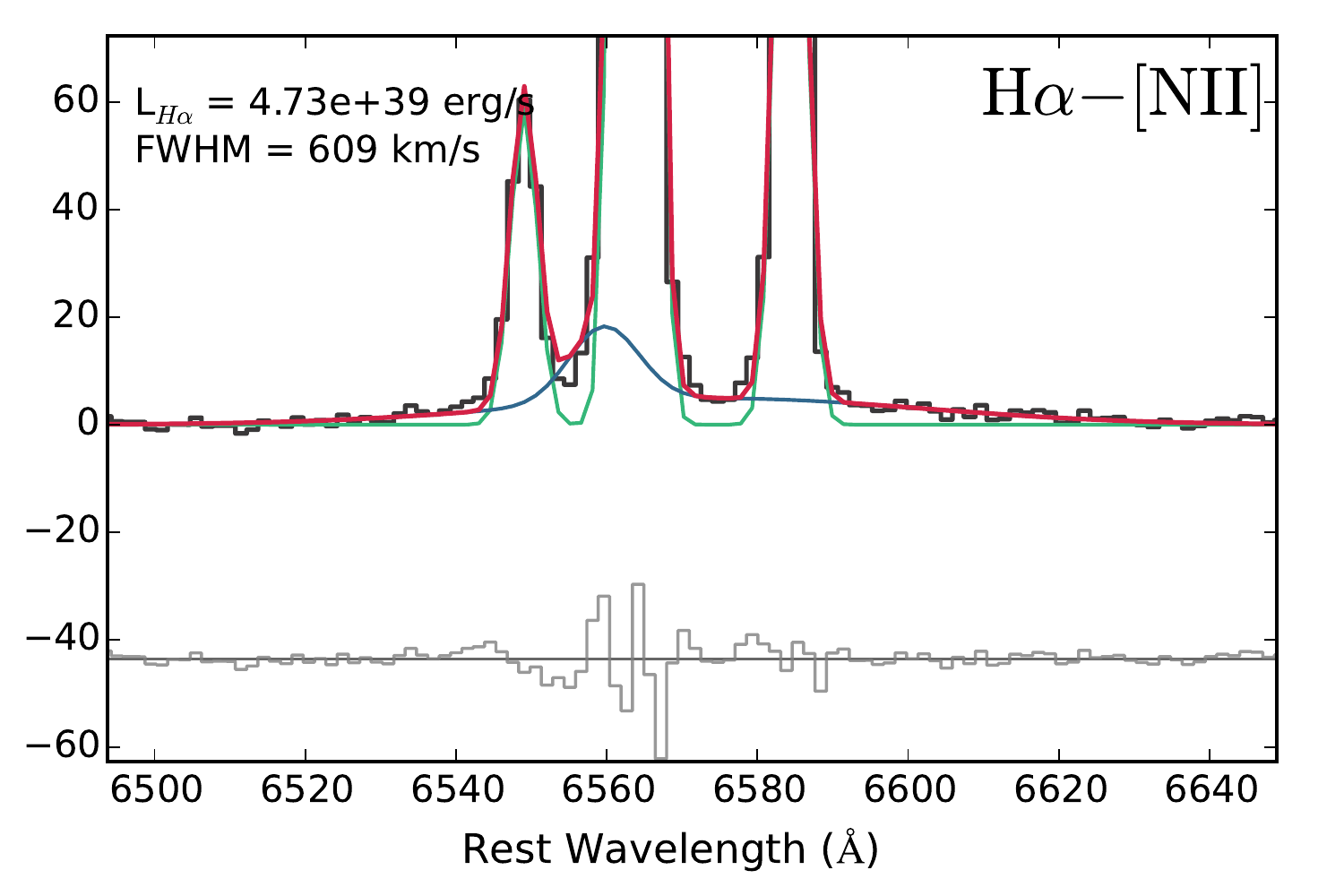}\\

\includegraphics[scale=0.44]{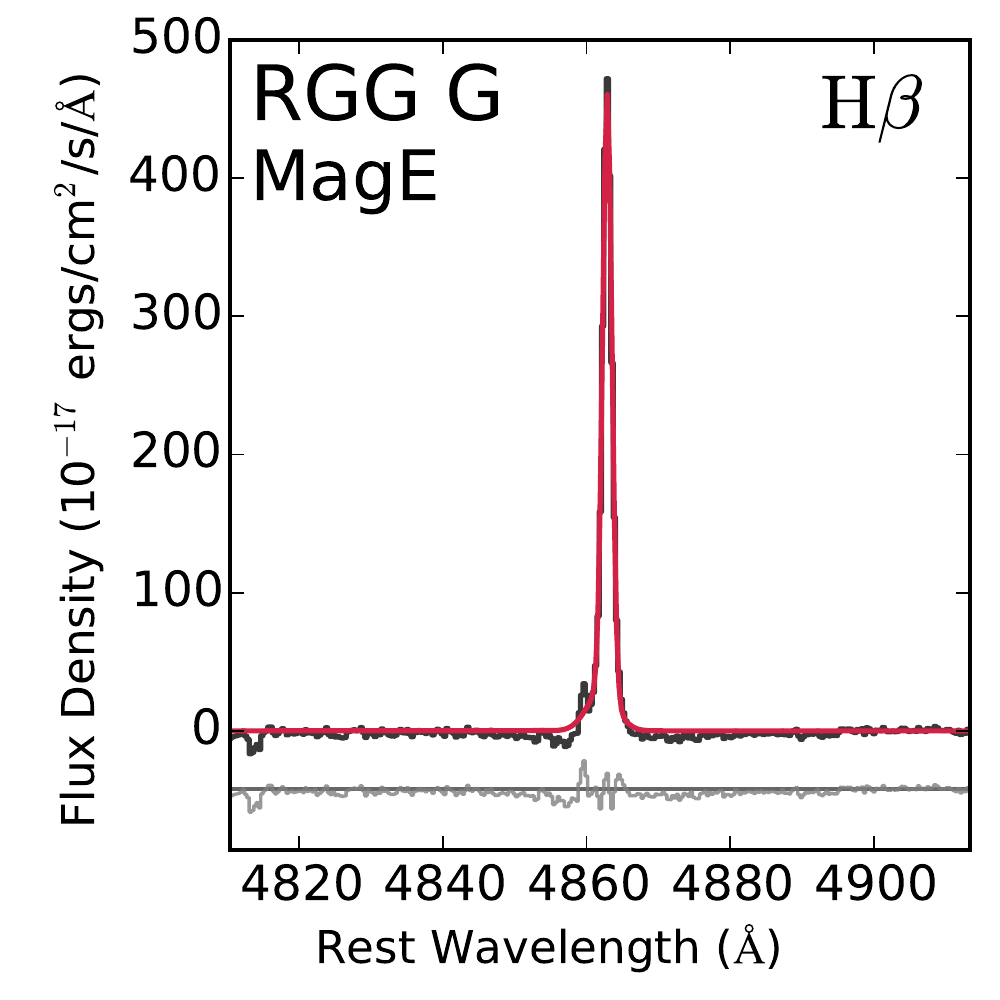}
\includegraphics[scale=0.44]{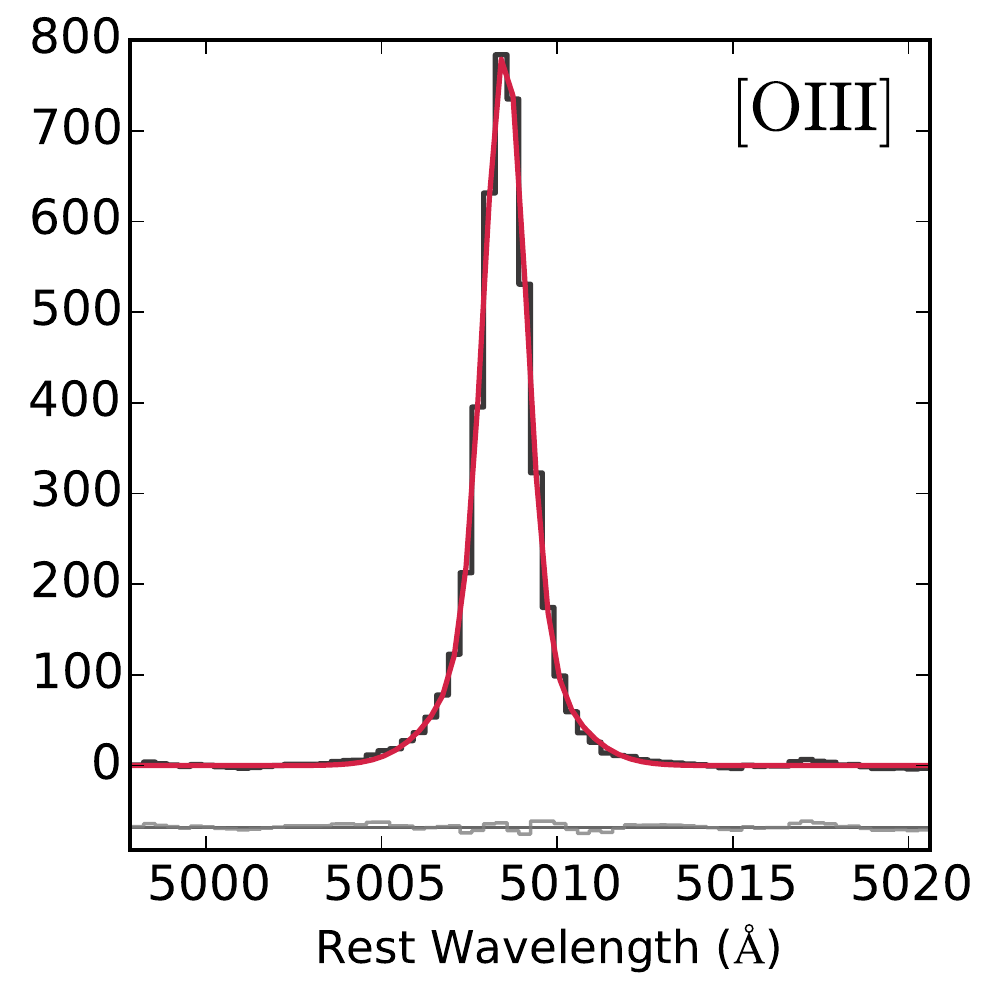}
\includegraphics[scale=0.44]{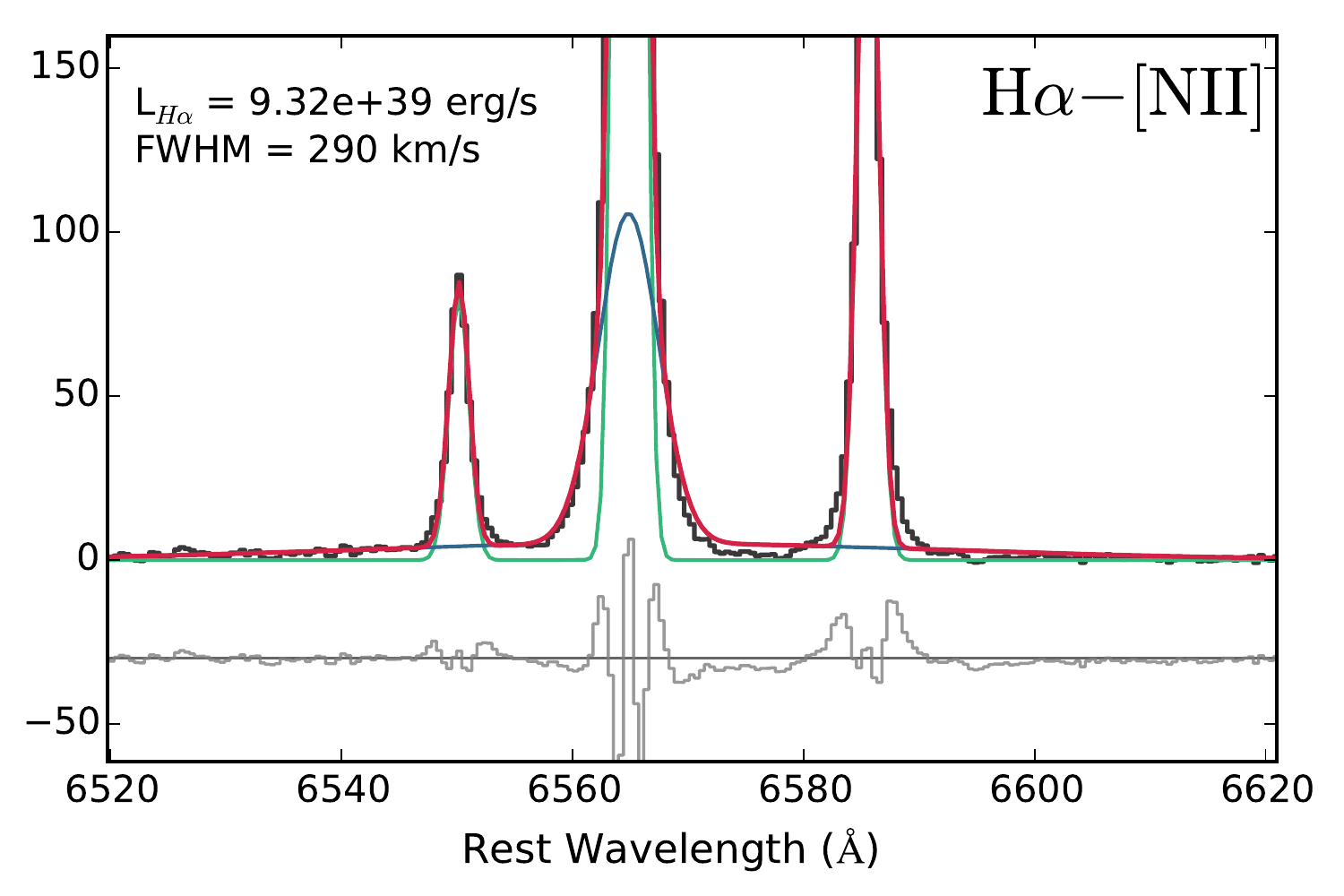}
\caption{These plots show the $\rm H\beta$, [OIII]$\lambda$5007, H$\alpha$, and [NII]$\lambda\lambda6718,6731$ lines for each observation taken of RGG G (NSA 13496). Description is same as for Figure~\ref{nsa15952}. We place this object in the ``transient" category.}
\label{nsa13496}
\end{figure*}

\begin{figure*}
\centering
\includegraphics[scale=0.44]{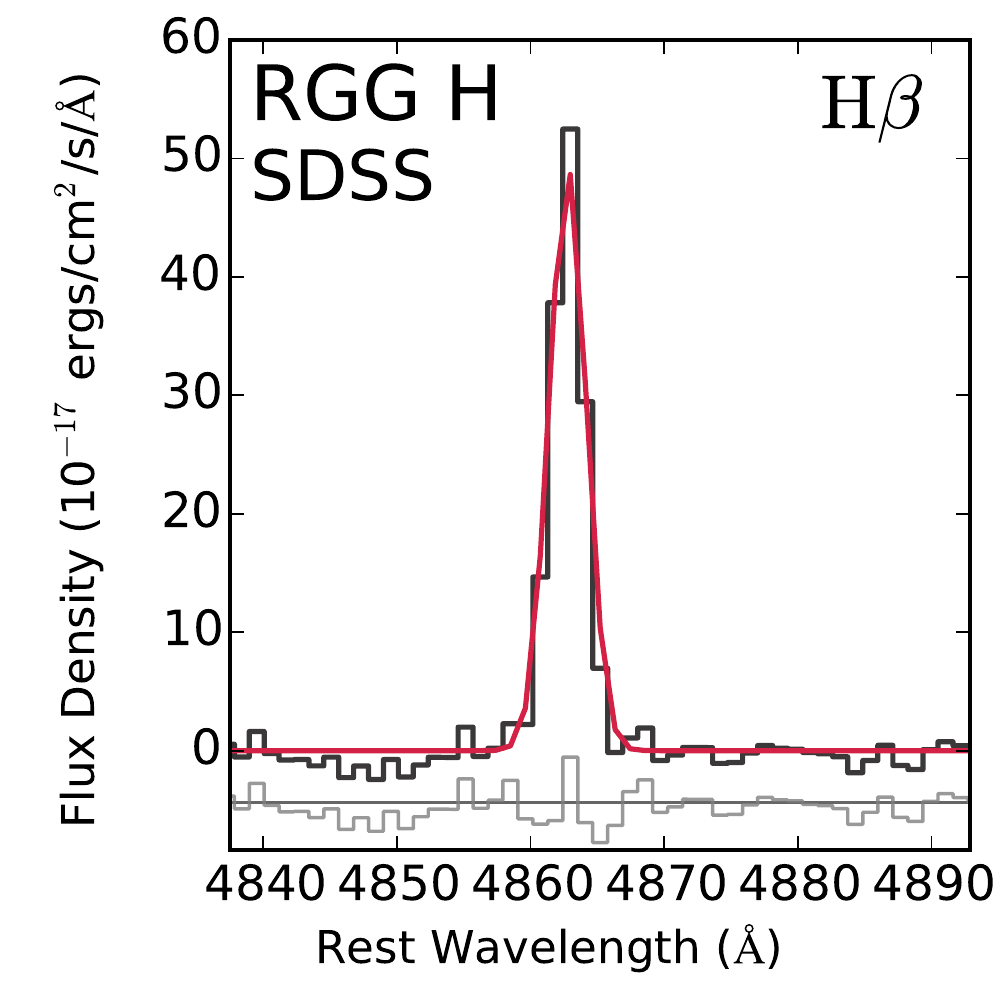}
\includegraphics[scale=0.44]{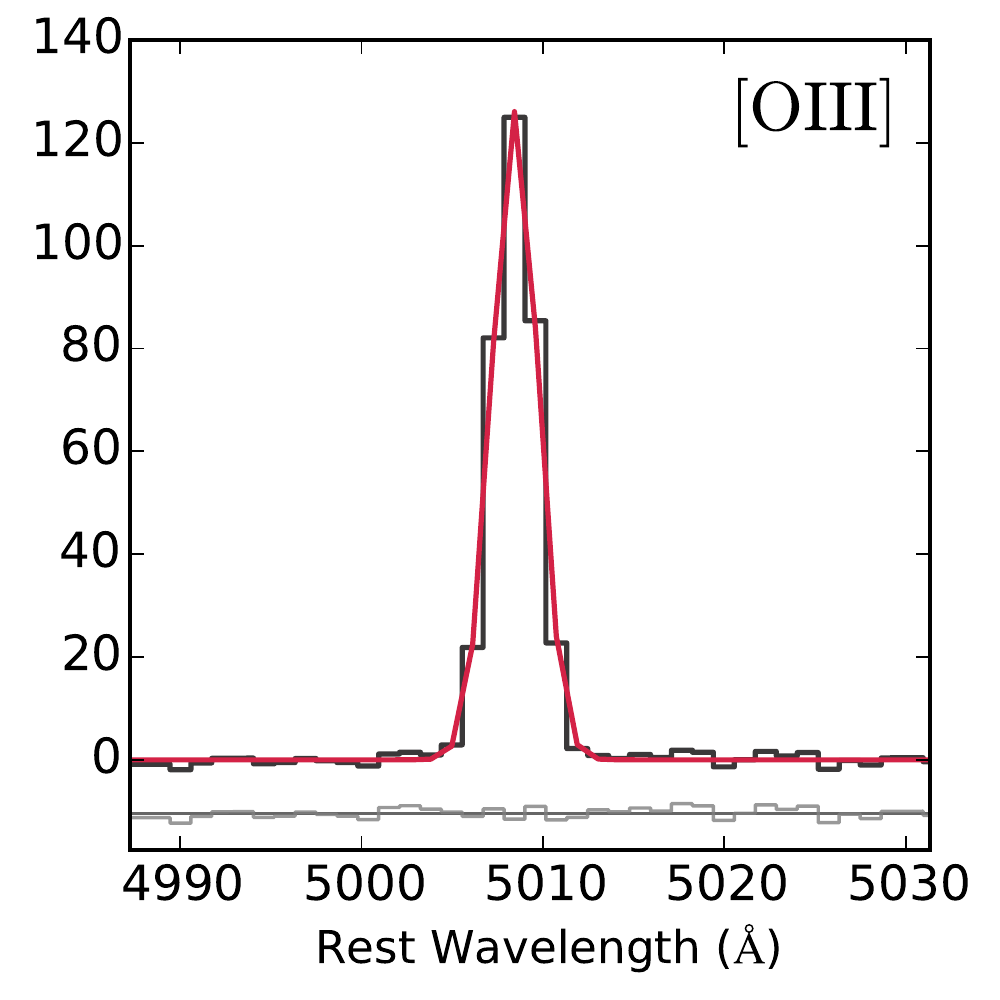}
\includegraphics[scale=0.44]{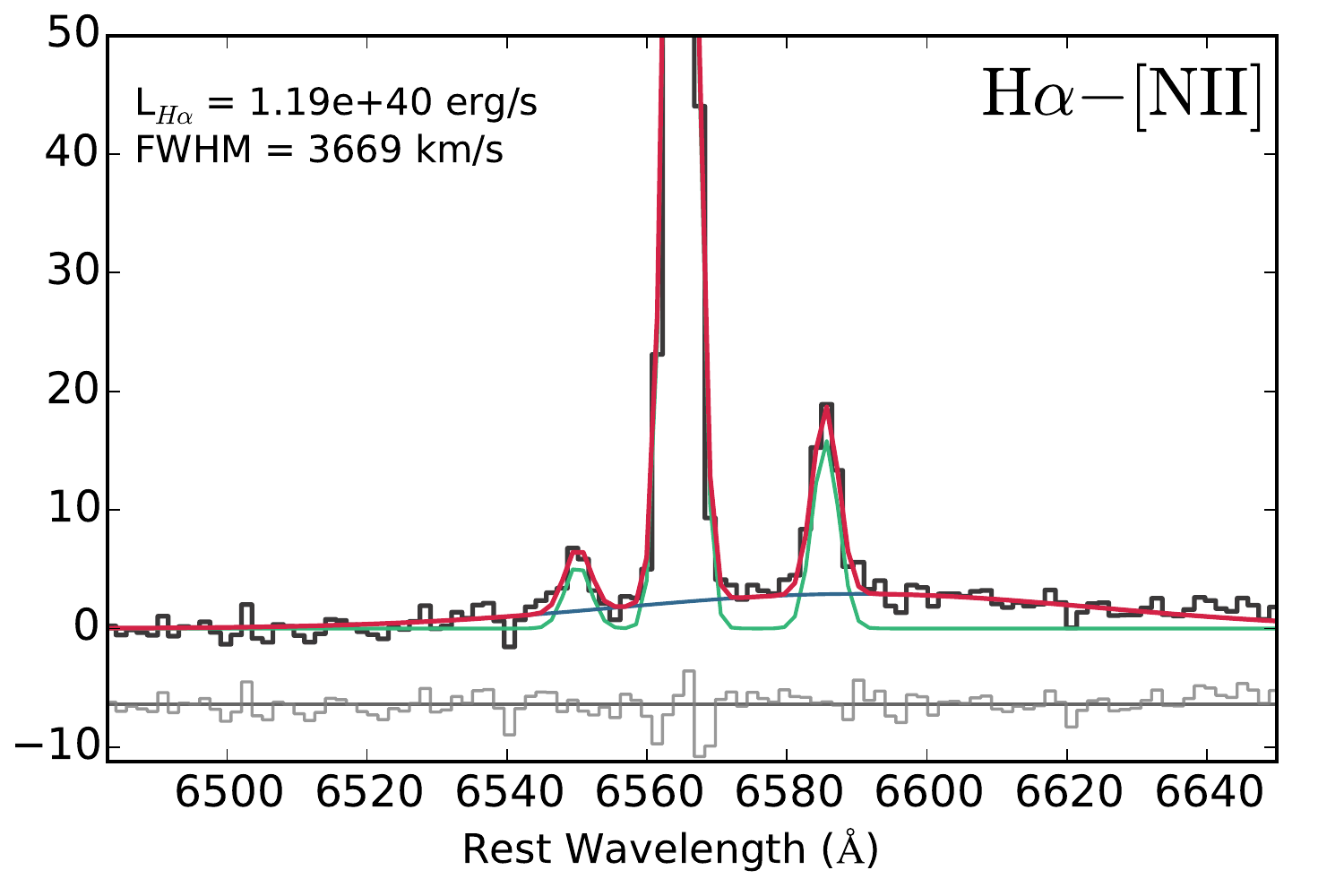}\\

\includegraphics[scale=0.44]{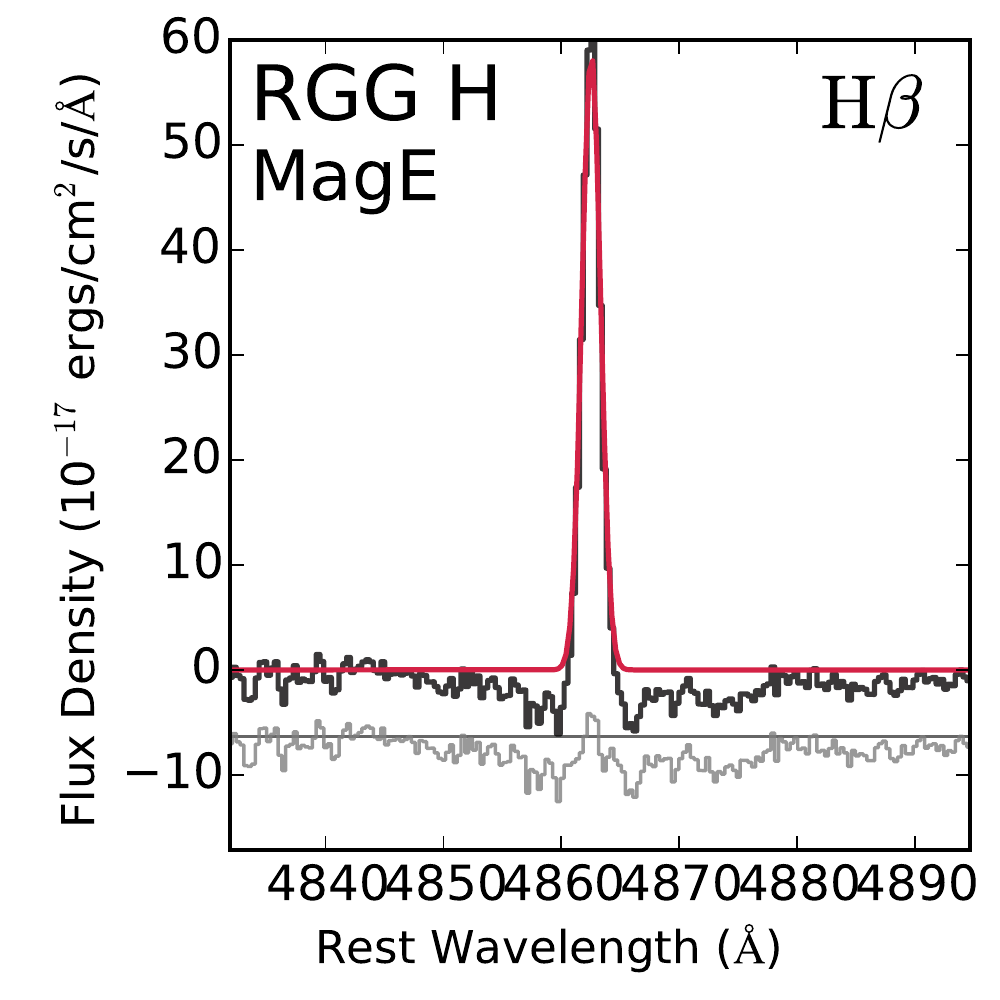}
\includegraphics[scale=0.44]{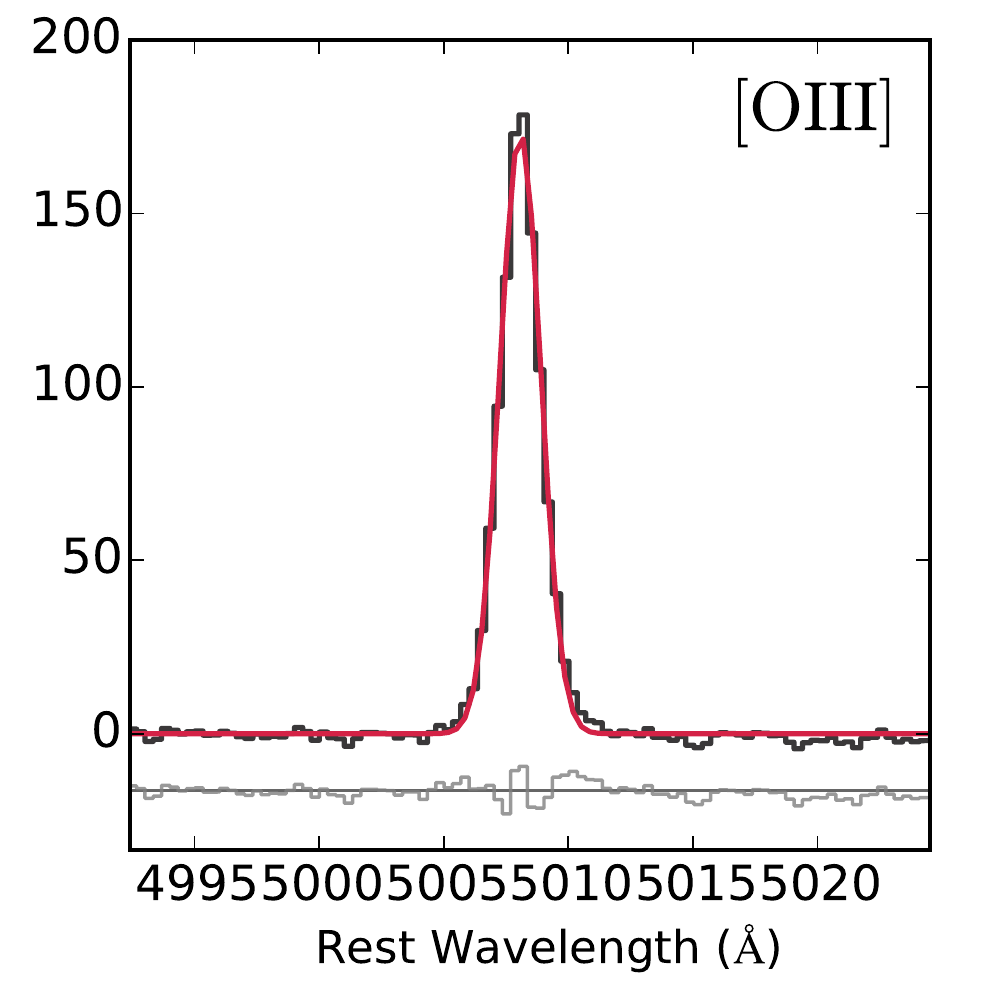}
\includegraphics[scale=0.44]{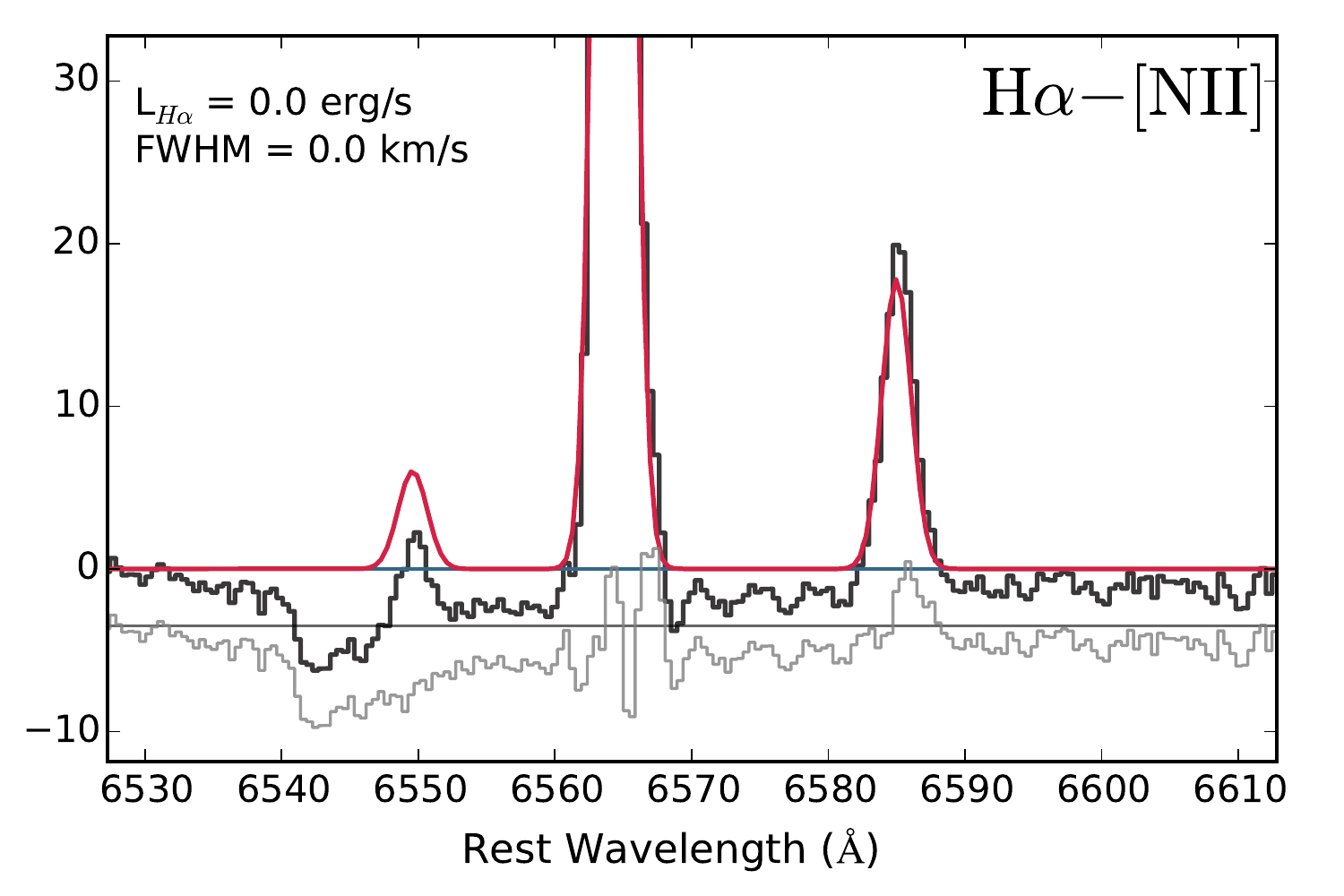}
\caption{These plots show the $\rm H\beta$, [OIII]$\lambda$5007, H$\alpha$, and [NII]$\lambda\lambda6718,6731$ lines for each observation taken of RGG H (NSA 74914). Description is same as for Figure~\ref{nsa15952}. We place this object in the ``transient" category.}
\label{nsa74914}
\end{figure*}

\begin{figure*}
\centering
\includegraphics[scale=0.44]{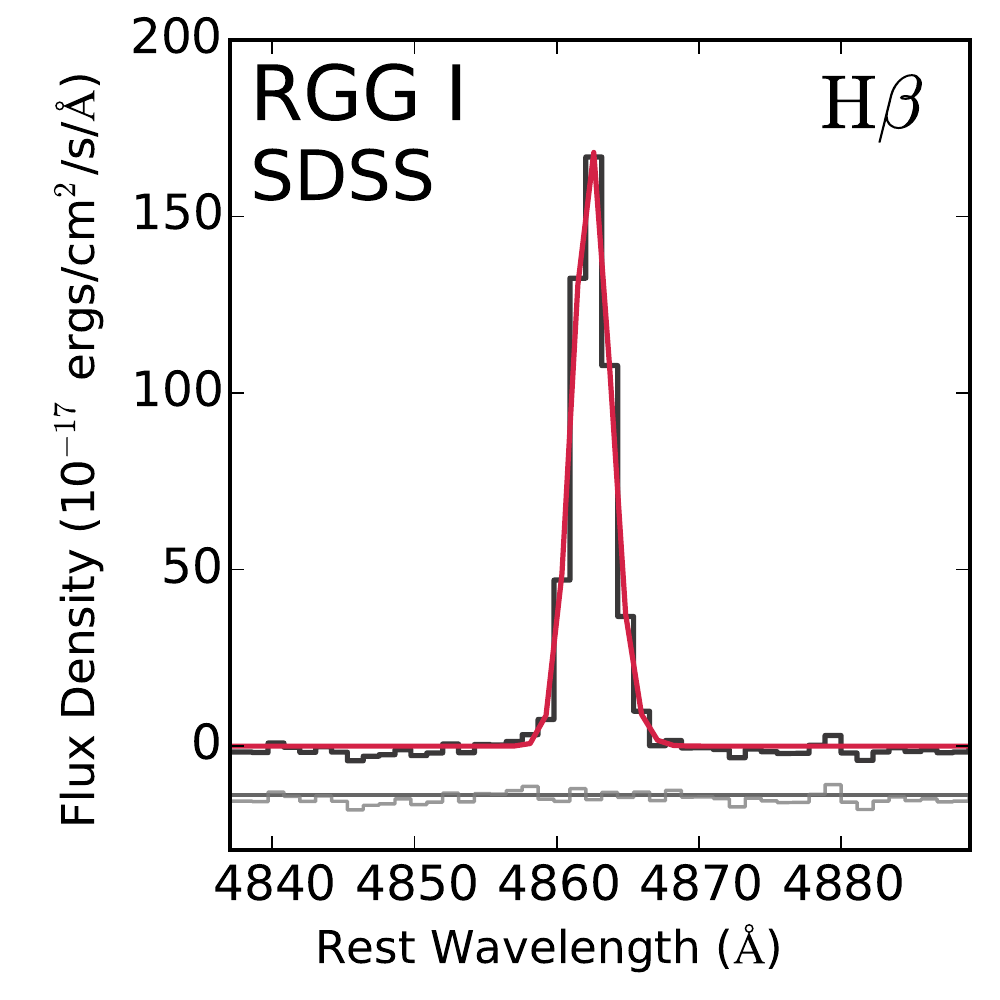}
\includegraphics[scale=0.44]{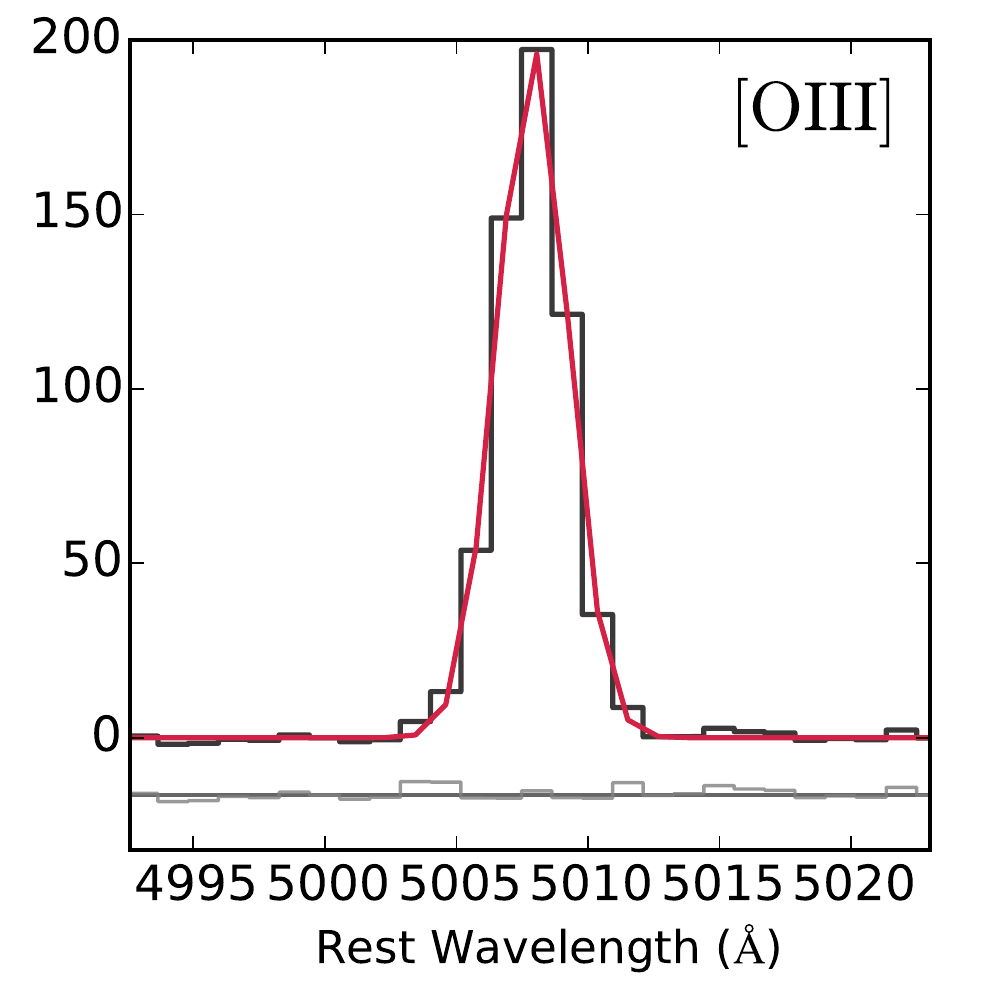}
\includegraphics[scale=0.44]{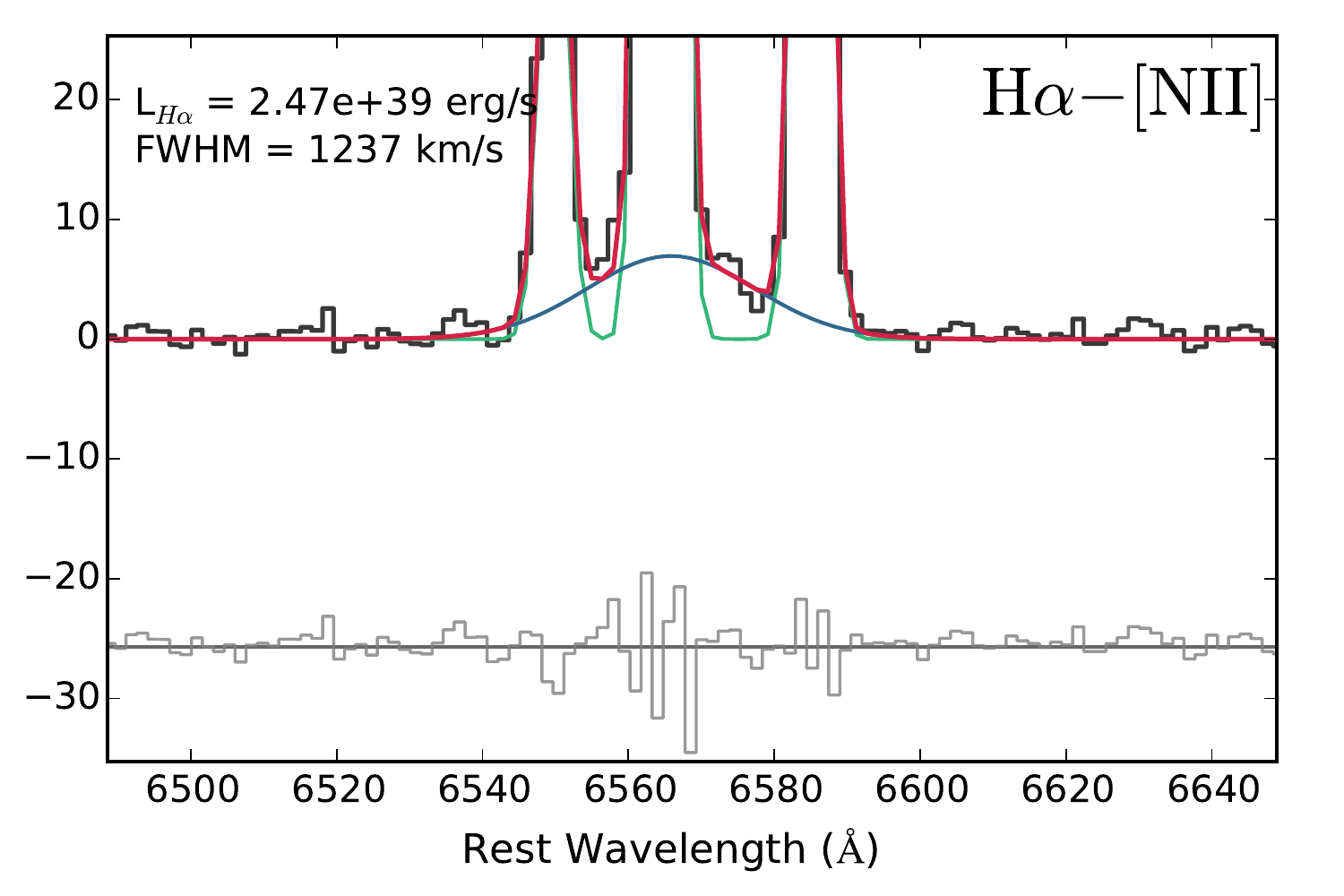}\\

\includegraphics[scale=0.44]{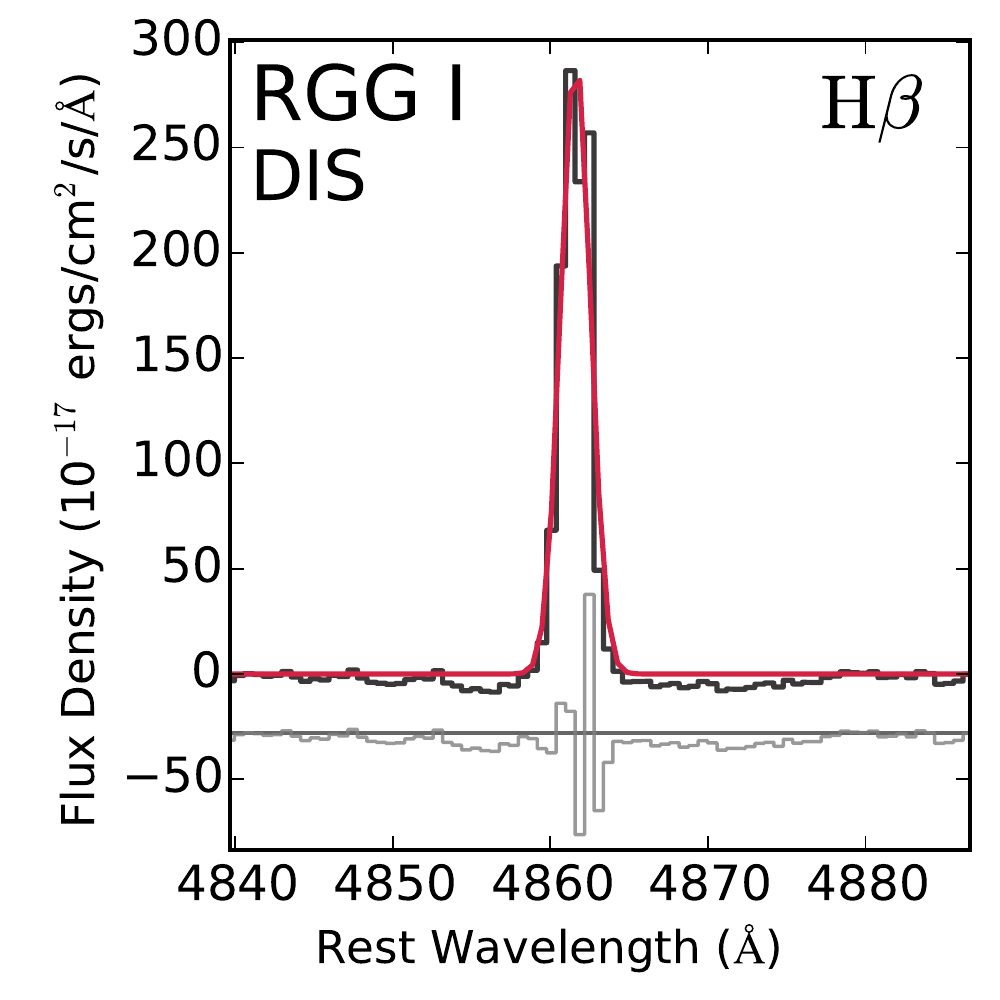}
\includegraphics[scale=0.44]{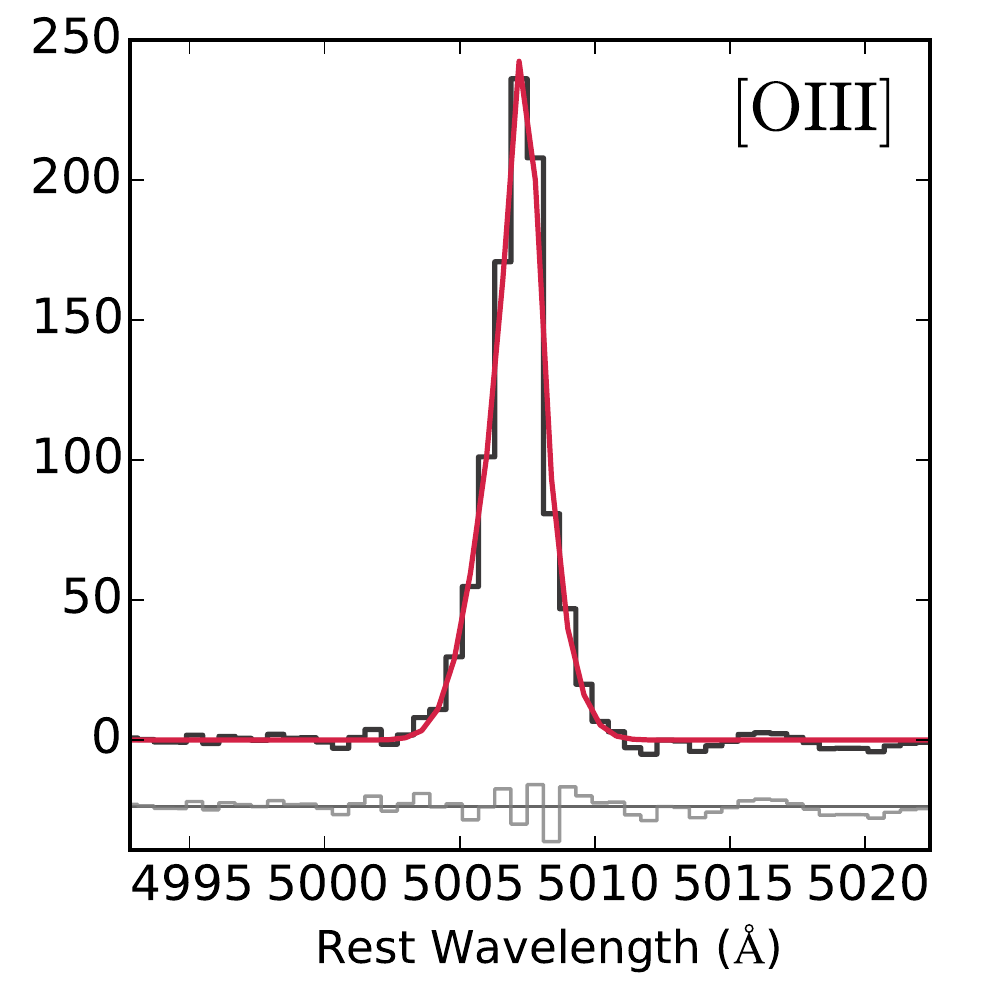}
\includegraphics[scale=0.44]{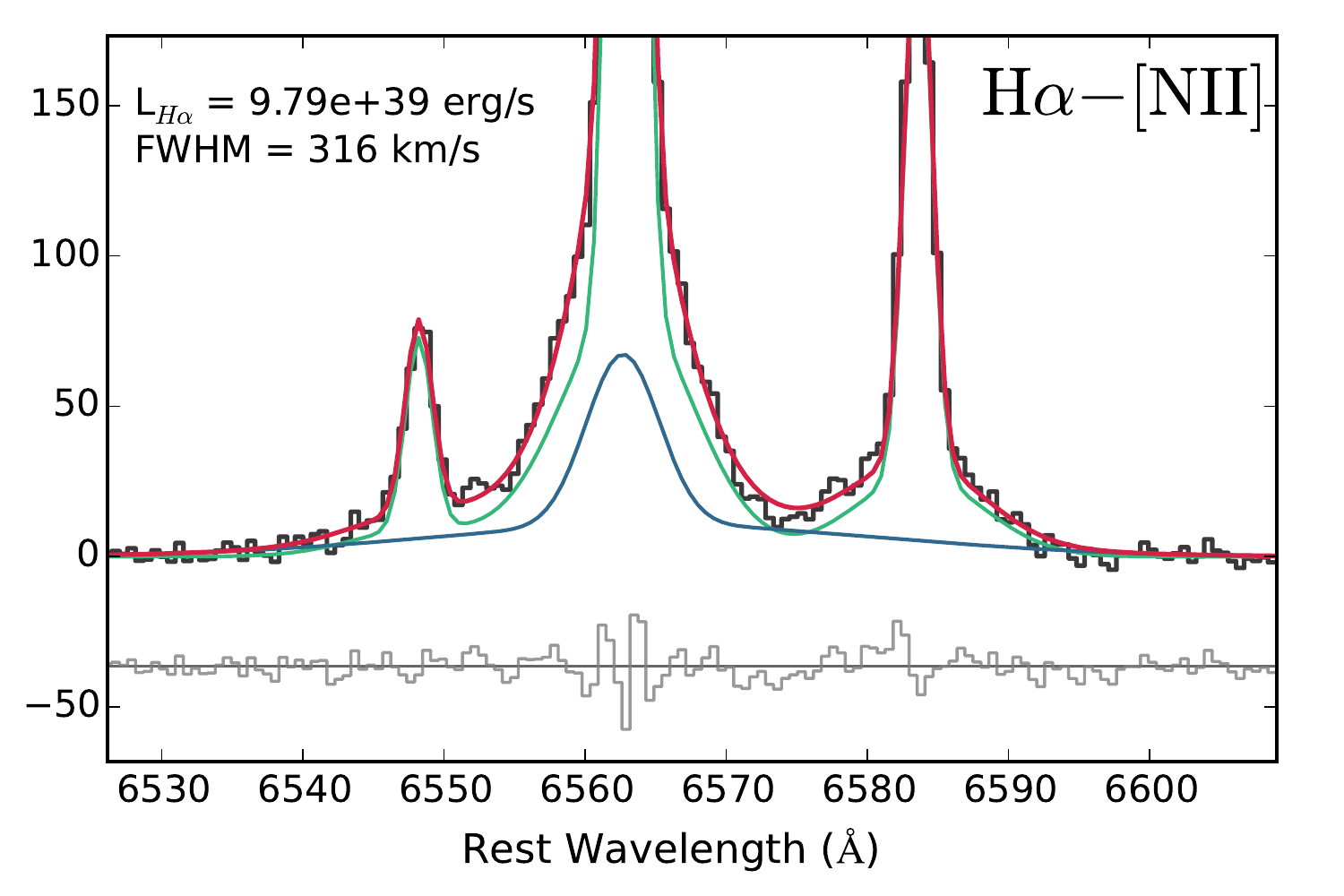}

\includegraphics[scale=0.44]{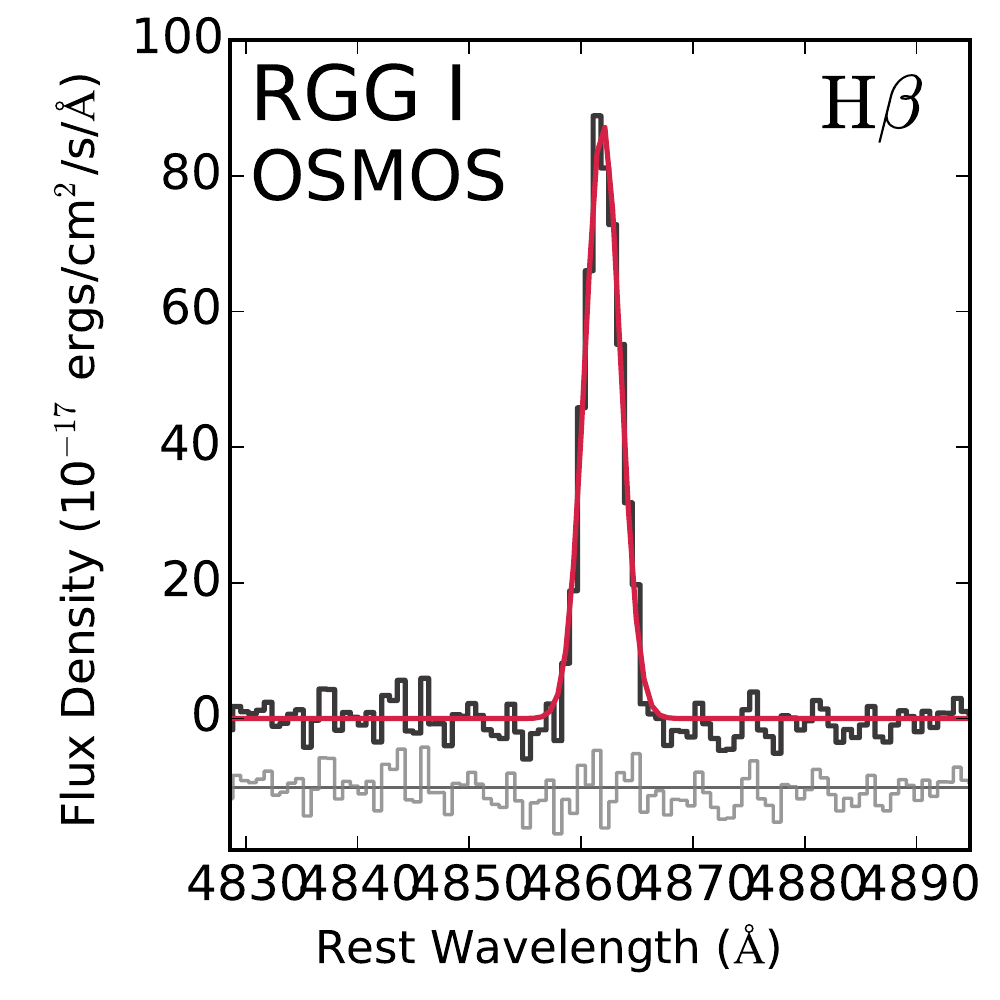}
\includegraphics[scale=0.44]{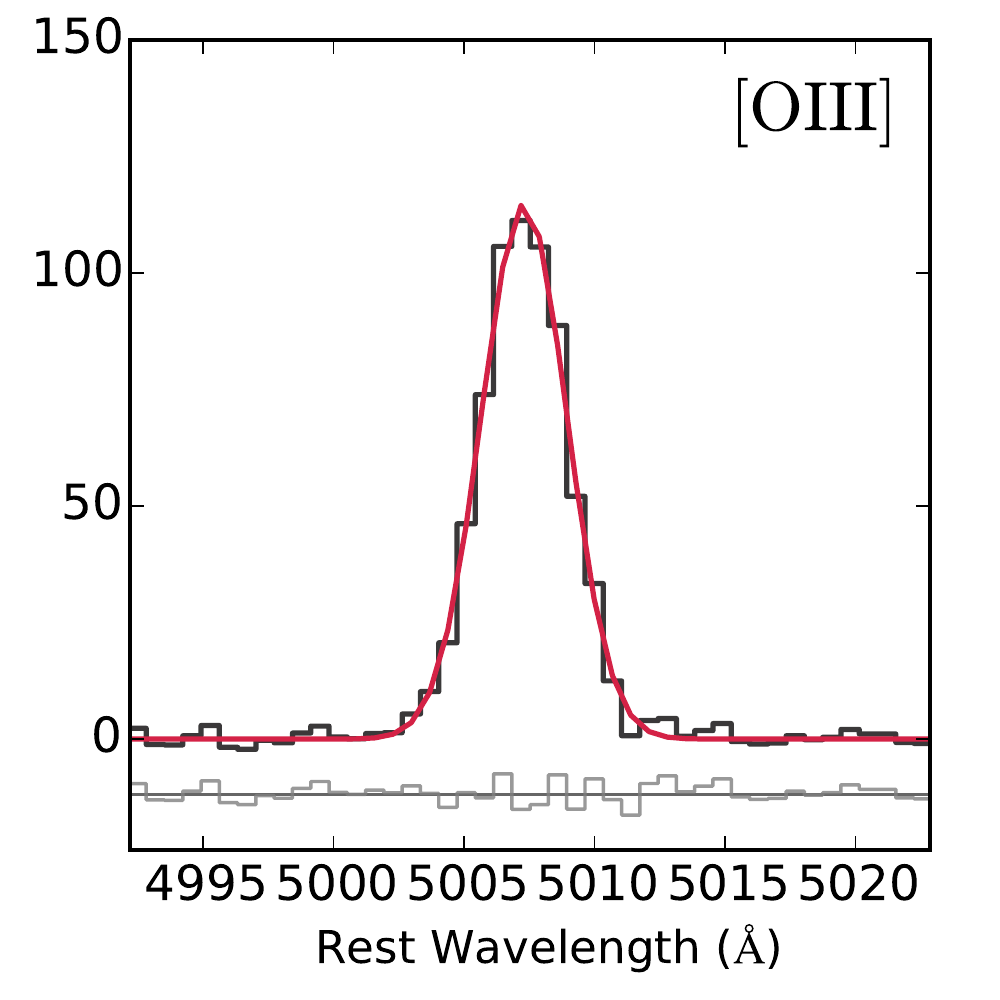}
\includegraphics[scale=0.44]{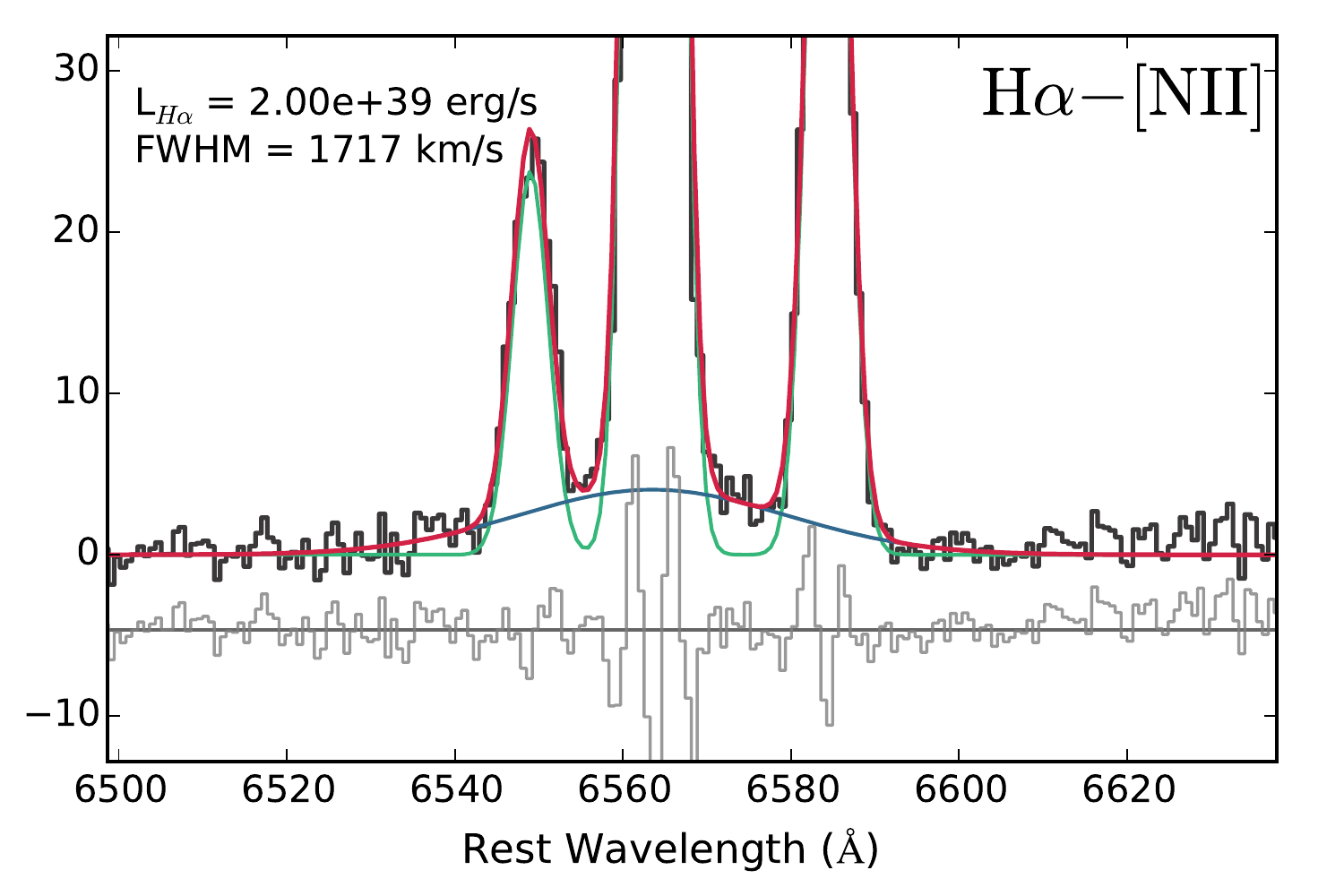}
\caption{These plots show the $\rm H\beta$, [OIII]$\lambda$5007, H$\alpha$, and [NII]$\lambda\lambda6718,6731$ lines for each observation taken of RGG I (NSA 112250). Description is same as for Figure~\ref{nsa15952}. We place this object in the ``ambiguous" category.}
\label{nsa112250}
\end{figure*}

\begin{figure*}
\centering
\includegraphics[scale=0.44]{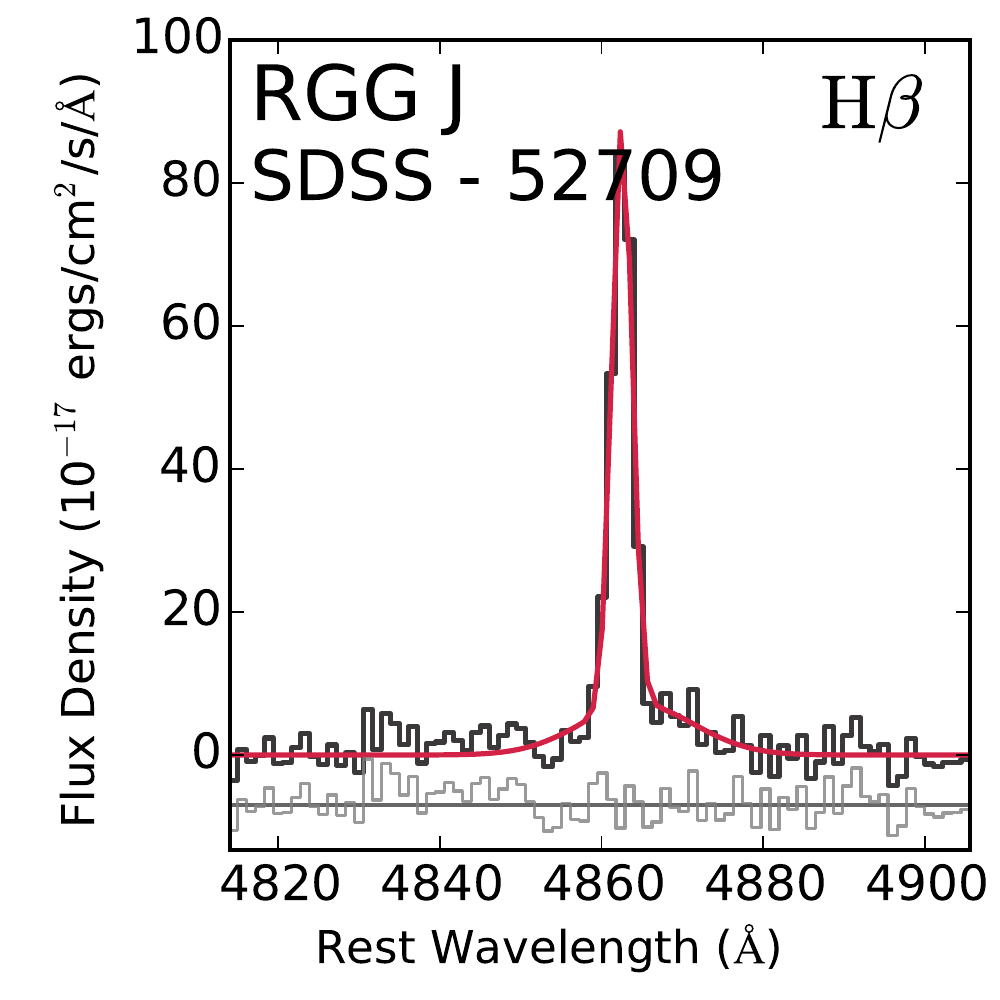}
\includegraphics[scale=0.44]{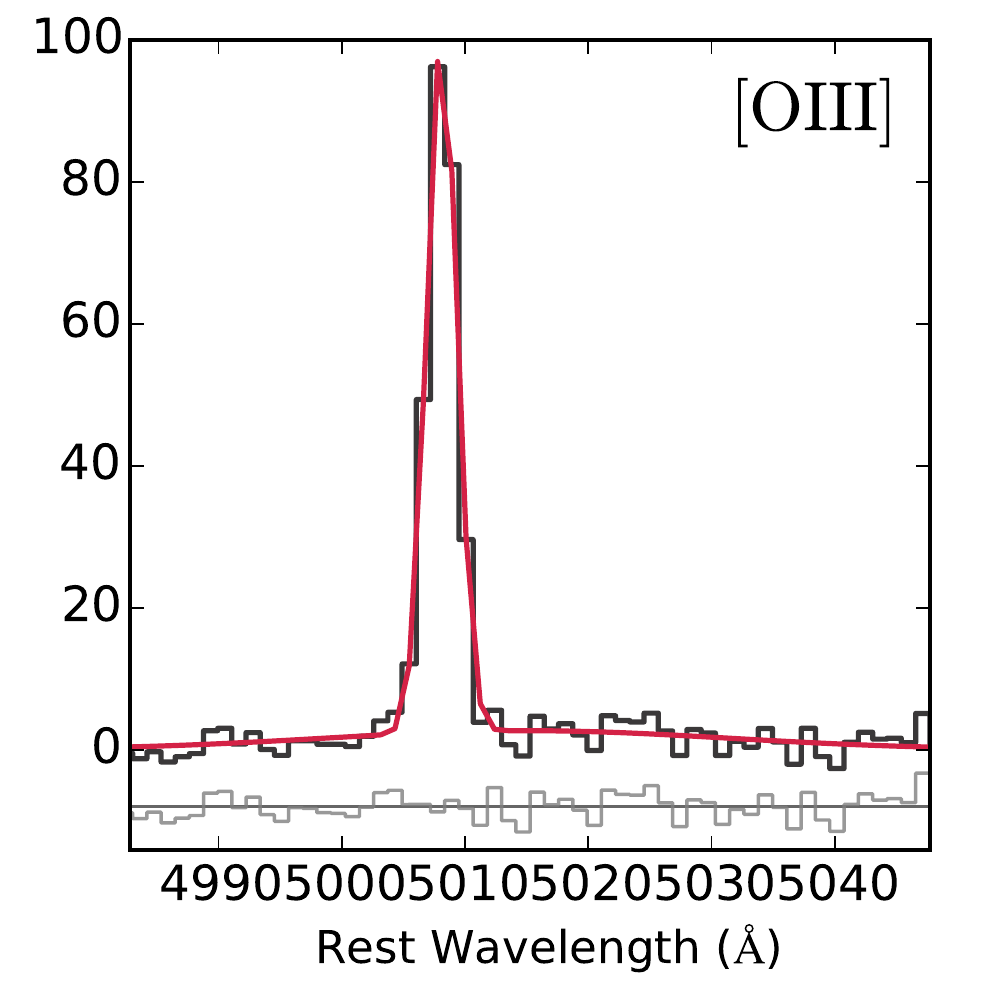}
\includegraphics[scale=0.44]{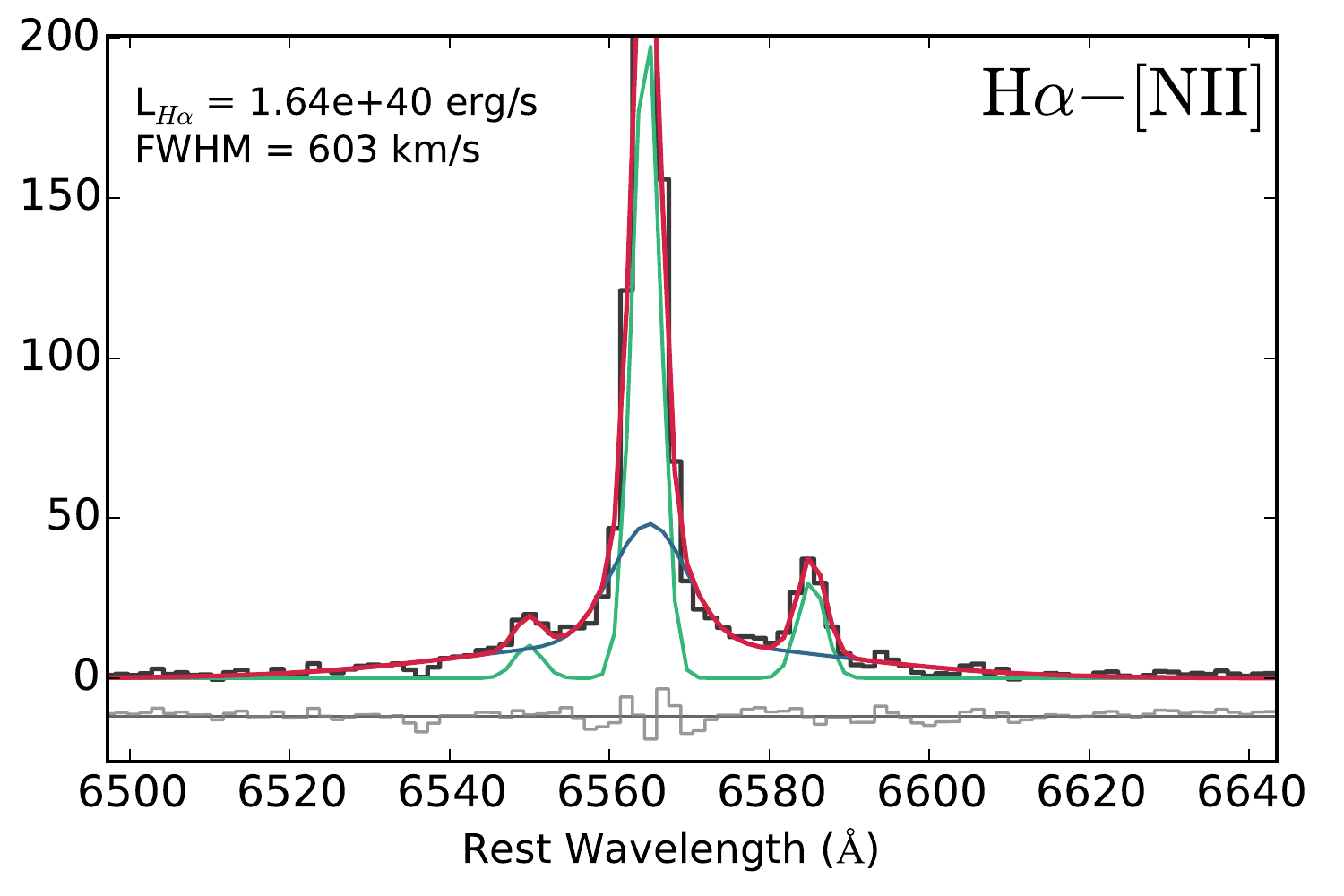}\\

\includegraphics[scale=0.44]{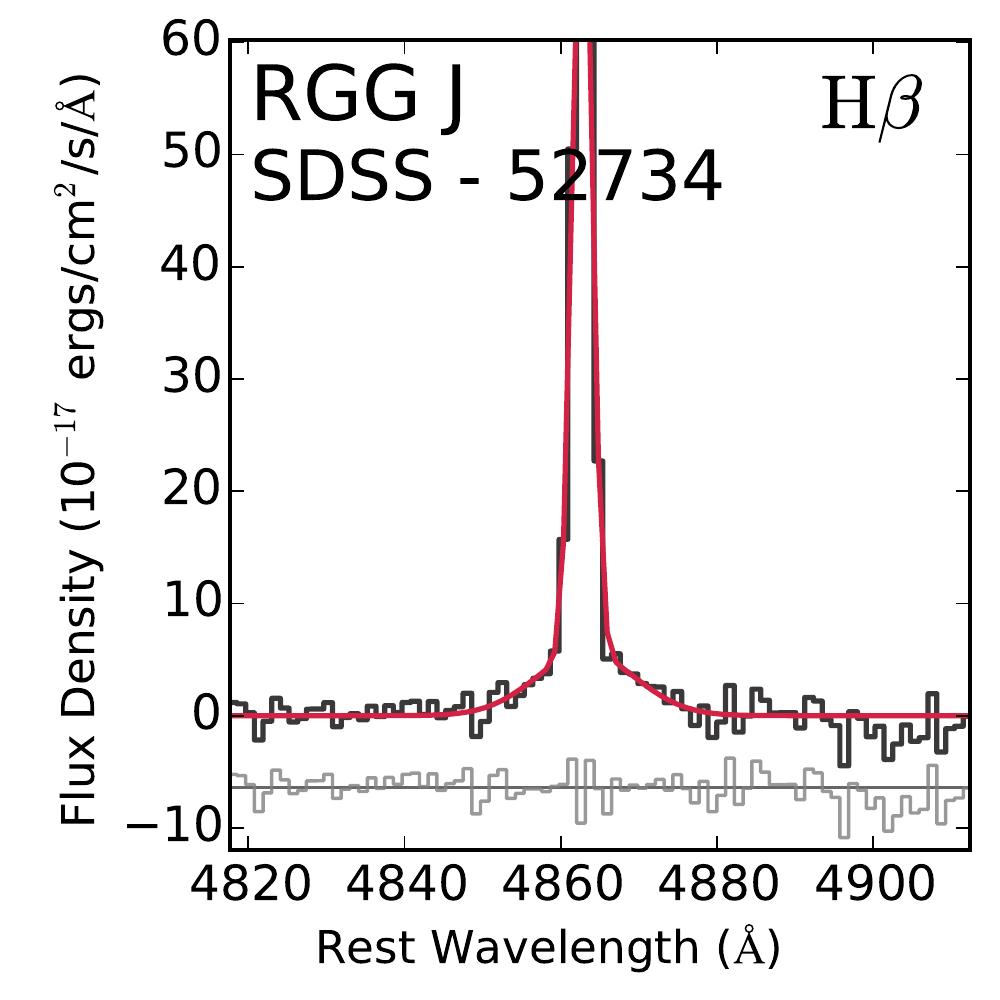}
\includegraphics[scale=0.44]{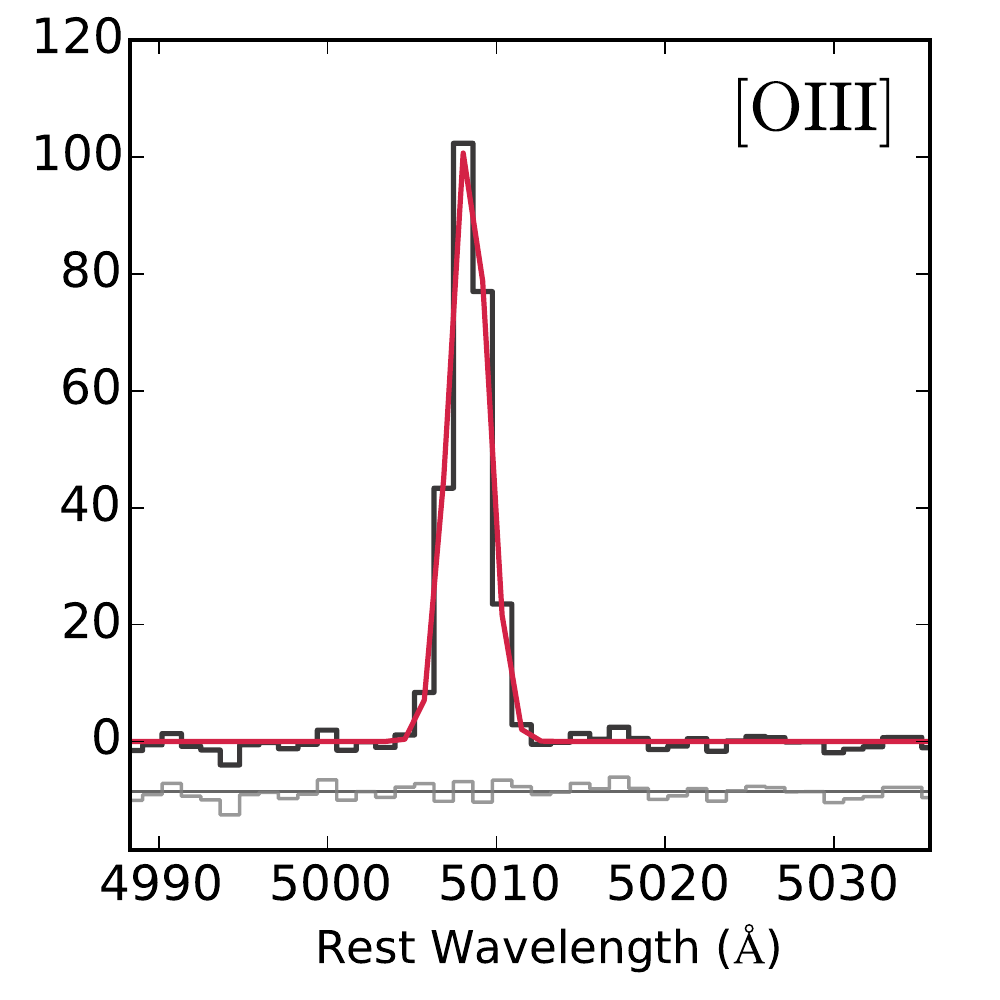}
\includegraphics[scale=0.44]{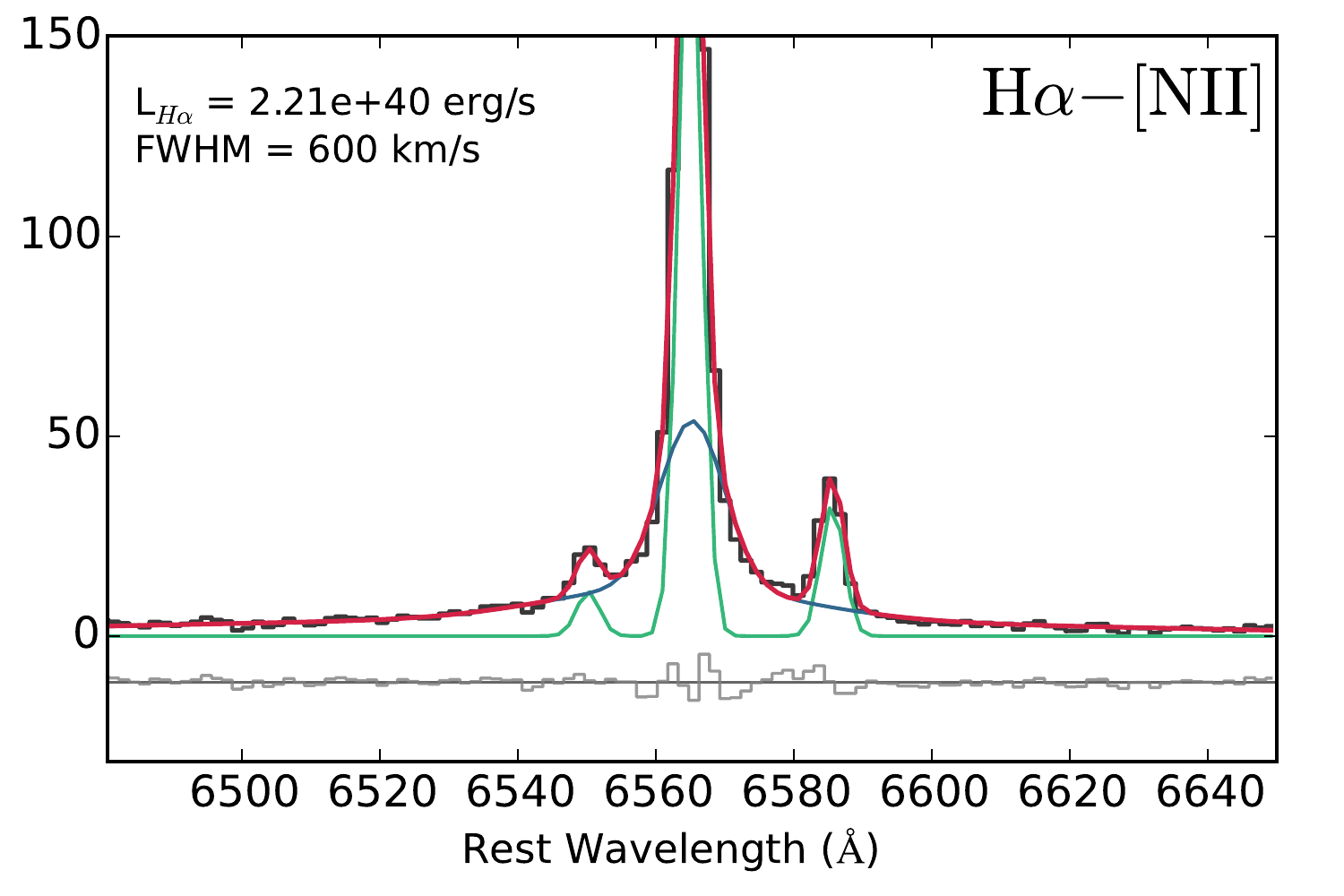}\\

\includegraphics[scale=0.44]{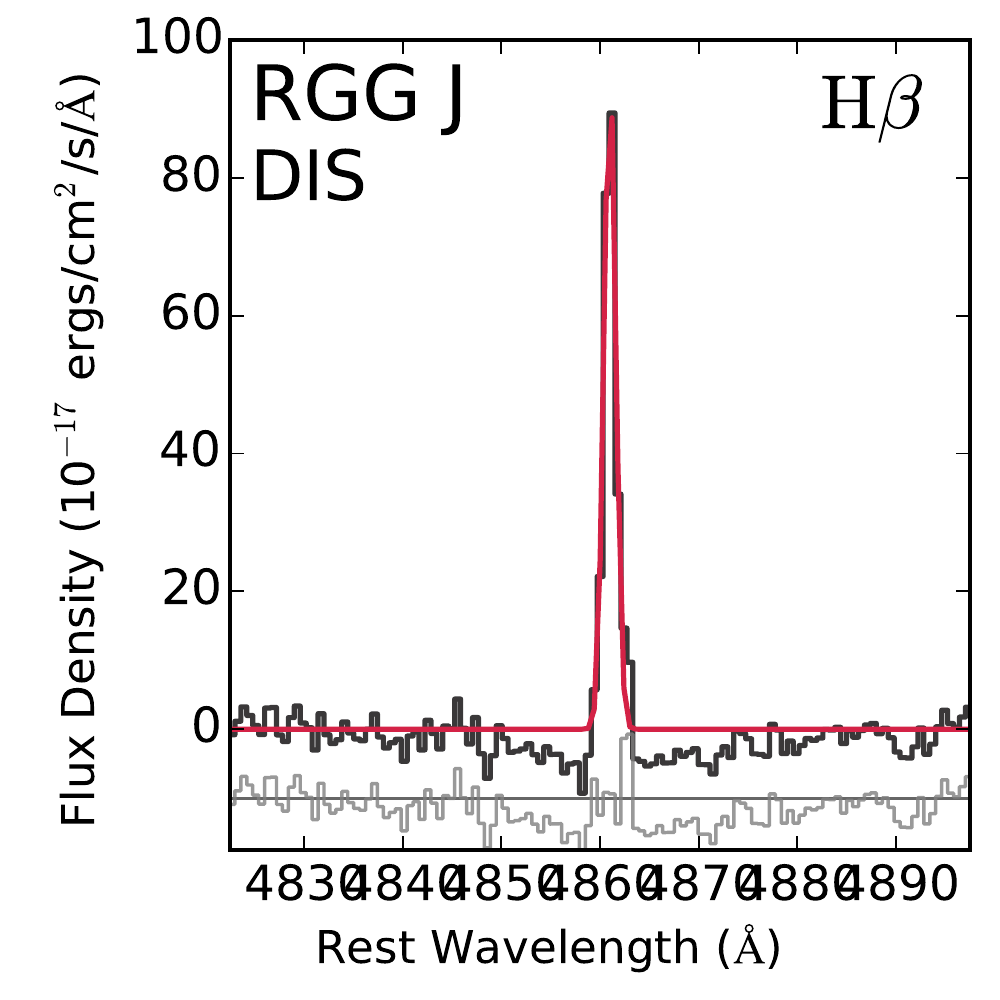}
\includegraphics[scale=0.44]{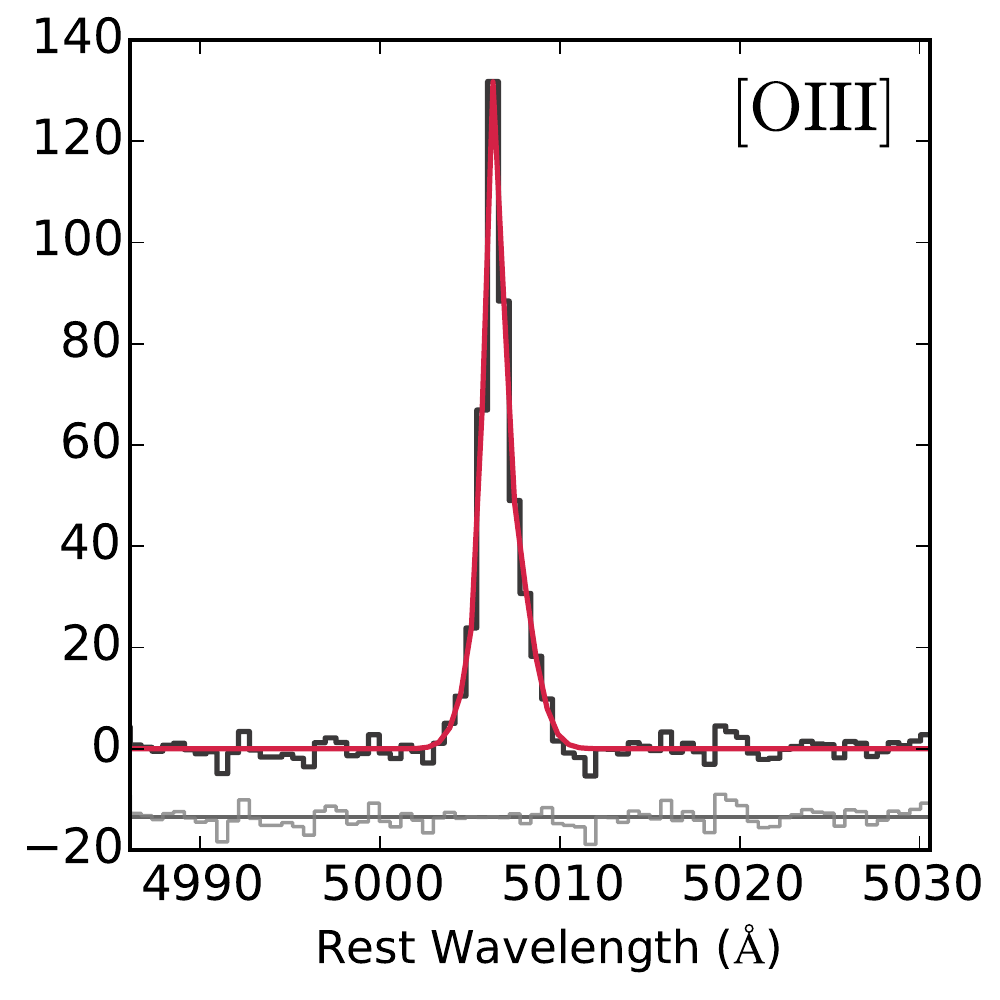}
\includegraphics[scale=0.44]{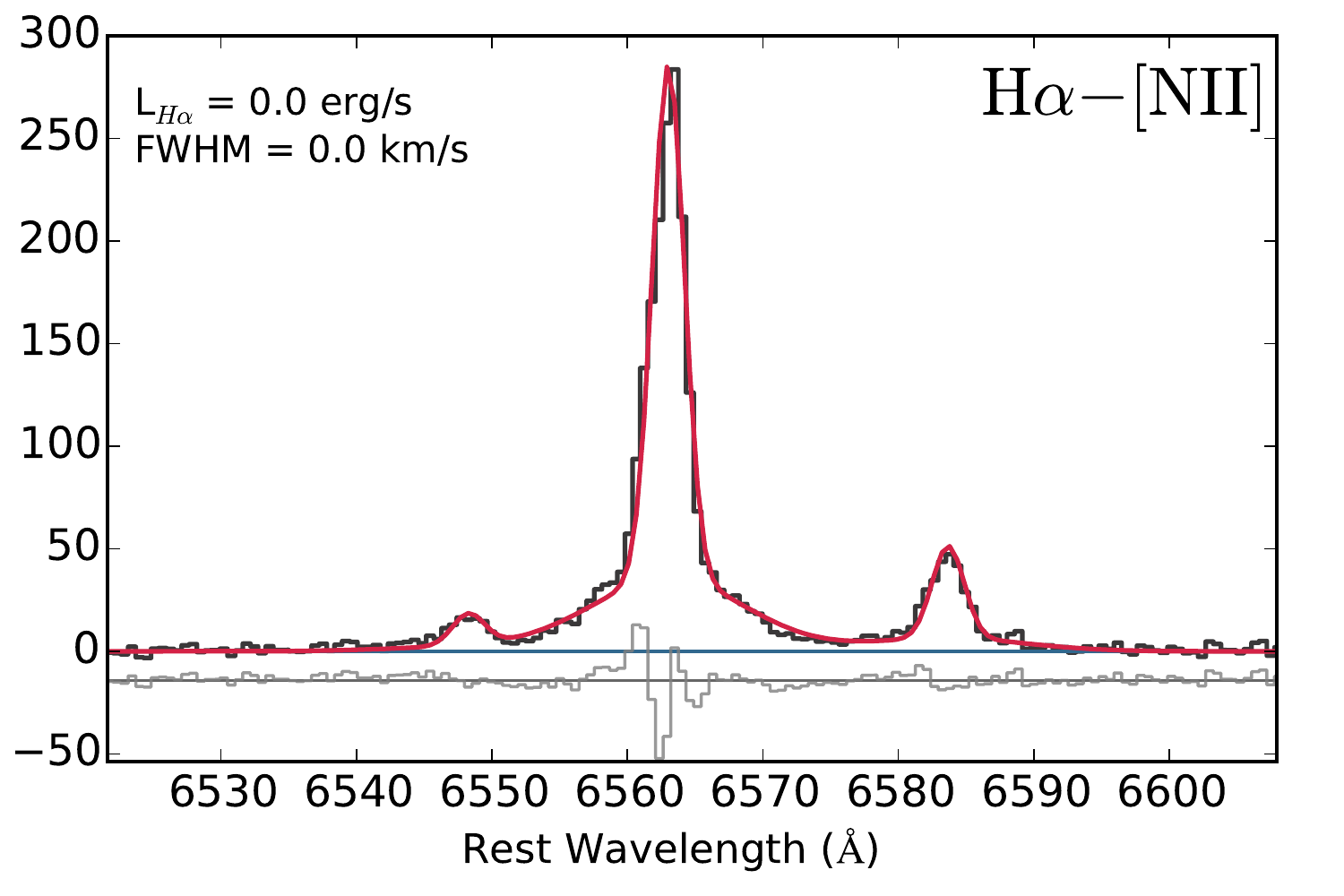}
\caption{These plots show the $\rm H\beta$, [OIII]$\lambda$5007, H$\alpha$, and [NII]$\lambda\lambda6718,6731$ lines for each observation taken of RGG J (NSA 41331). Description is same as for Figure~\ref{nsa15952}. This object has two SDSS observations (taken on MJD 52709 and 52734). We show both, and they are each labeled with their MJD. We place this object in the ``transient" category.}
\label{nsa41331}
\end{figure*}

\begin{figure*}
\centering
\includegraphics[scale=0.44]{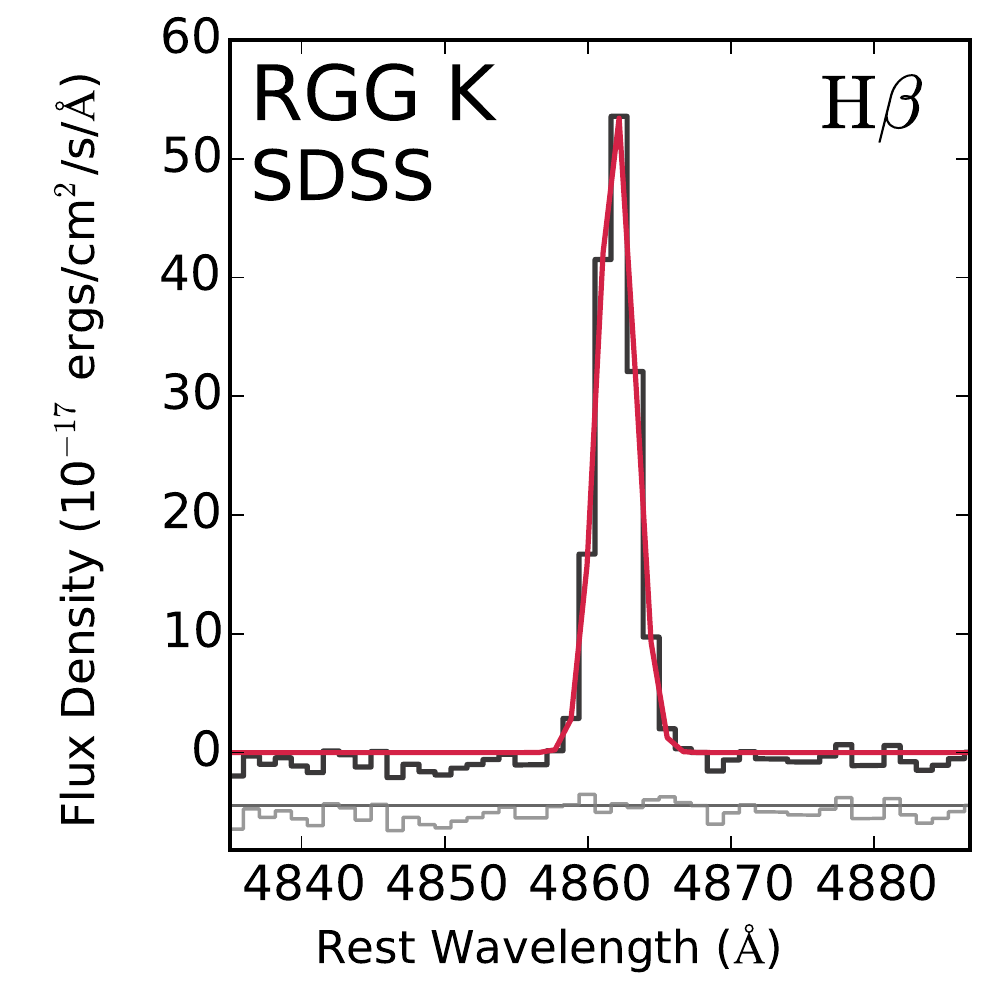}
\includegraphics[scale=0.44]{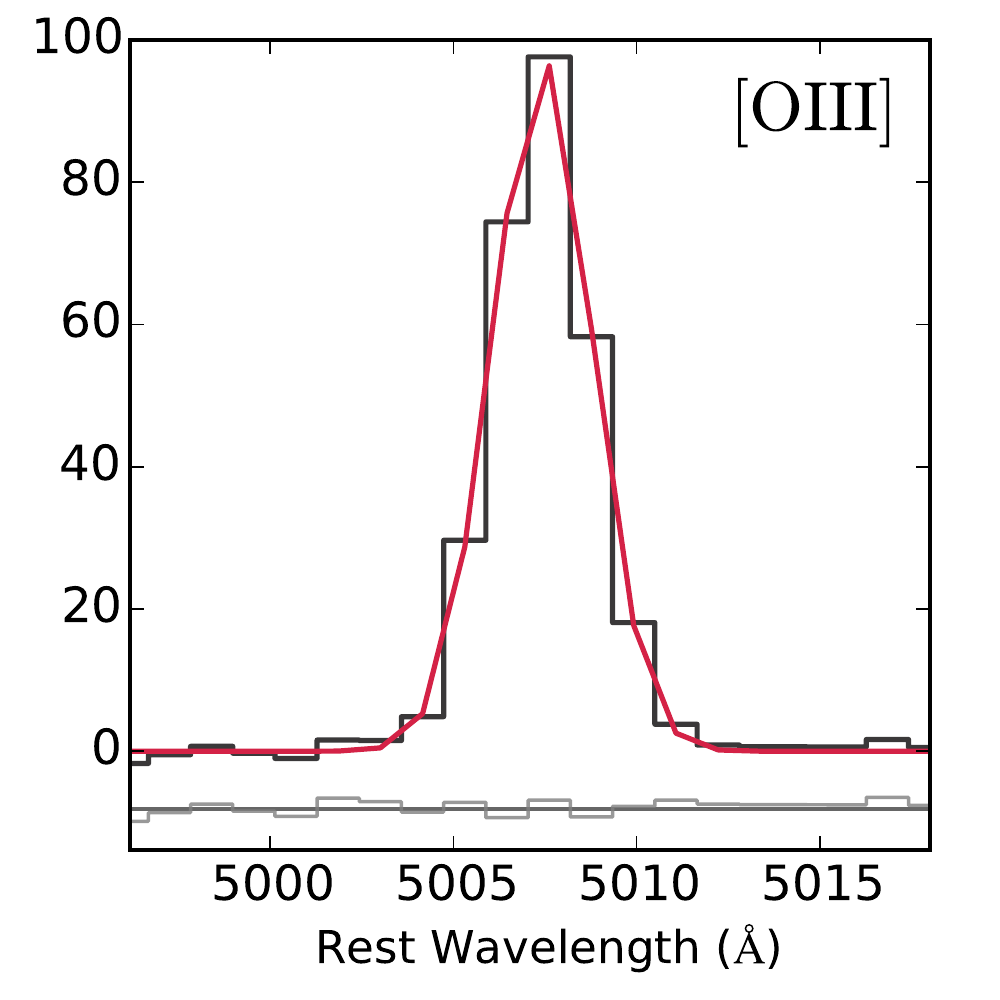}
\includegraphics[scale=0.44]{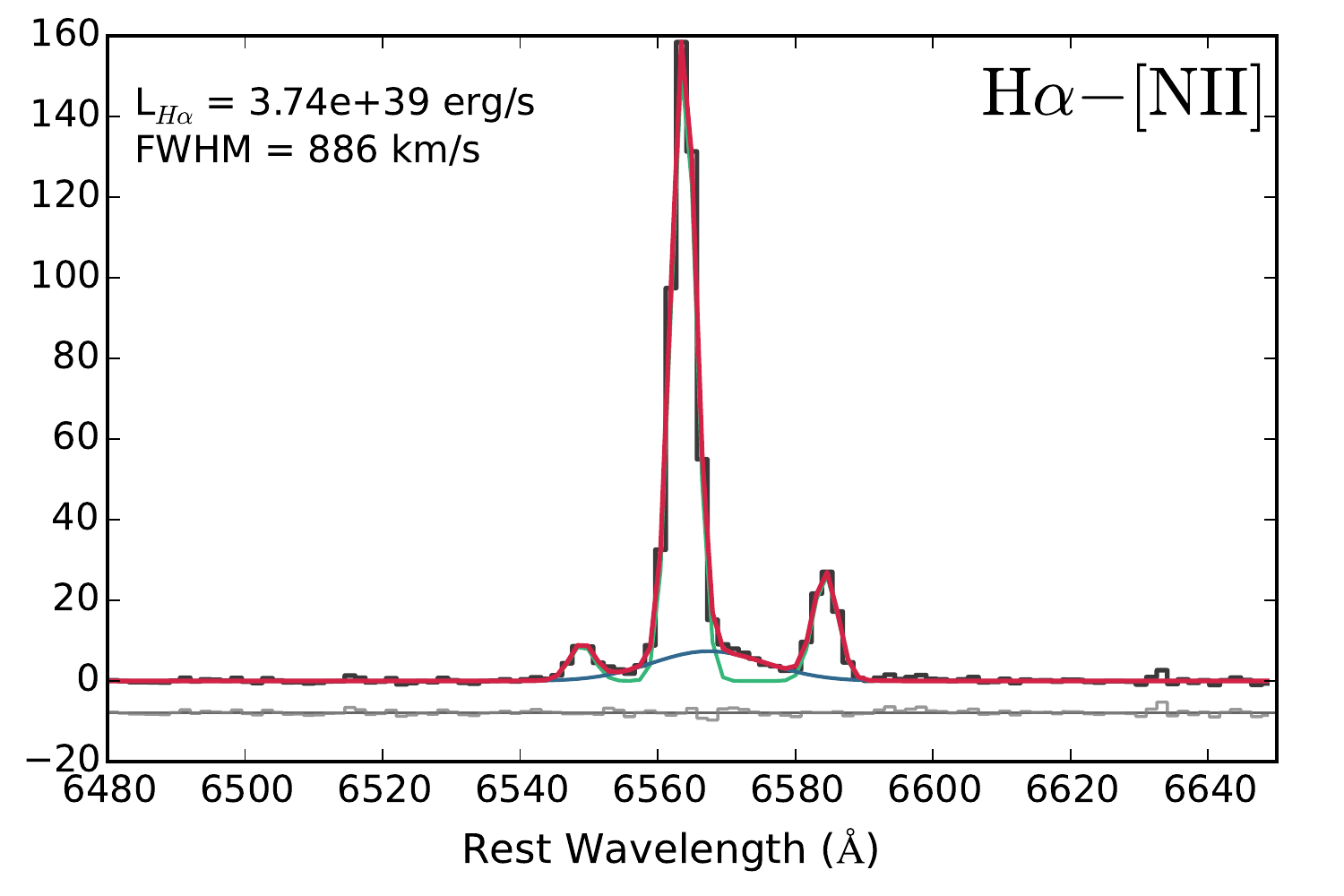}\\

\includegraphics[scale=0.44]{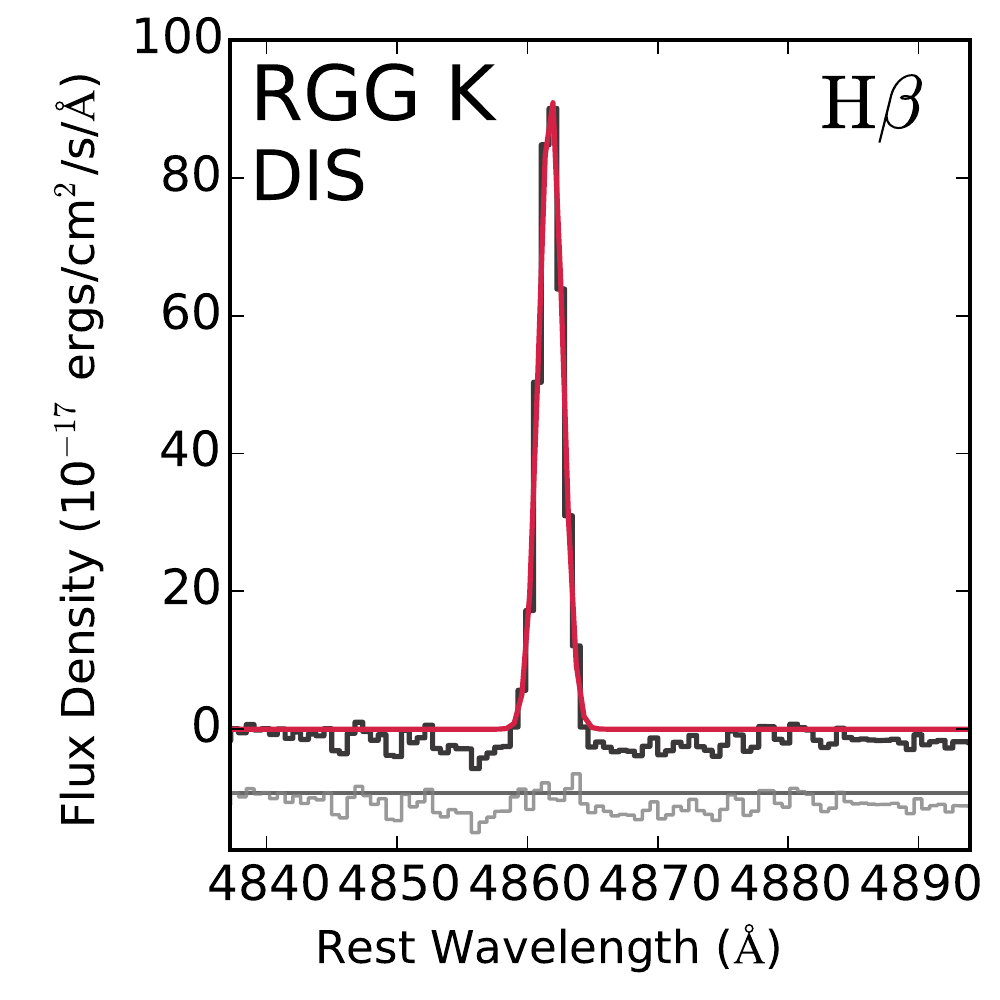}
\includegraphics[scale=0.44]{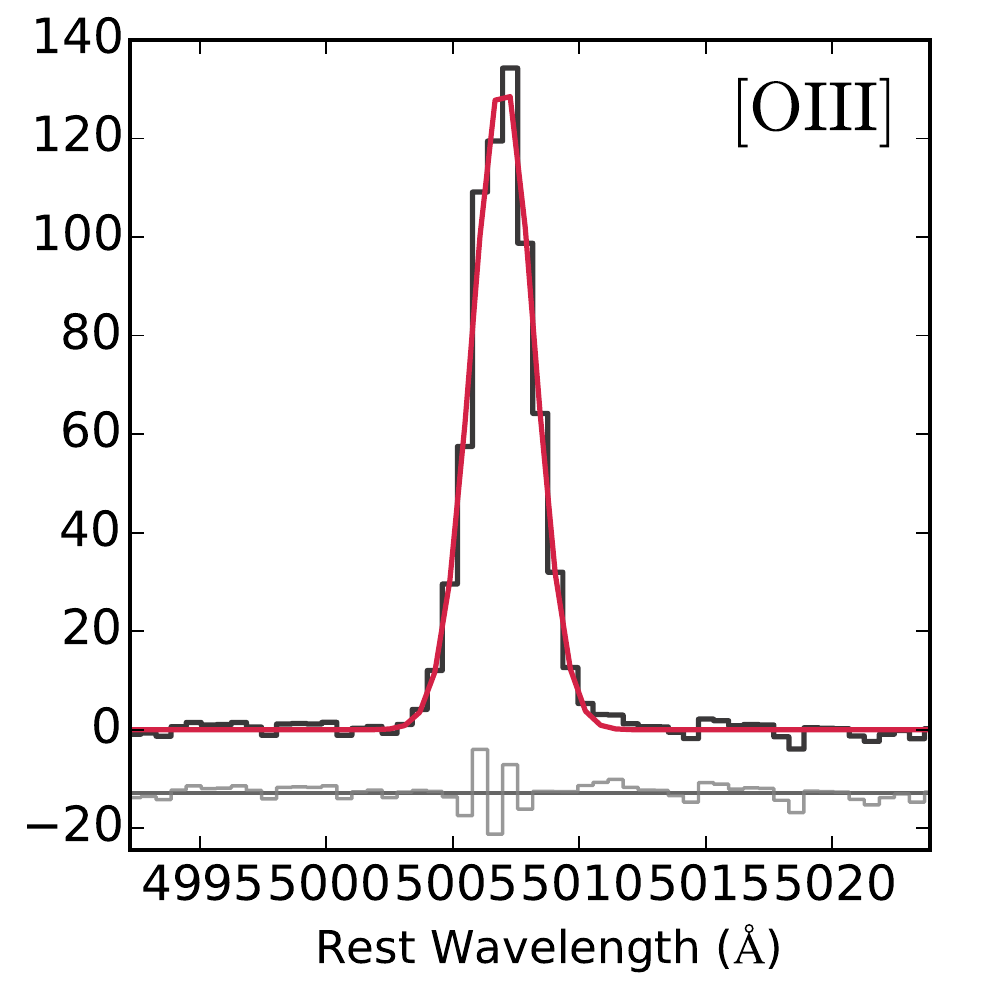}
\includegraphics[scale=0.44]{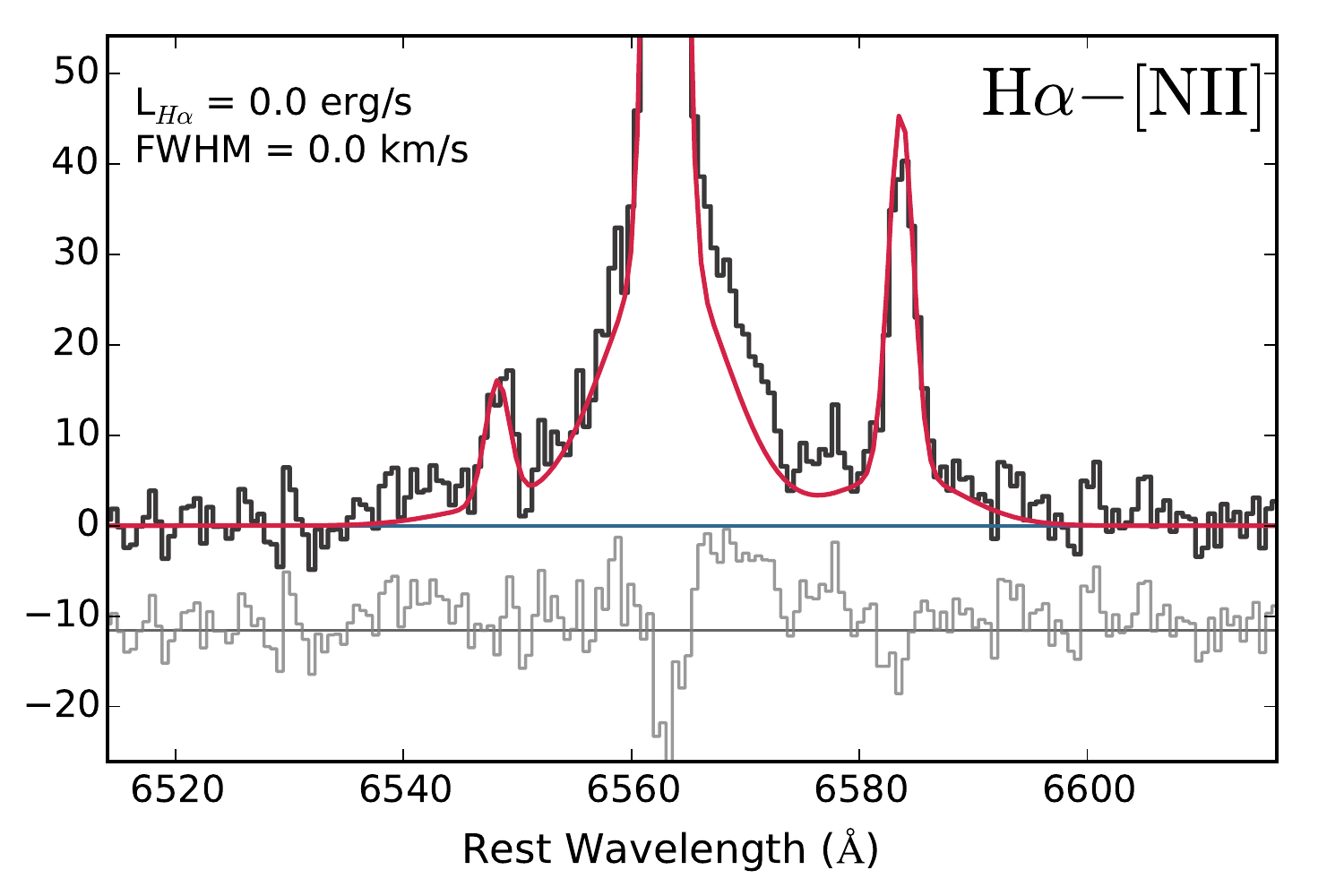}

\includegraphics[scale=0.44]{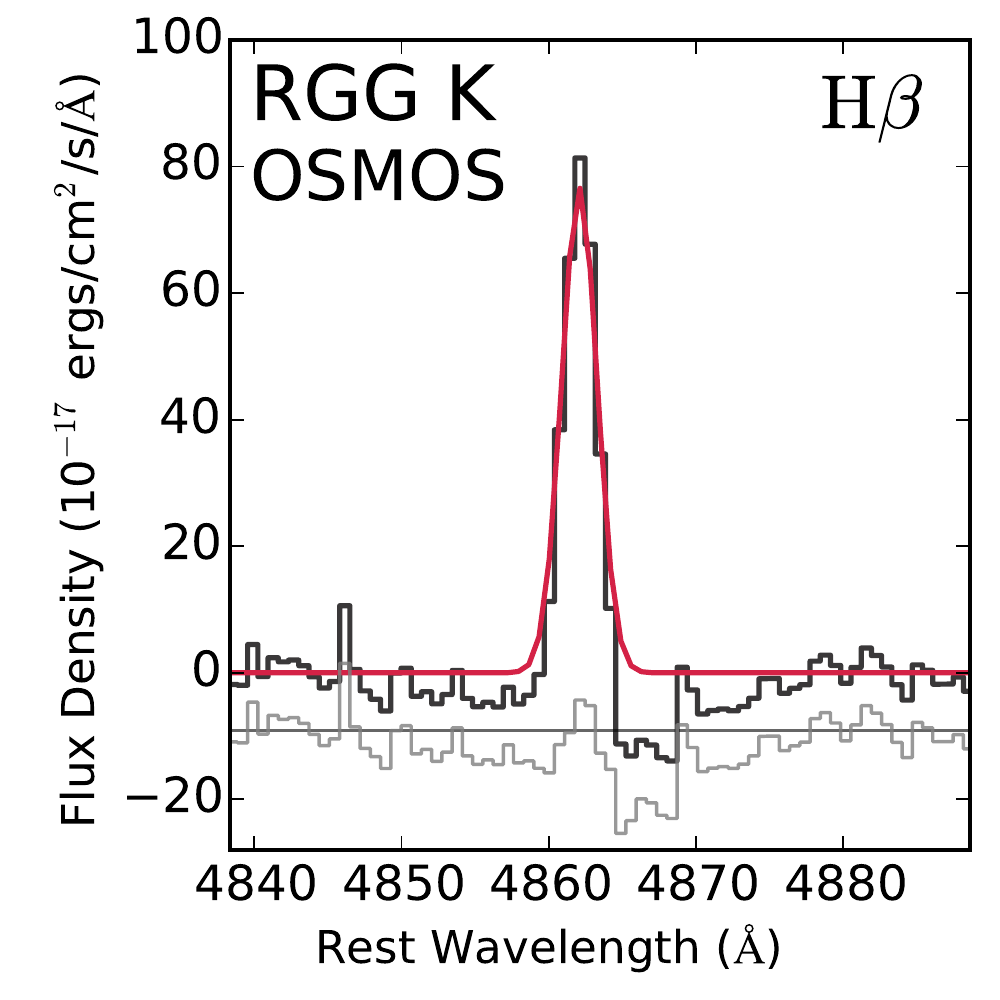}
\includegraphics[scale=0.44]{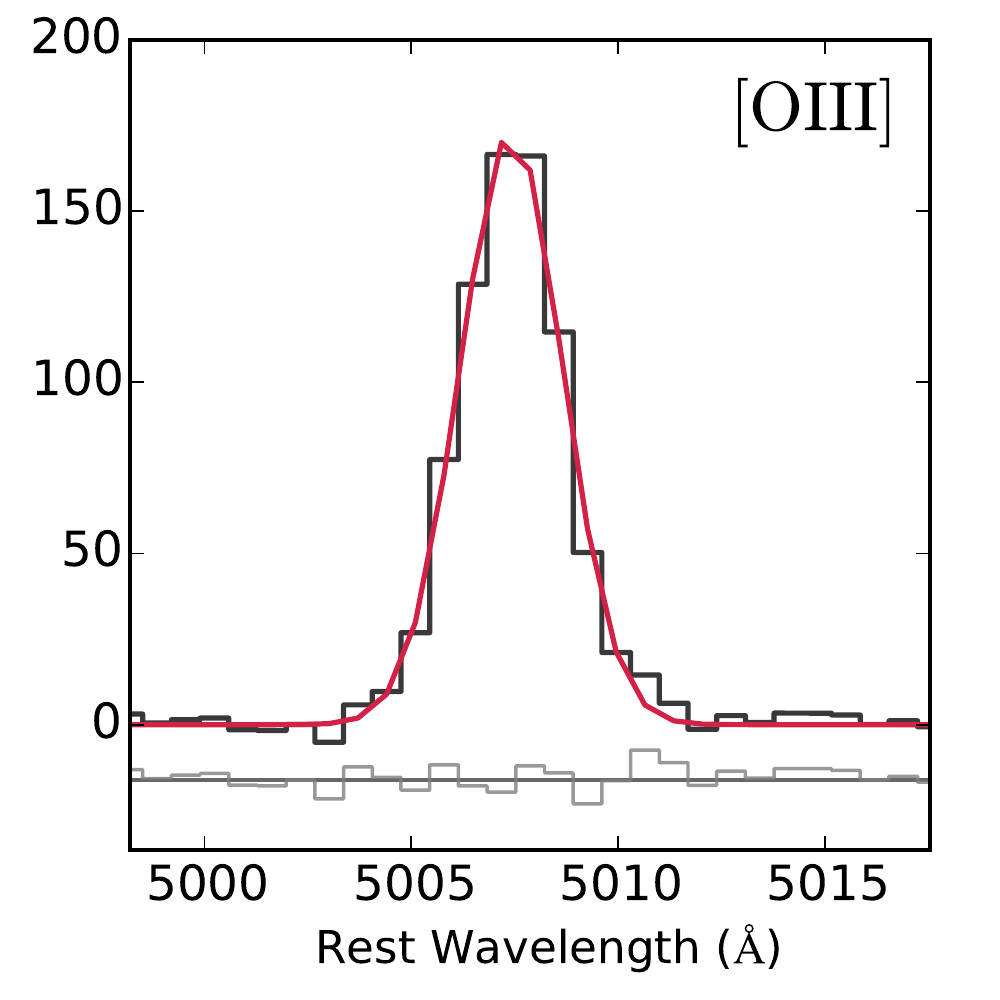}
\includegraphics[scale=0.44]{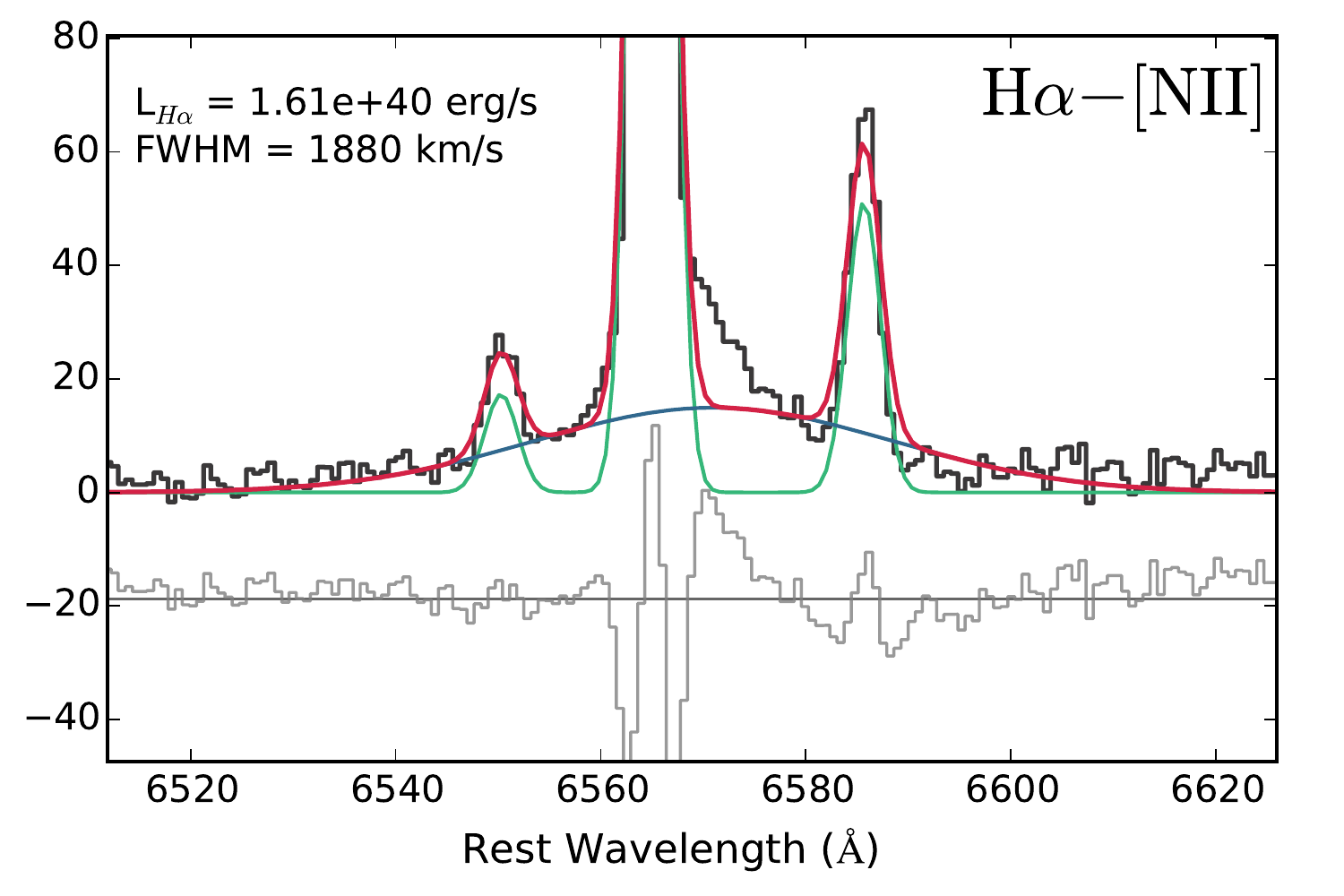}
\caption{These plots show the $\rm H\beta$, [OIII]$\lambda$5007, H$\alpha$, and [NII]$\lambda\lambda6718,6731$ lines for each observation taken of RGG K (NSA 91579). Description is same as for Figure~\ref{nsa15952}. We place this object in the ``ambiguous" category.}
\label{nsa91579}
\end{figure*}

\begin{figure*}
\centering

\includegraphics[scale=0.44]{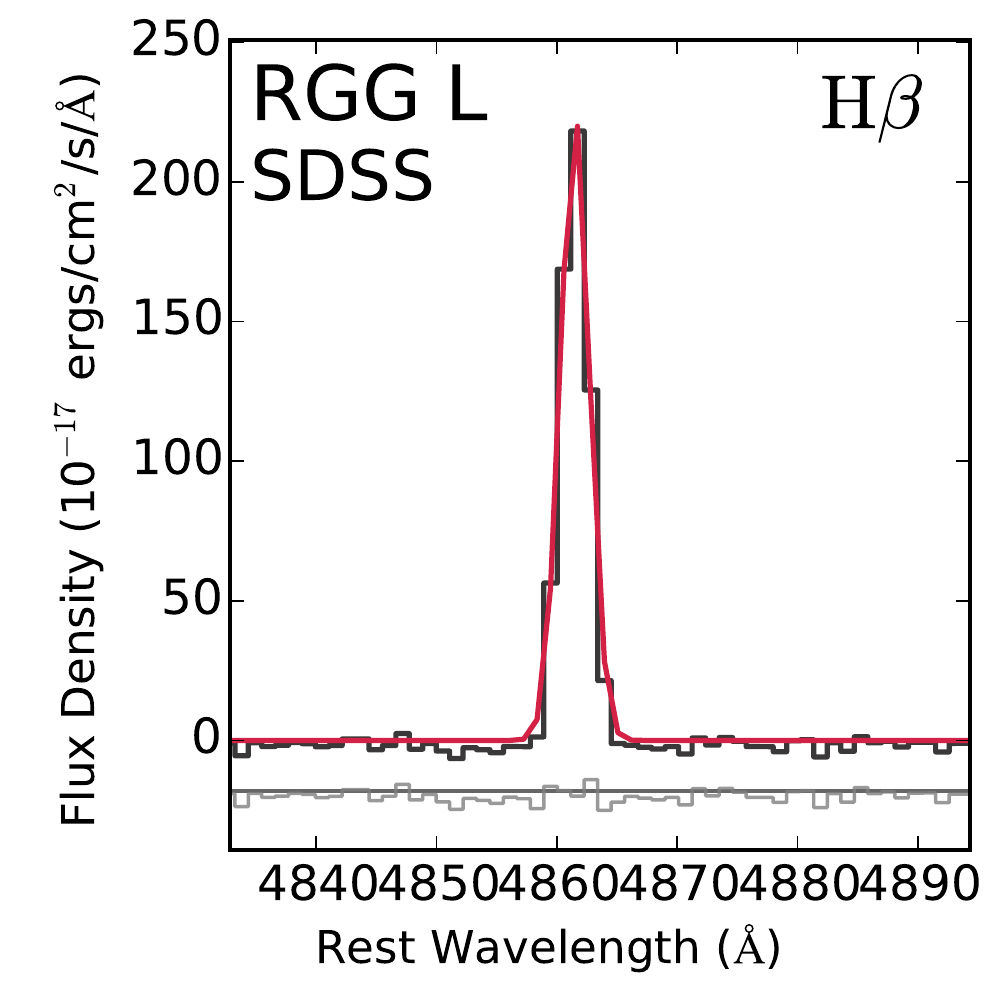}
\includegraphics[scale=0.44]{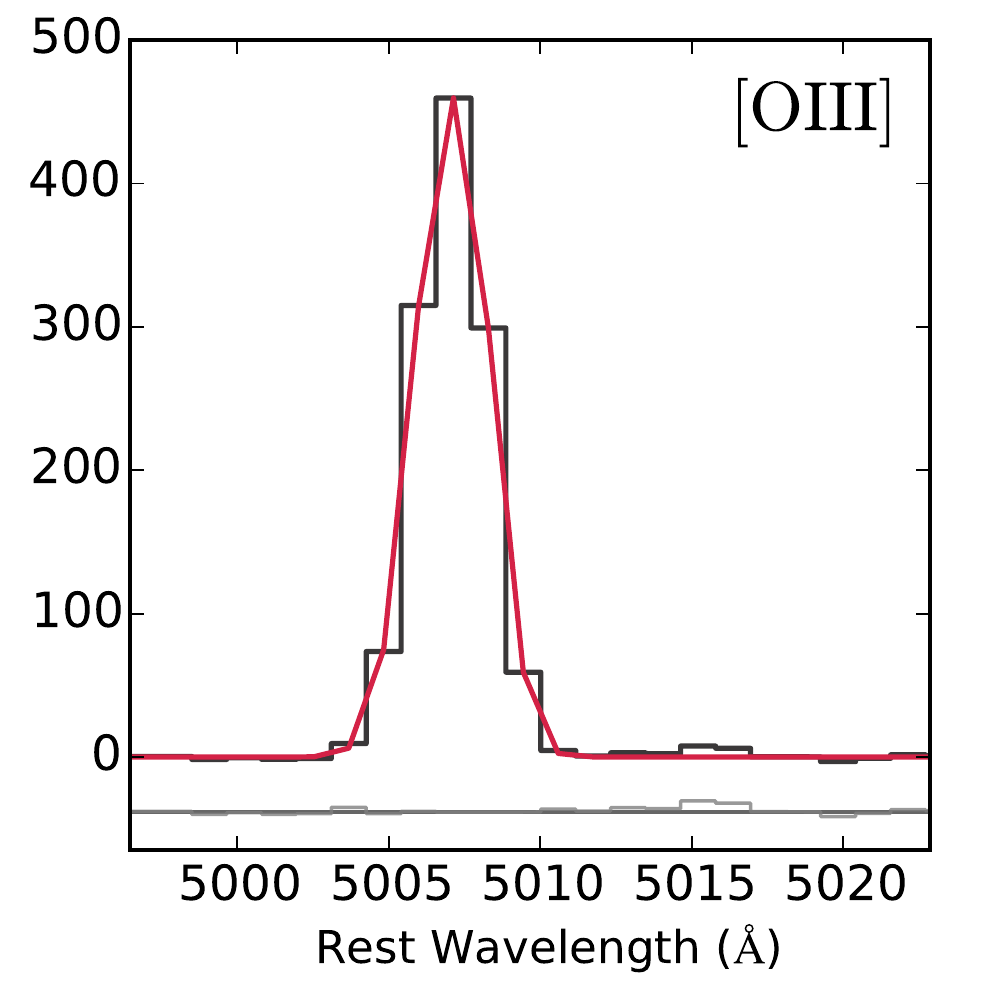}
\includegraphics[scale=0.44]{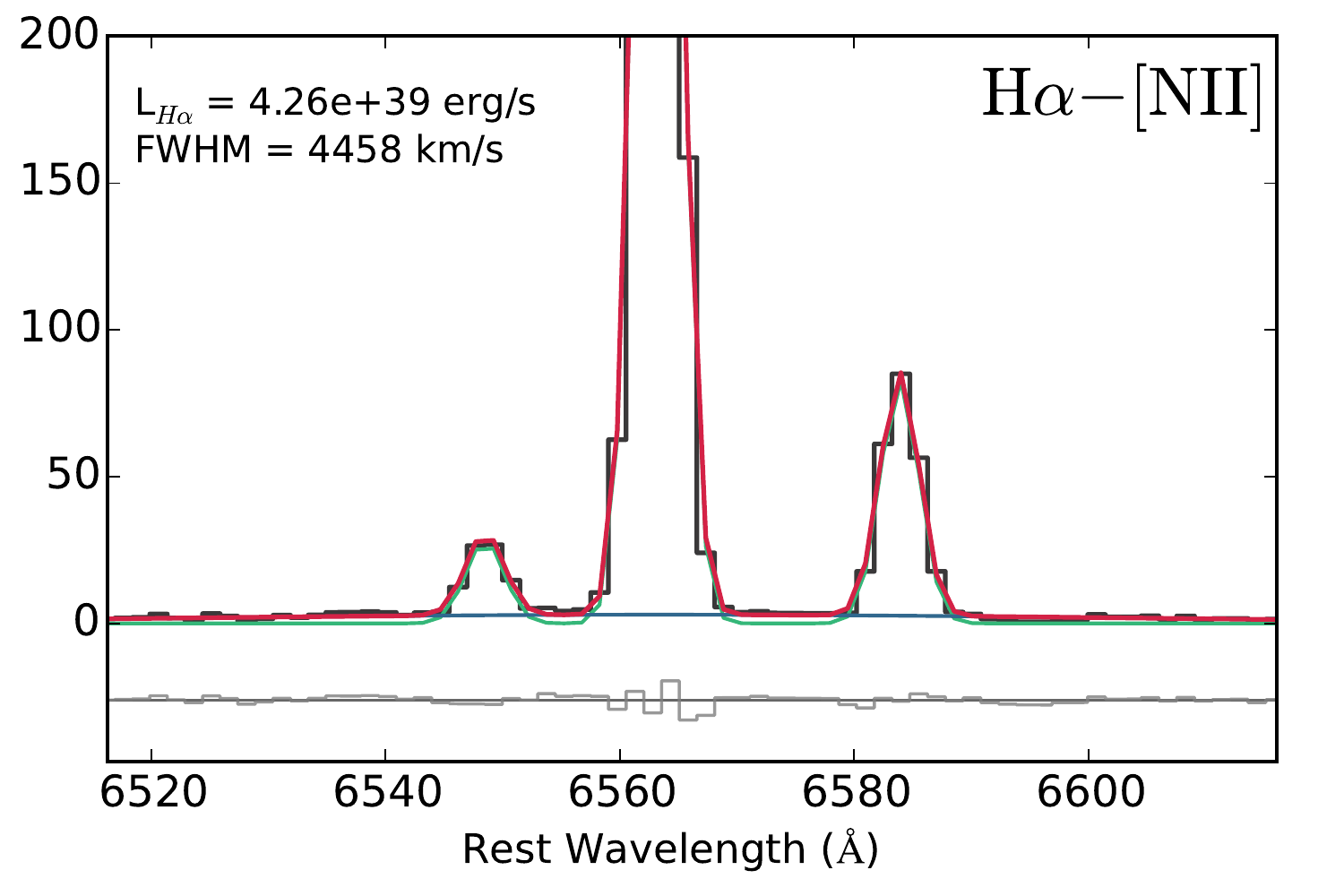}\\

\includegraphics[scale=0.44]{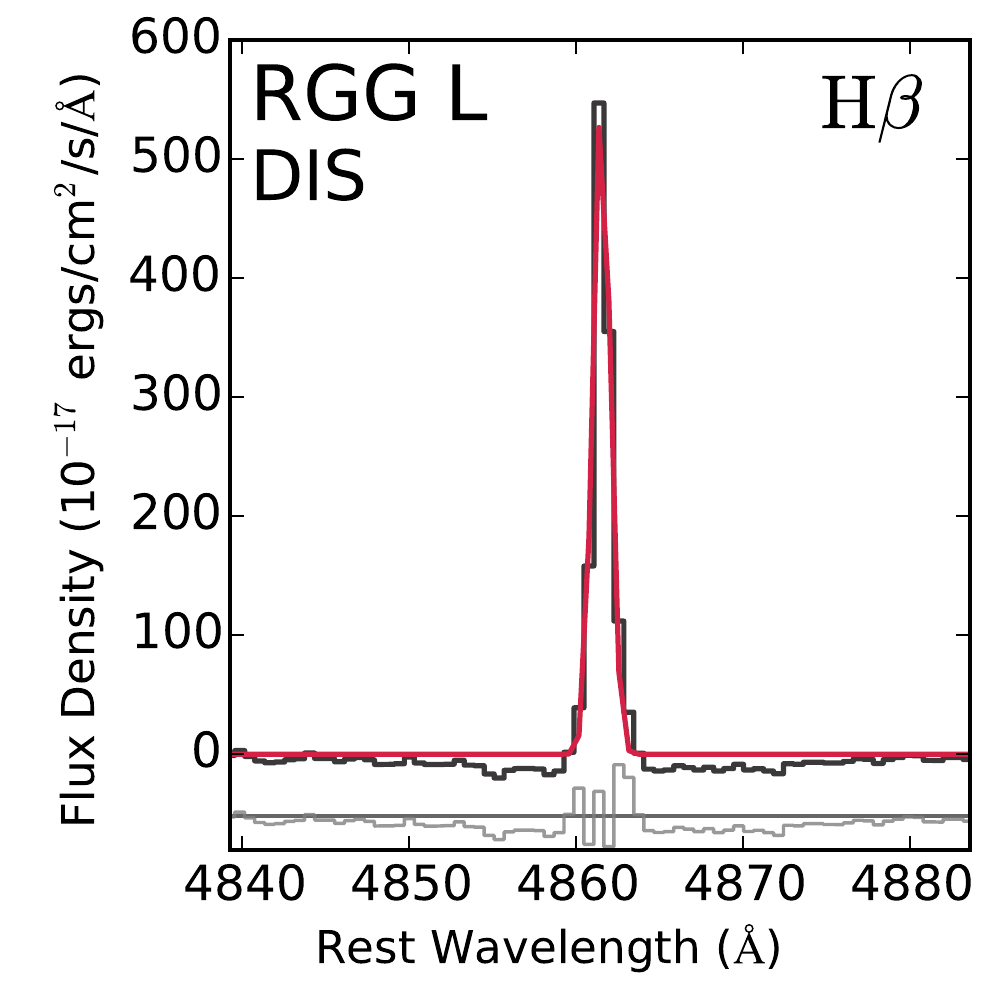}
\includegraphics[scale=0.44]{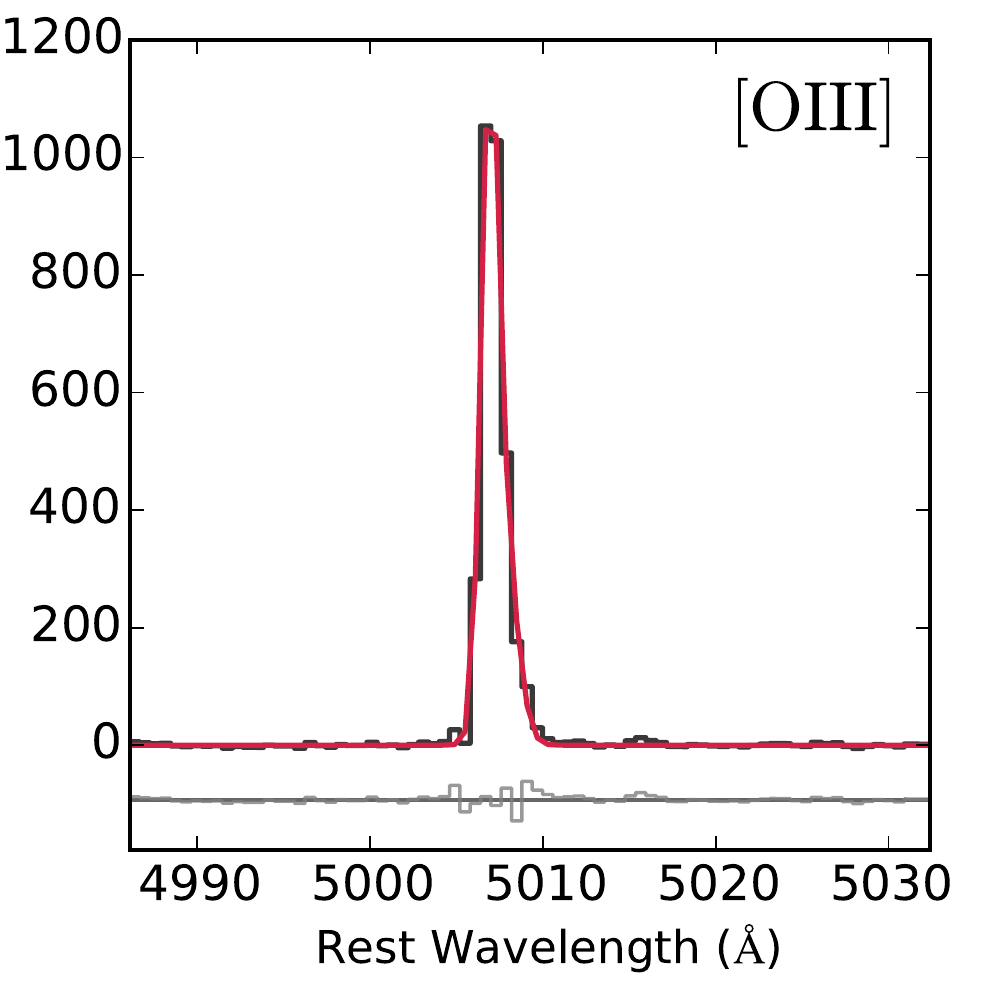}
\includegraphics[scale=0.44]{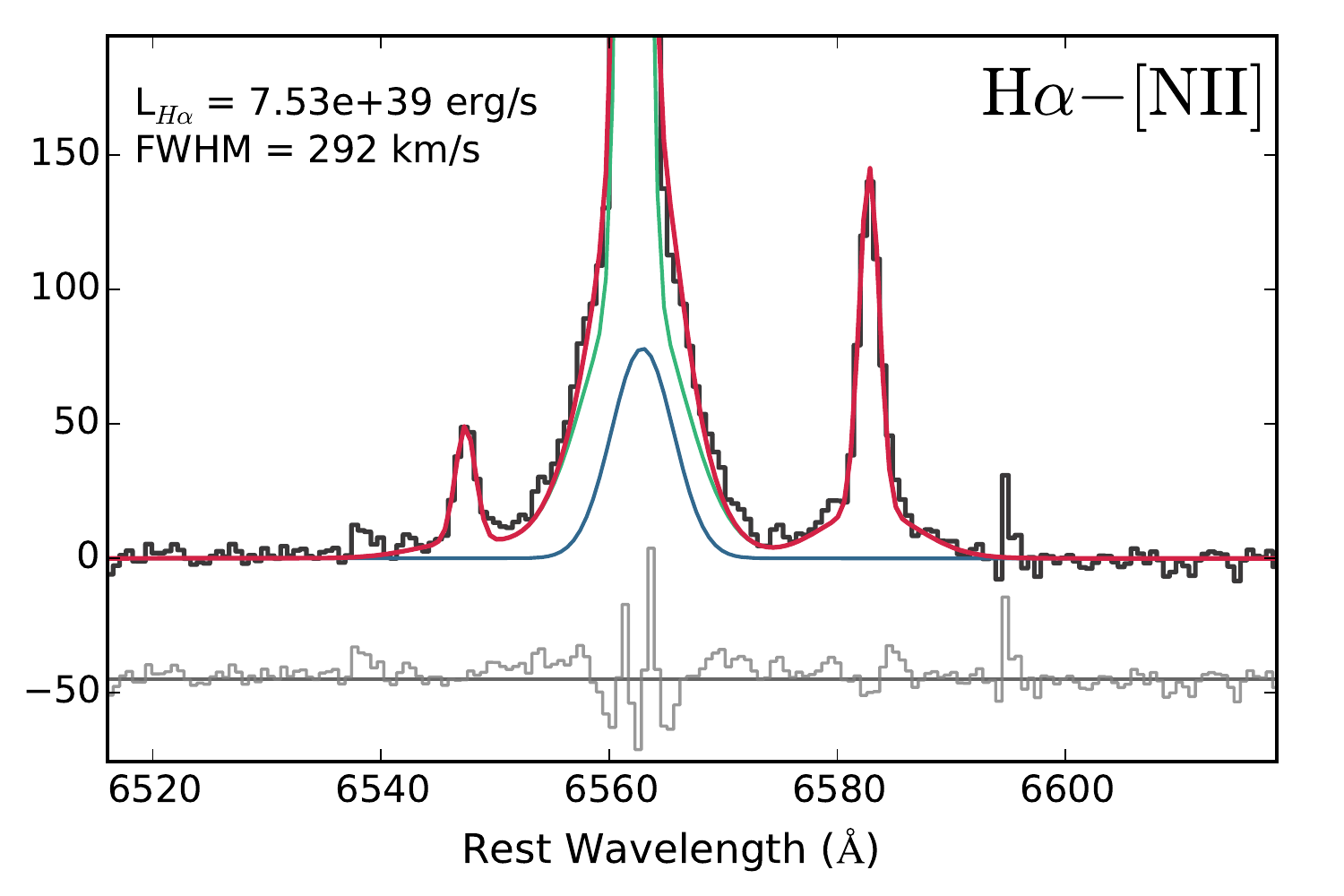}\\
\caption{These plots show the $\rm H\beta$, [OIII]$\lambda$5007, H$\alpha$, and [NII]$\lambda\lambda6718,6731$ lines for each observation taken of RGG L (NSA 33207). Description is same as for Figure~\ref{nsa15952}. We place this object in the ``transient" category.}
\label{nsa33207}
\end{figure*}

\begin{figure*}
\centering
\includegraphics[scale=0.44]{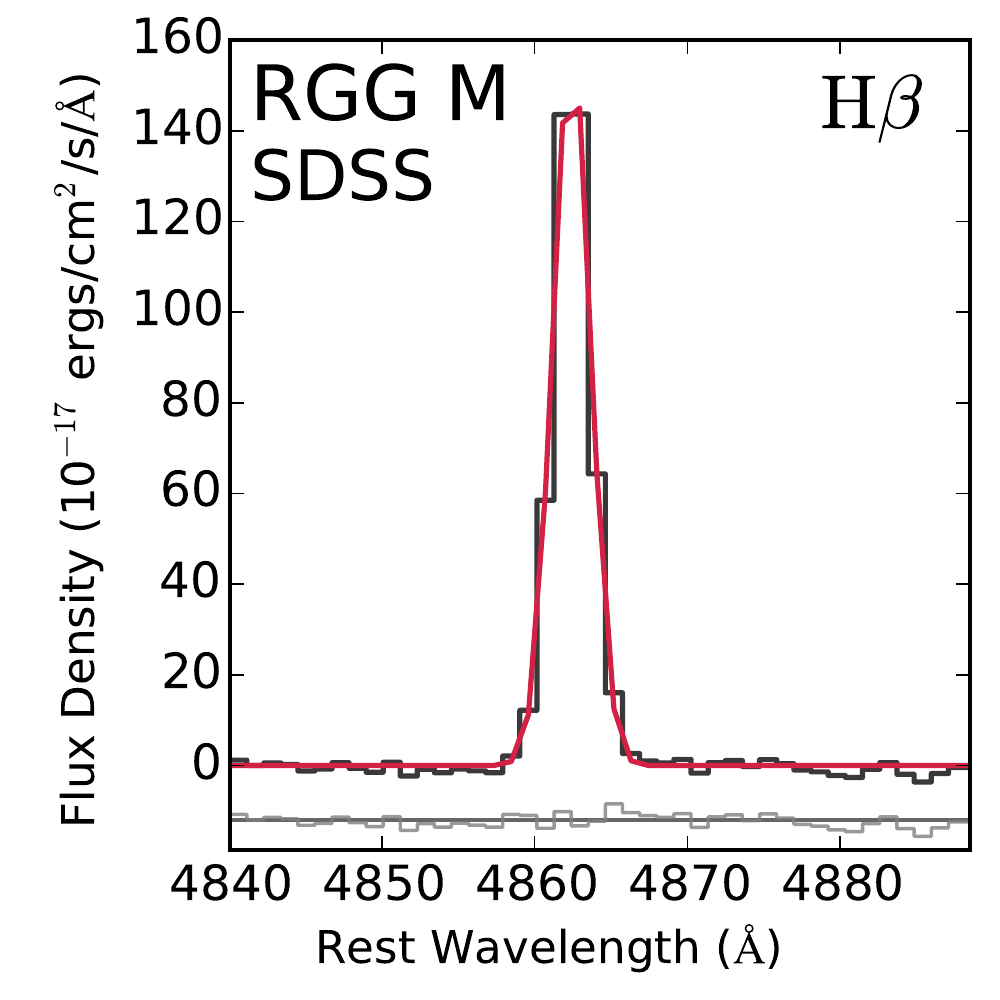}
\includegraphics[scale=0.44]{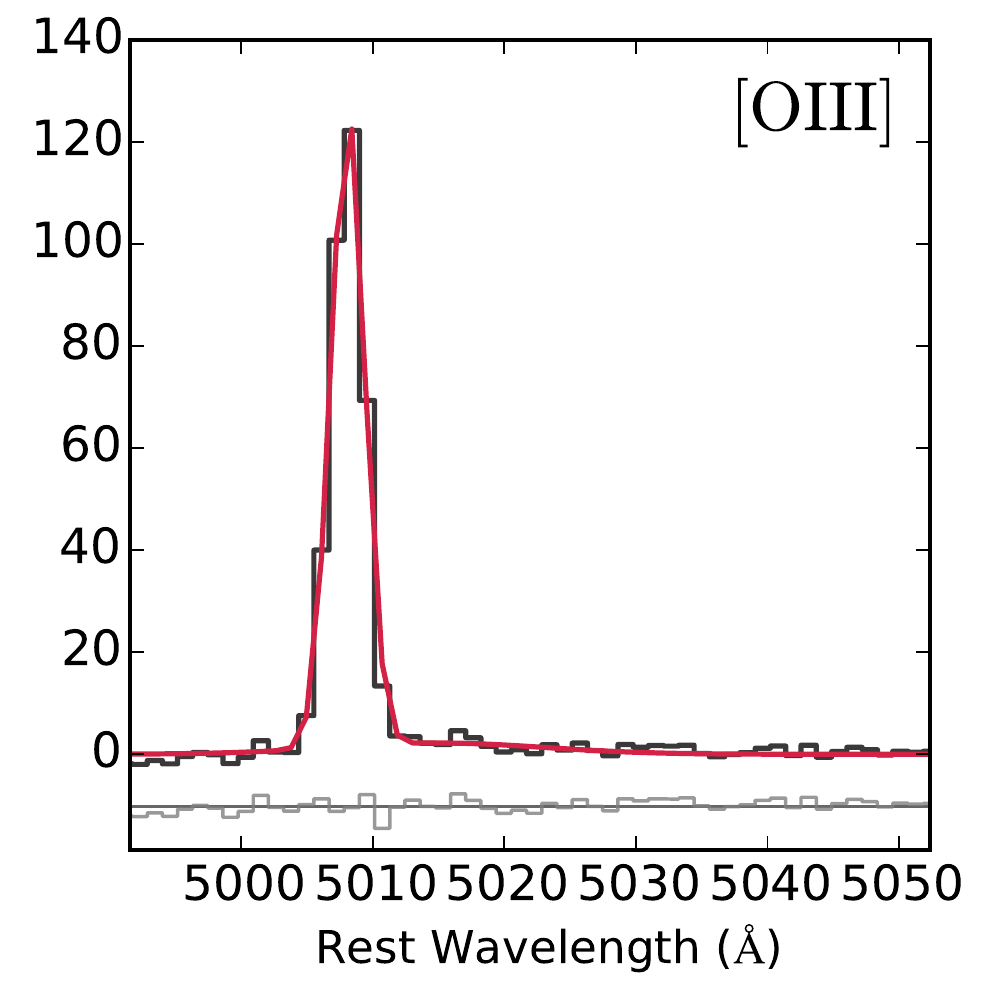}
\includegraphics[scale=0.44]{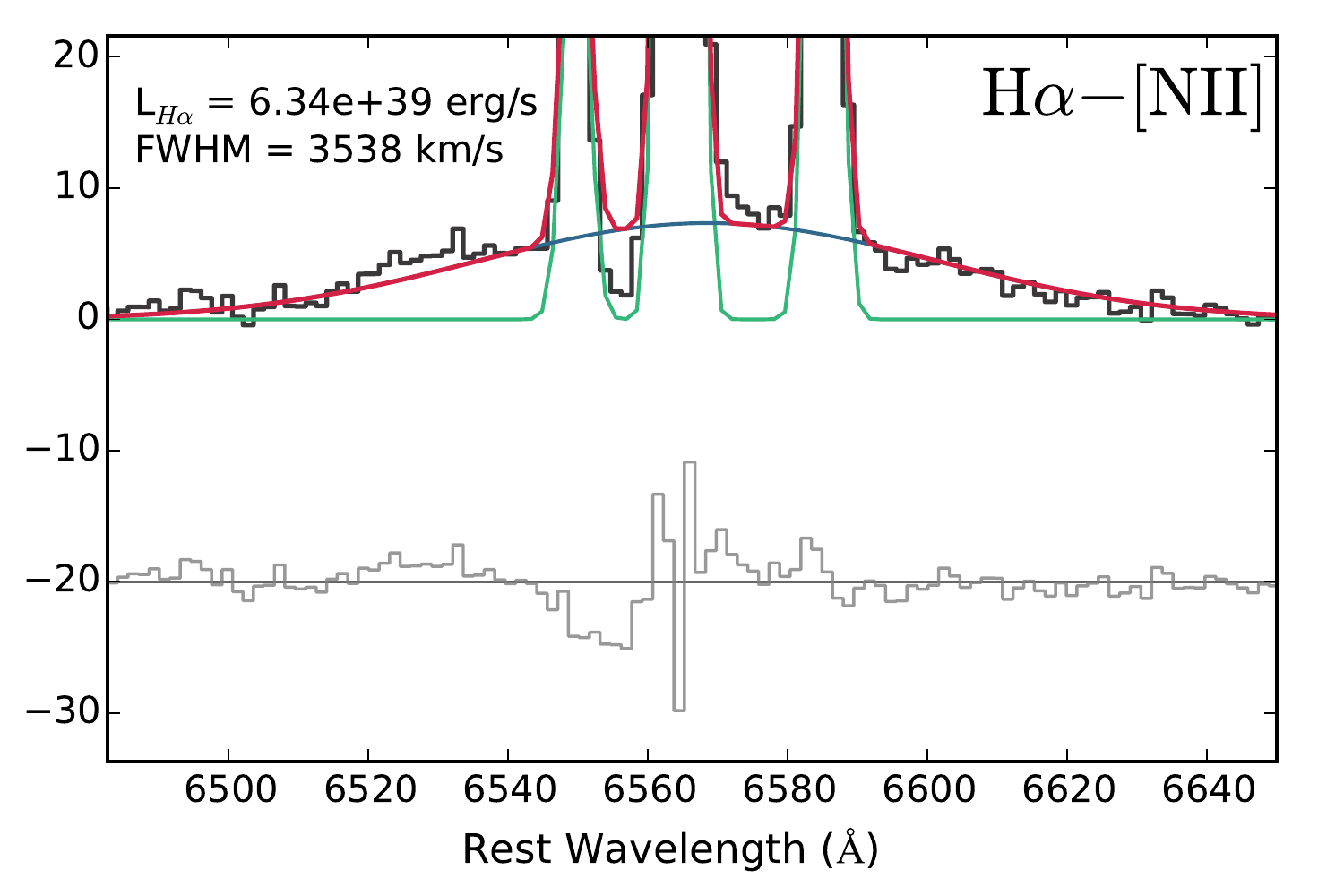}\\

\includegraphics[scale=0.44]{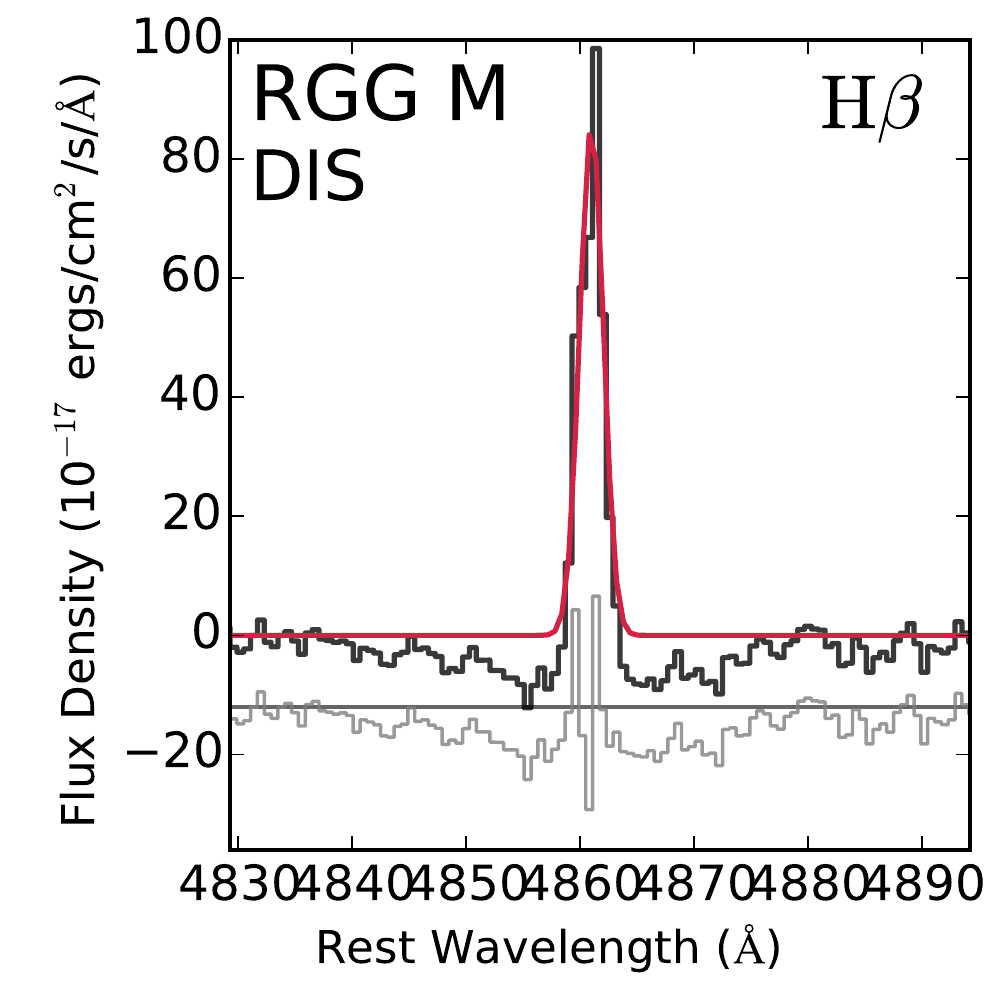}
\includegraphics[scale=0.44]{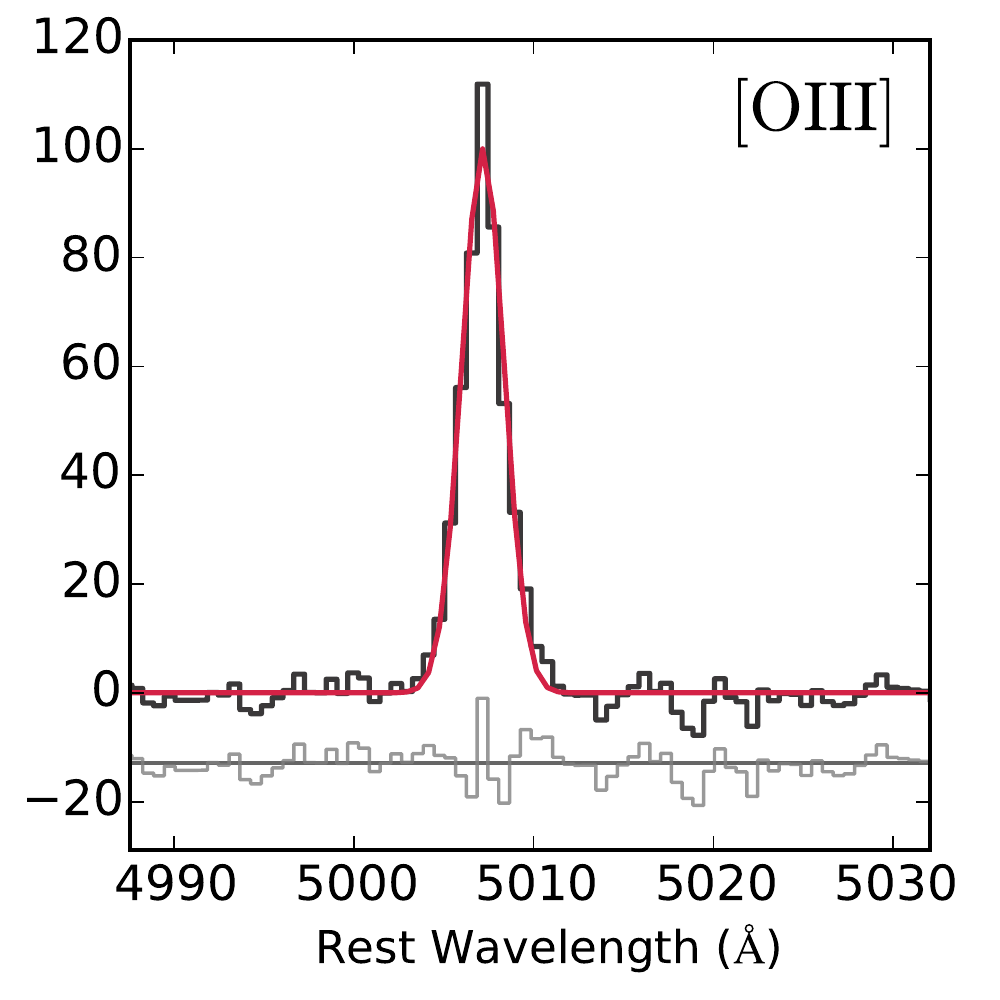}
\includegraphics[scale=0.44]{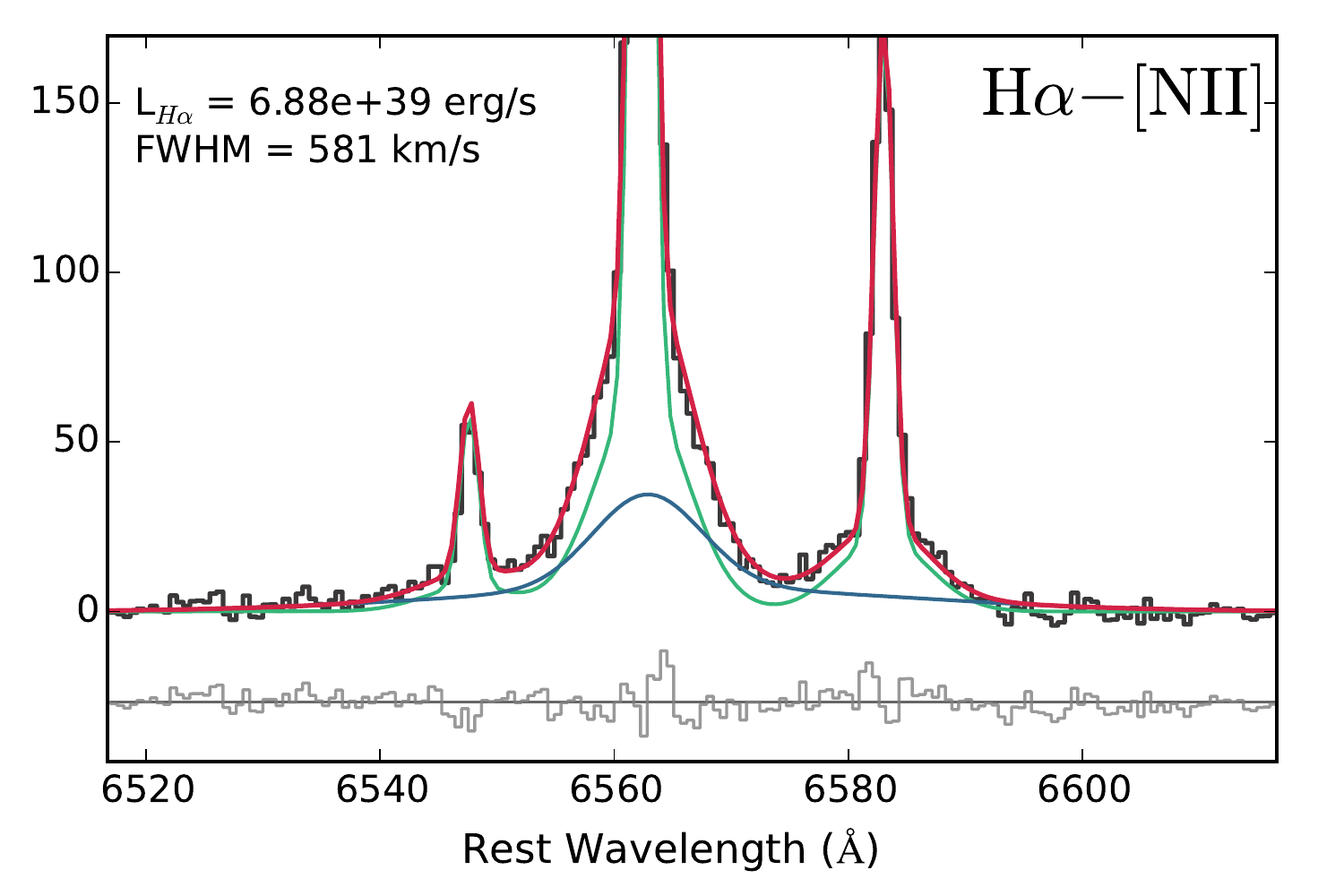}

\includegraphics[scale=0.44]{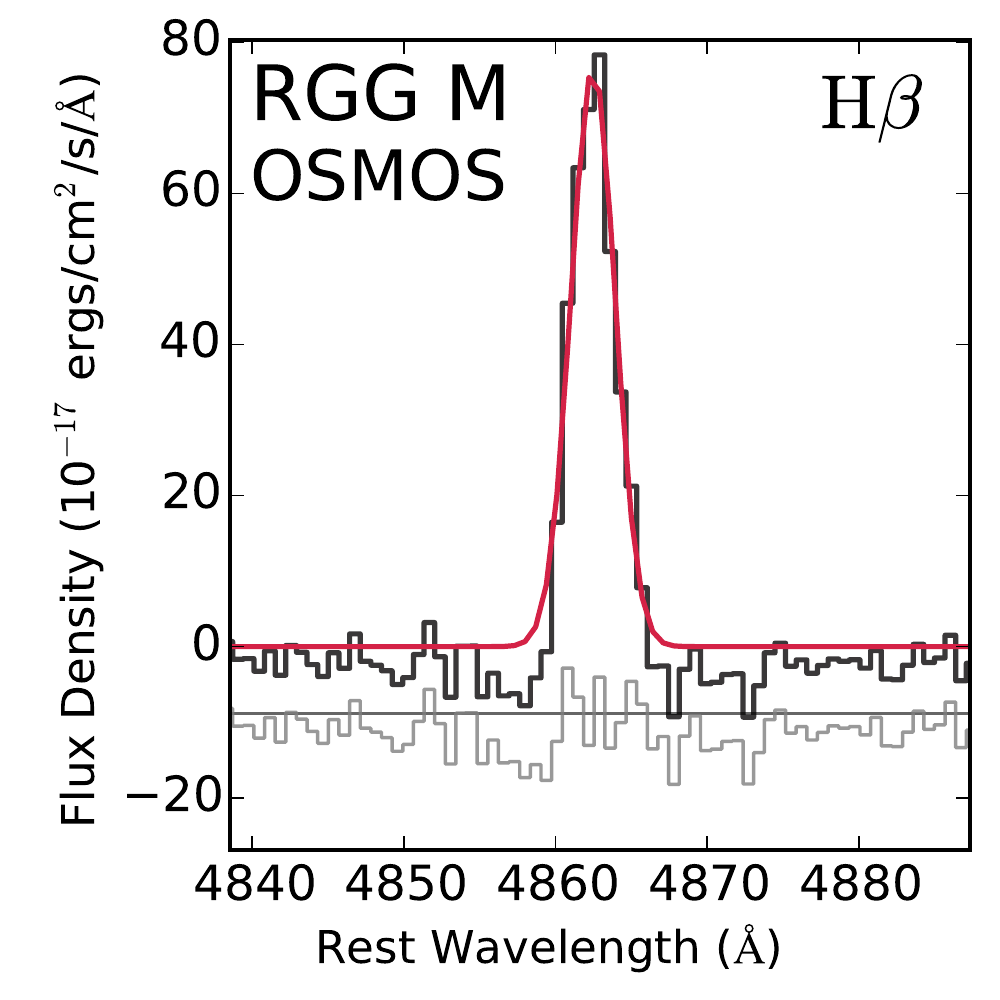}
\includegraphics[scale=0.44]{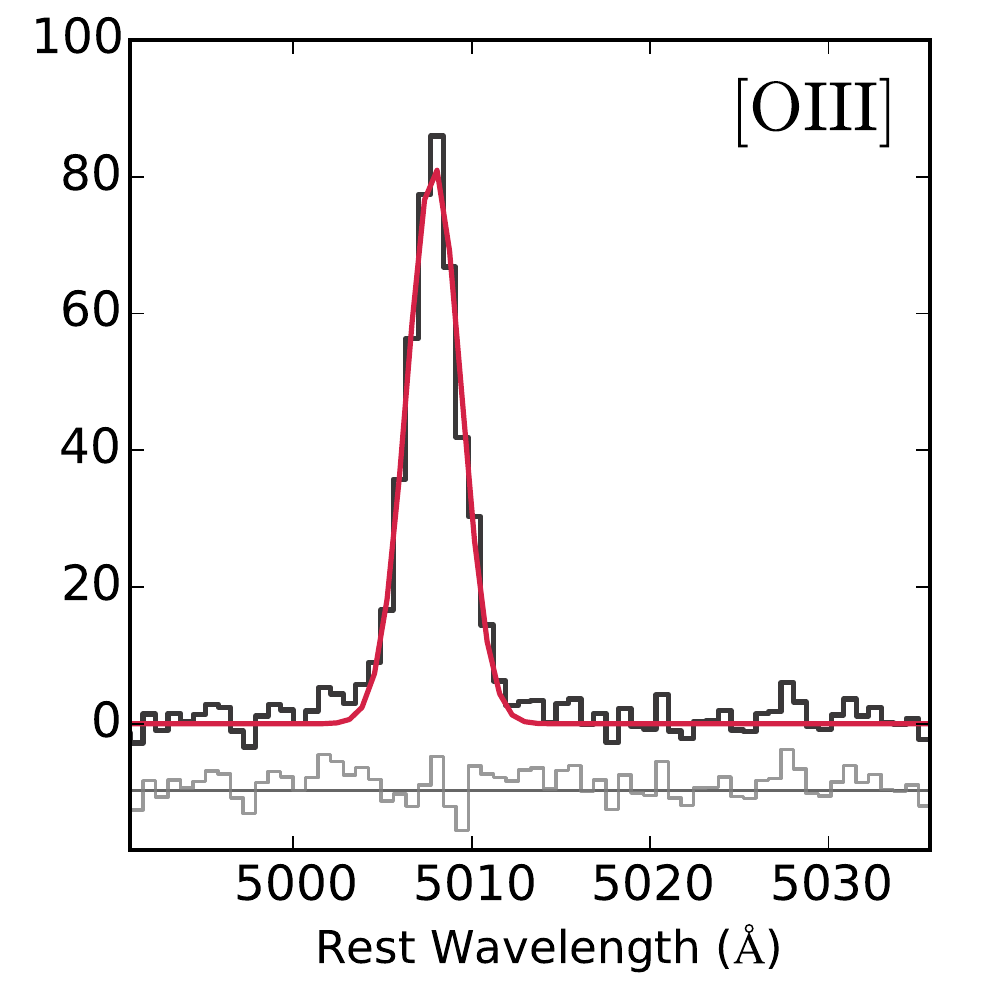}
\includegraphics[scale=0.44]{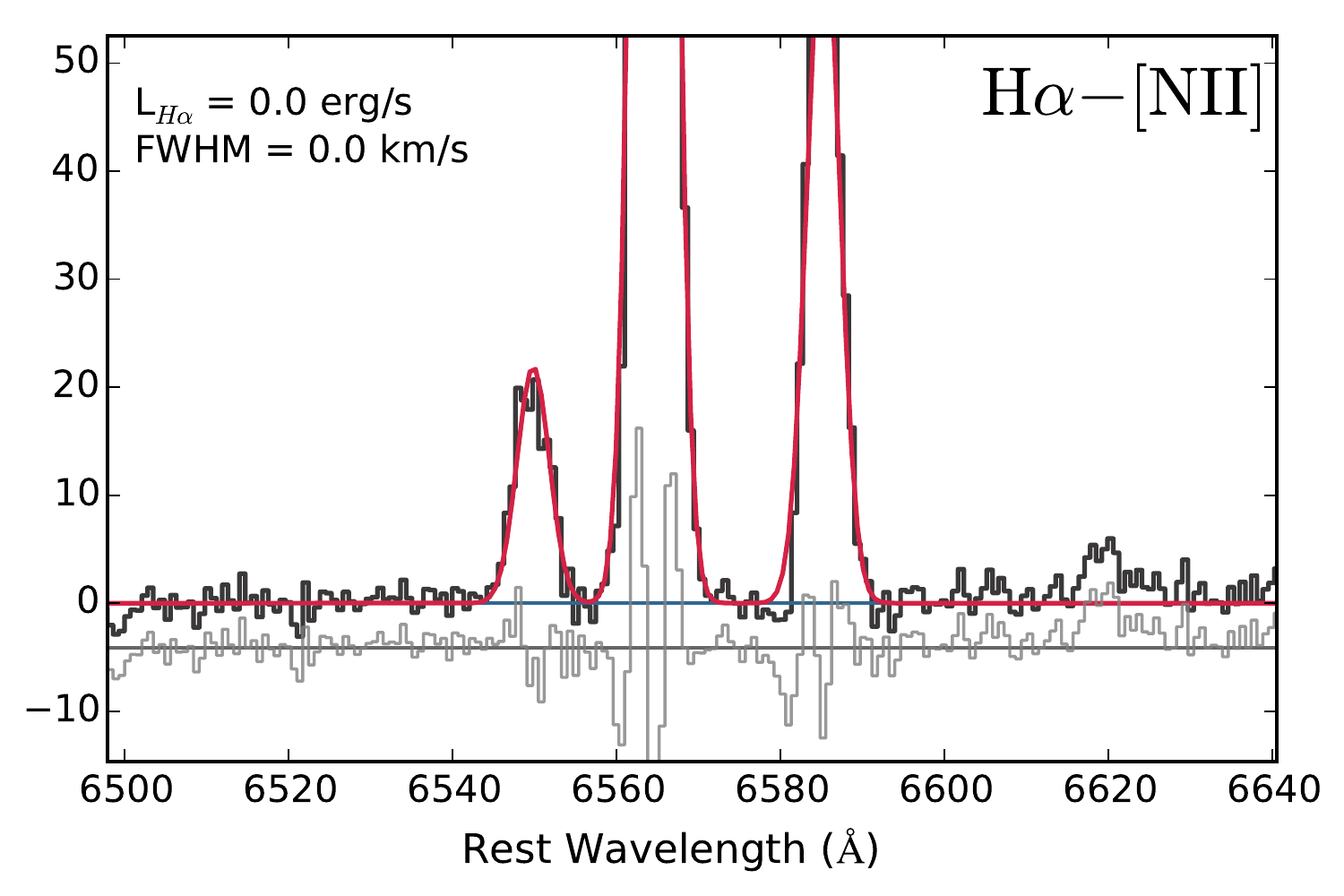}
\caption{These plots show the $\rm H\beta$, [OIII]$\lambda$5007, H$\alpha$, and [NII]$\lambda\lambda6718,6731$ lines for each observation taken of RGG M (NSA 119311). Description is same as for Figure~\ref{nsa15952}. We place this object in the ``transient" category.}
\label{nsa119311}
\end{figure*}

\begin{figure*}
\centering
\includegraphics[scale=0.44]{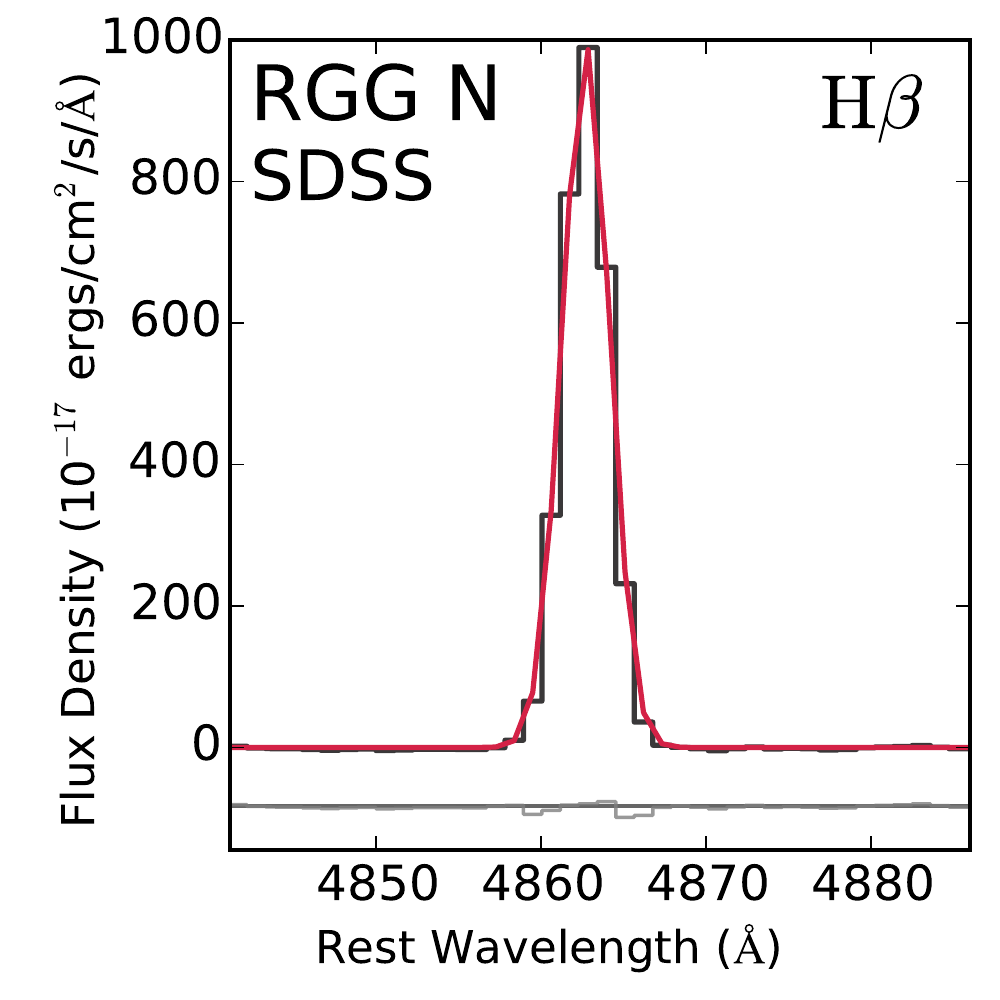}
\includegraphics[scale=0.44]{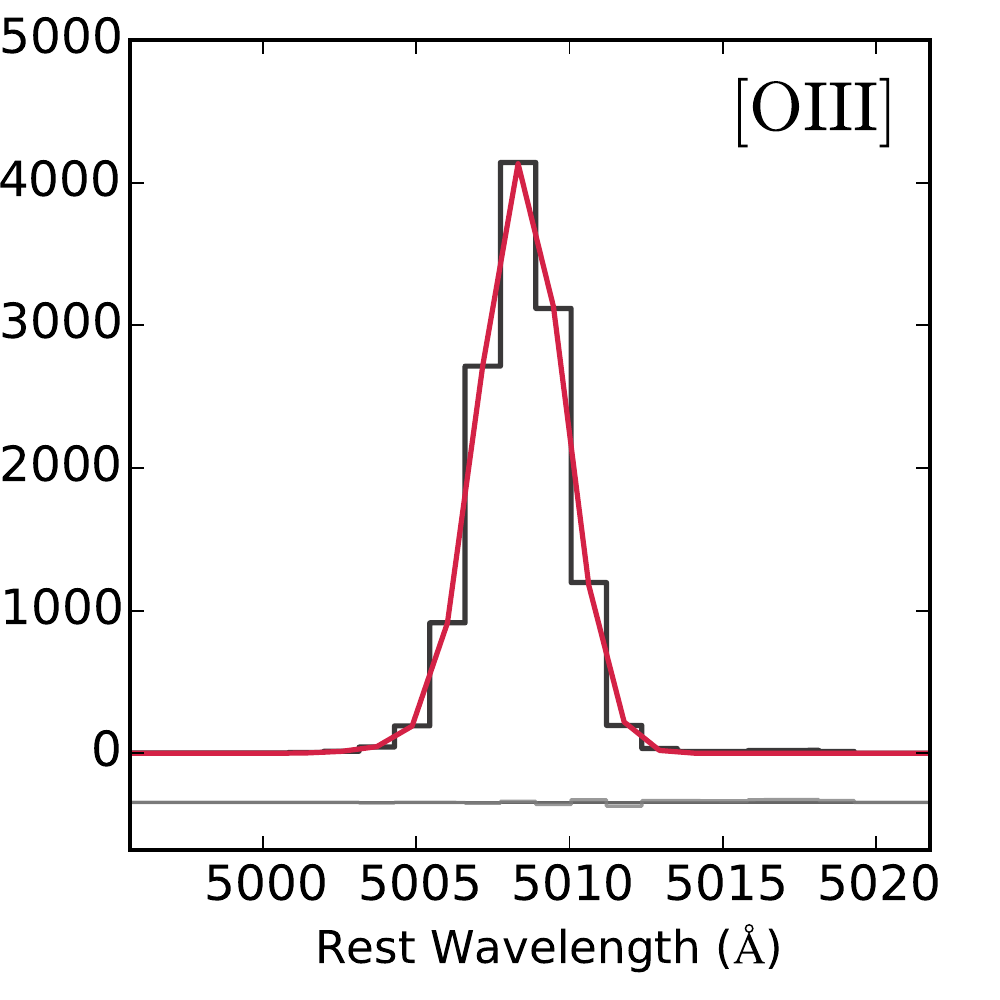}
\includegraphics[scale=0.44]{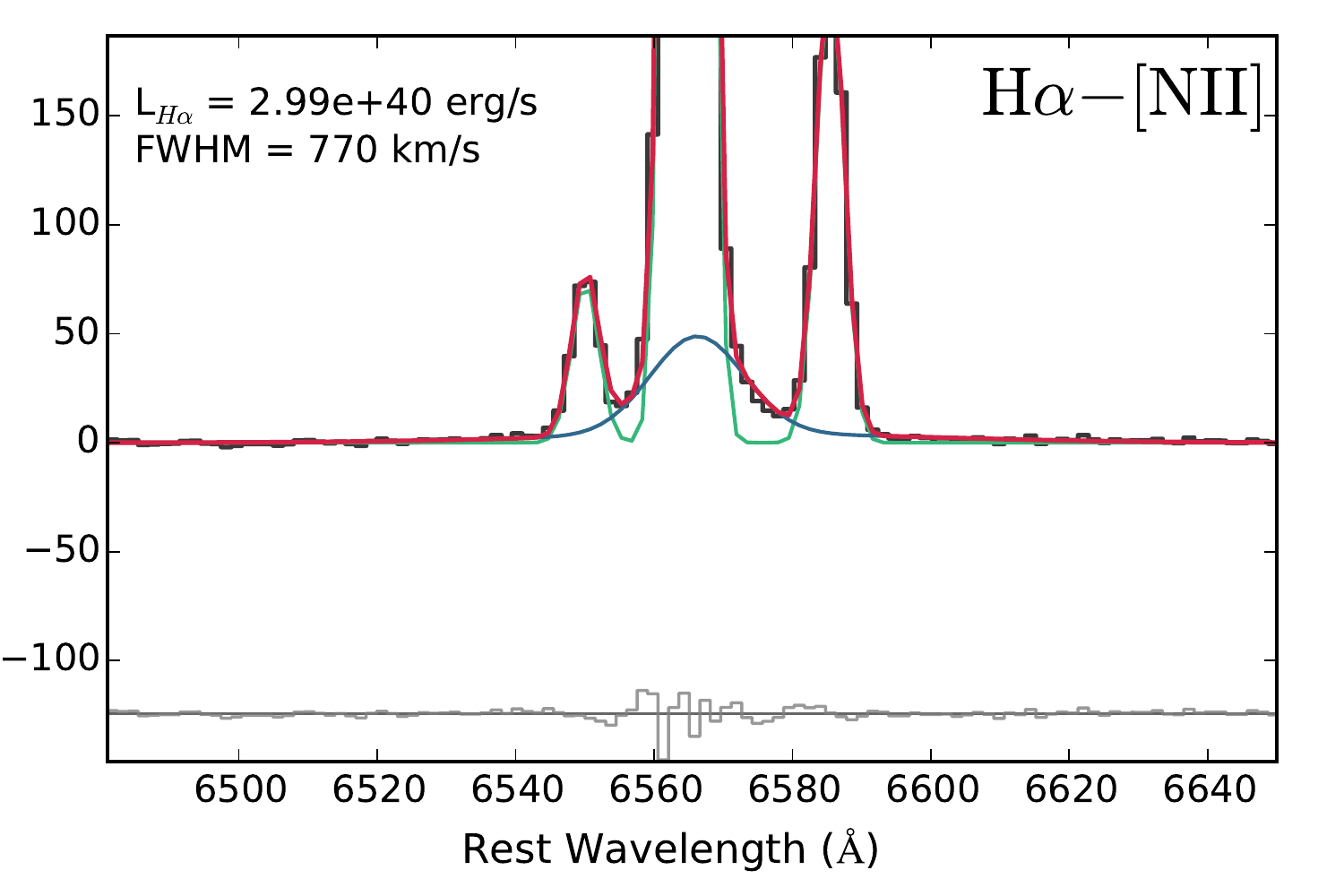}\\

\includegraphics[scale=0.44]{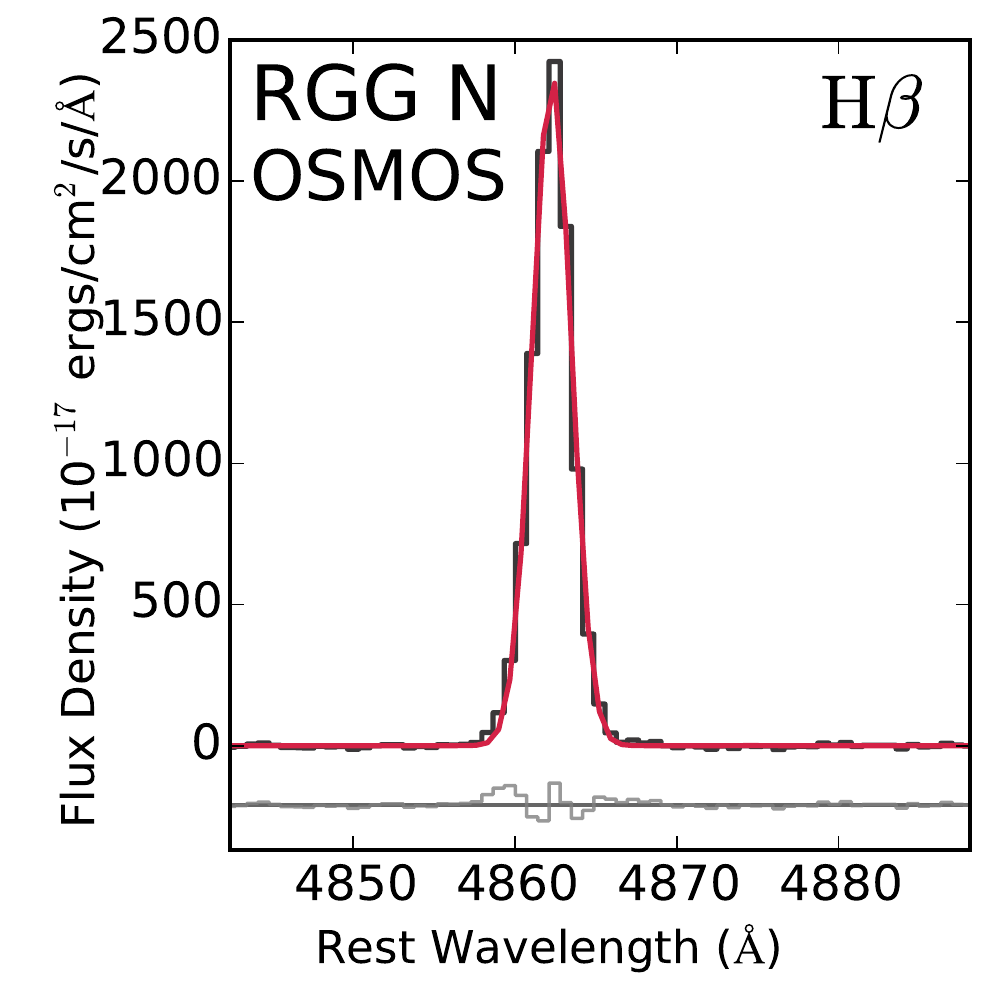}
\includegraphics[scale=0.44]{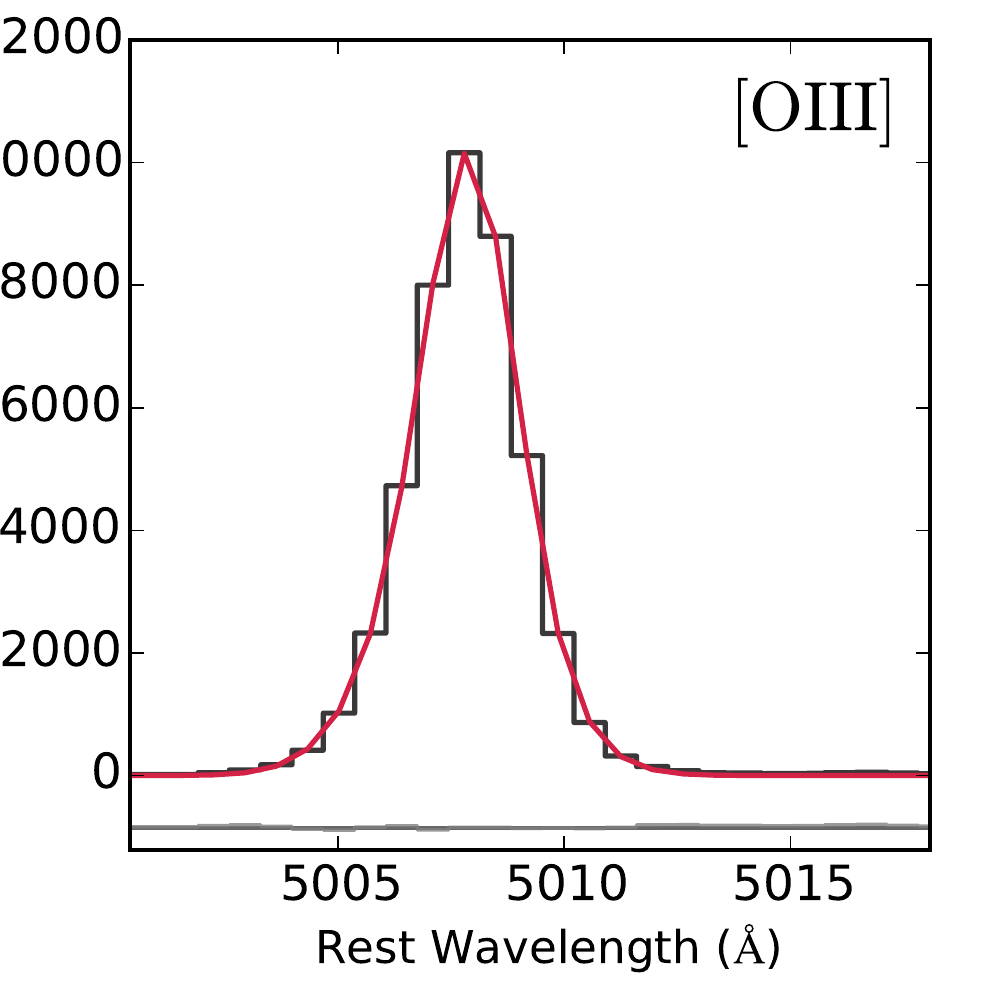}
\includegraphics[scale=0.44]{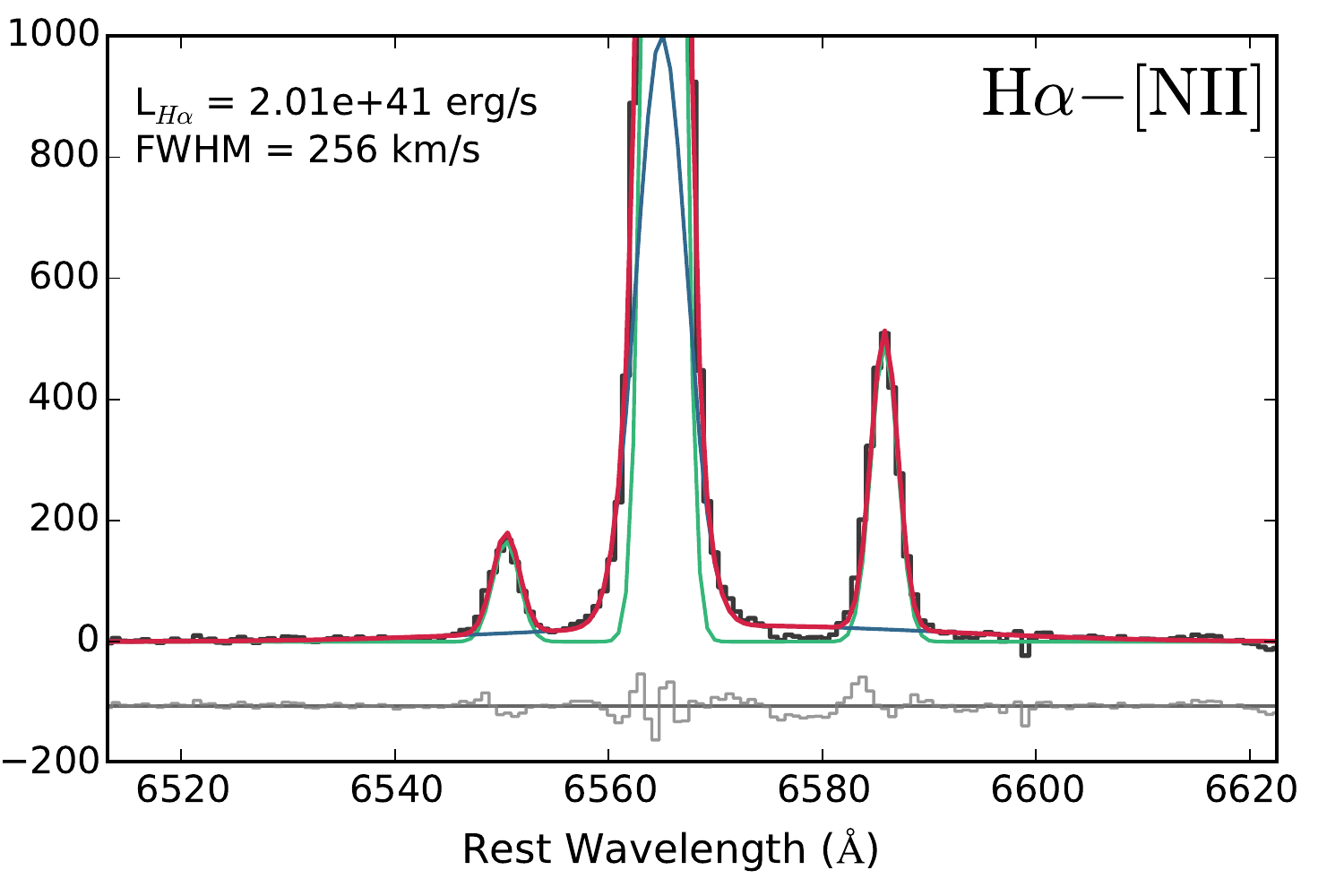}
\caption{These plots show the $\rm H\beta$, [OIII]$\lambda$5007, H$\alpha$, and [NII]$\lambda\lambda6718,6731$ lines for each observation taken of RGG N (NSA 88972). Description is same as for Figure~\ref{nsa15952}. We place this object in the ``transient" category.}
\label{nsa88972}
\end{figure*}

\begin{figure*}
\centering
\includegraphics[scale=0.44]{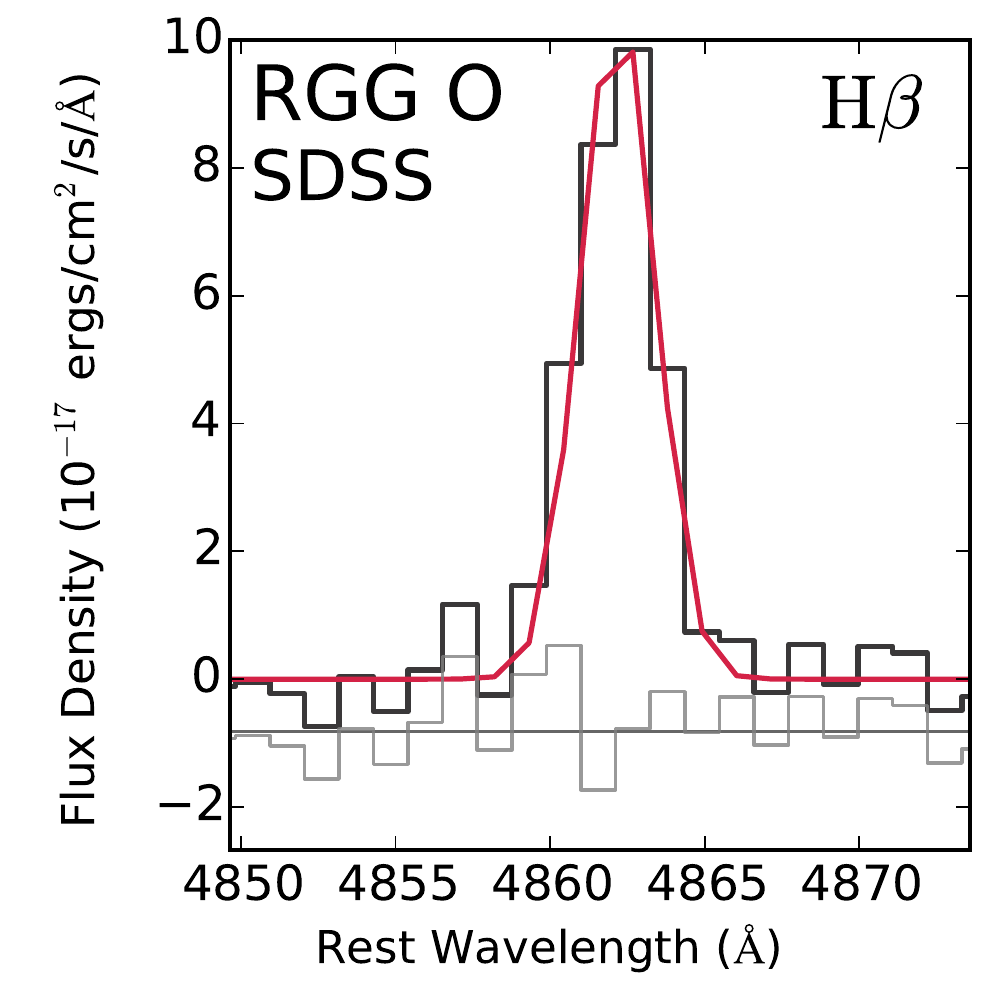}
\includegraphics[scale=0.44]{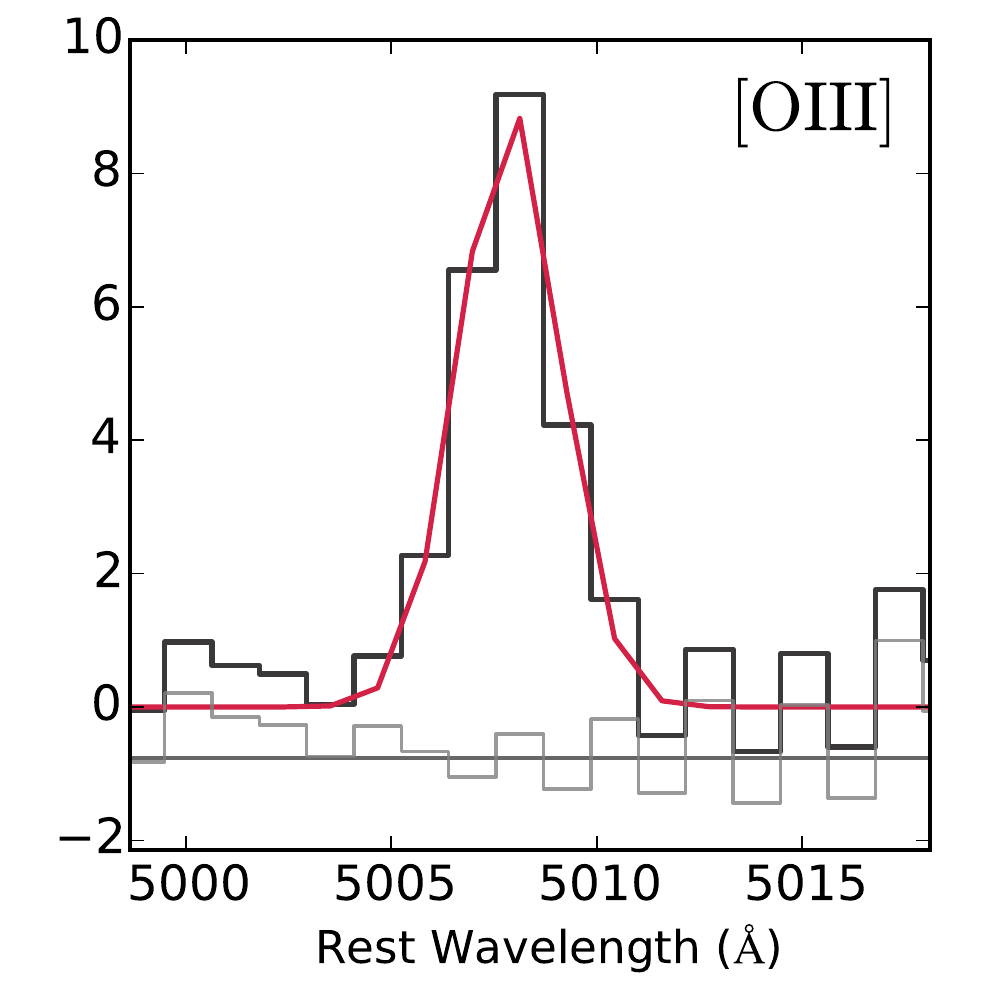}
\includegraphics[scale=0.44]{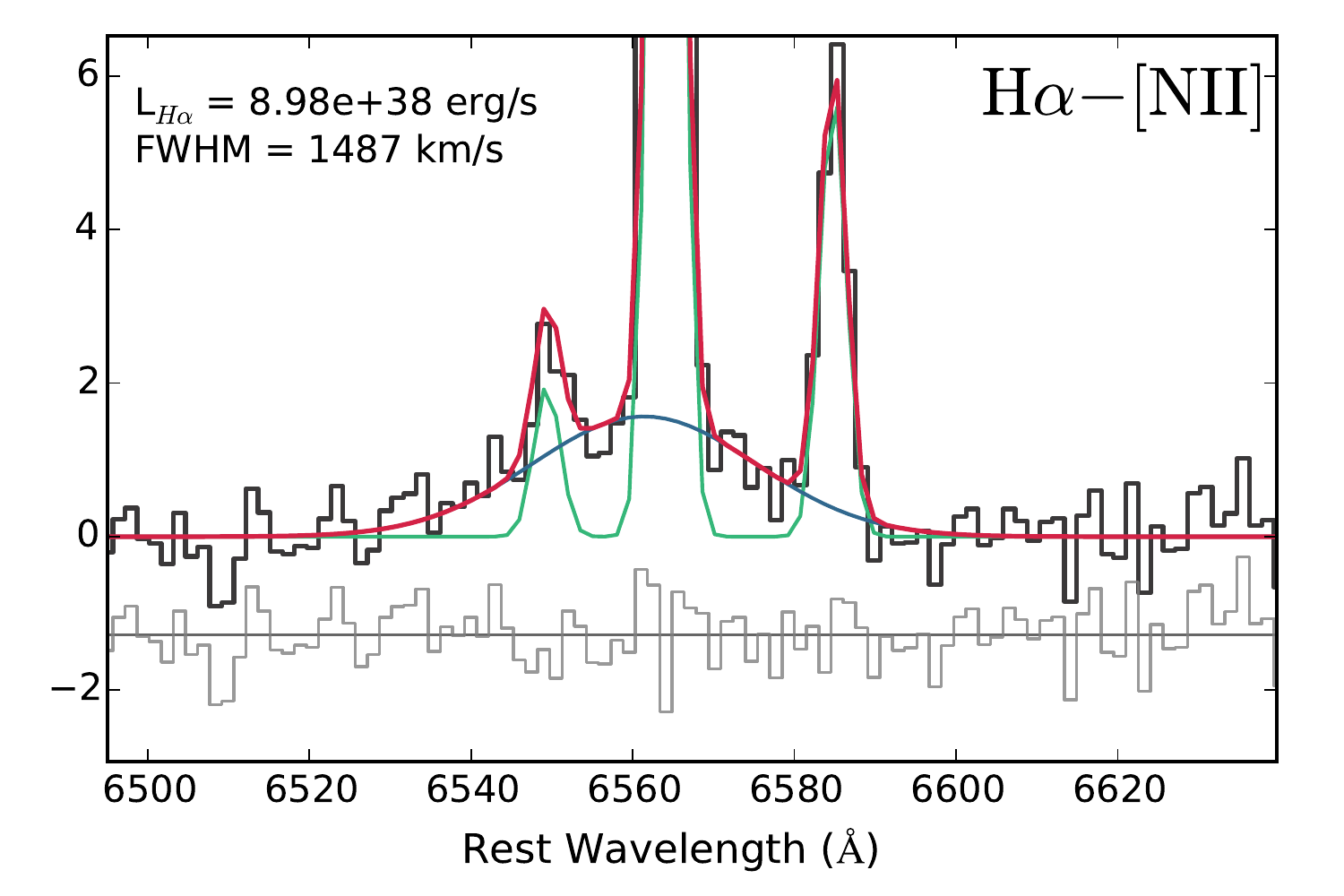}\\

\includegraphics[scale=0.44]{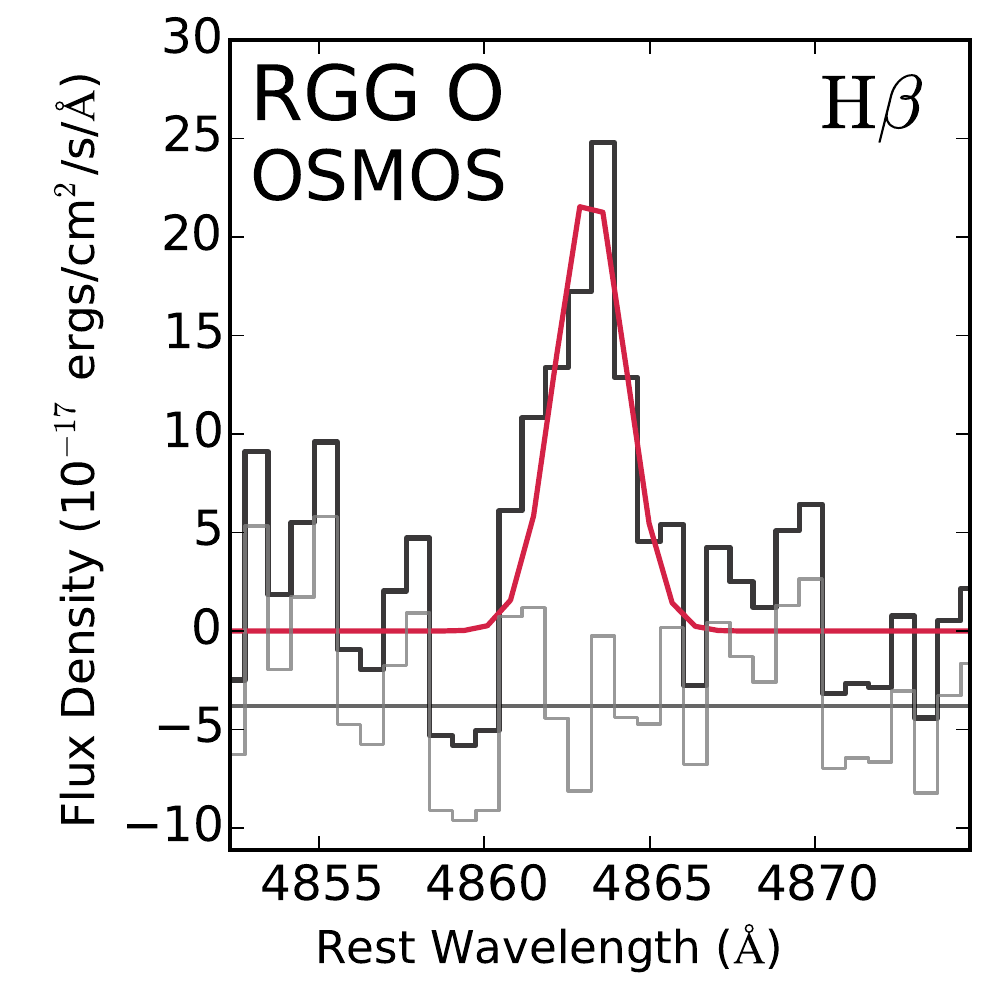}
\includegraphics[scale=0.44]{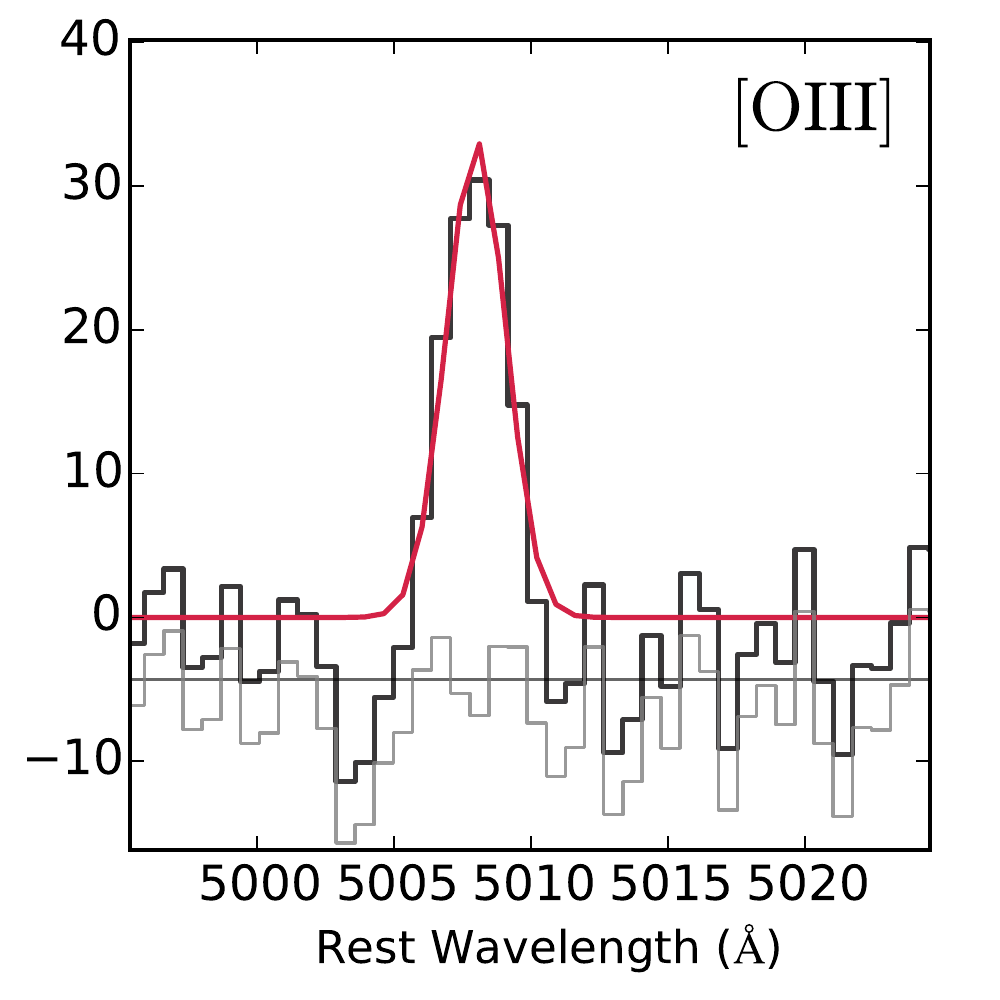}
\includegraphics[scale=0.44]{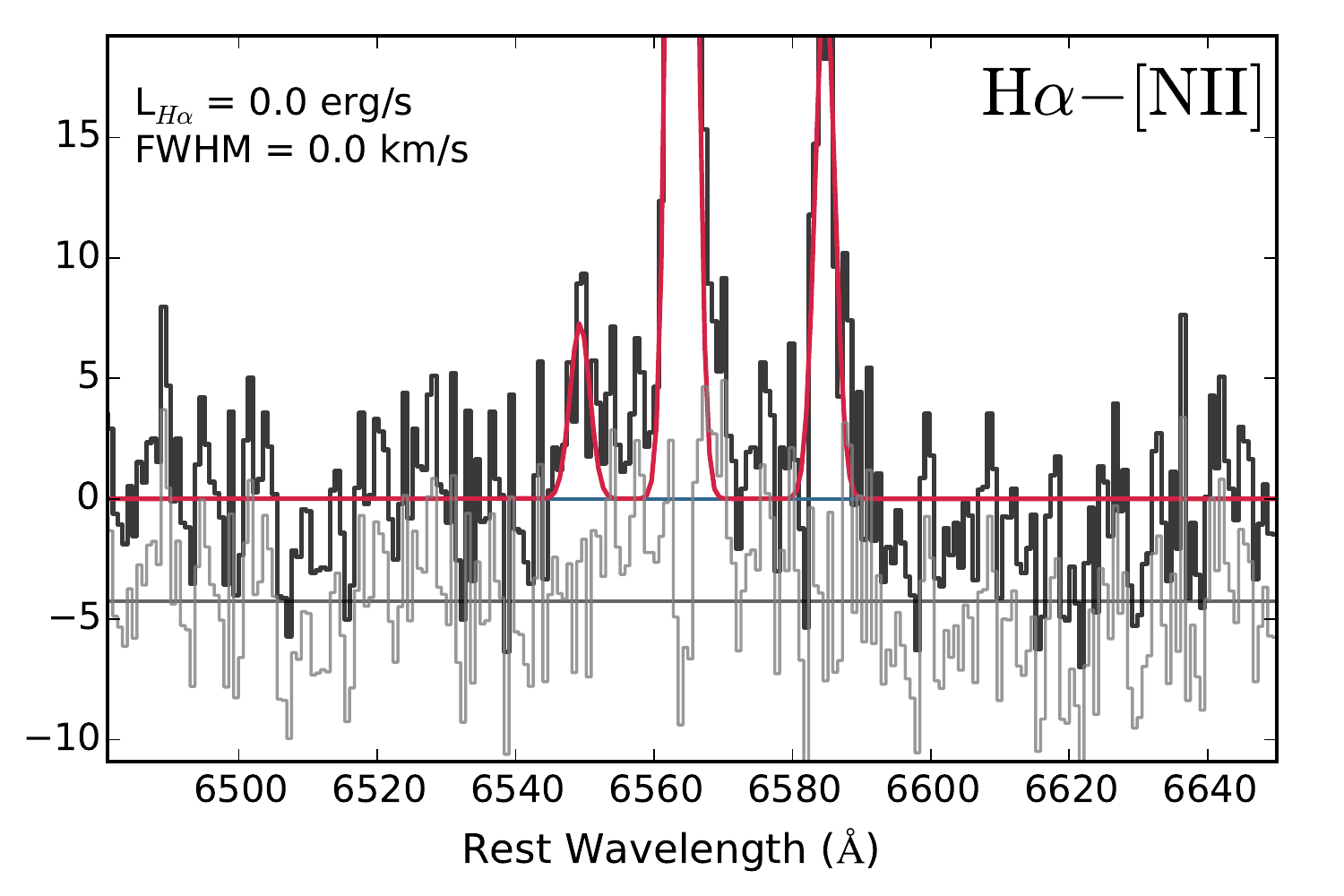}
\caption{These plots show the $\rm H\beta$, [OIII]$\lambda$5007, H$\alpha$, and [NII]$\lambda\lambda6718,6731$ lines for each observation taken of RGG O (NSA 104565. Description is same as for Figure~\ref{nsa15952}. We place this object in the ``transient" category.}
\label{nsa104565}
\end{figure*}

\begin{figure*}
\centering

\includegraphics[scale=0.44]{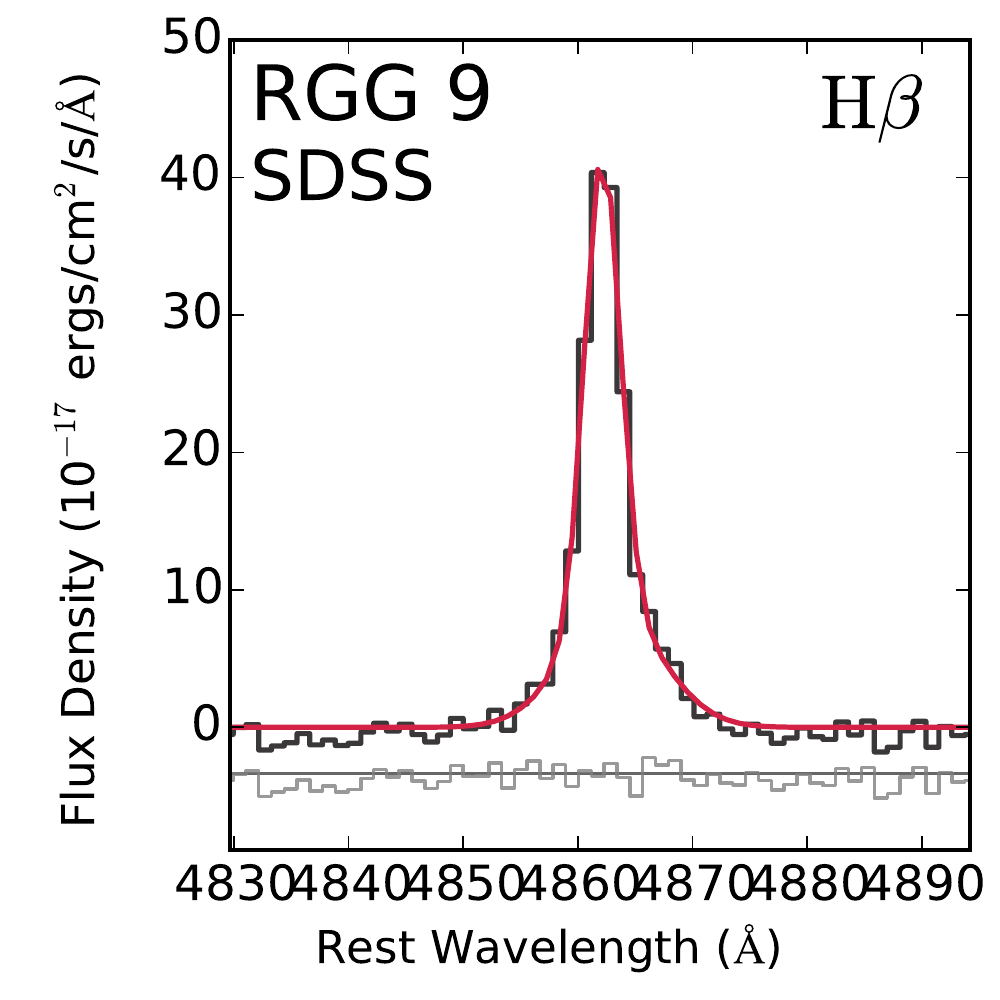}
\includegraphics[scale=0.44]{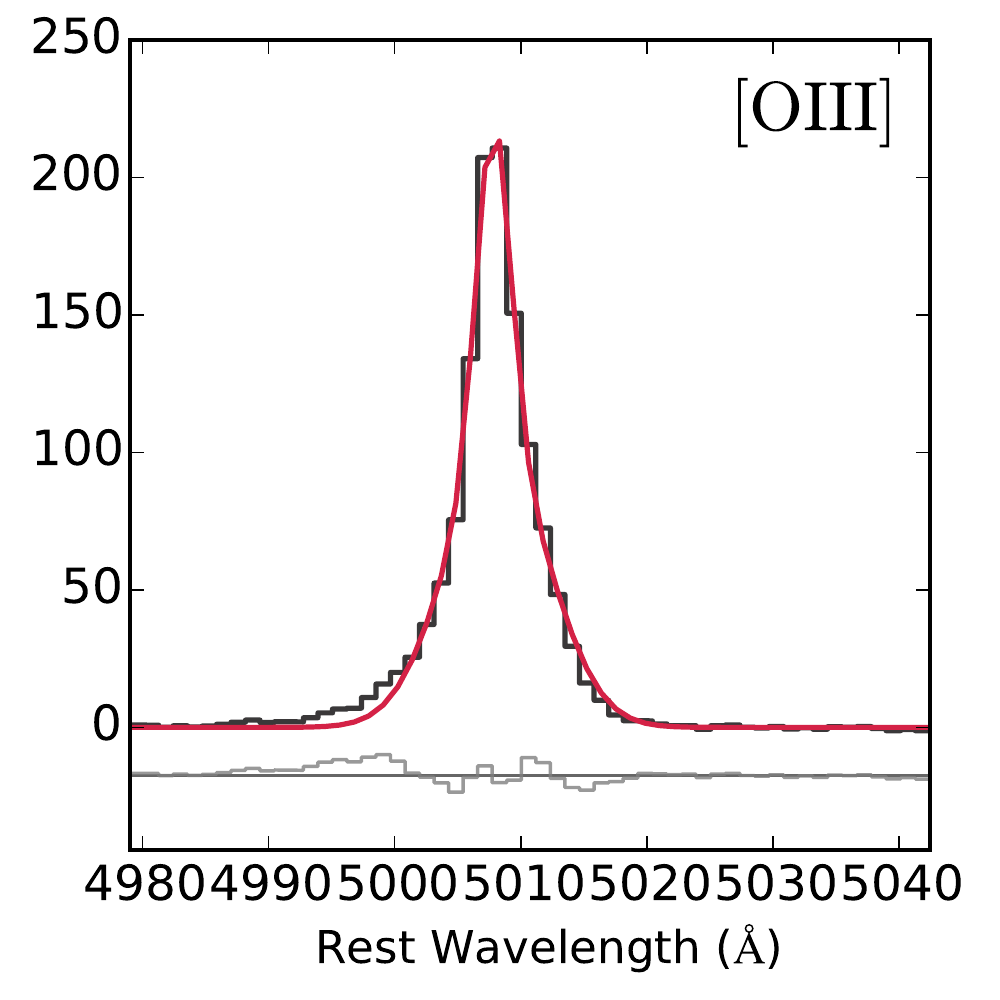}
\includegraphics[scale=0.44]{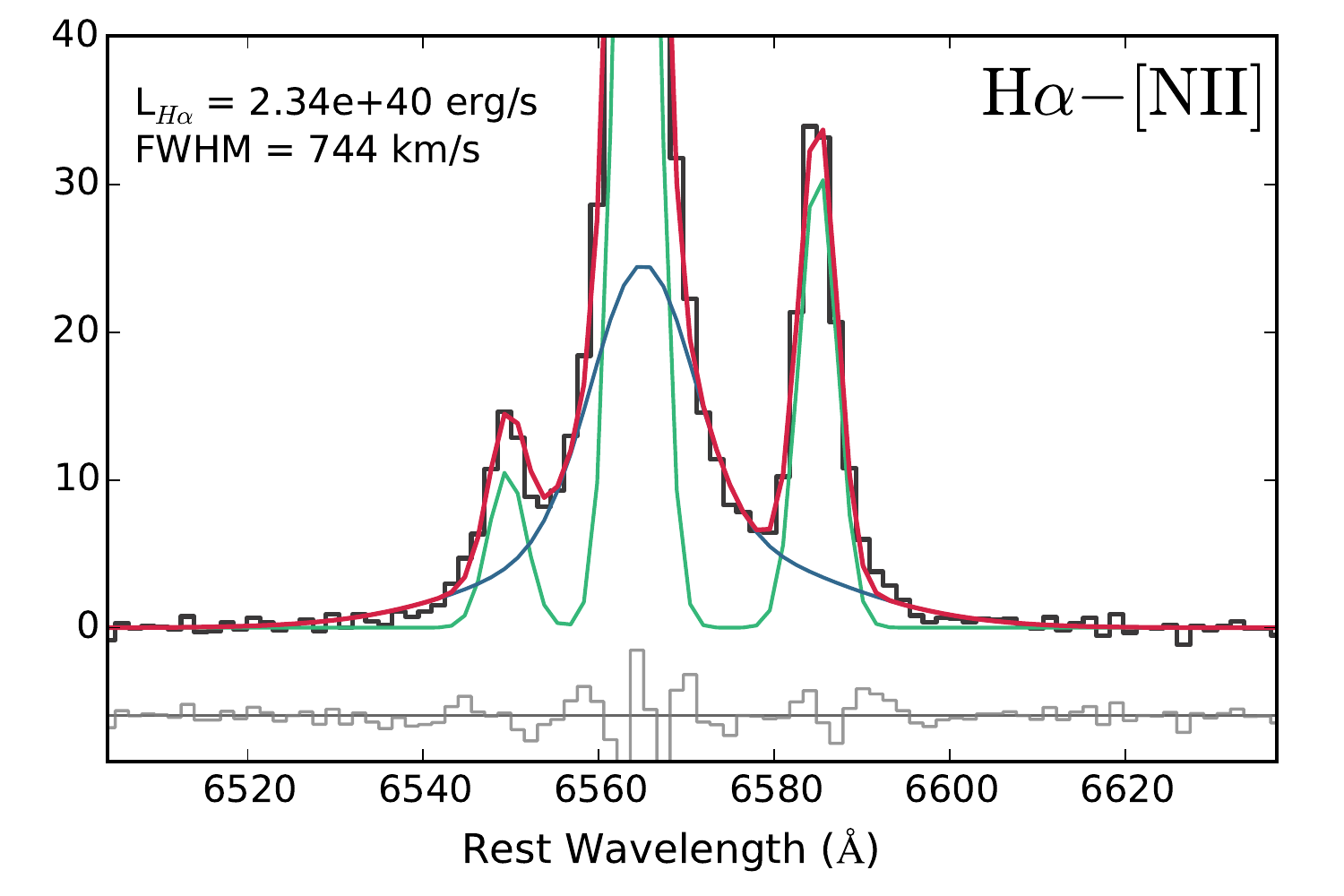}\\

\includegraphics[scale=0.44]{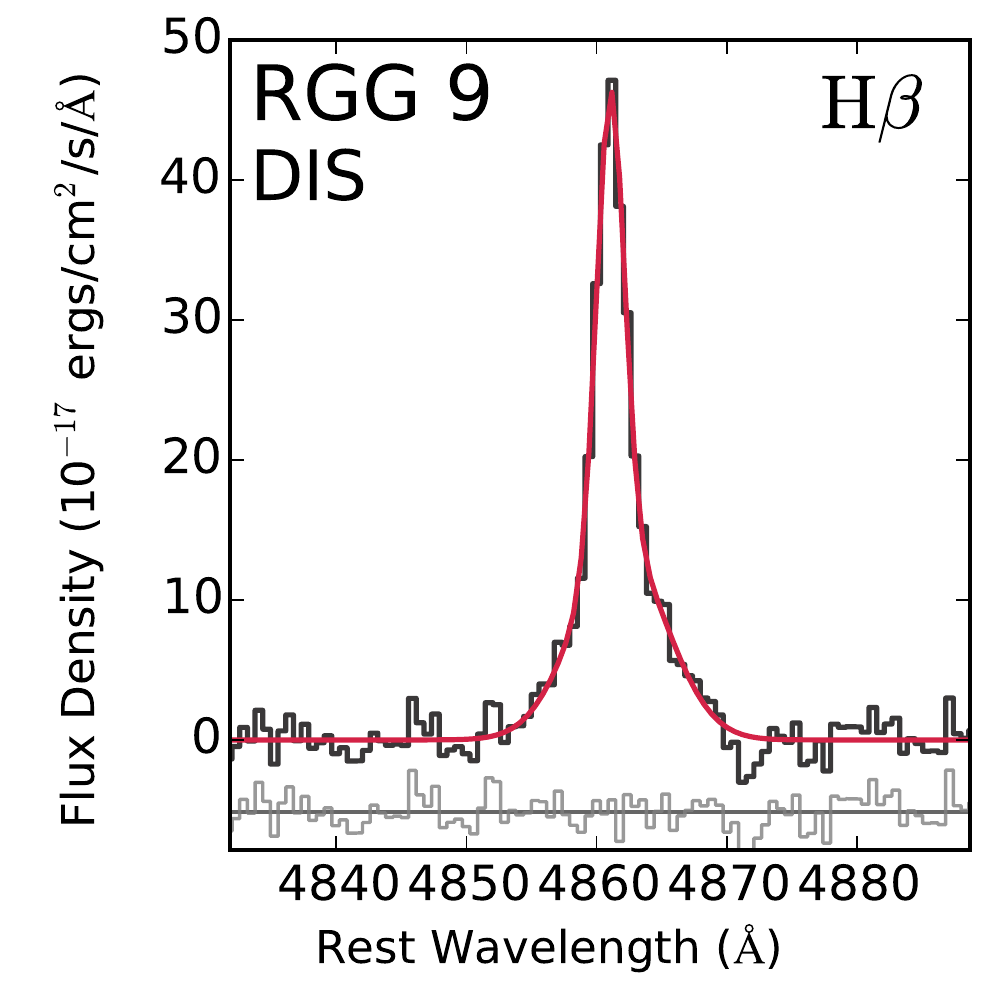}
\includegraphics[scale=0.44]{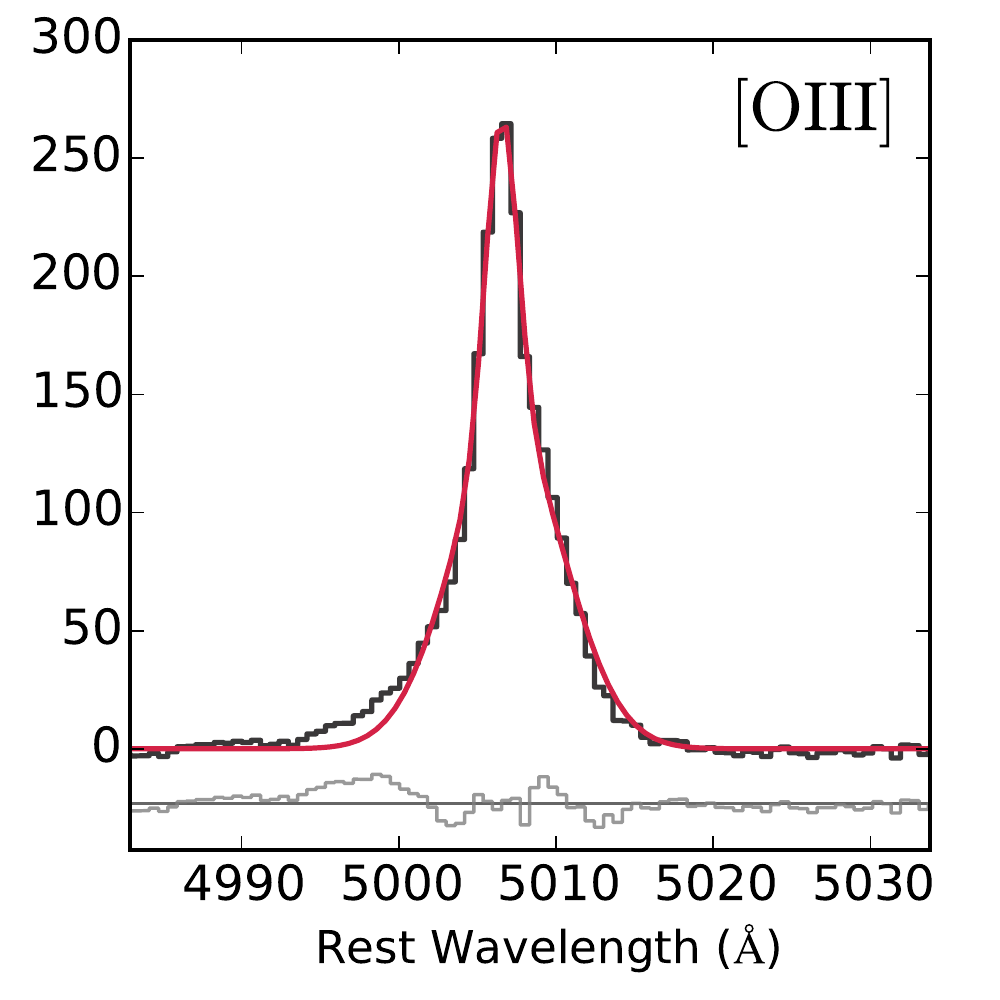}
\includegraphics[scale=0.44]{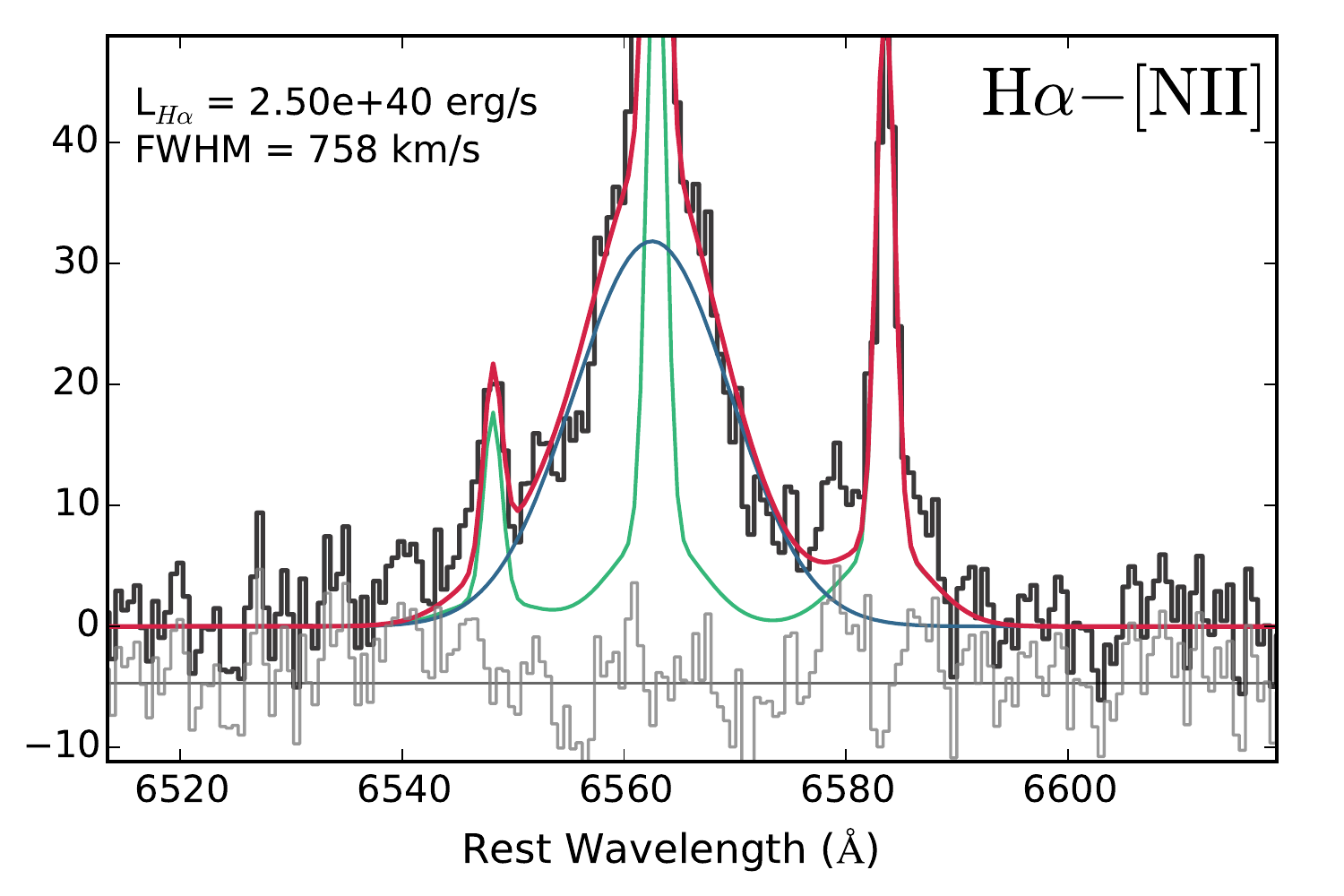}\\
\caption{These plots show the $\rm H\beta$, [OIII]$\lambda$5007, H$\alpha$, and [NII]$\lambda\lambda6718,6731$ lines for each observation taken of RGG 9 (NSA 10779). Description is same as for Figure~\ref{nsa15952}. We place this object in the ``persistent broad H$\alpha$" category.}
\label{nsa10779}
\end{figure*}

\begin{figure*}
\centering

\includegraphics[scale=0.44]{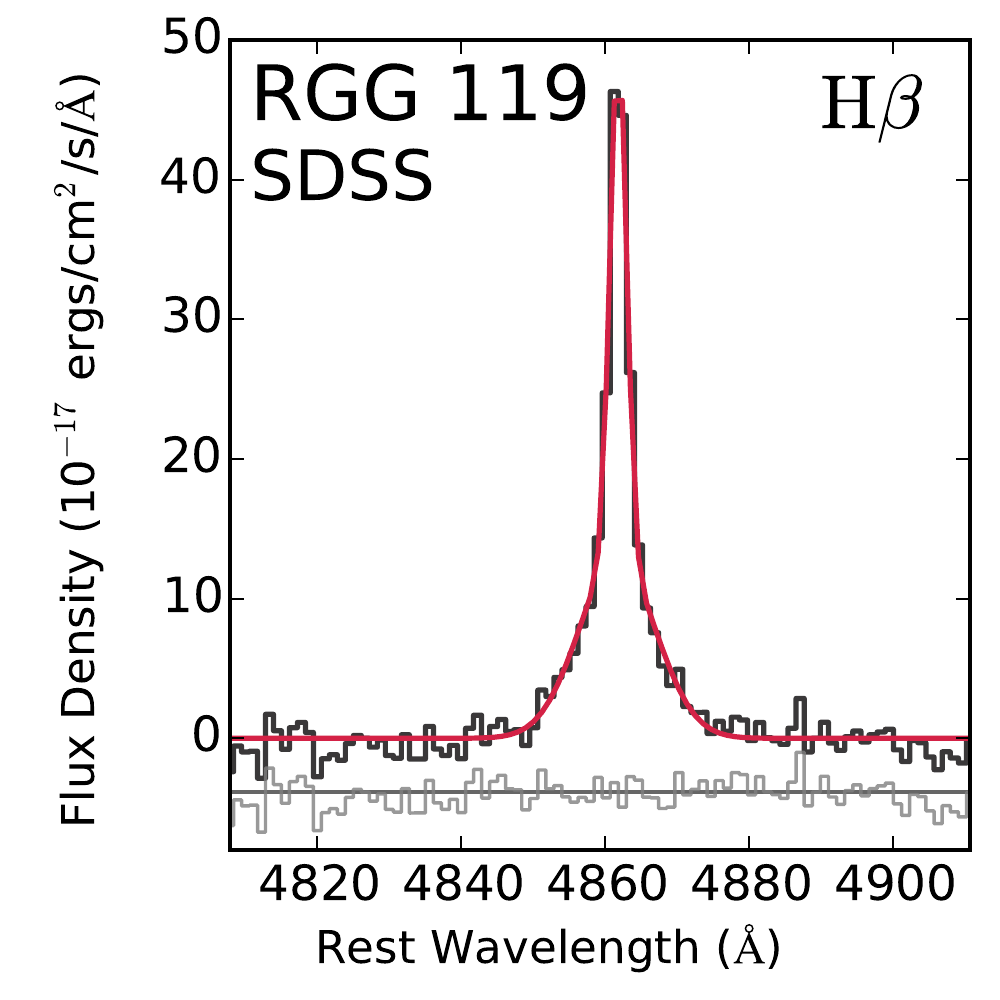}
\includegraphics[scale=0.44]{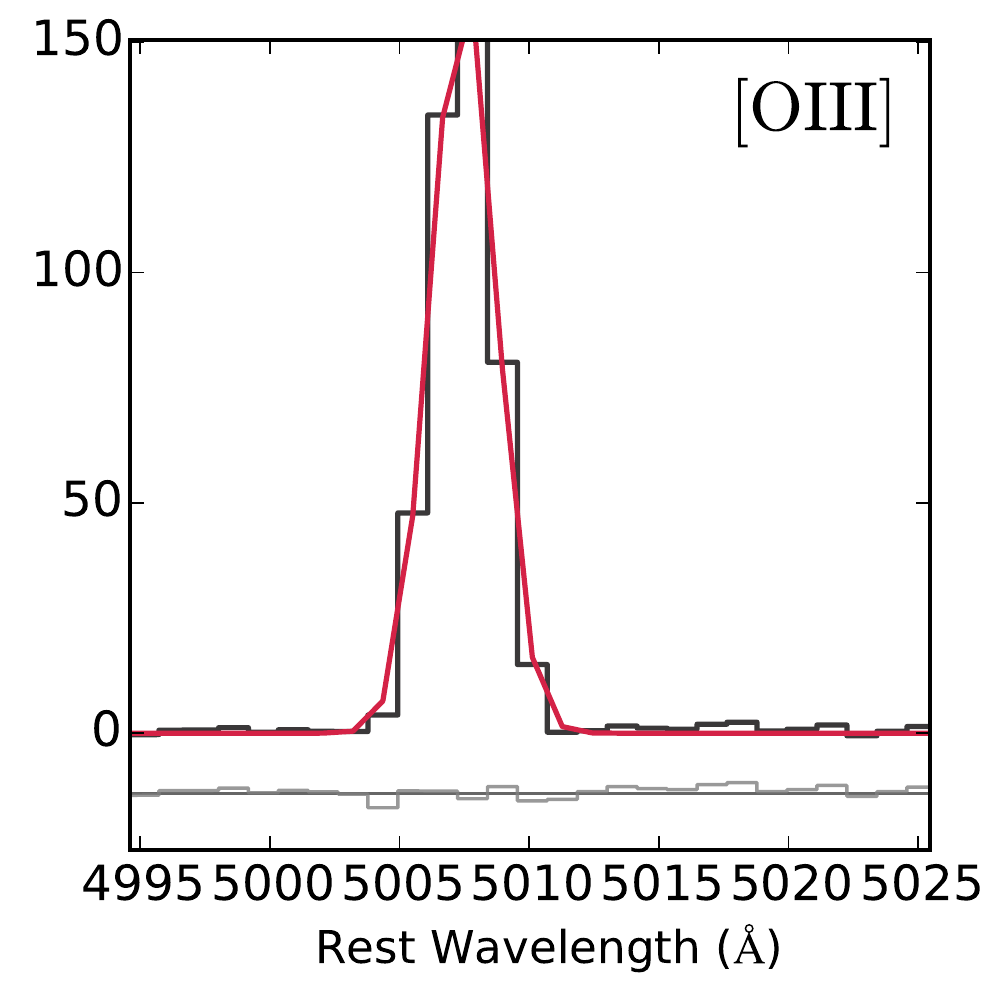}
\includegraphics[scale=0.44]{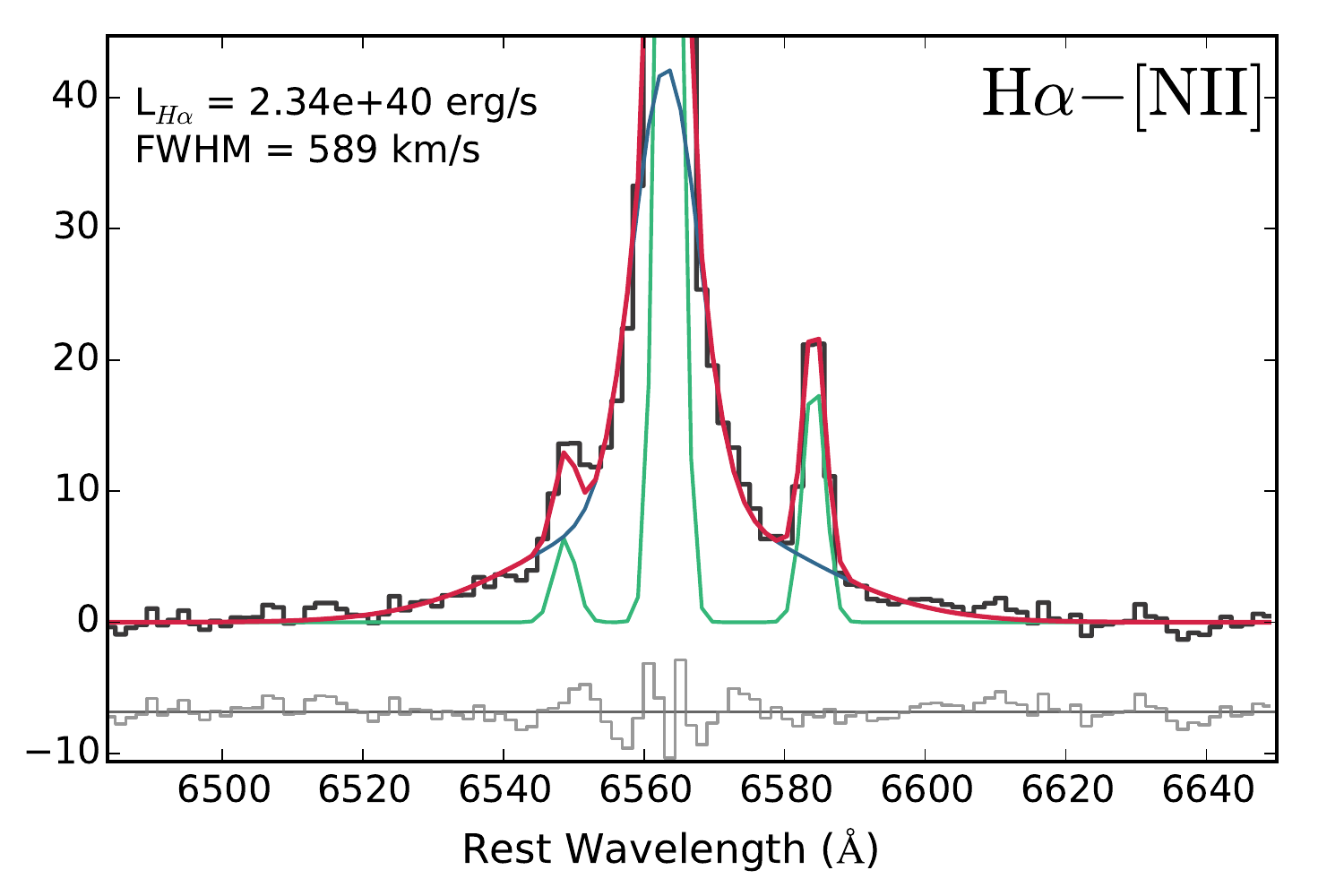}\\

\includegraphics[scale=0.44]{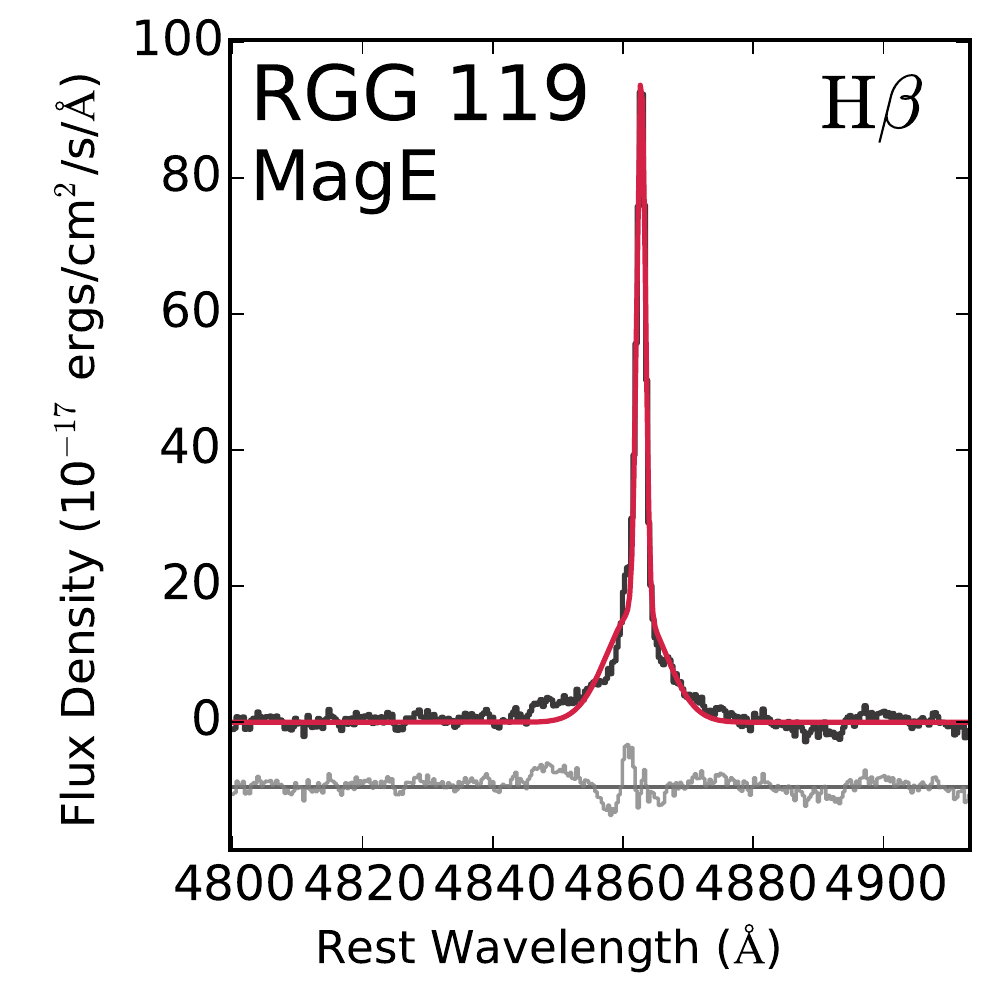}
\includegraphics[scale=0.44]{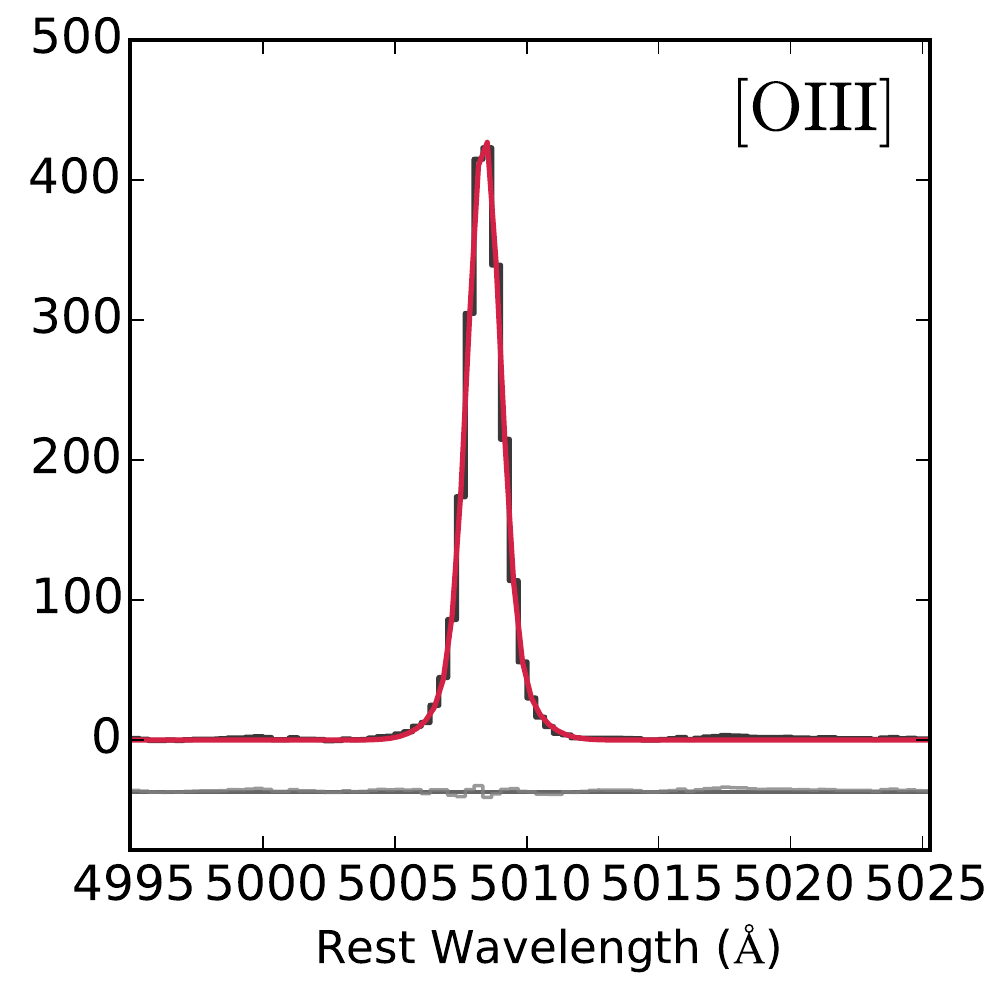}
\includegraphics[scale=0.44]{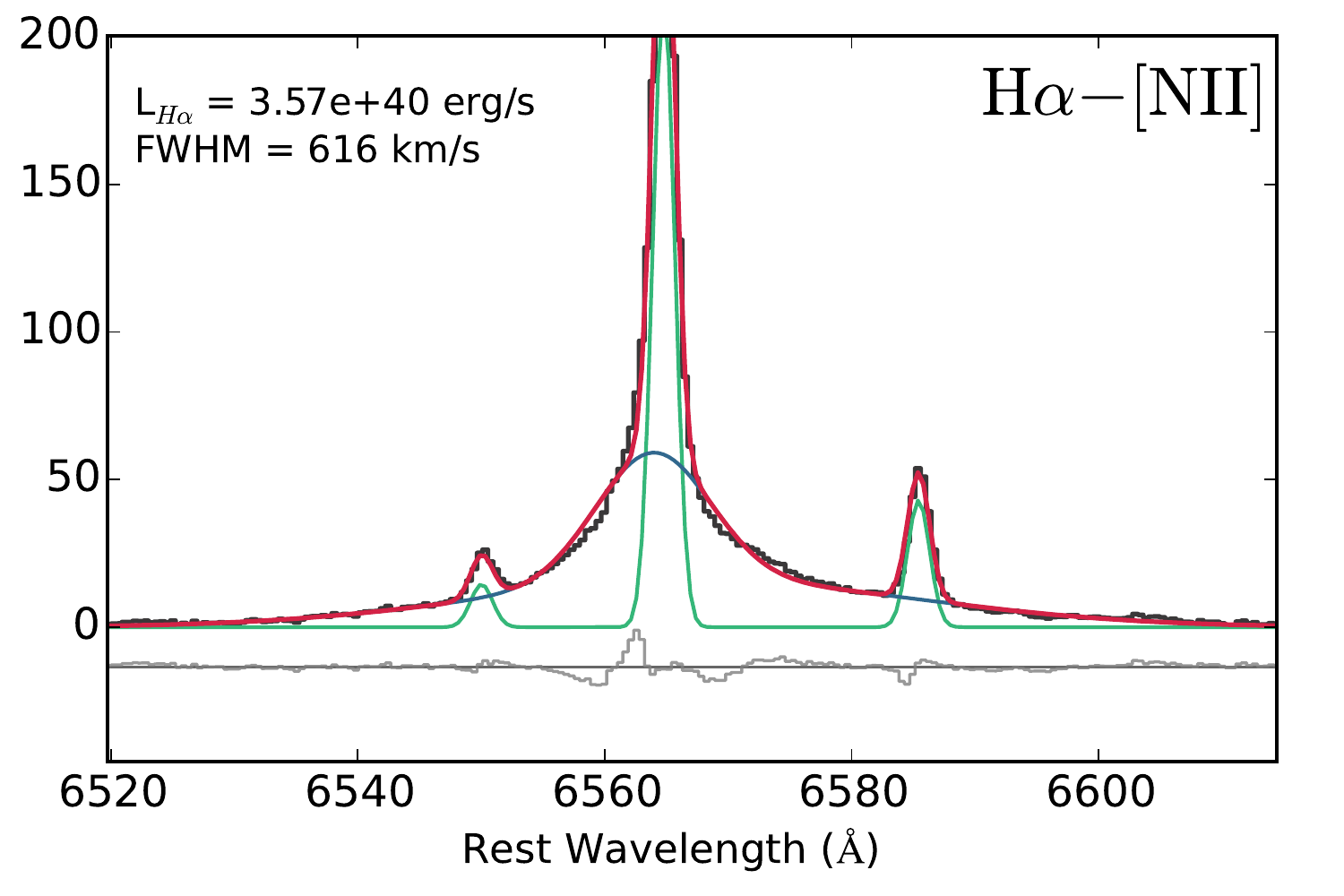}\\

\includegraphics[scale=0.44]{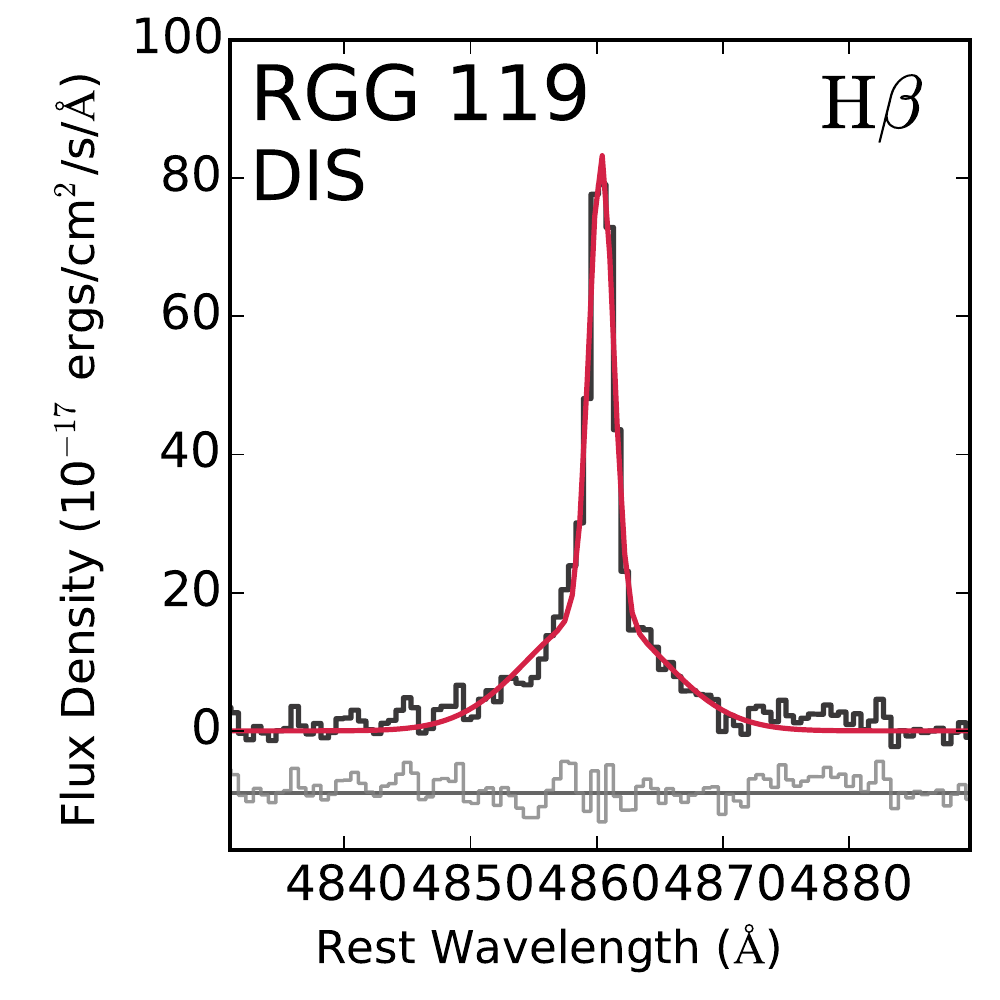}
\includegraphics[scale=0.44]{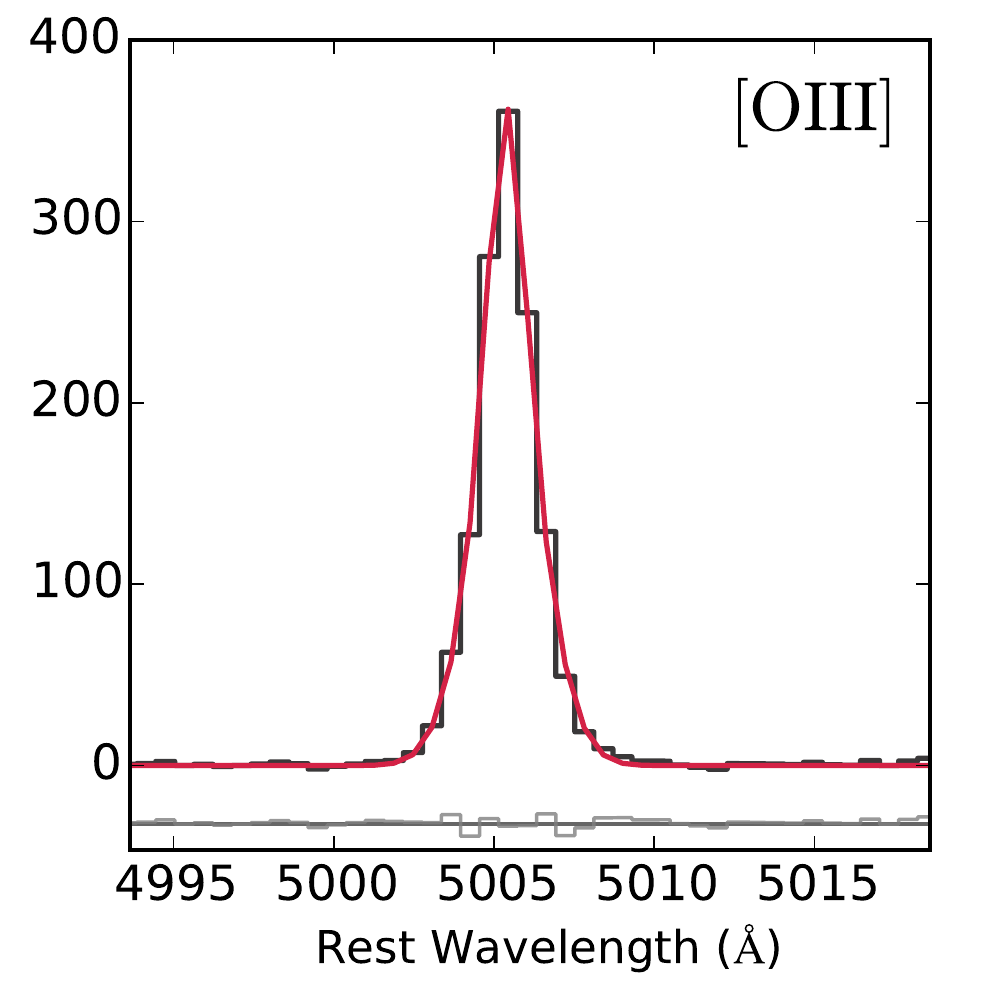}
\includegraphics[scale=0.44]{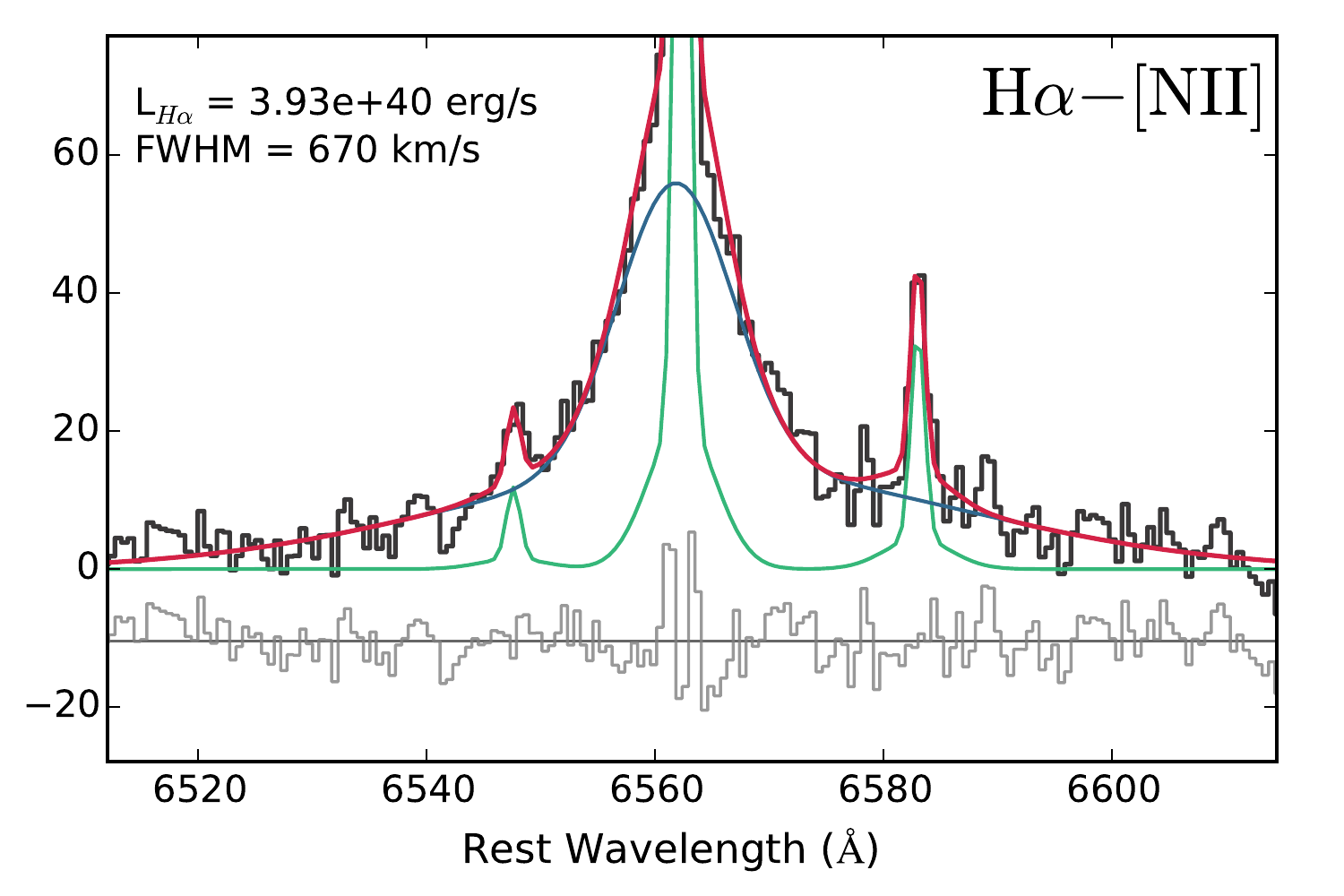}\\
\caption{These plots show the $\rm H\beta$, [OIII]$\lambda$5007, H$\alpha$, and [NII]$\lambda\lambda6718,6731$ lines for each observation taken of RGG 119 (NSA 79874). Description is same as for Figure~\ref{nsa15952}. We place this object in the ``persistent broad H$\alpha$" category.}
\label{nsa79874}
\end{figure*}

\clearpage

\end{appendix}

\end{document}